\documentclass[aps,pra,twocolumn,showpacs,groupedaddress,nofootinbib]{revtex4}
\usepackage{graphicx}
\usepackage{amsmath}
\usepackage{longtable}
\usepackage{bm}

\begin{document}

\title{Theory and applications of atomic and ionic polarizabilities}
\author{J. Mitroy}
\affiliation{School of Engineering, Charles Darwin University, Darwin NT 0909, Australia}
\author{M. S. Safronova}
\affiliation{Department of Physics and Astronomy, University of Delaware, Newark, Delaware, 19716, USA}
\author{Charles W. Clark}
\affiliation{Joint Quantum Institute, National Institute of Standards and Technology and the University of Maryland,
Gaithersburg, Maryland, 20899-8410, USA}

\date{\today}

\begin{abstract}
Atomic polarization phenomena impinge upon a number of areas and processes in physics.  The dielectric constant and
refractive index of any gas are examples of macroscopic properties that are largely determined by the dipole
polarizability.  When it comes to microscopic phenomena, the existence of alkaline-earth anions and the recently
discovered ability of positrons to bind to many atoms are predominantly due to the polarization interaction. An
imperfect knowledge of atomic polarizabilities is presently looming as the largest source of uncertainty in the new
generation of optical frequency standards. Accurate polarizabilities for the group I and II atoms and ions of the
periodic table have recently become available by a variety of techniques.  These include refined many-body perturbation
theory and coupled-cluster  calculations sometimes combined with precise experimental data for selected transitions,
microwave spectroscopy of Rydberg atoms and ions, refractive index measurements in microwave cavities, \textit{ab
initio} calculations of atomic structures using explicitly correlated wave functions, interferometry with atom beams,
and velocity changes of laser cooled atoms induced by an electric field.  This review examines existing theoretical
methods of determining atomic and ionic polarizabilities, and discusses their relevance to various applications with
particular emphasis on cold-atom physics and the metrology of atomic frequency standards.

\end{abstract}

\maketitle

\section{Introduction}

By the time Maxwell presented his article on a {\em Dynamical Theory of the Electromagnetic Field} \cite{maxwell1864a},
it was understood that bulk matter had a composition of particles of opposite electrical charge, and that an applied
electric field would rearrange the distribution of those charges in an ordinary object.  This rearrangement could be
described accurately even without a detailed microscopic understanding of matter.  For example, if a perfectly
conducting sphere of radius $r_0$ is placed in a uniform electric field ${\bf F}$, simple potential theory shows that
the resulting electric field at a position ${\bf r}$ outside the sphere must be ${\bf F} -\nabla ({\bf F}\cdot {\bf r}
r_0^3/r^3)$.  This is equivalent to replacing the sphere with a point electric dipole,
\begin{equation}
{\bf d} = \alpha {\bf F} , \label{defined}
\end{equation}
where $\alpha = r_0^3$ is the dipole {\it polarizability} of the sphere\footnote{For notational convenience, we use the
Gaussian system of electrical units, as discussed in subsection A below.  In the Gaussian system, electric
polarizability has the dimensions of volume.}. An arbitrary applied electric field can be decomposed into multipole
fields of the form ${\bf F}^k_q({\bf r}) = - F^k_q \nabla (r^k {\bf C}^{k}_q({\bf \hat{r}}))$, where ${\bf
C}^{k}_q({\bf \hat{r}})$ is a spherical tensor \cite{edmonds96}.  Each of these will induce a multipole moment of
$F^k_q r_0^{2k+1}$ in the conducting sphere, corresponding to a multipole polarizability of $\alpha^k = r_0^{2k+1}.$
Treatment of the electrical polarizabilities of macroscopic bodies is a standard topic of textbooks on electromagnetic
theory, and the only material properties that it requires are dielectric constants and conductivities.

Quantum mechanics, on the other hand, offers a fundamental description of matter, incorporating the effects of electric
and magnetic fields on its elementary constituents, and thus enables polarizabilities to be calculated from first
principles.  The standard framework for such calculations, perturbation theory, was first laid out by Schr\"{o}dinger
\cite{schrodinger26b} in a paper that reported his calculations of the Stark effect in atomic hydrogen.  A system of
particles with positions ${\bf r}_i$ and electric charges $q_i$ exposed to a uniform electric field, $({\bf F} = F
\hat{{\bf F}})$, is described by the Hamiltonian

\begin{equation}
H = H_0 - F \hat{{\bf F}} \cdot {\bf d}, \label{Hfield}
\end{equation}

\noindent where $H_0$ is the Hamiltonian in the absence of the field, and ${\bf d}$ is the dipole moment operator,

\begin{equation}
{\bf d} = \sum_i q_i {\bf r}_i.
\end{equation}

Treating the field strength, $F = |{\bf F}|$, as a perturbation parameter, means that the energy and wave function can
be expanded as
\begin{eqnarray}
|\Psi \rangle & = & |\Phi_0 \rangle + F | \Phi_1 \rangle  + F^2 | \Phi_2 \rangle + \ldots  \\
\label{pertvar1} E & = & E_0 + FE_1 + F^2 E_2 + \ldots
\end{eqnarray}

The first-order energy $E_1 = 0$ if $|\Phi_0 \rangle$ is an eigenfunction of the parity operator.  In this case,
$|\Phi_1 \rangle$ satisfies the equation

\begin{equation}
(H_0 - E_0) | \Phi_1 \rangle = - {\bf \hat{F}} \cdot {\bf d} | \Phi_0 \rangle \ . \label{first0}
\end{equation}
From the solution to Eq. (\ref{first0}), we can find the expectation value

\begin{eqnarray}
\langle \Psi | {\bf d} |\Psi \rangle & = &
 F (\langle \Phi_0 | {\bf d} |\Phi_1 \rangle + \langle \Phi_1 | {\bf d} |\Phi_0 \rangle) \nonumber \\
& = & \bar{\alpha} {\bf F} \label{dipole_moment} \label{pertvar2}
\end{eqnarray}
where $\bar{\alpha}$ is a matrix.  The second-order energy is given by

\begin{equation}
E_2 =  - \frac{1}{2} {\bf F} \cdot \bar{\alpha}{\bf F} . \label{secondO}
\end{equation}

Although Eq. (\ref{first0}) can be solved directly, and in some cases in closed form, it is often more practical to
express the solution in terms of the eigenfunctions and eigenvalues of $H_0$, so that  Eq. (\ref{secondO}) takes the
form

\begin{eqnarray}
E_2 &=& -\sum_{n}  \frac{ |\langle \Psi_0 | {\bf d} \cdot {\bf F}| \Psi_{n} \rangle |^2
 }{E_{n} - E_0 } .  \label{eq10}
\label{E2sum}
\end{eqnarray}
This sum over all stationary states shows that calculation of atomic polarizabilities is a demanding special case of
the calculation of atomic structure.  The sum extends in principle over the continuous spectrum, which sometimes makes
substantial contributions to the polarizability.

\begin{table*}[t]
\caption[]{List of data tables.} \label{tab:1}
\begin{ruledtabular}
\begin{tabular}{llllc}
  \multicolumn{1}{l}{Table} &  \multicolumn{1}{l}{System}& \multicolumn{1}{c}{Atoms and Ions} & \multicolumn{1}{c}{States} & \multicolumn{1}{c}{Data} \\
  \hline
Table \ref{tab:3} & Noble gases & He, Ne, Ar, Kr, Xe, Rn, Li$^+$, Na$^+$, K$^+$, Rb$^+$, Cs$^+$, Fr$^+$& ground & $\alpha_0$\\
 & &  Be$^{2+}$, Mg$^{2+}$, Ca$^{2+}$, Sr$^{2+}$, Ba$^{2+}$, Ra$^{2+}$&  & \\[0.5pc]
Table~\ref{alkali} & Alkali atoms &Li, Na, K, Rb, Cs, Fr & ground $ns$, $np$ &  $\alpha_0$, $\alpha_2$\\ [0.5pc] Table
\ref{alkaliions} & Alkali ions & Be$^{+}$, Mg$^{+}$, Ca$^{+}$, Sr$^{+}$, Ba$^{+}$, Ra$^{+}$ &ground &
                        $\alpha_0$ \\ [0.5pc]
Table \ref{alkalitensor} & Monovalent   & Li, Na, K, Ca$^+$, Rb, Sr$^+$ & excited & $\alpha_0$, $\alpha_2$ \\ [0.5pc]
Table \ref{starkalkali} & Alkali atoms & Resonance transition: Li, Na, K, Rb, Cs & $ns$, $np$& $\Delta\alpha_0$\\
 Table~\ref{sodium} & Alkali atom& Na & ground &$\alpha_0$ \\ [0.5pc]
Table \ref{cesium} & Alkali atom & Cs & 26 states & $\alpha_0$, $\alpha_2$ \\ [0.5pc]
 Table \ref{alkaline} & Group II type &Be, Mg, Ca, Sr, Ba, Ra, Al$^+$, Si$^{2+}$, Zn, Cd, Hg, Yb& ground, $nsnp
 \, ^3P_0$&$\alpha_0$\\ [0.5pc]
 Table \ref{miscellaneous} & Miscellaneous & Al, Ga, In, Tl, Si, Sn, Pb, Ir, U, Cu, Ag, Au, & ground& $\alpha_0$\\
   &   & Al$^{+}$, Si$^{3+}$, P$^{3+}$, Kr$^{6+}$, Cu$^+$, Ag$^+$, Hg$^+$, Yb$^+$, Zn$^+$ & & \\[0.5pc]
Table \ref{excitedalkaline}&Miscellaneous  & Ca, Sr, Ba, Zn, Cd, Hg, Yb, Al, Tl,  Yb$^+$ & excited&$\alpha_0$, $\alpha_2$\\[0.5pc]
Table \ref{starkother} & Miscellaneous&Li, Na,  Cs, Mg, Ca, Ba, Hg, Ga, Tl, Yb$^+$&&$\Delta\alpha_0$\\
Table \ref{BBshifts} & Miscellaneous& Mg, Ca, Sr, Yb, Zn, Cd, Hg,  & clock & $\Delta\nu_{\rm BBR}$\\
                     &              & Ca$^+$, Sr$^+$, Hg$^+$, Yb$^+$, Al$^+$, In$^+$ & transition &                \\
    Table \ref{BBRMicrowave}& Monovalent &Li, Na, K, Rb, Cs, Ba$^+$, Yb$^+$, Hg$^+$& ground hyperfine& BBR\\
Table \ref{C6} & Alkali atoms &Li, Na, K, Rb, Cs, Fr & ground & $C_6$ \\
\end{tabular}
\end{ruledtabular}
\end{table*}

Interest in the subject of polarizabilities of atomic states has recently been elevated by the appreciation that the
accuracy of next-generation atomic time and frequency standards, based on optical transitions
\cite{madej01a,udem02a,diddams04a,gill03a,gill05a,margolis09a}, is significantly limited by the displacement of atomic
energy levels due to universal ambient thermal fluctuations of the electromagnetic field: blackbody radiation (BBR)
shifts \cite{gallagher79a,itano82a,hollberg84a,porsev06b}. This phenomenon brings the most promising approach to a more
accurate definition of the unit of time, the second, into contact with deep understanding of the thermodynamics of the
electromagnetic radiation field.

Description of the interplay between these two fundamental phenomena is a major focus of this review, which in earlier
times might have seemed a pedestrian discourse on atomic polarizabilities.  The precise calculation of atomic
polarizabilities also has implications for quantum information processing and optical cooling and trapping schemes.
Modern requirements for precision and accuracy have elicited renewed attention to methods of accurate first-principles
calculations of atomic structure, which recently have been increased in scope and precision by developments in
methodology, algorithms, and raw computational power.  It is expected that the future will lead to an increased
reliance on theoretical treatments to describe the details of atomic polarization.  Indeed, at the present time, many
of the best estimates of atomic polarizabilities are derived from a composite analysis which integrates experimental
measurements with first principles calculations of atomic properties.

There have been a number of reviews and tabulations of atomic and ionic polarizabilities
\cite{dalgarno62a,teachout71a,miller77a,miller88a,wijngaarden96a,bonin97a,delone99a,schwerdtfeger06a,gould05a,lundeen05a,miller07a,lupinetti06b}.
 Some of these reviews, e.g.
\cite{miller77a,miller88a,gould05a,lundeen05a} have largely focussed upon experimental developments while others
\cite{bonin97a,schwerdtfeger06a,lupinetti06b} have given theory more attention.

In the present review,  the strengths and limitations of different theoretical techniques are discussed in detail given
their expected importance in the future. Discussion of experimental work is mainly confined to presenting a compilation
of existing results and very  brief overviews of the various methods.  The exception to this is the interpretation of
resonance excitation Stark ionization spectroscopy \cite{lundeen05a} since issues pertaining to the convergence of the
perturbation analysis of the polarization interaction are important here.   The present review is confined to
discussing the polarizabilities of low lying atomic and ionic states despite the existence of a body of research on
Rydberg states \cite{gallagher05a}. High-order polarizabilities  are not considered except in those circumstances where
they are specifically relevant to ordinary polarization phenomenon. The influence of external electric fields on energy
levels comprises part of this review as does the nature of the polarization interaction between charged particles with
atoms and ions.  The focus of this review is on developments related to contemporary topics such as development of
optical frequency standards, quantum computing, and study of fundamental symmetries. Major emphasis of this review is
to provide critically evaluated data on atomic polarizabilities.
 Table~\ref{tab:1} summarizes the data presented in this review to facilitate the search for particular information.

\subsection{Systems of units}

Dipole polarizabilities are given in a variety of units, depending on the
context in which they are determined.  The most widely used unit for
theoretical atomic physics is atomic units (a.u.), in which, $e$, $m_e$, $4 \pi
\epsilon_0$ and the reduced Planck constant $\hbar$ have the numerical value
$1$. The polarizability in a.u. has the dimension of volume, and its numerical
values presented here are thus expressed in units of $a_0 ^3$, where $a_0
\approx 0.052918$ nm is the Bohr radius.    The preferred unit systems for
polarizabilities determined by experiment are $\AA^3$, kHz/(kV/cm)$^2$,
cm$^3$/mol or C$\cdot$m$^2$/N where C$\cdot$m$^2$/N is the SI unit.  In this
review, almost all polarizabilities are given in a.u. with uncertainties in the
last digits (if appropriate) given in parentheses. Conversion factors between
the different units are listed in Table \ref{conversion}.  The last line of the
table gives conversion factors from SI units to the other units. For example,
the atomic units for $\alpha$ can be converted to SI units by multiplying by
0.248832.

Stark shift experiments which measure the change in photon frequency of an atomic transition as a function of electric
field strength are usually reported as a Stark shift coefficient in units of kHz/(kV/cm)$^2$.  The polarizability
difference is twice the size of the Stark shift coefficient, as in equation (\ref{secondO}).

\begin{table*}[th]
\caption[]{ Factors for converting polarizabilities between different unit systems.  The table entries give the
multiplying factor needed to convert the row entry to the corresponding column entry.  The last column in the Table is
the polarizability per mole and is often called the molar polarizability. The conversion factors from SI units to other
units are given in the last line. Here, $h$ is the Planck's constant, $\epsilon_0$ is the electric constant, $a_0$ is
the Bohr radius, and $N_A$ is the Avogadro constant.} \label{conversion}
\begin{ruledtabular}
\begin{tabular}{lccccc}
                &    a.u.            &   $\AA^3$                  & kHz/(kV/cm)$^2$            & C$\cdot$m$^2$/V   &   cm$^3$/mol \\ \hline
a.u.            &    1                & 0.1481847                  &  $0.2488319 $  &  $1.648773 \! \times \! 10^{-41}$ &  0.3738032    \\
$\AA^3$         &  6.748335           & 1                          &  $1.679201 $  &  $1.112650 \! \times \! 10^{-40}$ &  2.522549  \\
kHz/(kV/cm)$^2$   & $4.018778 $ & $0.5955214 $   &  1      &  $1.509190 \! \times \! 10^{40}$  & $ 1.502232 $ \\
Cm$^2$/V        & $6.065100 \! \times \! 10^{40}$ & $8.987552 \! \times \! 10^{39}$  & $6.626069 \! \times \! 10^{-39}$ & 1   &  $ 2.267154 \! \times \! 10^{40}$ \\
cm$^3$/mol  &  2.675205       &   0.3964244                 &  0.6656762   & $ 4.410816 \! \times \! 10^{-41}$         &   1   \\
Conversion from SI &$1/(4 \pi \epsilon_0 a_0^3)$&$10^{30}/(4 \pi \epsilon_0)$&$10^{-7} h$& 1&$10^{6} N_A /(3
\epsilon_0)$\\
\end{tabular}
\end{ruledtabular}
\end{table*}

\section{Atomic Polarizabilities and field-atom interactions }

\subsection{Static electric polarizabilities}
\subsubsection{Definitions of scalar and tensor polarizabilities}

The overall change in energy of the atom can be evaluated within the framework
of second-order perturbation theory. Upon reduction, the perturbation theory
expression given by Eq.(\ref{eq10}) leads to a sum-over-states formula for the
static scalar electric-dipole polarizability which is expressed most compactly
in terms of oscillator strengths as
\begin{equation}
\alpha_0 = \sum_n \frac{f_{gn}}{(\Delta E_{ng})^2} \ . \label{alpha1}
\end{equation}
In this expression, $f_{gn}$ is the absorption oscillator strength for a dipole transition from level $g$ to level $n$,
defined  in a $J$-representation as \cite{fuhr95a}
\begin{equation}
f_{gn} =  \frac {2 |\langle \psi_{g} \parallel  r {\bf C}^{1}({\bf \hat{r}}) \parallel \psi_{n} \rangle|^2 \Delta
E_{ng}} {3 (2J_g+1)}   , \label{fvaldef}
\end{equation}
where  $\Delta E_{ng} = E_{n} - E_{g}$ and ${\bf C}^{1}({\bf \hat{r}})$ is the spherical tensor of rank 1
\cite{edmonds96}. The definition of the oscillator strength in $LS$ coupling is transparently obtained from
Eq.~(\ref{fvaldef}) by replacing the total angular momentum by the orbital angular momentum.

The polarizability for a state with non-zero angular momentum $J$ depends on the magnetic projection $M$:
\begin{equation}
\alpha = \alpha_{0} + \alpha_{2} \frac{3M^2-J(J+1)}{J(2J-1)}. \label{alphaM}
\end{equation}
The quantity $\alpha_{0}$ is called the scalar polarizability  while $\alpha_{2}$ is the tensor polarizability in J
representation.

The scalar part of the polarizability can be determined using Eq.~(\ref{alpha1}). In terms of the reduced matrix
elements of the electric-dipole operator, the scalar polarizability $\alpha_{0}$ of an atom in a state $\psi$ with total
angular momentum $J$ and energy $E$ is also written as
\begin{equation}
 \alpha_{0} = \frac{2}{3(2J+1)} \sum_{n}
   \frac{ |\langle \psi \|r {\bf C}^1( {\bf{\hat r}})\| \psi_n  \rangle|^2 } {E_{n}-E}.
\label{alpha0J}
\end{equation}

The tensor polarizability $\alpha_{2}$  is defined as
\begin{eqnarray}\label{alpha2JJ}
&& \alpha_{2}=4\left(\frac{5J(2J-1)}{6(J+1)(2J+1)(2J+3)}\right)^{1/2}\\[0.2cm]
  &&\times \sum_{n}
  (-1)^{J+J_n}
  \left\{ \begin{array}{ccc}
   J & 1 & J_n \\
   1 & J & 2
   \end{array} \right\}
  \frac{ |\langle \psi \| r {\bf C}^1( {\bf{\hat r}})\| \psi_n|  \rangle^2 } {E_{n}-E}.\nonumber
\end{eqnarray}
It is useful in some cases to calculate polarizabilities in strict $LS$ coupling. In  such cases \cite{mitroy04b}, the
tensor polarizability $\alpha_{2,L}$ for a state with orbital angular momentum L is given by
\begin{eqnarray}
\alpha_{2,L} &=& \sum_n \Biggl[ \begin{pmatrix} L & 1 & L_n \\
-L & 0 & L \end{pmatrix}^2
-\frac{1}{3(2L+1)}  \Biggl] \nonumber \\[0.2cm]
 &\times & \frac{ 2 | \langle\psi \parallel  \ r {\bf C}^1( {\bf{\hat r}})
\parallel \psi_n\rangle |^2 } {E-E_n} \ .
\label{alpha2LL}
\end{eqnarray}

 The tensor polarizabilities $\alpha_{2}$ and $\alpha_{2,L}$  in the $J$ and $L$ representations, respectively,
  are  related by
 \begin{eqnarray}
\alpha_{2}&=&\alpha_{2,L}(-1)^{S+L+J+2} (2J+1) \left\{ \begin{array}{ccc}
   S & L & J \\
   2 & J & L
   \end{array} \right\} \nonumber \\
   &\times&
\left( \begin{array}{ccc}
   J & 2 & J \\
   -J & 0 & J
   \end{array} \right)
\left( \begin{array}{ccc}
   L & 2 & L \\
   -L & 0 & L
   \end{array} \right)^{-1}. \label{alphaLJ}
\end{eqnarray}
For $L=1$ and $J=3/2$, Eq.~(\ref{alphaLJ}) gives $\alpha_{2}=\alpha_{2,L}$. For $L=1$, $S=1$ and $J=1$, Eq.~(\ref{alphaLJ})
gives $\alpha_{2}=-\alpha_{2,L}/2$. For $L=2$, $\alpha_{2}=(7/10) \alpha_{2,L}$ for $J=3/2$ and $\alpha_{2}=
\alpha_{2,L}$ for $J=5/2$. We use the shorter  $\langle \psi \|D\| \psi_n  \rangle$ designation for the reduced
electric-dipole matrix elements instead of $\langle \psi \| r {\bf C}^1( {\bf{\hat r}})\| \psi_n  \rangle$ below.

Equation (\ref{alpha2JJ}) indicates that spherically symmetric levels (such as
the $6s_{1/2}$ and $6p_{1/2}$ levels of cesium) only have a scalar
polarizability. However, the hyperfine states of these levels can have
polarizabilities that depend  upon the hyperfine quantum numbers $F$ and $M_F$.
The relationship between $F$ and $J$ polarizabilities is discussed in
Ref.~\cite{arora07c}.  This issue is discussed in more detail in the section on
BBR shifts.

There are two distinctly different broad approaches to the calculation of atomic polarizabilities. The
``sum-over-states'' approach uses a straightforward interpretation of Eq.~(\ref{eq10}) with the contribution from each
state $\Psi_n$ being determined individually, either from a first principles calculation or from interpretation of
experimental data. A second class of approaches solves inhomogeneous equation (\ref{first0}) directly. We refer to this
class of approaches as direct methods, but note that there are many different implementations of this strategy.

\subsubsection{The sum-over-states method}
\label{sumoverstates}

The sum-over-states method utilizes expression such as Eqs.~(\ref{alpha1}, \ref{alpha0J} - \ref{alpha2LL}) to determine
the polarizability. This approach is widely used for systems with one or two valence electrons since the polarizability
is often dominated by transitions to a few low lying excited states.
 The sum-over-states approach can be used with
oscillator strengths (or electric-dipole matrix elements) derived from experiment or atomic structure calculations. It
is also possible to  insert high-precision experimental values of these quantities into an otherwise theoretical
determination of the total polarizability.
 For such monovalent or divalent systems, it
is computationally feasible to explicitly construct a set of intermediate states that is effectively complete.  Such an
approach is computationally more difficult to apply for atoms near the right hand side of the periodic table since the
larger dimensions involved would preclude an explicit computation of the entire set of intermediate state wave
functions.

For monovalent atoms, it is convenient to separate the total polarizability of
an atom into the core polarizability $\alpha_{\text{core}}$  and the valence
part defined by Eq.~(\ref{alpha0J}).  The core contribution actually has two
components, the polarizability of the ionic core and a small change due to the
presence of the valence electron \cite{arora07b}.  For the alkali atoms, the
valence part of the ground state polarizability  is completely dominated by the
contribution from the lowest excited state.  For example, the $5s-5p_{1/2}$ and
$5s-5p_{3/2}$ transitions contribute more than 99\% of the Rb valence
polarizability \cite{safronova99a}. The Rb$^+$ core polarizability contributes
about 3\%. Therefore, precision experimental measurements of the transition
rates for the dominant transitions can also be used to deduce accurate values
of the ground state polarizability. However, this is not the case for some
excited states where several transitions may have large contributions and
continuum contributions may be not negligible.

This issue is illustrated using the polarizability of the $5p_{1/2}$ state of the Rb atom \cite{arora07b}, which is
given by
\begin{eqnarray}
\label{5p}
 \alpha_{0}(5p_{1/2})& = &\frac{1}{3} \sum_{n}
   \frac{ |\langle ns \| D\| 5p_{1/2}  \rangle|^2 } {E_{ns}-E_{5p_{1/2}}}  \\
&+&\frac{1}{3} \sum_{n}
   \frac{ |\langle nd_{3/2} \| D\| 5p_{1/2}  \rangle|^2 } {E_{nd_{3/2}}-E_{5p_{1/2}}}+\alpha_{\textrm{core}} \nonumber
\end{eqnarray}

We present a solution to the  Eq.~(\ref{5p}) that combines first principles
calculations with experimental data. The strategy to produce a high-quality
recommended value with this approach is to calculate as many terms as realistic
or feasible using the high-precision atomic structure methods. Where
experimental high-precision data are available (for example, for the $5s-5p$
transitions) they are used in place of theory, assuming that the expected
theory uncertainty is higher than that of the experimental values. The
remainder that contains contributions from highly-excited states is generally
evaluated using (Dirac-Hartree-Fock) DHF or random-phase approximation (RPA)
methods. In our example, the contribution from the very high discrete ($n>10$)
and continuum states is about 1.5\% and cannot be omitted in a precision
calculation. Table ~\ref{table-rb} lists the dipole matrix elements and energy
differences required for evaluation of
 Eq.~(\ref{5p}) as well as the individual contributions to the polarizability.
Experimental values from \cite{volz96a} are used for the $5s-5p_j$ matrix
elements, otherwise the matrix elements are obtained from the all-order
calculations of Ref.~\cite{arora07b} described in
Section~\ref{all-order-section}. Absolute values of the matrix elements are
given. Experimental energies from \cite{NIST2,moore71b} are used. Several
transitions give significant contributions. This theoretical number agrees with
experimental measurement within the uncertainty. The comparison with experiment
is discussed in Section~\ref{results}.

\begin{table}
\caption{\label{table-rb}  The contributions (in a.u.) to the scalar polarizability of the Rb atom in the $5p_{1/2}$
state \cite{arora07b}. The uncertainties in each term are enclosed in parenthesis. The corresponding energy differences
 $\Delta E=E_{n}-E_{5p_{1/2}}$ \cite{moore71b} are given in cm$^{-1}$, which can be converted to atomic
units by division by 219474.6.  Experimental values from \cite{volz96a} are used for absolute values of the $5s-5p_j$
matrix elements, otherwise the matrix elements are obtained from all-order calculations of Ref.~\cite{arora07b}.}
\begin{ruledtabular}
\begin{tabular}{lcrr}
\multicolumn{1}{l}{Contribution}& \multicolumn{1}{c}{$|\langle  n\|D\|5p_{1/2} \rangle|$}&
 \multicolumn{1}{c}{$\Delta E$ }&
\multicolumn{1}{c}{$\alpha_0(5p_{1/2})$}\\
\hline
 $5p_{1/2}-5s$      &     4.231   &  -12579   &  -104.11(15)          \\
 $5p_{1/2}-6s$      &     4.146   &    7554   &   166.5(2.2)          \\
 $5p_{1/2}-7s$      &     0.953   &   13733   &     4.835(16)      \\
 $5p_{1/2}-8s$      &     0.502   &   16468   &     1.120(7)          \\
 $5p_{1/2}-9s$      &     0.331   &   17920   &     0.448(3)          \\
 $5p_{1/2}-10s$     &     0.243   &   18783   &     0.230(2)          \\
 $5p_{1/2}-11s$     &     0.189   &   19338   &     0.135(1)          \\
$5p_{1/2}-(12-\infty )s$                    &         &        &    1.9(0.2)    \\[0.5pc]
  $5p_{1/2}-4d_{3/2}$&    8.017    &   6777   &   694(30)       \\
  $5p_{1/2}-5d_{3/2}$&    1.352    &  13122   &  10.2(9)       \\
  $5p_{1/2}-6d_{3/2}$&    1.067    &  16108   &   5.2(1.1)       \\
  $5p_{1/2}-7d_{3/2}$&    0.787    &  17701   &   2.6(4)        \\
  $5p_{1/2}-8d_{3/2}$&    0.605    &  18643   &   1.4(2)        \\
  $5p_{1/2}-9d_{3/2}$&    0.483    &  19243   &   0.89(10)       \\
 $5p_{1/2}-(10-\infty )d_{3/2} $             &     &        &    10.5(10.5)     \\[0.1pc]
$\alpha_{\text{core}}            $              &          &        &    9.08(45)     \\[0.1pc]
  Total                                        &                 &        & 805(31)   \\
\end{tabular}
\end{ruledtabular}
\end{table}

\subsubsection{Direct  methods}

From a conceptual viewpoint, the finite-field method represents one of the
simplest ways to compute the polarizability. In this approach, one solves the
Schr{\"o}dinger equation using standard techniques for the perturbed
Hamiltonian given by Eq.~(\ref{Hfield}) for a variety of values of $F$. The
polarizability is then extracted from the dipole moment or the energy
eigenvalues of the perturbed Hamiltonian.  This usually entails doing a number
of calculations at different discrete field strengths. This approach is
generally used to obtain polarizabilities in coupled-cluster calculations (see,
for example, Refs.~\cite{lim04a,lupinetti05a}). We note that linearized
coupled-cluster calculations are implemented very differently, and
sum-over-states is used for the polarizability calculations
\cite{safronova08a}. These differences between coupled-cluster calculations are
discussed in Section~\ref{methods}.

Another direct approach to calculating polarizability is the
perturbation-variation method \cite{hibbert77a}. The perturbation-variation
approach has been outlined in the introduction as Eqs.~(\ref{pertvar1}) to
(\ref{pertvar2}). The unperturbed state, $|\Phi_0 \rangle$ and perturbed state,
$|\Phi_1 \rangle$ would be written as a linear combinations of basis states.
Equations (\ref{first0}) and (\ref{pertvar2}) then reduce to sets of matrix
equations. A general technique for  solving the inhomogeneous equation
(\ref{first0}) has been described by Dalgarno and Lewis in
Ref.~\cite{dalgarno55a}.

Exact solutions to Eqs.~(\ref{pertvar1}) - (\ref{pertvar2}) are possible for atomic hydrogen and hydrogenic ions.  The
non-relativistic solutions were first obtained independently in 1926 by Epstein \cite{epstein26}, Waller
\cite{waller26a}, and Wentzel \cite{wentzel26}; the relativistic case remains a subject of current research interest
\cite{yakhontov03, szmytkowski04,szmytkowski06,jentschura08}.  The nonrelativistic equations are separable in parabolic
coordinates, and the polarizability of a hydrogenic ion of nuclear charge Z in the state $|n_1 n_2 m>$ is (in a.u.)

\begin{equation}
\alpha = \frac{n^4}{8Z^4}[17 n^2 -3(n_1-n_2)^2 - 9 m^2 +19] a_0^3,
\label{hparab}
\end{equation}
where $n_1, n_2$ are parabolic quantum numbers \cite{bethe77a}, $m$ is the
projection of the orbital angular momentum onto the direction of the electric field,
and $n = n_1 + n_2 + |m| + 1$ is the principal quantum number.  A convenient special
case is $n = |m|+1$, which corresponds to the familiar circular states of hydrogen
in spherical coordinates, with orbital angular momentum $l = |m| = n-1$; for these
states, $\alpha = (|m|+2)(|m| + 9/4)a_0^3$.

\begin{figure} [h]
\includegraphics[width=8.50cm,angle=0]{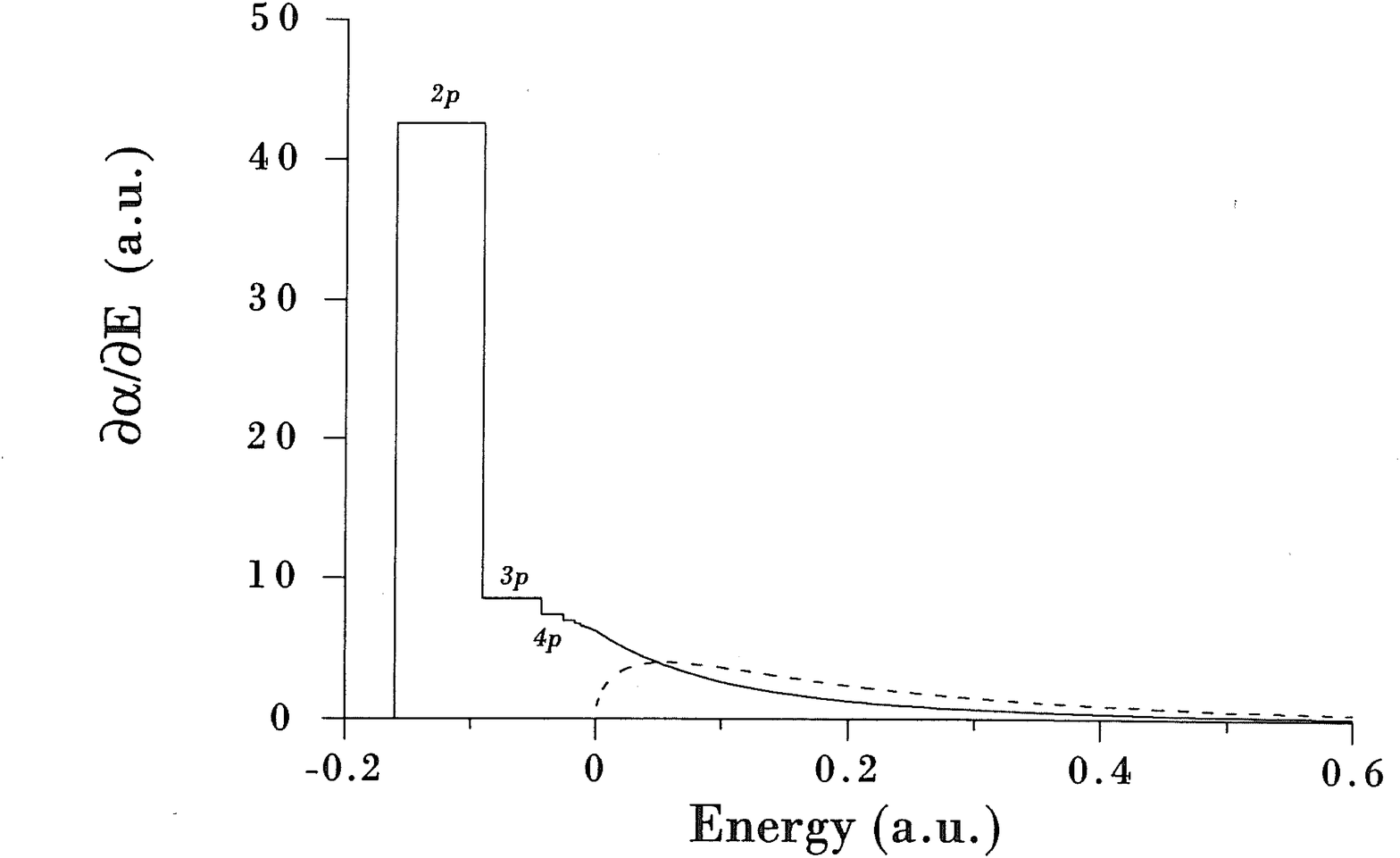}
\vspace{0.02cm} \caption[]{ \label{histo} Solid line: histogram representation of
the sum-over-states contributions to the polarizability of H 1$s$.  Following Ref.
\cite{fano68a}, the contributions of discrete states (e.g. 2$p$) are spread over the
inverse density of states, to show continuity with the continuum contributions near
energy $E=0$.  The polarizability, $\alpha$, is equal to the area under this curve.
Dashed line: the same construction, for an electron bound to a one-dimensional
delta-function potential with energy $E = -1/2$ a.u.  From \cite{clark90}.
}
\end{figure}

For the H 1$s$ ground state exposed to an electric field ${\bf F}=F {\bf{\hat{z}}}$, the
solution to Eq. (\ref{first0}) is (in a.u.)

\begin{eqnarray}
\Phi_0  & = & e^{-r}/\sqrt{\pi} ,\\
\label{Hgroundstate}
\Phi_1  & = & -z(1+r/2) \Phi_0,
\label{Hpertstate}
\end{eqnarray}

\noindent from which $\alpha = (9/2) a_0^3$.  Note that although $\Phi_1$ of Eq.~(\ref{Hpertstate}) is a $p$ state, it
is much more compact than any of the discrete $np$ eigenstates of H.  Thus building up $\Phi_1$ by the sum-over-states
approach requires a significant contribution from the continuous spectrum of H.  This is depicted in Fig. \ref{histo},
which employs the histogram construction of Fano and Cooper \cite{fano68a} to show the connection between discrete and
continuum contributions to the sum over states.  About 20\% of the polarizability of H 1$s$ comes from the continuum.

Clearly, the direct solution of the Schr{\"o}dinger equation for an atom in the presence of an electric field and
subsequent determination of the polarizability is formally equivalent to the sum-over-states approach described in the
previous subsection. However, it is useful to comment on how this equivalence is actually seen in  calculations for
many-electron atoms. For example, random-phase-approximation (RPA) results for polarizabilities of closed-shell atoms
\cite{johnson83a} that were obtained by direct solution of inhomogeneous equation  are the same (up to numerical
uncertainty of the calculations) as sum-over-state RPA results obtained using formula
\begin{equation}
   \alpha_{\textrm{core}} = \frac{2}{3} \sum_{ma}
   \frac{ |\langle \psi_a \| D^{\textrm{DHF}}\| \psi_m  \rangle \langle \psi_a \|
    D^{\textrm{RPA}}\| \psi_m  \rangle |}{E_{m}-E_{a}},
    \label{alphaRPA}
\end{equation}
where $\langle \psi_a \| D^{\textrm{DHF}}\| \psi_m  \rangle$ is reduced matrix
element of dipole operator obtained in the DHF approximation and the $\langle
\psi_a \| D^{\textrm{RPA}}\| \psi_m\rangle$ matrix elements include RPA terms
using many-body perturbation theory as discussed, for example, in
\cite{johnson96a}. The index  $a$ refers to all core orbitals, while the  $m$
includes all other orbitals.  The sum-over-states can be calculated with a
finite basis set \cite{johnson88a}, and such an approach intrinsically includes
the continuum states when complete sum over the entire basis set is carried
out. When the contributions from highly-excited states are significant, it
becomes difficult to account for these terms accurately within the framework of
the sum-over-states approach. Direct method automatically accounts for these
states and this problem does not arise. However, it becomes difficult and
cumbersome to include corrections to the dipole operator beyond RPA. The method
implemented in \cite{johnson83a} is different from the finite field approach
and does not involve performing a number of calculations at different discrete
field strengths.

In most high-precision calculations, the determination of polarizabilities follows the calculation of wave functions or
quantities that represent the wave functions (such as excitation coefficients). The type of approach used for this
initial calculation generally determines whether polarizabilities are determined by Eqs.~(\ref{first0}) or by
sum-over-states method. For example, relativistic linearized coupled-cluster approach \cite{safronova08a} is formulated
in a way that does not explicitly generate numerical wave functions on a radial grid, and all quantities are expressed
in terms of excitation coefficients. Therefore, the polarizabilities are calculated by the sum-over-states method using
resulting high-quality dipole matrix elements and energies. In the case of methods that combine relativistic
configuration interaction and perturbation theory [CI+MBPT], it is natural to determine polarizabilities by directly
solving the inhomogeneous equation. In this case, it is solved in the valence space with the effective operators that
are determined using MBPT \cite{kozlov99a}. The ionic core polarizability is calculated separately in this approach.
The effective dipole operator generally includes RPA corrections, with other corrections calculated independently.

The direct and sum-over-states approaches can also be merged in a hybrid approach. One strategy is to perform a direct
calculation using the best available techniques, and then replace the transition matrix elements for the most important
low-lying states with those from a higher level theory. This hybrid method is discussed further in the sections on the
CI+MBPT and CI+all-order methods.

\subsection{The frequency-dependent polarizability}

So far, we have described the polarizability for static fields.  The numerical value of the
polarizability changes when the atom is immersed in an alternating (AC) electromagnetic field.
To second order, one writes $\Delta E = -\frac{1}{2}
\alpha(\omega) F^2 + \ldots$. The valence part of the scalar frequency dependent polarizability,
usually called the dynamic polarizability, is calculated using
the sum-over-states approach with a straightforwardly modified version of Eq.~(\ref{alpha0J}):
\begin{equation}
 \alpha_{0}(\omega) = \frac{2}{3(2J+1)} \sum_{n}
   \frac{\Delta E |\langle \psi \| D\| \psi_n  \rangle|^2 } {(\Delta E)^2 - \omega^2}.
\label{alphao}
\end{equation}
Eq.~(\ref{alphao}) assumes that $\omega$  is at least a few linewidths away from resonant frequencies defined by
$\Delta E = E_n - E$. As noted previously, atomic units are used throughout this paper, and $\hbar=1$. The core part of
the polarizability  may also be corrected for frequency dependence in random phase approximation  by similarly
modifying the formula (\ref{alphaRPA}). Static values may be used for the core contribution in many applications since
the frequencies of interest, (i.e. corresponding to commonly used lasers) are very far from the excitation energies of
the core states. The calculations of the ground and excited state frequency-dependent polarizabilities of the
alkali-metal atoms are described in detail in Refs.~\cite{safronova06c} and \cite{arora07c}, respectively.
 It is essentially the same as the calculation of the static polarizability described in Section~\ref{sumoverstates},
only for $\omega \neq 0$.

 The expression for the tensor polarizability given by
Eq.~(\ref{alpha2JJ}) is modified in the same way, i.e. by replacing
\begin{equation}
\frac{1}{\Delta E} \rightarrow \frac{\Delta E}{\Delta E^2 - \omega^2}.
\end{equation}

There has been more interest recently in the determination of
frequency-dependent polarizabilities due to the need to know various ``magic
wavelengths'' \cite{katori99b} for the development of optical frequency
standards and other applications. At such wavelengths, the frequency-dependent
polarizabilities of two states are the same, and the AC Stark shift of the
transition frequency between these two states is zero. An example of the
calculation of frequency-dependent polarizabilities and  magic wavelengths is
given in Section~\ref{magicwav}. Experimentally determined magic wavelengths
may also be used to gauge the accuracy of the theory.

\section{Measurements of polarizabilities and related quantities}

Experimental measurements of atomic and ionic polarizabilities are somewhat rarer than theoretical determinations.
There are two types of measurements, those which directly determine the  polarizability, and those which determine
differences in polarizabilities of two states from Stark shift of atomic transitions.

For the most part, we make brief comments on the major experimental techniques and refer the reader to primarily
experimental reviews \cite{miller88a,bonin97a,gould05a,lundeen05a} for further details.

\subsection{$f$-sum rules}

This approach makes use of Eqs.~(\ref{alpha1}-\ref{alpha2LL}).  Many of the most interesting atoms used in cold atom
physics typically have only one or two valence electrons. The ground state polarizability of  these atoms is dominated
by a single low-lying transition.  As mentioned in Section~\ref{sumoverstates}, 97\% of the total value of Rb ground
state polarizability comes from  $5s \to 5p$  transition. In the case of Na, about 99.4$\%$ of the valence
polarizability and 98.8$\%$ of the total polarizability of sodium arises from the $3s \to 3p$ resonant transition.

Composite estimates of the polarizability using both experimental and theoretical inputs are possible.  One type of
estimate would use experimental oscillator strengths to determine the valence polarizability.  This could be combined
with a core contribution obtained by other methods to estimate the total polarizability. Another approach replaces the
most important matrix elements in a first-principles calculation by high precision experimental values
\cite{derevianko99a,porsev06a}. Various types of experiments may be used to determine particular matrix elements,
including photo-association experiments \cite{bouloufa09a}, lifetime, oscillator strengths, or Stark shift measurements
\cite{arora07b} with photoassociation experiments generally giving the most reliable matrix elements. This hybrid
method may provide values accurate to better than 0.5\% in certain cases \cite{derevianko99a}.

\subsection{Dielectric constant}
The dielectric constant $K$ of an atomic or molecular gas is related to the dipole polarizability, $\alpha$, by the
identity
\begin{equation}
\alpha = \frac{K - 1}{4 \pi N},
\end{equation}
where $N$ is the atomic number density.  The technique has only been applied to
the rare gas atoms, and the nitrogen and oxygen atoms by the use of a shock
tube.  Results for the rare gases typically achieve precisions of $0.01 - 0.1
\%$. Examples are reported in Table~\ref{tab:3}.

\subsection{Refractive index}

The frequency-dependent refractive index of a gas  $n(\omega)$, is related to
the polarizability by the expression
\begin{equation}
\alpha(\omega) = \frac{n(\omega)- 1}{2 \pi N} ,
\end{equation}
where $N$ is the atomic number density.  The static dipole polarizability, $\alpha(0)$, can be extracted from the
frequency-dependent polarizability $\alpha(\omega)$ by the following technique.

The energy denominator in Eq.~(\ref{alphao}) can be expanded when the frequency is smaller than the frequency of the
first excitation giving
\begin{equation}
\alpha(\omega) = \alpha(0) + \omega^2 S(-4) + \omega^4 S(-6) \ldots  \\
\label{alphaw2}
\end{equation}
The $S(-q)$ factors are the Cauchy moments of the oscillator strength distribution and are defined by
\begin{equation}
S(-q) =  \sum_{n}  \frac{ f_{gn}} {\Delta  E_{ng}^q }.   \\
\label{cauchy}
\end{equation}
Specific Cauchy moments arise in a number of atomic physics applications, as reviewed by Fano and Cooper
\cite{fano68a}. For example, the Thomas-Reiche-Kuhn sum rule states that $S(0)$ is equal to the number of electrons in
the atom. The $S(-3)$ moment is related to the non-adiabatic dipole polarizability \cite{kleinman68a,dalgarno68a}.

The general functional dependence of the polarizability at low frequencies is given by Eq.~(\ref{alphaw2})
\cite{dalgarno60a,langhoff69a}.  The achievable precision for the rare gases is 0.1$\%$ or better
\cite{langhoff69a,schmidt07a}.  Experiments on the vapours of Zn, Cd and Hg gave polarizabilities with uncertainties of
$1 - 10 \%$ \cite{goebel95a,goebel96a}.
\subsection{Deflection of an atom beam by electric fields}

The beam deflection experiment is conceptually simple.  A collimated atomic beam is directed through an interaction
region containing an inhomogeneous electric field.  While the atom is in the interaction region, the electric field
${\bf F}$ induces a dipole moment on the atom.  Since the field is not uniform, a force proportional to the gradient of
the electric field and the induced dipole moment results in the deflection of the atomic beam.  The polarizability is
deduced from the deflection of the beam.  The overall uncertainty in the derived polarizabilities is between $5 - 10\%$
\cite{hall74a}. Therefore, this method is mainly useful at the present time for polarizability measurements in atoms
inaccessible by any other means.

\subsection{The $E$-$H$ balance method}

In this approach, the $E$-$H$ balance configuration applies an inhomogeneous
electric field and an inhomogeneous magnetic field in the interaction region
\cite{molof74a}.  The magnetic field acts on the magnetic moment of the atom
giving a magnetic deflection force in addition to the electric deflection. The
experiment is tuned so that the electric and magnetic forces are in balance.
The polarizability can be determined since the magnetic moments of many atoms
are known. Uncertainties range from $2\%$ to $10\%$
\cite{molof74a,miller76a,schwartz74a}.

\subsection{Atom interferometry}

The interferometry approach splits the beam of atoms so that one path  sends a beam through a parallel plate capacitor
while the other goes through a field free region.  An interference pattern is then measured when the beams are
subsequently merged and detected.   The polarizability is deduced from the phase shift of the beam passing through the
field free region.  So far, this approach has been used to measure the polarizabilities of helium (see \cite{cronin09a}
for a discussion of this measurement), lithium \cite{miffre06a}, sodium \cite{ekstrom95a, holmgren10a}, potassium
\cite{holmgren10a}, and rubidium \cite{holmgren10a} achieving uncertainties of $0.35 - 0.8\%$.

It has been suggested that multi-species interferometers could possibly determine the polarizability ratio $R =
\alpha_{X}/\alpha_{{\rm Y}}$ to $10^{-4}$ relative accuracy \cite{cronin09a}.  Consequently, a measurement of $R$ in
conjunction with a known standard, say lithium, could lead to a new level of precision in polarizability measurements.
Already the Na:K and Na:Rb polarizability ratios have been measured with a precision of 0.3$\%$ \cite{holmgren10a}.

\subsection{Cold atom velocity change}

The experiment of Amini and Gould \cite{amini03a} measured the kinetic energy gained
 as cold cesium atoms  were
launched from a magneto-optical trap into a region with a finite electric
field. The kinetic energy gained only depends on the final value of the
electric field.  The experimental arrangement actually measures  the  time of
return for cesium atoms to fall back after they are launched into a region
between a set of parallel
 electric-field plates. The only such
experiment  reported so far gave a very precise estimate of the Cs ground state
 polarizability, namely $\alpha_0 =
401.0(6)$ a.u.. This approach can in principle be applied to measure the
polarizability of many other atoms with a
precision approaching 0.1$\%$ \cite{gould05a}.

\subsection{Other approaches}

The deflection of an atomic beam by pulsed lasers has been used to obtain the
dynamic polarizabilities of rubidium and uranium
\cite{kadarkallen92a,kadarkallen94a}. The dynamic polarizabilities of some
metal atoms sourced from an exploding wire have been  measured
interferometrically \cite{hu02a,sarkisov06a}.  These approaches measure
polarizabilities to an accuracy of 5-20$\%$.

\subsection{Spectral analysis for ion polarizabilities}
\label{resis}

 The polarizability of an ion can in principle be extracted from the energies of non-penetrating  Rydberg
series of the parent system\cite{born24a,waller26a,mayer33a}.  The polarizability of the ionic core leads to a shift in
the $(n,L)$ energy levels away from their hydrogenic values.

Consider a charged particle interacting with an atom or ion at large distances.  To zeroth order, the interaction
potential between a highly excited electron and the residual ion is just
\begin{equation}
V(r) = \frac{Z-N}{r} \ ,
\label{eatom}
\end{equation}
where $Z$ is the nuclear charge and $N$ is the number of electrons. However, the outer electron perturbs the atomic
charge distribution. This polarization of the electron charge cloud leads to an attractive polarization potential
between the external electron and the atom. The Coulomb interaction in a multipole expansion with $\mid {\bf r} \mid >
\mid {\bf x} \mid$, is written as
\begin{equation}
\frac{1}{\mid {\mathbf r}-{\mathbf x} \mid} = \sum_{k}{\mathbf C}^k({\mathbf x}) \cdot {\mathbf C}^k({\mathbf r})
\frac{x^k}{r^{k+1}}. \label{coulomb}
\end{equation}
Applying second-order perturbation theory leads to the adiabatic polarization potential between the charged particle
and the atom, e.g.
\begin{equation}
V_{\rm pol}(r) = - \sum_{k=1}^{\infty} \frac{\alpha^{Ek}}{2r^{2k+2}}. \label{vpol}
\end{equation}
The quantities $\alpha^{Ek}$ are the  multipole polarizabilities defined as
\begin{equation}
\alpha^{Ek} = \sum_n \frac{f^{(k)}_{gn}}{(\Delta E_{gn})^2} \ . \label{alphak}
\end{equation}

In this notation, the electric-dipole polarizability is written as $\alpha^{E1}$,
 and $f^{(k)}_{gn}$ is the absorption
oscillator strength for a multipole transition from $g \longrightarrow n$.
  Equation (\ref{vpol}), with its
leading term involving the dipole polarizability is not absolutely convergent in $k$ \cite{dalgarno56a}.  At any finite
$r$, continued summation of the series given by Eq.~(\ref{vpol}), with respect to $k$, will eventually result in a
divergence in the value of the polarization potential.

Equation (\ref{vpol}) is modified by non-adiabatic corrections \cite{kleinman68a,dalgarno68a}.  The non-adiabatic
dipole term is written as
\begin{equation}
V_{\rm non-ad} = \frac{6 \beta_0}{2r^6} \ ,
\label{nonad}
\end{equation}
where the non-adiabatic dipole polarizability, $\beta_0$ is defined
\begin{equation}
\beta_0 = \sum \frac{f^{(1)}_{gn}}{2(\Delta E_{gn})^3} \ .
\label{betad}
\end{equation}
The non-adiabatic interaction is repulsive for atoms in their ground states.  The polarization interaction includes
further adiabatic, non-adiabatic and higher order terms that contribute at the $r^{-7}$ and $r^{-8}$, but there has
been no systematic study of what could be referred to as the non-adiabatic expansion of the polarization potential.

When the Rydberg electron is in a state that has negligible overlap with the core (this is best achieved with the
electron in high angular momentum orbitals), then the polarization interaction usually provides the dominant
contribution to this energy shift.  Suppose the dominant perturbation to the long-range atomic interaction is
\begin{equation}
V_{\rm pol}(r) = - \frac{C_4}{r^4} - \frac{C_6}{r^6}, \label{edlen1}
\end{equation}
where $C_4 = \alpha_0/2$ and $C_6 = (\alpha_0 - 6\beta)/2$.  Equation (\ref{edlen1}) omits the $C_7/r^7$ and $C_8/r^8$
terms that are included in a more complete description \cite{drachman82b,drachman95a,mitroy09a}. The energy shift due
to an interaction of this type can be written
\begin{equation}
\frac{\Delta E}{\Delta \langle r^{-4} \rangle} =
C_4 + C_6 \frac{\Delta \langle r^{-6} \rangle}{\Delta \langle r^{-4} \rangle} \ ,
\label{edlen2}
\end{equation}
where $\Delta E$ is usually the energy difference between two Rydberg states.  The expectation values $\Delta \langle
r^{-6} \rangle$ and $\Delta \langle r^{-4} \rangle$ are simply the differences in the radial expectations of the two
states.  These are easily evaluated using the identities of Bockasten \cite{bockasten74a}.  Plotting $\frac{\Delta
E}{\Delta \langle r^{-4} \rangle}$ versus $\frac{\Delta \langle r^{-6} \rangle}{\Delta \langle r^{-4} \rangle}$ yields
$C_4$ as the intercept and $C_6$ as the gradient.  Such a graph is sometimes called a polarization plot.

Traditional spectroscopies such as discharges or laser excitation find it difficult to excite atoms into Rydberg states
with $L > 6$. Exciting atoms into states with $L > 6$ is best done with resonant excitation Stark ionization
spectroscopy (RESIS) \cite{lundeen05a}. RESIS spectroscopy first excites an atomic or ionic beam into a highly-excited
state, and then uses a laser to excite the system into a very highly-excited state which is Stark ionized.

While this approach to extracting polarizabilities from Rydberg series energy shifts is appealing, there are a number
of perturbations that act to complicate the analysis.  These include relativistic effects $\Delta E_{\rm rel}$, Stark
shifts from ambient electric fields $\Delta E_{\rm ss}$, second-order effects due to relaxation of the Rydberg electron
in the field of the polarization potential $\Delta E_{\rm sec}$ \cite{drake91a,swainson92b,mitroy08k}, and finally the
corrections due to the $C_7/r^7$ and $C_8/r^8$ terms, $\Delta E_7,\Delta E_8$, and $\Delta E_{8L}$.  Therefore,  the
energy shift between two neighbouring Rydberg states is
\begin{eqnarray}
\Delta E & = & \Delta E_4 + \Delta E_6 + \Delta E_7 + \Delta E_8 + \Delta E_{8L}  \nonumber \\
         & + & \Delta E_{\rm rel} + \Delta E_{\rm sec} + \Delta E_{\rm ss}.
\label{DeltaE1}
\end{eqnarray}
One way to solve the problem is to simply subtract these terms from
the observed energy shift, e.g.
\begin{eqnarray}
\frac{\Delta E_{c}}{\Delta \langle r^{-4} \rangle} &=&
  \frac{\Delta E_{\rm obs}}{\Delta \langle r^{-4} \rangle}  \nonumber \\
 &-& \left( \frac{\Delta E_{\rm rel} + \Delta E_{\rm sec} + \Delta E_{ss}} {\Delta \langle r^{-4} \rangle} \right) \nonumber \\
 &-& \left( \frac{\Delta E_{\rm 7} + \Delta E_{\rm 8} + \Delta E_{8L}} {\Delta \langle r^{-4} \rangle} \right) \ .
\label{DeltaEc1}
\end{eqnarray}
and then deduce $C_4$ and $C_6$ from the polarization plot of
the corrected energy levels \cite{mitroy09a}.

\subsection{Stark shift measurements of polarizability differences}

The Stark shift experiment predates the formulation of quantum
 mechanics in its modern form \cite{stark13a}.  An atom
is immersed in an electric field, and the shift in wavelength of one
of its spectral lines is measured as a function of
the field strength. Stark shift experiments effectively measure the difference between the polarizability of the two
atomic states involved in the transition. Stark shifts can be measured for both static and dynamic electric fields.
While there have been many Stark shift measurements, relatively few have achieved an overall precision of 1$\%$ or
better.

While the polarizabilities can generally be extracted from the Stark shift measurement, it is useful to compare the
experimental values directly with theoretical predictions where high precision is achieved for both theory and
experiment. In this review, comparisons of the theoretical static polarizability differences for the resonance
transitions involving the alkali atoms with the corresponding Stark shifts are provided in Section~\ref{results}. Some
of the alkali atom experiments report precisions between 0.01 and 0.1 a.u.
\cite{hunter88a,hunter91a,hunter92a,bennett99a}. The many Stark shift experiments involving Rydberg atoms
\cite{wijngaarden99a} are not detailed here.

Selected Stark shifts for some non-alkali atoms that are of interest for
applications described in this review are discussed in Section~\ref{results} as
well. The list is restricted to low-lying excited states for which high
precision Stark shifts are available. When compared with the alkali atoms,
there are not that many measurements and those that have been performed have
larger uncertainties.

The tensor polarizability of an open shell atom can be extracted from the difference in polarizabilities between the
different magnetic sub-levels. Consequently, tensor polarizabilities do not rely on absolute polarizability
measurements and can be extracted from Stark shift measurements by tuning the polarization of a probe laser.  Tensor
polarizabilities for a number of states of selected systems are discussed in Section~\ref{results}.

One unusual experiment was a measurement of the AC energy shift ratio for the
$6s$ and $5d_{3/2}$ states of Ba$^+$ to an accuracy of 0.11$\%$ \cite{sherman05a}.
This experiment does not give polarizabilities, and is mainly valuable as an
additional constraint upon calculation \cite{iskrenova08a}.

\subsection{AC Stark shift measurements}

There are  few experimental measurements of AC Stark shifts at optical frequencies.   Two recent examples would be the
determination of the Stark shift for the Al$^+$ clock transition \cite{rosenband06a} and the Li $2s$-$3s$ Stark shifts
\cite{sanchez09a} at the frequencies of the pump and probe laser of a two-photon resonance transition  between
the two states. One difficulty in the interpretation of AC Stark shift experiments is the lack of precise knowledge
about the overlap of the laser beam with atoms in the interaction region. This is also a complication in the analysis
of experiment on deflection of atomic beams by lasers \cite{kadarkallen92a,kadarkallen94a}.

\section{Practical calculation of atomic polarizabilities}
\label{methods}

There have been numerous theoretical studies  of atomic and ionic polarizabilities in the last several decades. Most
methods used to determine atomic wave functions and energy levels can be adapted to generate polarizabilities.  These
have been divided into a number of different classes that are listed below.  We give a brief description of each
approach. It should be noted that the list is not exhaustive, and the emphasis here has been on those methods that have
achieved the highest accuracy or those methods that have been applied to a number of different atoms and ions.

\subsection{Configuration interaction}

 The configuration interaction (CI) method \cite{hibbert75a} and its variants are widely used for atomic
structure calculations owing to  general applicability of the CI method.  The CI wave function is written as a linear
combination of configuration state functions
\begin{equation}
\Psi_{\textrm{CI}} = \sum_{i} c_{i} \Phi_i,
\end{equation}
i.e. a linear combination of Slater determinants from a model subspace \cite{dzuba96b}.
 Each configuration is constructed with consideration
given to anti-symmetrization, angular momentum and parity requirements. There is a great deal of variety in how the CI
approach is implemented.  For example, sometimes the exact functional form of the orbitals in the excitation space is
generated iteratively during successive diagonalization of the excitation basis. Such a scheme is called the
multi-configuration Hartree-Fock (MCHF) or multi-configuration self consistent field (MCSCF) approach
\cite{fischer97a}. The relativistic version of MCHF is referred to as multi-configuration Dirac-Fock (MCDF) method
\cite{grant07}.

The CI approach has a great deal of generality since there are no restrictions imposed upon the virtual orbital space
and classes of excitations beyond those limited by the computer resources.  The method is particularly useful for open
shell systems which contain a number of strongly interacting configurations. On the other hand, there can be a good
deal of variation in quality between different CI calculations for the same system, because of the flexibility of
introducing additional configuration state functions.

The most straightforward way to evaluate polarizability within the framework of the CI method it to use a direct
approach by solving the inhomogeneous equation (\ref{first0}). RPA corrections to the dipole operator can be
incorporated using the effective operator technique described in Section~\ref{ci+mbpt}. It is also possible to use
CI-generated matrix elements and energies to evaluate sums over states. The main drawback of the CI method is its loss
of accuracy for heavier systems. It becomes difficult to include a sufficient number of configurations for heavier
systems to produce accurate results even with modern computer facilities. One solution of this problem is to use a
semi-empirical core potential (CICP method) described in the next subsection. Another, \textit{ab initio} solution,
involves construction of the effective Hamiltonian using either many-body perturbation theory (CI+MBPT) or all-order
linearized coupled-cluster method (CI+all-order) and carrying out CI calculations in the valence sector. These
approaches are described in the last two sections of this chapter.

\subsection{CI calculations with a semi-empirical core potential (CICP)}

The \textit{ab initio} treatment of core-valence correlations greatly increases the complexity of any structure
calculation. Consequently, to include this physics in the calculation, using a semi-empirical approach is an attractive
alternative for an atom with a few valence electrons \cite{laughlin88a,muller84,mitroy03f}.

In this method, the active Hamiltonian for a system with two valence electrons is written as
\begin{eqnarray}
H  &=&  \sum_{i=1}^2 \left( -\frac {1}{2} \nabla_i^2 + V_{\rm dir}({\bf r_i})
     + V_{\rm exc}({\bf r_i}) +  V_{\rm p1}({\bf r_i}) \right ) \nonumber \\
   &+&  \frac{1}{r_{12}} + V_{\rm p2}({\bf r_1},{\bf r}_2).
\end{eqnarray}

The $V_{\rm dir}$ and $V_{\rm exc}$ represent the direct and exchange interactions with the core electrons. In some
approaches, these terms are represented by model potentials, \cite{victor76,norcross76a,santra04a}. More refined
approaches evaluate $V_{\rm dir}$ and $V_{\rm exc}$  using core wave functions calculated with the Hartree-Fock (or
Dirac-Fock) method \cite{muller84,migdalek78a,mitroy03f}. The one-body polarization interaction $V_{\rm pol}(r)$ is
semi-empirical in nature and can be written in its most general form as an $\ell$-dependent potential, e.g.
\begin{equation}
V_{\rm p1}({\bf r})  =  -\sum_{\ell m} \frac{\alpha g_{\ell}^2(r)}{2 r^4}
                    |\ell m \rangle \langle \ell m|,
\label{polar1}
\end{equation}
where  $\alpha$ is the static dipole polarizability of the core and $g_{\ell}^2(r)$ is a cutoff function that
eliminates the $1/r^4$ singularity at the origin.  The cutoff functions usually include an adjustable parameter that is
tuned to reproduce the binding energies of the valence states.  The two-electron
 or di-electronic polarization potential is
written
\begin{equation}
V_{\rm p2}({\bf r}_i,{\bf r}_j) = - \frac{\alpha} {r_i^3 r_j^3} ({\bf r}_i\cdot{\bf r}_j)g(r_i)g(r_j)\ . \label{polar2}
\end{equation}
There is variation between expressions for the core polarization potential, but what is described above is fairly
representative.  One choice for the cutoff function is $g_{\ell}^2(r) = 1-\exp\bigl(-r^6/\rho_{\ell}^6 \bigr)$
\cite{mitroy03f}, but other choices exist.

A complete treatment of the core-polarization corrections also implies that corrections have to be made to the
multipole operators \cite{hameed68a,hameed72a,muller84,mitroy03f}. The modified transition operator is obtained from
the mapping
\begin{equation}
r^k {\bf C}^k({\bf r})  \to g_{\ell}(r) r^k {\bf C}^k({\bf r}).
\end{equation}
Usage of the modified operator is essential to the correct prediction of the oscillator strengths.  For example, it
reduces the K$(4s \to 4p)$ oscillator strength by 8$\%$ \cite{muller84}.

One advantage of this configuration interaction plus core-polarization (CICP) approach is in reducing the
size of the calculation.  The elimination of the core from active
consideration permits very accurate solutions of the Schr{\"o}dinger equation for the valence electrons.
Introduction of the core-polarization potentials, $V_{p1}$ and $V_{p2}$, introduces an additional
source of uncertainty into the calculation.  However, this additional small source of uncertainty
is justified by the almost complete elimination of computational uncertainty in the solution of
the resulting simplified Schr{\"o}dinger equation.

The CICP approach only gives the polarizability of the valence electrons. Core polarizabilities are typically quite
small for the group I and II atoms, e.g. the cesium atom has a large core polarizability of about 15.6~$a_0^3$
\cite{zhou89a}, but this represents only 4$\%$ of the total ground state polarizability of 401 $a_0^3$ \cite{amini03a}.
Hence, usage of moderate accuracy core polarizabilities sourced from theory or experiment will lead to only small
inaccuracies in the total polarizability.

Most implementation of the  CICP approach to the calculation of polarizabilities have been within a non-relativistic
framework. A relativistic variant (RCICP) has recently been applied to zinc, cadmium, and mercury \cite{ye08a}.  It
should be noted that even non-relativistic calculations incorporate relativistic effects to some extent.  Tuning the
core polarization correction to reproduce the experimental binding energy partially incorporates relativistic effects
on the wave function.

\subsection{Density functional theory}

Approaches based on Density Functional Theory (DFT) are not expected to give
polarizabilities as accurate as those coming from the refined \textit{ab
initio} calculations described in the following sections. Polarizabilities from
DFT calculations are most likely to be useful for systems for which large scale
\textit{ab initio} calculations are difficult, e.g. the transition metals. DFT
calculations are often much less computationally expensive than \textit{ab
initio} calculations. There have been two relatively extensive DFT compilations
\cite{doolen84a,chu04a} that have reported dipole polarizabilities for many
atoms in the periodic table.

\subsection{Correlated basis functions}

The accuracy of atomic structure calculations can be dramatically improved by the use of basis functions which
explicitly include the electron-electron coordinate.  The most accurate calculations reported for atoms and ions with
two or three electrons have typically been performed with exponential basis functions including the inter-electronic
coordinates as a linear factor. A typical Hylleraas basis function for lithium would be
\begin{equation}
\chi = r_1^{j_1} r_2^{j_2} r_{3}^{j_3} r_{12}^{j_{12}} r_{13}^{j_{13}} r_{23}^{j_{23}} \exp \left(-\alpha r_1 - \beta
r_2 - \gamma r_3 \right). \label{Hyl}
\end{equation}
Difficulties with performing the multi-center integrals have effectively precluded the use of such basis functions for
systems with more than three electrons. Within the framework of the non-relativistic Schr{\"o}dinger equation,
calculations with Hylleraas basis sets achieve accuracies of 13 significant digits \cite{pachucki01a} for
polarizability of two-electron systems and 6 significant digits for the polarizability of three-electron systems
\cite{yan96a,tang09a}. Inclusion of relativistic and quantum electrodynamic (QED) corrections to the polarizability of helium has been carried
out in Refs.~\cite{pachucki01a,lach04a}, and the resulting final value is accurate to  7 significant digits.

 Another correlated basis set that has recently found increasingly widespread use
  utilizes the
explicitly correlated gaussian (ECG).  A typical spherically symmetric explicitly correlated gaussian for a
three-electron system is  written as \cite{suzuki98a}
\begin{equation}
\chi = \exp \left(-\sum_{i=1}^3 \alpha_i r_i^2 -\sum_{i<j} \beta_{ij} r_{ij}^2 \right ). \label{ECG}
\end{equation}
The multi-center integrals that occur in the evaluation of the Hamiltonian can be generally reduced to analytic
expressions that are relatively easy to compute.  Calculations using correlated gaussians do not achieve the same
precision as Hylleraas forms,  but are still capable of achieving much higher precision than orbital based calculations
provided the parameters $\alpha_i$ and $\beta_{ij}$ are well optimized \cite{suzuki98a,komasa02a}.

\subsection{Many-body perturbation theory}

The application of many-body perturbation theory (MBPT)  is discussed in this
section
 in the context of the Dirac
equation.   While MBPT has been applied with the non-relativistic Schr{\"o}dinger equation, many recent applications
most relevant to this review have been using a relativistic Hamiltonian.

 The point of departure for the discussions of
relativistic many-body perturbation theory (RMBPT) calculations is the {\em no-pair} Hamiltonian obtained from QED by
\citet{brown51a}, where the contributions from negative-energy (positron) states are projected out. The no-pair
Hamiltonian can be written in second-quantized form as $H=H_0+V$, where
\begin{align}
H_0 & = \sum_i \epsilon_i [a_i^\dagger a_i]\, , \label{np1}\\
V & = \frac{1}{2} \sum_{ijkl}\left( g_{ijkl} + b_{ijkl}\right)
[ a^\dagger_ia^\dagger_j a_l a_k ]\label{np2} \\
  & + \sum_{ij} \left(V_\text{DHF} + B_\text{DHF} - U\right)_{ij} [a^\dagger_ia_j
  ]\nonumber,
\end{align}
and a c-number term that just  provides an additive constant to the energy of the atom has been omitted.

In Eqs.~(\ref{np1} - \ref{np2}), $a^\dagger_i$ and $a_i$ are creation and annihilation operators for an electron state
$i$, and the summation indices range over electron bound and scattering states only. Products of operators enclosed in
brackets, such as $[ a^\dagger_ia^\dagger_j a_l a_k ]$, designate normal products with respect to a closed core. The
core DHF potential is designated by $V_\text{DHF}$ and its Breit counterpart is designated by
$B_\text{DHF}$.   The quantity $\epsilon_i$ in Eq.~(\ref{np1}) is the eigenvalue of the Dirac equation. The quantities
$g_{ijkl}$ and $b_{ijkl}$ in Eq.~(\ref{np2}) are two-electron Coulomb and Breit matrix elements, respectively
\begin{align}
g_{ijkl}& = \left< ij \left| \frac{1}{r_{12}} \right| kl\right>\, ,
\\
b_{ijkl}& = - \left< ij \left|\ \frac{{\bm \alpha}_1\cdot{\bm \alpha}_2 + ({\bm \alpha}_1\cdot \hat{\bm r}_{12})({\bm
\alpha}_2\cdot \hat{\bm r}_{12})} {2r_{12}} \ \right| kl\right>,
\end{align}
where $\bm \alpha$ are Dirac matrices.

For neutral atoms, the Breit interaction is often a small perturbation that can be ignored compared to the Coulomb
interaction. In such cases, it is particularly convenient to choose the starting potential $U(r)$ to be the core DHF
potential $U = V_{\textrm{DHF}}$,
\begin{equation}
(V_\text{DHF})_{ij} =\sum_{a} \left[g_{iaja}-g_{iaaj}\right],
\end{equation}
since with this choice, the second term in Eq.~(\ref{np2}) vanishes. The index  $a$ refers to all core orbitals. The
Breit $(B_\text{DHF})_{ij}$ term is defined as
\begin{equation}
(B_\text{DHF})_{ij} =\sum_{a} \left[b_{iaja}-b_{iaaj}\right].
\end{equation}

For monovalent atoms, the lowest-order wave function is written as
\begin{equation}
|\Psi_v^{(0)}\rangle  = a^\dagger_v | 0_c\rangle\, ,\label{vwf}
\end{equation}
where $| 0_c\rangle = a^\dagger_a a^\dagger_b \cdots  a^\dagger_n |0\rangle$ is the closed core wave function,
$|0\rangle$ being the vacuum wave function, and $a_v^\dagger$ being a valence-state
 creation operator. The indices  $a$
and $b$ refer to core orbitals.

The perturbation expansion for the wave function leads immediately to a perturbation expansion for matrix elements.
Thus, for the one-particle operator written in the second-quantized form as
\begin{equation}
\label{z}
 Z = \sum_{ij} z_{ij} a^\dagger_i a_j ,
\end{equation}
perturbation theory leads to an order-by-order expansion for  the matrix
element of $Z$ between states $v$ and $w$ of an atom with one valence electron:
\begin{equation}
 \langle \Psi_w | Z | \Psi_v \rangle = Z^{(1)}_{wv} + Z^{(2)}_{wv} + \cdots,
\end{equation}
The first-order matrix element is given by the DHF value in the present case
\begin{equation}
 Z^{(1)}_{wv} = z_{wv}.
\end{equation}

The second-order expression for the  matrix element of a one-body operator $Z$
in a Hartree-Fock potential is given by
\begin{equation}
        Z^{(2)}_{wv}=\sum_{am} \frac {z_{am} \widetilde{g}_{wmva}}{\epsilon_{av}-\epsilon_{mw}}
           +\sum_{am} \frac {z_{ma} \widetilde{g}_{wavm}}{\epsilon_{wa}-\epsilon_{mv}},
\end{equation}
where $\epsilon_{wa}=\epsilon_{w}+\epsilon_{a}$.
 The summation index $a$  ranges over states in the closed
core, and the summation index $m$  ranges over the excited states. The complete third-order MBPT expression for the
matrix elements of monovalent systems was given in Ref.~\cite{johnson96a}. The monumental task of deriving and
evaluating the complete expression for the fourth-order matrix elements has been carried out for Na in
Ref.~\cite{cannon04a}.

The polarizabilities are obtained using  a sum-over-state approach by combining
the resulting matrix elements and either experimental or theoretical energies.
The calculations are carried out with a finite basis set, resulting in a finite
sum in the sum-over-state expression that it is equivalent to the inclusion of
all bound states and the continuum. Third-order MBPT calculation of
polarizabilities is described in detail, for example, in
Ref.~\cite{safronova09a} for Yb$^+$.

The relativistic third-order many-body perturbation  theory generally gives
good results for electric-dipole (E1) matrix elements of lighter systems in the
cases when the correlation corrections are not unusually large. For example,
the third-order value of the Na $3s-3p_{1/2}$ matrix element agrees with
high-precision experiment to 0.6\% \cite{safronova08a}. However, the
third-order values for the $6s-6p_{1/2}$ matrix element in Cs and $7s-7p_{1/2}$
matrix element in Fr differ from the experimental data by 1.3\% and 2\%,
respectively \cite{safronova08a}. For some small matrix elements, for example
$6s-7p$ in Cs, third-order perturbation theory gives much poorer values. As a
result, various methods that are equivalent to summing dominant classes of
perturbation theory terms to all orders have to be used to obtain precision
values, in particular when sub-percent accuracy is required.

 The relativistic all-order correlation potential method that enables
  efficient treatment of dominant core-valence
 correlations was developed in Ref.~\cite{dzuba89b}. It was used to study
 fundamental symmetries in heavy atoms
 and to calculate atomic properties of alkali-metal atoms and isoelectronic ions
  (see, for example, Refs.~\cite{dzuba01a,dzuba02a} and references therein).
     In the correlation potential method for monovalent systems, the calculations generally
     start from the relativistic Hartree-Fock
     method in the $V^{N-1}$ approximation. The correlations are incorporated by
     means of a correlation potential $\Sigma$ defined in such a way that  its
     expectation value over a valence electron
     wave function is equal to the RMBPT expression for the correlation correction
     to the energy of the electron.
Two classes of higher-order corrections are generally included in the
correlation potential: the screening of
     the Coulomb interaction between a valence electron and a core electron by
     outer electrons,
     and hole-particle interactions. Ladder diagrams were included to all orders
     in Ref.~\cite{dzuba08a}.
     The correlation potential is used to build a new set of single-electron
      states for subsequent evaluation of
     various matrix
      elements using the random-phase approximation.
     Structural radiation and the normalization corrections to matrix elements
     are also incorporated.
     This approach was used to evaluate black-body radiation shifts in microwave
     frequency standards in Refs.~\cite{angstmann06a,angstmann06b}
      (see Section~\ref{BBR-mic}).

Another class of the all-order approaches based on the coupled-cluster method
is discussed in the next subsection.

\subsection{Coupled-cluster methods}
\label{all-order-section}


 In the coupled-cluster method, the
exact many-body wave function is represented in the form
\cite{coester60a}
\begin{equation}
|\Psi \rangle = \exp(S) |\Psi^{(0)}\rangle, \label{cc}
\end{equation}
where $|\Psi^{(0)}\rangle$ is the lowest-order atomic wave function. The operator $S$ for an N-electron atom consists
of ``cluster'' contributions from one-electron, two-electron, $\cdots$, N-electron excitations of the lowest-order wave
function $|\Psi^{(0)}\rangle$: $S=S_1+S_2+ \dots +S_N $. In the single-double approximation of the coupled-cluster
(CCSD) method, only single and double excitation terms with $S_1$ and $S_2$ are retained. Coupled-cluster calculations
which use a relativistic Hamiltonian are identified by a prefix of R, e.g. RCCSD.

The exponential in Eq.~(\ref{cc}), when expanded in terms of the $n$-body
excitations $S_n$, becomes
\begin{equation}
|\Psi \rangle = \left\{ 1+S_1+S_2+S_3+\frac{1}{2} S_1^2 + S_1 S_2 +
  \cdots \right\} |\Psi^{(0)}\rangle.
\label{cc1}
\end{equation}

Actual implementations of the coupled-cluster approach and subsequent determination of polarizability vary
significantly with the main source of variation being the inclusion of triple excitations or non-linear terms and use
of different basis sets. These differences account for some discrepancies between different coupled-cluster
calculations for the same system. It is common for triple excitations to be included perturbatively. In this review,
all coupled-cluster calculations that include triples in some way are labelled as CCSDT (or RCCSDT, RLCCSDT)
calculations with no further distinctions being made.

We can generally separate coupled-cluster calculations of polarizabilities into two groups, but note that details of
calculations vary between different works. Implementations of the CCSDT method in the form typically used for the
quantum chemistry calculations use gaussian type orbital  basis sets.  Care should be taken to explore the dependence
of the final results on the choice and size of the basis set. The dependence of the dipole polarizability values on the
quality of the basis set used has been discussed, for example, in Ref.~\cite{lim04a}. In those calculations, the
polarizabilities
 are generally calculated using the finite-field approach \cite{lim04a,lim05a,lupinetti05a}. Consequently, such
CC calculations are not restricted to monovalent systems, and RCC calculations of polarizabilities of divalent systems
have been reported in Refs.~\cite{sadlej91a,schafer07a,lim04a}.

The second type of relativistic coupled-cluster calculations is carried out using the linearized variant of the
coupled-cluster method (referred to as the relativistic all-order method in most references), which was first developed
for atomic physics calculations and applied to He in Ref.~\cite{blundell89b}. The extension of this method to
 monovalent systems was introduced in Ref.~\cite{blundell89a}. We refer to this
approach as the RLCCSD or RLCCSDT method \cite{safronova08a}. We note that RLCCSDT method includes only valence triples
using perturbative approach.  As noted above, all CC calculations that include triples in some way are labelled as
CCSDT. The RLCCSDT method uses finite
 basis set of B-splines rather than gaussian orbitals. The B-spline basis sets are effectively complete for each partial wave,
  i.e. using a
larger basis set will produce  negligible changes in the results. The partial waves with $l=0-6$ are generally used.
Third-order perturbation theory is used to account for higher partial waves where necessary.  Very large basis sets are
used, typically a total of 500 - 700 orbitals are included for monovalent systems. Therefore, this method avoids the
basis set issues generally associated with other coupled-cluster calculations. The actual algorithm implementation is
distinct from standard quantum chemistry codes as well.

In the linearized coupled-cluster approach,  all non-linear terms
are omitted and the wave function takes the form
\begin{equation}
|\Psi \rangle = \left\{1+S_1+S_2+S_3 + \cdots +S_N \right\}
|\Psi^{(0)}\rangle\, . \label{cc2}
\end{equation}
The inclusion of the nonlinear terms within the framework of this method is described in Ref.~\cite{pal07a}.
Restricting the sum in Eq.~(\ref{cc2}) to single, double, and valence triple excitations yields the expansion for the
wave function of a monovalent atom in state $v$:

\begin{eqnarray}
|\Psi_v \rangle &= &\left[ 1 + \sum_{ma} \, \rho_{ma} a^\dagger_m a_a +
\frac{1}{2} \sum_{mnab} \rho_{mnab} a^\dagger_m a^\dagger_n a_b a_a +
 \right.  \nonumber \\
& +& \left. \sum_{m \neq v} \rho_{mv} a^\dagger_m a_v + \sum_{mna} \rho_{mnva}
a^\dagger_m a^\dagger_n a_a a_v \right. \nonumber \\
& + &\left. \frac{1}{6}\sum_{mnrab} \rho_{mnrvab} a^\dagger_m a^\dagger_n a^\dagger_r a_b a_a a_v \right]|
\Psi_v^{(0)}\rangle\label{eqall} ,
\end{eqnarray}
where the indices $m$, $n$, and $r$ range over all possible virtual states while indices $a$ and $b$ range over all
occupied core states. The quantities $\rho_{ma}$, $\rho_{mv}$ are single-excitation coefficients for core and valence
electrons and $\rho_{mnab}$ and $\rho_{mnva}$ are double-excitation coefficients for core and valence electrons,
respectively,  $\rho_{mnrvab}$ are the triple valence excitation coefficients.  For the monovalent systems, $U$ is
generally taken to be  the \textit{frozen-core} $V^{\textrm{N-1}}$ potential, $U=V_{\textrm{DF}}$.

We refer to results obtained with this approach as RLCCSDT, indicating inclusion of single, double, and partial triple
excitations. The triple excitations are generally included perturbatively. Strong cancellations between groups of
smaller terms, for example non-linear terms and certain triple excitation terms have been found in
Ref.~\cite{porsev06c}. As a result, additional inclusion of certain classes of terms may not necessarily lead to more
accurate values.

The matrix elements for any one-body operator $Z$ given in second-quantized form by Eq.~(\ref{z})
 are obtained within the framework of the linearized coupled-cluster method as
\begin{equation}
Z_{wv}=\frac{\langle \Psi_w |Z| \Psi_v \rangle}{\sqrt{\langle \Psi_v | \Psi_v
\rangle \langle \Psi_w | \Psi_w \rangle}},  \label{eqr}
\end{equation}
where $|\Psi_v\rangle$ and $|\Psi_w\rangle$ are given by the expansion (\ref{eqall}). In the SD approximation, the
resulting expression for the numerator of Eq.~(\ref{eqr}) consists of the sum of the DHF matrix element $z_{wv}$ and 20
other terms that are linear or quadratic functions of the excitation coefficients \cite{blundell89a}. The main
advantage of this
 method is its general applicability to calculation of  many atomic properties of ground and excited states: energies, electric and magnetic
 multipole
 matrix elements and other transition properties such as oscillator strengths and lifetimes,
  $A$ and $B$ hyperfine constants, dipole and quadrupole polarizabilities, parity-nonconserving matrix
 elements, electron electric-dipole-moment   (EDM) enhancement factors, $C_3$ and $C_6$ coefficients, etc.

The all-order method yields results for the properties of alkali atoms \cite{safronova99a} in excellent agreement with
experiment. The application of this method to the calculation of alkali  polarizabilities (using a sum-over-state
approach) is described in detail in Refs.~\cite{safronova99a,derevianko99a,arora07b,arora07c}.

In its present form  described above, the  RLCCSDT method is only applicable to the calculation of polarizabilities of
monovalent systems. The work on combining the RLCCSDT approach with the CI method to create a method that is more
general is currently in progress \cite{safronova09c} and is described in Section \ref{ci+all}.

\subsection{Combined CI and many-body perturbation theory}
\label{ci+mbpt}

Precise calculations for atoms with several valence electrons require an accurate treatment of valence-valence
correlations.  While finite-order MBPT is a powerful technique for atomic systems with weakly interacting
configurations, its accuracy can be limited when the wave function has a number of strongly interacting configurations.
One example occurs for the alkaline-earth atoms where there is strong mixing between the $ns^2$ and $np^2$
configurations of $^1$S symmetry. For such systems, an approach combining both aspects has been developed by Dzuba {\em
et al.} \cite{dzuba96b} and later applied to the calculation of atomic properties of many other systems
\cite{porsev01a,kozlov99a,porsev06a,porsev08a,dzuba06a,dzuba10a}. This composite approach to the calculation of atomic
structure is often abbreviated as CI+MBPT (we use RCI+MBPT designations in this review to indicate that the method is
relativistic).

For systems with more than one valence electron, the precision of the CI method is drastically limited  by the sheer
number of the configurations that should be included. As a result, the  core-core and core-valence correlations might
only receive a limited treatment, which can lead to a significant loss of accuracy.  The RCI+MBPT approach provides a
complete treatment of core correlations to a limited order of perturbation theory.  The RCI+MBPT approach uses
perturbation theory to construct an effective core Hamiltonian, and then a CI calculation is performed to generate the
valence wave functions.

The no-pair Hamiltonian given by Eqs.~(\ref{np1}) and (\ref{np2}) separates into a sum of the one-body and two-body
interactions,
\begin{equation}
H = H_1 + H_2,
\end{equation}
where $H_2$ contains the Coulomb (or Coulomb + Breit) matrix elements $v_{ijkl}$. In the RCI+MBPT approach, the
one-body term $H_1$ is modified to include a correlation potential $\Sigma_1$ that accounts for part of the
core-valence correlations, $H_1 \rightarrow H_1+\Sigma_1$. Either the second-order expression for $\Sigma_1^{(2)}$ or
all-order chains of such terms can be used (see, for example, 
Ref.~\cite{dzuba96b}).  The two-body Coulomb interaction term in $H_{2}$ is modified by including the two-body part of
core-valence interaction that represents screening of the Coulomb interaction by valence electrons; $H_2 \rightarrow
H_2+\Sigma_2$. The quantity $\Sigma_2$ is calculated in second-order MBPT \cite{dzuba96b}.  The CI method is then used
with the modified $H_\text{eff}$ to obtain improved energies and wave functions.

The polarizabilities are determined using the direct approach (in the valence sector) by solving the inhomogeneous
equation in the valence space, approximated from Eq.~(\ref{first0}). For state  $v$ with total angular momentum $J$ and
projection $M$, the corresponding equation is written as \cite{kozlov99a}
\begin{equation}
(E_v - H_{\textrm{eff}})|\Psi(v,M^{\prime})\rangle = D_{\textrm{eff}} |\Psi_0(v,J,M)\rangle.
\end{equation}
The wave function $\Psi(v,M^{\prime})$ is composed of parts that have angular momenta of $J^{\prime}=J,J \pm 1$. This
then permits the scalar and tensor polarizability of the state $|v,J,M\rangle$ to be determined \cite{kozlov99a}.

 The
construction of $H_\textrm{eff}$ was described in the preceding paragraphs. The
effective dipole operator $D_{\textrm{eff}}$ includes random phase
approximation (RPA) corrections and several smaller MBPT corrections described
in \cite{JETP}. Non-RPA corrections may be neglected in some cases
\cite{kozlov99a}.  There are several variants of the RCI+MBPT method that
differ by the corrections included in the effective operators
$H_{\textrm{eff}}$ and $D_{\textrm{eff}}$, the functions used for the basis
sets, and versions of the CI code. In some implementations of the RCI+MBPT, the
strength of the effective Hamiltonian is rescaled to improve agreement with
binding energies. However, this procedure may not necessarily improve the
values of polarizabilities.

 The contributions from the dominant
transitions may be separated and replaced by more accurate experimental matrix elements when appropriate. Such a
procedure is discussed in detail in Ref.~\cite{porsev08a}. This hybrid RCI+MBPT approach
\cite{porsev06a,porsev06b,hachisu08a} has been used to obtain present recommended values for the polarizabilities of
the $ns^2$ and $nsnp~^3P_0$ states of Mg, Ca, Sr, Hg, and Yb needed to evaluate the blackbody radiation shifts of the
relevant optical frequency standards.

\subsection{Combined CI and all-order method}
\label{ci+all}

The RCI+MBPT approach described in the previous section includes only a limited
number of the core-valence excitation terms (mostly in second order) and
deteriorates in accuracy for heavier, more complicated systems.
 The linearized coupled-cluster approach described in Section~\ref{all-order-section} is designed to treat core-core and core-valence
correlations with high accuracy. As noted above, it is restricted in its present form to the calculation of
properties of monovalent systems. Direct extension of this method to even divalent systems faces two major problems.

First, use of the Rayleigh-Schr\"{o}dinger RMBPT for heavy systems with more than one valence electron leads to a
non-symmetric effective Hamiltonian and to the problem of ``intruder states'' \cite{lindroth04}. Second, the complexity
of the all-order formalism for matrix elements increases rapidly with the number of valence electrons.  The direct
extensions of the all-order approach to more complicated systems is impractical. For example, the expression for
all-order matrix elements in divalent systems contains several hundred terms instead of the twenty terms in the
corresponding monovalent expression. However, combining  the linearized coupled-cluster approach (also referred to as
the all-order method) with CI method eliminates many of these difficulties. This method (referred to as CI+all-order)
was developed in Ref.~\cite{safronova09c} and tested on the calculation of energy levels of  Mg, Ca, Sr, Zn, Cd, Ba,
and Hg. The prefix R is used to indicate the use of the relativistic Hamiltonian.

In the RCI+all-order approach, the effective Hamiltonian is constructed using fully converged all-order excitations
coefficients $\rho_{ma}$, $\rho_{mnab}$, $\rho_{mv}$, $\rho_{mnva}$, and $\rho_{mnvw}$ (see section
\ref{all-order-section} for designations). The  $\rho_{mnvw}$ coefficients do not arise in the monovalent all-order
method, but are straightforwardly obtained from the above core and core-valence coefficients.
 As a result,
the core-core and core-valence sectors of the correlation corrections for systems with few valence electrons are
treated with the same accuracy as in the all-order approach for the monovalent systems. The CI method is used to treat
valence-valence correlations and to evaluate matrix elements and polarizabilities.

The RCI+all-order method employs a  variant of the Brillouin-Wigner many-body perturbation theory, rather than
Rayleigh-Schr\"{o}dinger perturbation theory. In the Brillouin-Wigner variant of MBPT, the effective Hamiltonian is
symmetric and accidentally small denominators do not arise ~\cite{safronova09c}. Comparisons of the RCI+MBPT and
RCI+all-order binding energies for the ground and excited states of a number of two-electron systems reveal that the
RCI+all-order energies are usually more accurate by at least a factor of three \cite{safronova09c}.

The preliminary calculations of polarizabilities values in Ca and Sr  indicate better agreement of the RCI+all-order
\textit{ab initio} results with recommended values from Ref.~\cite{porsev06b} in comparison with the RCI+MBPT approach.

\begin{table*}[t]
\caption[]{Ground state polarizabilities $\alpha_0$ (in atomic units) of noble gases and isoelectronic ions.
 Uncertainties in the last digits are given in parentheses. References are given
in square brackets. Method abbreviations: DC - dielectric constant, RI - refractive index, SA - spectral analysis, RRPA
- relativistic random-phase approximation, (R)CCSDT - (relativistic) coupled-cluster calculations. The RCCR12
calculation is a CCSDT calculation
 which allows for explicitly correlated electron pairs.
 $^a$See text for further discussion of He polarizability calculations, $^b$Finite mass Hylleraas alculation incorporating relativistic effects from an RCI calculation as an additive correction,$^c$PNO-CEPA (pseudo-natural orbital coupled
electron pair approximation). } \label{tab:3}
\begin{ruledtabular}
\begin{tabular}{lllllll}
  \multicolumn{1}{l}{\textbf{He}} & \multicolumn{1}{l}{\textbf{Ne}} & \multicolumn{1}{l}{\textbf{Ar}} &
   \multicolumn{1}{l}{\textbf{Kr}} & \multicolumn{1}{l}{\textbf{Xe}} &
   \multicolumn{1}{l}{\textbf{Rn}}&\multicolumn{1}{l}{Method [Ref.]}
   \\ \hline
  1.322&2.38 &10.77 & 16.47 & 26.97 & & Th. RRPA \cite{johnson83a}\\
  1.383763 & 2.6648  & 11.084 & & & & Th.  CCSDT \cite{soldan01a} \\
     & 2.697  & 11.22 & 16.80 & 27.06 & 33.18  &Th.   RCCSDT \cite{nakajima01a}\\
     & 2.665 \cite{hald03a}&   11.085(6) \cite{lupinetti05a} & & & & Th. CCSDT  \\
     & 2.6557 &   11.062  & 17.214 & 28.223 & & Th. MBPT \cite{thakkar92a}  \\
     & 2.668(6) \cite{franke01a}&  & & & & Th. RCCR12  \\
 1.38376079(23)$^{a,b}$ \cite{lach04a}   &                              &                            && & & Th.  \\
   1.383223(67) \cite{gugan80a,gugan80b}&2.670(3)  \cite{orcutt67a}&11.081(5) \cite{orcutt67a} &16.766(8) \cite{orcutt67a}&&&Expt. DC \\
    1.3838 &2.6680 & 11.091 & 16.740 & 27.340 & & Expt. RI \cite{langhoff69a}\\
   1.384  & 2.663 & 11.080  & 16.734 &27.292 && Expt. RI \cite{dalgarno60a} \\
 1.383759(13) \cite{schmidt07a} & & 11.083(2) \cite{newell65a}&&&&Expt. RI  \\
  \hline
  \multicolumn{1}{l}{\textbf{Li$^+$}} & \multicolumn{1}{l}{\textbf{Na$^+$}} & \multicolumn{1}{l}{\textbf{K$^+$}} &
   \multicolumn{1}{l}{\textbf{Rb$^+$}} & \multicolumn{1}{l}{\textbf{Cs$^+$}} &
   \multicolumn{1}{l}{\textbf{Fr$^+$}}&\multicolumn{1}{l}{}
   \\ \hline
   0.192486$^b$ \cite{bhatia97a,johnson96b}& 0.9947$^c$  \cite{muller84} & 5.354$^c$ \cite{muller84} & & & & Th.\\
  0.1894&0.9457&5.457& 9.076&15.81&& Th.   RRPA \cite{johnson83a}\\
       & 1.00(4) & 5.52(4) &  9.11(4) & 15.8(1) & 20.4(2) &  Th.  RCCSDT \cite{lim02a}\\
    0.1883(20) \cite{cooke77a}&0.978(10) \cite{opik67a}     &5.47(5) \cite{opik67a} &9.0
   \cite{johansson60a}&15.544(30) \cite{safinya80a}&&Expt.  SA\\
                            &1.0015(15) \cite{freeman76a} &&
   &15.759 \cite{curtis81b}&& Expt. SA \\
                             & 0.9980(33) \cite{gray88a}   &&
   &15.644(5) \cite{zhou89a,weber87a}&& Expt. SA\\
\hline
  \multicolumn{1}{l}{\textbf{Be$^{2+}$}} & \multicolumn{1}{l}{\textbf{Mg$^{2+}$}} & \multicolumn{1}{l}{\textbf{Ca$^{2+}$}} &
   \multicolumn{1}{l}{\textbf{Sr$^{2+}$}} & \multicolumn{1}{l}{\textbf{Ba$^{2+}$}} &
   \multicolumn{1}{l}{\textbf{Ra$^{2+}$}}&\multicolumn{1}{l}{}
   \\ \hline
   0.05182&0.4698&3.254&  5.813& 10.61&& Th. RRPA \cite{johnson83a} \\
   0.052264$^b$ \cite{bhatia97a,johnson96b} & 0.4814$^c$ \cite{muller84} & 3.161$^c$ \cite{muller84}  & & & & Th. \\
            &      &  3.262 &  5.792   &10.491  &13.361 & Th.  RCCSDT \cite{lim04a} \\
 & 0.489(5) \cite{opik67a}&3.26(3)   \cite{opik67a}&&&&Expt.  SA\\
 & 0.486(7) \cite{bockasten56a}&&&&&Expt. SA\\
\end{tabular}
\end{ruledtabular}
\end{table*}

\begin{table*}[t]
\caption[]{ Ground and $np_j$ excited state polarizabilities (in a.u.) of alkali atoms. Scalar ($\alpha_0$) and tensor
($\alpha_2$) polarizabilities are given for the $np_{3/2}$ states. Static polarizabilities for the $np_{1/2}$ and
$np_{3/2}$ states are the same for the non-relativistic Hylleraas and CICP calculations. Uncertainties in the last
digits are given in parentheses. References are given in square brackets. Method abbreviations: EH - $E\text{-}H$
balance or beam-deflection, sum-rule - hybrid $f$-sum rules with experimental data for primary contribution, SA -
spectral analysis, CI - configuration interaction, CICP - CI calculations with a semi-empirical core potential, MBPT -
many-body perturbation theory, RLCCSDT - linearized CCSD method with partial triple contributions. All values in the
sum-rule row explicitly include a core polarizability. $^a$Non-relativistic Hylleraas calculation for $^{\infty}$Li,
$^b$Hylleraas calculations for $^{7}$Li that includes estimate of relativistic effects, $^c$CI, $^d$Hybrid-RLCCSD data
for the alkali ground states from \cite{derevianko99a} are listed as recommended ``sum-rule'' data, $^e$interferometry,
$^f$interferometry ratio, $^g$cold atom velocity change experiments.} \label{alkali}
\begin{ruledtabular}
\begin{tabular}{llllllll}
 \multicolumn{1}{l}{}& \multicolumn{1}{l}{\textbf{Li}} & \multicolumn{1}{l}{\textbf{Na}} & \multicolumn{1}{l}{\textbf{K}} & \multicolumn{1}{l}{\textbf{Rb}} & \multicolumn{1}{l}{\textbf{Cs}} &
 \multicolumn{1}{l}{\textbf{Fr}} &  \multicolumn{1}{l}{Method}  \\
  \multicolumn{1}{l}{$\alpha_0$} &\multicolumn{1}{l}{$2s$} & \multicolumn{1}{l}{$3s$}& \multicolumn{1}{l}{$4s$} & \multicolumn{1}{l}{$5s$}& \multicolumn{1}{l}{$6s$} & \multicolumn{1}{l}{$7s$}
\\ \hline
      &164.112(1)$^a$ \cite{tang09a}&164.50$^{c}$ \cite{hamonou07a} & & & 398.2(9)$^{**}$ \cite{safronova04b} & & Th.\\
      & 164.11(3)$^b$ \cite{tang10a} &  & & & & & Th. Hyl.\\
 & 164.21 \cite{zhang07a} & 162.8  \cite{mitroy03f}& 290.0 \cite{mitroy03f} & 315.7 \cite{mitroy03f} &&&Th.  CICP\\
 && 165.50 \cite{maroulis04a} & 301.28 \cite{lim99a} &&&& Th. CCSD\\
  & 163.74 \cite{lim99a}&162.9(6) \cite{thakkar05a}&291.12 \cite{lim05a}&316.17 \cite{lim05a}&396.02 \cite{lim05a}&315.23 \cite{lim05a}&Th. RCCSDT \\
 &&163.0 &289.1 &316.4 &401.5  &315.1& Th. RLCCSD \cite{derevianko99a}\\
 &164.08  \cite{johnson08a}&&289.3 \cite{safronova08b} & & 398.4(7) \cite{iskrenova07a}&313.7 \cite{safronova07a}&Th. RLCCSDT \\
 &164(3) &159(3)& 293(6)&319(6)&402(8)&&Expt. EH \cite{molof74a}\\
&164.2(1.1)$^e$ \cite{miffre06a}&162.7(8)$^e$ \cite{ekstrom95a} & 290.8(1.4)$^f$ \cite{holmgren10a} & 318.8(1.4)$^f$ \cite{holmgren10a}&401.0(6)$^g$ \cite{amini03a} &&Expt. \\
  &&162.6(3)&290.2(8)&318.6(6)&399.9(1.9)& 317.8(2.4) &Sum-rule$^d$ \cite{derevianko99a}\\
 \hline
  \multicolumn{1}{l}{$\alpha_0$} &\multicolumn{1}{l}{$2p_{1/2}$} & \multicolumn{1}{l}{$3p_{1/2}$}& \multicolumn{1}{l}{$4p_{1/2}$} & \multicolumn{1}{l}{$5p_{1/2}$}& \multicolumn{1}{l}{$6p_{1/2}$} & \multicolumn{1}{l}{}&
\multicolumn{1}{l}{}\\ \hline
& 126.9458(3) \cite{tang09a} &&&&&&Th. $^a$Hyl.\\
 &126.95 \cite{zhang07a}&360.7  \cite{zhang07c}&615.3 \cite{zhang07c}& 854.4  \cite{zhang07c}&&&Th. CICP  \\
&126.980 \cite{johnson08a}&& 604.1 \cite{safronova08b}&805(31) \cite{arora07b}&1338(54)  \cite{iskrenova07a}&&Th. RLCCSDT \\
&& 359.7 &605 & 807 &&& Th. RCI+MBPT \cite{zhu04a} \\
\hline
  \multicolumn{1}{l}{$\alpha_0$} & \multicolumn{1}{l}{$2p_{3/2}$} & \multicolumn{1}{l}{$3p_{3/2}$}&\multicolumn{1}{l}{$4p_{3/2}$} & \multicolumn{1}{l}{$5p_{3/2}$}& \multicolumn{1}{l}{$6p_{3/2}$} &
  \multicolumn{1}{l}{}&
\\ \hline
 &126.995  \cite{johnson08a}&& 614.1 \cite{safronova08b} & & 1648(58)  \cite{iskrenova07a}&&Th. RLCCSDT \\
 &&361.4 &616 &870&&&Th. RCI+MBPT \cite{zhu04a} \\
\hline
  \multicolumn{1}{l}{$\alpha_2$} &\multicolumn{1}{l}{$2p_{3/2}$} & \multicolumn{1}{l}{$3p_{3/2}$}&
   \multicolumn{1}{l}{$4p_{3/2}$} & \multicolumn{1}{l}{$5p_{3/2}$}& \multicolumn{1}{l}{$6p_{3/2}$} & \multicolumn{1}{l}{}
&\\ \hline
&1.6214(3)  \cite{tang09a}&&&&&&Th. Hyl.\\
 &1.6627 \cite{zhang07a}  & $-$87.89 \cite{zhang07c}&$-$107.9 \cite{zhang07c}&$-$160.5 \cite{zhang07c}&&&Th. CICP \\
&& $-$88.0 &  $-$111 & $-$171   &&&Th. RCI+MBPT \cite{zhu04a} \\
& 1.59 \cite{johnson08a}&& $-$107.9 \cite{safronova08b}&& $-$261(13) \cite{iskrenova07a}&&Th. RLCCSDT \\
 & 1.64(4) \cite{windholz92a}& $-$88.3(4)  \cite{windholz89a} & $-$107(2) \cite{krenn97a}  & $-$163(3)  \cite{krenn97a}&$-$261(8) \cite{hunter88a}&& Expt.\\
 & & $-$113(16) \cite{hannaford79a} & $-$110.9(2.8) \cite{kawamura09a}&& $-$262.4(1.5)
\cite{tanner88a}&&Expt.\\
\end{tabular}
\end{ruledtabular}
\end{table*}

\begin{table*}[t]
\caption[]{ Ground state polarizabilities (in a.u.) of alkali-like ions.
Uncertainties in
 the last digits are given in parentheses. References are
given in square brackets. Method abbreviations: SA - spectral analysis, RESIS - resonant excitation Stark ionization
spectroscopy, $^a$non-relativistic Hylleraas calculation for $^{\infty}$Be$^+$, $^b$Hylleraas calculations for
$^{9}$Be$^+$ that includes estimate of relativistic effects, $^c$$f$-sum rule for valence polarizability with
core-polarzation from \cite{werner76a} added.} \label{alkaliions}
\begin{ruledtabular}
\begin{tabular}{lllllll}
 \multicolumn{1}{l}{\textbf{Be}$^+$} & \multicolumn{1}{l}{\textbf{Mg}$^+$} & \multicolumn{1}{l}{\textbf{Ca}$^+$} &
   \multicolumn{1}{l}{\textbf{Sr}$^+$} & \multicolumn{1}{l}{\textbf{Ba}$^+$} &
  \multicolumn{1}{l}{\textbf{Ra}$^+$}&\multicolumn{1}{l}{Method} \\
\multicolumn{1}{l}{$2s$} & \multicolumn{1}{l}{$3s$}& \multicolumn{1}{l}{$4s$} & \multicolumn{1}{l}{$5s$}& \multicolumn{1}{l}{$6s$} &
     \multicolumn{1}{l}{$7s$}&
\\ \hline
 24.4966(1)$^a$ \cite{tang09b} & & & & & & Th. Hyl. \\
 24.489(4)$^b$ \cite{tang10a} & & & & & & Th. Hyl. \\
 24.495 \cite{wang94a}&35.66 \cite{hamonou07a}&&&&&Th. CI\\
 24.493 \cite{tang09b} &  34.99 \cite{mitroy09a}             & 75.49 \cite{mitroy08b}    & 89.9 \cite{mitroy08c}&&&Th. CICP\\
 & 35.05 \cite{mitroy09a}                  & 76.1(1.1) \cite{arora07a}     &91.3(9) \cite{jiang09a} &124.15 \cite{iskrenova08a}&106.5 \cite{safronova07a}&Th. RLCCSD \\
 &                             & 75.88      & 91.10      & 123.07 & 105.37 &Th. RCCSDT \cite{lim04a} \\
  & 33.80(50) \cite{lyons98a}               & 75.3  \cite{chang83a}&                        & 125.5(1.0) \cite{gallagher82a}&&Expt. SA \\
   & 35.04(3)  \cite{mitroy09a}          &                  &                     &124.30(16) \cite{snow05a} &&Expt. RESIS\\
  &         35.00(5) \cite{snow08a} &                           &                        &123.88(5) \cite{snow07b} &&Expt. RESIS \\
  & 35.10 \cite{theodosiou95a}& 74.11 \cite{theodosiou95a}&&&& $^c$$f$-sum rule\\
\end{tabular}
\end{ruledtabular}
\end{table*}

\begin{table*}[t]
\caption[]{Polarizability differences $\alpha_0(np_J)-\alpha_0(ns)$  (in a.u.) of the alkali atoms derived from Stark
shift measurements. Values are negative when the $np_J$ state polarizability is smaller than the ground state
polarizability.  Stark shifts for the $np_{1/2}$ and $np_{3/2}$ states are the same for the non-relativistic Hylleraas
and CICP methods. Uncertainties in the last digits are given in parentheses. References are given in square brackets.
The experimental values and Hylleraas calculations \cite{tang10a} are those reported for $^7$Li, the CICP and RLCCSDT
values are for $^{\infty}$Li.
 }
\label{starkalkali}
\begin{ruledtabular}
\begin{tabular}{lllllll}
 \multicolumn{1}{l}{\textbf{$^7$Li}} & \multicolumn{1}{l}{\textbf{Na}} & \multicolumn{1}{l}{\textbf{K}} & \multicolumn{1}{l}{\textbf{Rb}} & \multicolumn{1}{l}{\textbf{Cs}} &
   \multicolumn{1}{l}{Method}  \\
  \multicolumn{1}{l}{$2s$-$2p_{1/2}$} & \multicolumn{1}{l}{$3s$-$3p_{1/2}$}
  & \multicolumn{1}{l}{$4s$-$4p_{1/2}$} & \multicolumn{1}{l}{$5s$-$5p_{1/2}$}& \multicolumn{1}{l}{$6s$-$6p_{1/2}$} & \\ \hline
    $-$37.14(3) \cite{tang10a}  &     &       &       &        &      Th. Hylleraas   \\
    $-$37.26 \cite{mitroy03f,zhang07a}  &  197.9 \cite{mitroy03f,zhang07c}  &  325.3 \cite{mitroy03f,zhang07c} &   &   & Th. CICP   \\
   $-$37.104 \cite{johnson08a}    &   196.7  \cite{derevianko99a,zhu04a}  &   314.8  \cite{safronova08b}  &    488(4)$^b$   \cite{arora07b,zhu04a}   &   940(55)$^b$ \cite{iskrenova07a}   &    Th. RLCCSDT   \\
    $-$37.146(17) \cite{hunter91a}  &     &  316.68(4) \cite{miller94a} &   491.52(6) \cite{miller94a}  &  926.08(12) \cite{hunter92a} & Expt.   \\
    $-$37.11(33) \cite{windholz92a}  &   196.86(45) \cite{windholz85a}  &   315(3) \cite{krenn97a} &   &      & Expt.  \\ \hline
  \multicolumn{1}{l}{$2s$-$2p_{3/2}$} & \multicolumn{1}{l}{$3s$-$3p_{3/2}$}
  & \multicolumn{1}{l}{$4s$-$4p_{3/2}$} & \multicolumn{1}{l}{$5s$-$5p_{3/2}$}& \multicolumn{1}{l}{$6s$-$6p_{3/2}$} &  \\ \hline
  $-$37.089 MBPT \cite{johnson08a}   & 198.4  \cite{derevianko99a,zhu04a}  & 324.8 \cite{safronova08b}   &  554   \cite{derevianko99a,zhu04a}    &   1250(59) \cite{iskrenova07a}   &    Th. RLCCSDT   \\
    $-$37.30(42) \cite{windholz92a}  &  198.0(6) \cite{windholz89a}   &   322.3(3.2)  \cite{krenn97a} &    538.5(3.2) \cite{krenn97a}  &   1240.2(2.4) \cite{tanner88a} & Expt.   \\
                     &                             &                               &     &   1264(13) \cite{hunter88a} & Expt.   \\
\end{tabular}
\end{ruledtabular}
\end{table*}

\begin{table}[th]
\caption[]{Excited state scalar $\alpha_0$ and tensor $\alpha_2$ polarizabilities (in a.u.) of monovalent
systems.  All experimental values are derived from Stark shift experiments and the polarizability of the lower
state is added to the Stark shift to get the upper state polarizability. Uncertainties in the last digits are given in parentheses. References are
given in square brackets. $^a$Hylleraas basis functions, $^b$CICP, $^c$RLCCSDT, $^d$CA.  Polarizabilities marked with
an asterisk (*) were not published, but obtained from the matrix elements of \cite{zhang07c}.  } \label{alkalitensor}
\begin{ruledtabular}
\begin{tabular}{lllrr}
  \multicolumn{1}{l}{Atom} & State && \multicolumn{1}{c}{Expt.}  &  \multicolumn{1}{c}{Theory}  \\ \hline
\textbf{Li}&$3d_{3/2}$   &$\alpha_0$&   -15082(60)  \cite{ashby03a}   &       $-$14928$^a$ \cite{tang09a} \\
&  &&    &                                                          $-$15044$^b$ \cite{zhang07a} \\
&                &$\alpha_2$&   11626(68)  \cite{ashby03a}   &  11409$^a$ \cite{tang09a}      \\
                &&        &                                  &  11490$^b$ \cite{zhang07a}      \\
&$3d_{5/2}$   &$\alpha_0$&   -15159(32)  \cite{ashby03a}   &     $-$14928$^a$ \cite{tang09a} \\
&  &&    &                                                          $-$15044$^b$ \cite{zhang07a} \\
&              &$\alpha_2$&   16308(52)  \cite{ashby03a}   &  16298$^a$ \cite{tang09a}      \\
                &&          &                              &  16414$^b$  \cite{zhang07a}      \\[0.5pc]
\textbf{Na}&$5s_{1/2}$& $\alpha_0$&  21000(1200) \cite{harvey75a}  & 21780*$^b$  \cite{zhang07c}   \\
&$4d_{3/2}$   &$\alpha_0$& 624000(7000) \cite{harvey75a}& 633800*$^b$  \cite{zhang07c}\\
&             &$\alpha_2$&   $-$154700(2800)  \cite{harvey75a}  &     $-$148700*$^b$ \cite{zhang07c} \\
&$4d_{5/2}$   &$\alpha_0$& 627000(5000) \cite{harvey75a}& \\
&             &$\alpha_2$&   $-$213800(2000)  \cite{harvey75a}  &     $-$212400*$^b$ \cite{zhang07c}  \\[0.5pc]
\textbf{K}&$5p_{3/2}$    &$\alpha_0$ &                                      &  7118*$^b$ \cite{zhang07c}     \\
  &           &$\alpha_2$ &  $-$1057(161) \cite{schmeider71a}    &  $-$1019*$^b$ \cite{zhang07c}     \\[0.5pc]
\textbf{Ca}$^{+}$&$3d_{5/2}$&$\alpha_0$ &  & 32.73$^b$ \cite{mitroy08b}       \\
       &       &     &                             &32.0(1.1)$^c$ \cite{arora07a} \\
       &$3d_{5/2}$&$\alpha_2$ &  & $-$25.20$^b$ \cite{mitroy08b}       \\
       &       &     &                             &$-$24.5(4)$^c$ \cite{arora07a} \\[0.5pc]
\textbf{Rb}&$6p_{3/2}$   &$\alpha_2$&  $-$2090(80) \cite{khadjavi68a}   &  $-$2040$^d$  \cite{wijngaarden97a}  \\
&$6d_{3/2}$   &$\alpha_2$&  $-$42.2(28)  \cite{hogevorst75a}   & $-$559$^d$  \cite{wijngaarden97a}  \\
&$6d_{5/2}$   &$\alpha_2$&  3780(200)    \cite{hogevorst75a}   &  3450$^d$    \cite{wijngaarden97a}  \\
&$7p_{3/2}$   &$\alpha_2$&  $-$12900(800) \cite{svanberg72a}   &  $-$12500$^d$  \cite{wijngaarden97a}     \\[0.5pc]
\textbf{Sr}$^{+}$&$4d_{5/2}$&$\alpha_0$& &  61.77$^a$ \cite{mitroy08c}  \\
          &         &        & &  62.0(5)$^c$ \cite{jiang09a}    \\
        & $4d_{5/2}$ & $\alpha_2$ & &  $-$47.20$^a$ \cite{mitroy08c}  \\
          &         &        & &  $-$47.7(3)$^c$ \cite{jiang09a}    \\
\end{tabular}
\end{ruledtabular}
\end{table}

\begin{table}[th]
\caption[]{Selected theoretical and experimental ground state polarizabilities $\alpha_0$ (in a.u.) of sodium
atom.  Uncertainties in the last digits are given in parentheses. References are given in square brackets.
 HF - Hartree-Fock, PNO-CEPA - pseudonatural orbital configuration
expansion, CICP - CI calculations with a semi-empirical core potential, RLCCSDT - linearized CCSD method with
additional partial triple contributions included, EH - $E\text{-}H$ balance or beam-deflection.} \label{sodium}
\begin{ruledtabular}
\begin{tabular}{lcc}
Method   &   Year  & Value   \\ \hline
\multicolumn{3}{c}{Theory} \\
HF \cite{yoshimine64a}    &  1964   & 183   \\
HF \cite{muller84}        &  1984   & 189.2   \\
PNO-CEPA \cite{reinsch76a}    & 1976    & 165.02    \\
CICP \cite{maeder79a}        &  1979   & 162.6   \\
CICP \cite{muller84}         &  1984   & 162.4   \\
CICP \cite{mitroy03f}        &  2003   & 162.8      \\
CI  \cite{hamonou07a}        & 2007    & 164.50     \\
RLCCSD \cite{derevianko99a}     & 1999    &  163.0    \\
RCCSDT \cite{lim99a}     & 1999    &  164.89      \\
CCSDT \cite{maroulis01a}  & 2001    &  165.06      \\
RCCSDT \cite{soldan03a}     & 2003    &  166.3      \\
RCCSDT \cite{maroulis04a}  & 2004    &  165.5     \\
CCSDT \cite{thakkar05a}    & 2005    &  162.88(60)   \\
\multicolumn{3}{c}{Experiment}                            \\
$f$-sum \cite{dalgarno59a} &  1959 & 166  \\
EH \cite{hall74a}          &   1974 &  165(11)     \\
EH \cite{molof74a}          &   1974 &  159(3)    \\
Interferometry \cite{ekstrom95a}   &  1995 & 162.7(8)    \\
Hybrid $f$-sum \cite{derevianko99a} &  1999 & 162.6(3)  \\
Interferometry \cite{holmgren10a}   &  2010 & 162.7(1.3)    \\
\end{tabular}
\end{ruledtabular}
\end{table}

\begin{table*}
\caption{\label{cesium} Excited state scalar $\alpha_0$ and tensor $\alpha_2$ polarizabilities (in multiples of $1000$ a.u.) of the Cs atom. Uncertainties in the last digits are given in parentheses. References are given in square
brackets. Experimental values: $^b$derived from the Ref.~\cite{bennett99a} $7s-6s$ Stark shift measurement and the $6s$
result from \cite{amini03a}, $^c$ Ref.~\cite{gunawardena07a}, $^d$Refs.~\cite{wijngaarden94a,wijngaarden:1994},
$^e$Ref.~\cite{khvoshtenko68a}, $^f$Ref. \cite{fredriksson77a}, $^g$Ref.~\cite{wessel87a}, $^h$Ref.~\cite{xia:1997},
$^i$Ref.~\cite{auznish06a},  CA - Coulomb approximation, RLCCSDT - relativistic linearized coupled-cluster method with
single, double, and partial triple excitations.  }
\begin{ruledtabular}
\begin{tabular}{llllllll}
  $\alpha_0$  &         $ 7s$       & $8s$       &   $ 9s$      &  $10s$       & $11s$   & $12s $ &Ref. \\
 \hline
 &      6.238(41)  &  38.27(28) &  153.7(1.0)  &  478(3)      & 1246(8) &2866(30)&Th. RLCCSDT \cite{iskrenova07a}  \\
  &        6.14       &  37.9      &  153         &   475        & 1240  &   2840&Th. CA \cite{wijngaarden:1994}  \\
 &  6.238(6)$^b$& 38.06(25)$^c$ &          &   478.5(1.1)$^d$   &  1245(1)$^d$& 2867(2)$^d$&Expt.  \\ [0.3pc]
 \hline
 $\alpha_0$   &    $7p_{1/2}$ &  $8p_{1/2}$&   $9p_{1/2}$ &  $10p_{1/2}$& &&\\[0.3pc]
 \hline
  &        29.9(7)  &    223(2)  &     1021(7)  &    3499(19) & &&Th. RLCCSDT \cite{iskrenova07a} \\
     &    29.4     &    221     &     1020     &    3490&&& Th. CA \cite{wijngaarden:1994} \\
 &    29.6(6) &            &              &           & &&Expt. \cite{domelunksen83a} \\ [0.3pc]
 \hline
   &    $7p_{3/2}$ &  $8p_{3/2}$&    $9p_{3/2}$&  $10p_{3/2}$&&& \\[0.3pc]
 \hline
 $\alpha_0$&      37.5(8) &     284(3) &     1312(7)  &   4522(19)  &&&Th. RLCCSDT \cite{iskrenova07a}\\
      &     36.9    &     282    &     1310     &   4510&&&Th. CA \cite{wijngaarden:1994} \\
  &   37.9(8) &          &               &                            &     & &Expt. \cite{khvoshtenko68a} \\ [0.3pc]
 \hline
$\alpha_2$&      $-$4.41(17) &     $-$30.6(6) &     $-$135(2)  &    $-$451(5)    &&&Th. RLCCSDT \cite{iskrenova07a}  \\
  &      $-$4.28     &     $-$30.2    &     $-$134     &    $-$449 &&&Th. CA \cite{wijngaarden:1994}  \\
  &   $-$4.43(12)$^e$& $-$30.5(1.2)$^f$ & & &&&Expt.  \\
                       &  $-$4.33(17) & & & &&&Expt. \cite{khadjavi68a}\\
                       &  $-$4.00(8)  & & & &&& Expt. \cite{domelunksen83a}\\   [0.3pc]
         \hline
  &  $5d_{3/2}$ &  $6d_{3/2}$ &  $7d_{3/2}$&   $8d_{3/2}$ &   $9d_{3/2}$ &  $10d_{3/2}$ &\\ [0.3pc]
 \hline
  $\alpha_0$ &  $-$0.352(69) &   $-$5.68(45) &   $-$66.7(1.7)&   $-$369(5)   &  $-$1402(13)   &   $-$4234(32) &Th. RLCCSDT \cite{auznish07a,iskrenova07a} \\
   &  $-$0.418     &   $-$5.32     &   $-$65.2    &    $-$366      &  $-$1400       &   $-$4220&Th. CA \cite{wijngaarden:1994}\\
  &             &             &  $-$60(8)$^g$&              &   $-$1450(120)$^f$& $-$4185(4)$^h$& Expt.\\  [0.3pc]
 \hline
$\alpha_2$  &   0.370(28)&    8.77(36)  &   71.1(1.2)  &    339(4)    &     1189(10)$^i$   &   3416(26)&Th. RLCCSDT \cite{auznish07a,iskrenova07a}  \\
   &   0.380     &    8.62      &   70.4       &    336       &     1190       &    3410  &   Th. CA \cite{wijngaarden:1994}  \\
   &             &              &  74.5(2.0)$^i$&   332(16)$^f$&   1183(35)$^i$& 3401(4)$^h$&  Expt.   \\ [0.3pc]
\hline
  &  $5d_{5/2}$ &  $6d_{5/2}$ &  $7d_{5/2}$&   $8d_{5/2}$ &   $9d_{5/2}$ &  $10d_{5/2}$& \\ [0.2pc] \hline
 $\alpha_0$  &   $-$0.453(70) &   $-$8.37(55) &  $-$88.8(2.0)&    $-$475(5)   &   $-$1777(14)  &   $-$5316(38)  &Th. RLCCSDT \cite{auznish07a,iskrenova07a}\\
  &   $-$0.518    &   $-$7.95     &  $-$87.1     &    $-$472      &   $-$1770      &   $-$5300&Th. CA \cite{wijngaarden:1994} \\
 &             &             &   $-$76(8)$^g$&             & $-$2050(100)$^f$&   $-$5303(8)$^h$ & Expt. \\[0.3pc]
 \hline
  $\alpha_2$ &   0.691(40) &    17.33(50) &     142(2)   &       678(5) &      2386(13)  &    6869(34) &Th. RLCCSDT \cite{auznish07a,iskrenova07a}\\
   &  0.704      &    17.00     &     140      &       675    &      2380  &   6850 & Th. CA \cite{wijngaarden:1994}\\
    &            &              &   129(4)$^g$&    731(40)$^f$&     2650(140)$^f$& 6815(20)$^h$& Expt.\\
       &             &              &                &              &                 & 7110(360)          &Expt. \cite{fredriksson77a}  \\
\end{tabular}
\end{ruledtabular}
\end{table*}

\begin{table*}[t]
\caption[]{ Ground  and excited ($nsnp~^3P_0$) state scalar polarizabilities $\alpha_0$ (in a.u.) of group II
atoms and divalent ions. Uncertainties in the last digits are given in parentheses. References are given in square
brackets.
 The abbreviations conform to those used in Table \ref{tab:3} and \ref{alkali}. Hybrid-RCI+MBPT
include experimental data for some transitions.  $^a$RCCSDT, $^b$MBPT, $^c$Hybrid-RCI+MBPT data for the
alkaline-earth ground states from \cite{porsev06a} are listed as recommended ``sum-rule'' data, $^d$RCI+MBPT. }
\label{alkaline}
\begin{ruledtabular}
\begin{tabular}{lllllll}

  \multicolumn{1}{l}{\textbf{Be}} & \multicolumn{1}{l}{\textbf{Mg}} & \multicolumn{1}{l}{\textbf{Ca}} &
   \multicolumn{1}{l}{\textbf{Sr}} & \multicolumn{1}{l}{\textbf{Ba}} & \multicolumn{1}{l}{\textbf{Ra}}& \multicolumn{1}{l}{Method}
   \\
 \multicolumn{1}{l}{$2s^{2}$} & \multicolumn{1}{l}{$3s^{2}$}&
 \multicolumn{1}{l}{$4s^{2}$} & \multicolumn{1}{l}{$5s^{2}$}& \multicolumn{1}{l}{$6s^{2}$} & \multicolumn{1}{l}{$7s^{2}$}
\\ \hline
   37.755 \cite{komasa02a}&  &  &  &   &&Th. ECG \\
37.73(5)$^a$ \cite{tunega97a} & 71.7$^b$ \cite{archibong91a} & 157$^b$ \cite{archibong91a} &&&&Th.\\
 37.807  \cite{bendazzoli04a}& 70.90 \cite{hamonou08a}&171.7 \cite{glass87a} & &&&Th. CI\\[0.2pc]
 37.29 \cite{muller84} & 70.74 \cite{muller84}& 156.0 \cite{muller84}&&&& Th. CICP\\
 37.69 \cite{mitroy03f} & 71.35 \cite{mitroy03f}& 159.4 \cite{mitroy03f}& 201.2  \cite{mitroy03f}&&& Th. CICP\\[0.2pc]
&&158.00 \cite{lim04a} & 198.85 \cite{lim04a} & 273.9 \cite{lim04a} &248.56 \cite{lim04a}&Th. RCCSDT \\
 &&152 \cite{sadlej91a} & 190 \cite{sadlej91a}& 275.5 \cite{schafer07a} &&Th. RCCSDT \\[0.2pc]
37.76  \cite{porsev06a} & 71.33 \cite{porsev06a} & 159.0 \cite{porsev06a} & 202.0  \cite{porsev06a} & 272.1 \cite{porsev06a} &&Th. RCI+MBPT\\
&    &169(17) \cite{miller76a}& 186(15) \cite{schwartz74a} & 268(22) \cite{schwartz74a}&&Expt. EH\\
 & 74.9(2.7) \cite{reshetnikov08a}&  157.1(1.3) \cite{porsev06a}&197.2(2) \cite{porsev06a}& 273.5(2.0) \cite{porsev06a}&&Sum-rule$^c$\\[0.5pc]
 \hline
\multicolumn{1}{c}{$2s2p \ ^{3}P^o_0$}  & \multicolumn{1}{l}{$3s3p \  ^{3}P^o_0$}&\multicolumn{1}{l}{$4s4p \ ^{3}P^o_0$} & \multicolumn{1}{l}{$5s5p \ ^{3}P^o_0$}& \multicolumn{1}{l}{$6s6p \ ^{3}P^o_0$}&&
\\ \hline
39.02 \cite{mitroy04b} & 101.5 \cite{mitroy07e,mitroy08a}&295.3 \cite{mitroy08g}&&&&Th. CICP \\
&101.2(3)  \cite{porsev06b} & 290.3(1.5) \cite{porsev06b} & 458.3(3.6) \cite{porsev06b}& &&Th. Hybrid-RCI+MBPT \\
 && &457.0   \cite{porsev08a}&$-$13 \cite{kozlov99a}&&Th. RCI+MBPT\\[0.5pc]
 \hline
  \multicolumn{1}{l}{\textbf{Al}$^+$} & \multicolumn{1}{l}{\textbf{Si}$^{2+}$} & \multicolumn{1}{l}{\textbf{Zn}} &
   \multicolumn{1}{l}{\textbf{Cd}} & \multicolumn{1}{l}{\textbf{Hg}} & \multicolumn{1}{l}{\textbf{Yb}}&
   \\
\multicolumn{1}{l}{$3s^{2}$} & \multicolumn{1}{l}{$3s^{2}$}&
 \multicolumn{1}{l}{$4s^{2}$} & \multicolumn{1}{l}{$5s^{2}$}& \multicolumn{1}{l}{$5d^{10}6s^{2}$} & \multicolumn{1}{l}{$4d^{14}6s^{2}$}&
\\ \hline
24.2$^b$ \cite{archibong91a}&&&&&                          &Th.\\
24.14(12) \cite{mitroy09b}& 11.688 \cite{mitroy08k}& 38.12 \cite{ye08a} & 44.63 \cite{ye08a} & 31.32  \cite{ye08a}&& Th. CICP \\
 24.12 CI \cite{hamonou08a}&11.75 CI \cite{hamonou08a} &&&33.6$^d$ \cite{hachisu08a}& 111.3$^d$ \cite{porsev06b}&Th. \\
                           &                           &&&                      & 138.9 \cite{dzuba10a}&Th. RCI+MBPT\\
                           &                           &&&                      & 141(6) \cite{dzuba10a}&Th. Hybrid-RCI+MBPT\\
&& 39.2(8) \cite{goebel96b} & & & 140.4  \cite{thierfelder09a} & Th. RCCSDT \\
&&38.8(8) \cite{goebel96b}&49.65(1.49) \cite{goebel95a}&33.75 \cite{tang08a}&&Expt. RI\\
 &&&  &33.91 \cite{goebel96a}&&Expt. RI\\
 &11.666(4) \cite{komara05a} &&&&&Expt. RESIS\\
 &11.669(9) \cite{mitroy08k} &&&&&Expt. RESIS\\
 24.20(75) \cite{reshetnikov08a}  &&&&&&Sum-rule\\
\hline \multicolumn{1}{l}{$3s3p \ ^{3}P^o_0$} &&
 \multicolumn{1}{l}{$4s4p \ ^{3}P^o_0$} & \multicolumn{1}{l}{$5s5p \ ^{3}P^o_0$}& \multicolumn{1}{l}{$6s6p \ ^{3}P^o_0$} & \multicolumn{1}{l}{$6s6p \ ^{3}P^o_0$}&
\\ \hline
24.62(25) \cite{mitroy09b} &&67.69 \cite{ye08a} & 75.29 \cite{ye08a} & 55.32 \cite{ye08a}&&Th. CICP\\
&&&& 54.6 \cite{hachisu08a}& 315.9 \cite{dzuba10a} &Th. RCI+MBPT\\
&&&&&252(25) \cite{porsev99a} &Th. RCI+MBPT\\
&&&&&266(15) \cite{porsev06b}& Th. RCI+MBPT\\
&&&&&302(14) \cite{dzuba10a}& Th. RCI+MBPT\\
\end{tabular}
\end{ruledtabular}
\end{table*}

\begin{table*}[t]
\caption[]{ Ground state polarizabilities $\alpha_0$ (in a.u.) of other systems. Uncertainties in the last
digits are given in parentheses. References are given in square brackets. The average over magnetic projections is
given for atoms which do not have a spherically symmetric ground state. $^{a}$Third-order MBPT, $^{b}$RRPA, $^c$RCCSD,
$^d$RLCCSD, $^e$CICP, $^f$light deflection, $^{g}$RCI+MBPT, $^{h}$RESIS reanalysis using theoretical estimates of
higher order polarization corrections, $^{i}$spectral analysis, $^j$CI, $^k$CCSDT, $^{l}$f-sum rule, $^{m}$RMBPT
 }.
\label{miscellaneous}
\begin{ruledtabular}
\begin{tabular}{lllllll}
  \multicolumn{1}{l}{\textbf{Cu}} & \multicolumn{1}{l}{\textbf{Ag}} & \multicolumn{1}{l}{\textbf{Au}} &
     \multicolumn{1}{l}{\textbf{Zn}$^+$} & \multicolumn{1}{l}{\textbf{Hg}$^+$} & \multicolumn{1}{l}{\textbf{Yb}$^+$} & \multicolumn{1}{l}{Method} \\
\multicolumn{1}{l}{$4s$} & \multicolumn{1}{l}{$5s$}& \multicolumn{1}{l}{$6s$} & \multicolumn{1}{l}{$4s$}& \multicolumn{1}{l}{$6s$} & \multicolumn{1}{l}{$6s$}&
\\ \hline
 45.0   \cite{schwerdtfeger94a}&52.2 \cite{schwerdtfeger94a}&35.1   \cite{schwerdtfeger94a}& & &62.04$^{a}$ \cite{safronova09a} &Th. \\
 46.50  \cite{neogrady97a} &52.46  \cite{neogrady97a} & 36.06 \cite{neogrady97a} & 18.84 \cite{ilias99a} & 19.36 \cite{ilias99a} &&Th. RCCSDT  \\
  41.65  &46.17  &&&&&Th. CICP \cite{zhang08d} \\
 &&30(4)$^{l}$ \cite{henderson97a}& 15.4(5)$^{l}$ \cite{curtis95a} &&&Expt. \\
\hline
  \multicolumn{1}{l}{\textbf{Al}$^{2+}$} & \multicolumn{1}{l}{\textbf{Si}$^{3+}$}& \multicolumn{1}{l}{\textbf{P}$^{3+}$}& \multicolumn{1}{l}{\textbf{Kr}$^{6+}$}&
  \multicolumn{1}{l}{\textbf{Cu}$^+$} & \multicolumn{1}{l}{\textbf{Ag}$^+$}  & \\
\multicolumn{1}{l}{$3s$} & \multicolumn{1}{l}{$3s$}& \multicolumn{1}{l}{$3s^2$} & \multicolumn{1}{l}{$3d^{10}4s^2$}& \multicolumn{1}{l}{$3d^{10}$} & \multicolumn{1}{l}{$4d^{10}$}&
\\ \hline
  14.44 \cite{hamonou07a} & 7.50 \cite{hamonou07a} & 6.73 \cite{hamonou08a} &&&&Th. CI \\
  &7.399$^e$ \cite{mitroy09a} &&&5.36$^{b}$ \cite{johnson83a} & 8.829$^{b}$ \cite{johnson83a}&Th.\\
 &7.419$^d$ \cite{mitroy09a} && $2.555^{m}$  &6.57$^{c}$ \cite{neogrady96a} & 9.21$^{c}$ \cite{neogrady96a} & Th.  \\
  &7.426(12) \cite{snow07a}&                         &2.69(4) \cite{lundeen07b}&&&Expt.  RESIS \\
 &7.433(25)$^h$ \cite{snow07a,mitroy09a}& 6.312(10)$^i$ \cite{magnusson77a} & & & & Expt.   \\
  \hline
  \multicolumn{1}{l}{\textbf{Al}} & \multicolumn{1}{l}{\textbf{Ga}} & \multicolumn{1}{l}{\textbf{In}} &   \multicolumn{1}{l}{\textbf{Tl}} & &&   \\
\multicolumn{1}{l}{$3s^23p$} & \multicolumn{1}{l}{$4s^24p$}& \multicolumn{1}{l}{$5s^25p$} & \multicolumn{1}{l}{$6s^26p$}& \multicolumn{1}{l}{} & \multicolumn{1}{l}{}&
\\ \hline
 57.74$^k$ \cite{lupinetti05a}&&&49.2$^g$ \cite{kozlov01b} &&&Th.\\
 59.5$^j$  \cite{hibbert80a} &&&          &&&Th.\\
 & 49.9   \cite{fleig05a}&61.9   \cite{fleig05a}&51.6 \cite{fleig05a}&&&Th. RCCSDT \\
46.2(20) \cite{milani90a} &&68.7(8.1) \cite{guella84a} &51.3(5.4)  \cite{guella85a}&&& Expt. EH \\
\hline
  \multicolumn{1}{l}{\textbf{Si}} & \multicolumn{1}{l}{\textbf{Sn}} & \multicolumn{1}{l}{\textbf{Pb}} &
   \multicolumn{1}{l}{\textbf{Ir}} & \multicolumn{1}{l}{\textbf{U}} & &\\
\multicolumn{1}{l}{$3s^23p^2$} & \multicolumn{1}{l}{$5s^25p^2$}& \multicolumn{1}{l}{$6s^26p^2$} & &
\multicolumn{1}{l}{} & \multicolumn{1}{l}{}&
\\
   \hline
 37.0  \cite{hibbert80a}&&&&&&Th. CI\\
 37.17 \cite{lupinetti05a}&&&&&&Th. CCSD\\
37.3 \cite{thierfelder08a}&52.9 \cite{thierfelder08a}&47.3 \cite{thierfelder08a}&&&&Th. RCCSDT \\
   & 42.4(11.0)  \cite{thierfelder08a}&47.1(7.0) \cite{thierfelder08a}&54.0(6.7)  \cite{bardon84a}& 137.0(9.4)$^f$ \cite{kadarkallen94a} &&Expt. EH\\
\end{tabular}
\end{ruledtabular}
\end{table*}

\begin{table}[th]
\caption[]{Excited state scalar $\alpha_0$ and tensor $\alpha_2$  polarizabilities (in a.u.) of selected systems.
Uncertainties in the last digits are given in parentheses. References are given in square brackets. $^a$CICP,
$^b$RCI+MBPT, $^c$CCSDT, $^d$RCCSDT, $^e$one-electron model potential.} \label{excitedalkaline}
 \begin{ruledtabular}
\begin{tabular}{lllrr}
Atom      &State & & \multicolumn{1}{c}{Expt.}    & \multicolumn{1}{c}{Theory}       \\ \hline
\textbf{Ca}&$4s4p \ ^{1}P^o_1$   &$\alpha_0$ &                           & 242.4$^a$ \cite{mitroy08g}      \\
&  &  $\alpha_2$& $-$54.7(1.2) \cite{kreutztrager73a}      &  $-$55.54$^a$    \cite{mitroy08g}   \\
&$4s4p \ ^3P^o_1$    & $\alpha_2$& 12.9(3.2)  \cite{oppen70a}      &       14.2$^a$  \cite{mitroy08g}     \\
                     & &  &  10.54(6) \cite{yanagimachi02a}      &          \\
                     & &  &12.1(8) \cite{zeiske95a,yanagimachi02a}  &            \\
\textbf{Sr} &$5s5p \ ^1P^o_1$   &  $\alpha_2$ & $-$63.1(7.6) \cite{oppen71a}     &          \\
&                    &  $\alpha_2$ & $-$57.55(60) \cite{kreutztrager73a}   &            \\
  &$5s5p \ ^{3}P^o_1$ &$\alpha_0$ &    &  498.8$^b$    \cite{porsev08a}    \\
&   &  $\alpha_2$& $-$24.5(3.2) \cite{oppen69b}   &           \\
\textbf{Ba}&$6s6p \ ^{1}P^o_1$ &$\alpha_0$ &                           &  409$^b$   \cite{kozlov99a}   \\
&  &  $\alpha_2$& $-$43.08(40) \cite{kreutztrager73a}    &       $-$51$^b$   \cite{kozlov99a}    \\
                     & & &  $-$43.4(1.2) \cite{hese77a}    &                     \\
&$6s5d \ ^{1}D_2$   &  $\alpha_2$& 85.2(2.4) \cite{vanleeuwen83a}      &  81$^b$ \cite{kozlov99a}       \\
&$6p^{2} \ ^{3}P_2$   &  $\alpha_2$& $-$109.7(4) \cite{li04b}      &           \\
\textbf{Zn} & $4s4p \ ^3P^o_1$   &  $\alpha_2$ & 7.35(32) \cite{rinkleff79a}  &  6.73$^e$ \cite{robinson69a}    \\
\textbf{Cd} & $5s5p \ ^3P^o_1$   & $\alpha_2$ & 7.11(32) \cite{rinkleff79a}   &  6.30$^e$ \cite{robinson69a}     \\
                    & & & 5.10(24) \cite{legowski95a}    &             \\
                    & & & 5.35(16) \cite{legowski95a}    &             \\
\textbf{Hg}&$6s6p \ ^{3}P^o_1$ & $\alpha_0$&                &  60.6$^{b}$ \cite{hachisu08a}    \\
& &  $\alpha_2$& 6.31(24) \cite{khadjavi68a}    &                \\
                      && &  6.35(8)  \cite{sandle75a}    &                \\
                      &&&  6.34(6)  \cite{kaul72a}    &                \\
\textbf{Al}&$3s^23p \ ^2P^o_{3/2}$ &  $\alpha_0$   &      &   57.74$^{c}$   \cite{lupinetti05a}    \\
                                   &&  $\alpha_2$& $-$8.15(40)  \cite{martin68a}  &   $-$8.53$^{c}$  \cite{lupinetti05a}   \\
\textbf{Tl}& $6s^26p \ ^2P^o_{3/2}$& $\alpha_0$  &     &  81.2$^d$  \cite{fleig05a}\\
&          &              &                 &   79.6$^b$ \cite{kozlov01b}\\
&&  $\alpha_2$& $-$24.2(3)   \cite{peterson68a,gould76a}  & $-$24.56$^d$ \cite{fleig05a}        \\
&&            &                                           & $-$25.0$^b$ \cite{kozlov01b}        \\
\textbf{Yb}&$6s6p \ ^{1}P^o_1$ &$\alpha_0$ &                           &    501(200)$^b$ \cite{porsev99a}     \\
& &  $\alpha_2$& $-$57.4(5.6) \cite{rinkleff80a}              &   $-$118(60) \cite{porsev99a}     \\
&$6s6p \ ^{3}P^o_1$ & $\alpha_0$&                   &    278(15)$^b$ \cite{porsev99a}     \\
&  &  $\alpha_2$&  24.26(84) \cite{kulina82a}    &   24.3(1.5)$^b$ \cite{porsev99a}     \\
                      &&&  23.35(52) \cite{li95b} &             \\
\textbf{Yb}$^+$&$5d \ ^2D_{3/2}$  &  $\alpha_2$& $-$82.5(1.3) \cite{schneider05a}    &     \\
\end{tabular}
\end{ruledtabular}
\end{table}

\section{Benchmark comparisons of theory and experiment}
\label{results}
\subsection{Noble gases and isoelectronic ions}

 Theoretical
\cite{johnson83a,soldan01a,nakajima01a,lupinetti05a,thakkar92a,franke01a,yan96a,pachucki01a,bhatia92a,muller84,lim02a}
and experimental
\cite{gugan80a,gugan80b,orcutt67a,langhoff69a,dalgarno60a,newell65a,cooke77a,opik67a,johansson60a,safinya80a,freeman76a,curtis81b,gray88a,zhou89a,weber87a,bockasten56a}
values for the ground state polarizabilities of the noble gases and isoelectronic ions are listed in
Table~\ref{tab:3}. References are given in square
brackets. The reference is given at the end of the row when all data in this row come from the same work. Otherwise,
the references are listed together with the particular value. The following method abbreviations are used in the table:
DC - dielectric constant, RI - refractive index, SA - spectral analysis, RRPA - relativistic random phase
approximation, MBPT - many-body perturbation theory, (R)CCSDT - (relativistic) coupled-cluster method. If any triple
excitations are included, CCSDT
 abbreviation is used for coupled-cluster calculations, single-double coupled cluster calculations are labelled
 (R)CCSD.  The RCCR12 calculation \cite{franke01a} is a CCSDT calculation
 which allows for explicitly correlated electron pairs.
  The pseudo-natural orbital coupled electron pair approximation (PNO-CEPA) \cite{muller84} can be regarded as precursor of modern CCSD type models.
We first discuss the general trends of values for the noble gases as a whole, and then consider He in more detail separately.

The most precise calculations of the noble gas polarizabilities (apart from helium) have mostly been obtained with
coupled-cluster type calculations. As we noted in the previous sections, particular care has to be taken  to ensure
that the basis set used in CC calculations is of sufficiently high quality to obtain accurate values.
  One curious aspect about the noble gases is their insensitivity to
relativistic effects.  The relativistic correction to $\alpha_0$ is less than 1$\%$ for Ne, Ar, and Kr and is only
about 2$\%$ for Xe \cite{nakajima01a}.

One notable feature of Table~\ref{tab:3} is the good agreement of the RRPA \cite{johnson83a} with the much more
elaborate coupled-cluster and Hylleraas basis function calculations and experimental data. The difference between RRPA
values and other calculations/experimental value for neutral systems ranges from $10\%$ for Ne to $1.6\%$ for Kr (4\%
for He). The RRPA values \cite{johnson83a} improve significantly for the singly ionized systems and differ from other
values by $5\%$ for Na$^+$ and only $0.4\%$ for Rb$^+$.  The discrepancies are reduced further for doubly ionized
systems owing to the decrease in the relative contribution of the correlation corrections beyond RRPA.  Core
polarizabilities for the alkali and alkaline-earth atoms are important for the construction of CICP type models of
these atoms.  In addition, the RRPA calculations of the core polarizabilities are embedded into many calculations of
the polarizabilities of alkali and alkaline-earth ions (see, for example,
Refs.~\cite{derevianko99a,porsev06a,porsev06b}).

\subsubsection{Helium}

The helium atom is of particular interest since it allows for the most precise calculations and benchmark tests of
theory and experiment. Within the framework of the non-relativistic Schr{\"o}dinger equation with infinite-nuclear-mass
Hamiltonian, the He polarizability value obtained using a modified version of the generalized Hylleraas basis set
\cite{korobov01a} is 1.383192174455(1) a.u. \cite{pachucki01a}, achieving accuracy of 13 significant digits. This
value is in agreement with 1996 calculation of \cite{yan96a}.

The  finite mass effects increase the polarizability by about 0.00062~a.u., with the mass polarization effect
accounting for 0.000049~a.u. resulting in the $^4$He nonrelativistic value of
1.38380999 a.u. \cite{pachucki01a,cencek01a,lach04a}. The $\alpha^2$ relativistic corrections contribute
-0.00008035(2)~a.u. \cite{pachucki01a,cencek01a,lach04a}. The $\alpha^3$ QED corrections  with exception of the terms
containing electric-field derivative of the Bethe logarithm were calculated in Ref.~\cite{pachucki01a} to give
0.0000305~a.u.. These latter terms were calculated in \cite{lach04a}, together with the estimates of the $\alpha^4$,
$\alpha^2m_e/M_{\textrm{He}}$, and $\alpha^3m_e/M_{\textrm{He}}$, yielding the final value of $^4$He polarizability of
1.38376079(23)~a.u. listed in Table~\ref{tab:3}.

A non-relativistic coupled-cluster calculation in the infinite mass limit carried out in Ref.~\cite{soldan01a} provides
a detailed study of the dependence of the CCSDT results on the choice of the basis set and tests of basis set
convergence. The values obtained with different uncontracted, even-tempered basis sets
varied in the fifth significant
digit. Their final value of $\alpha(^{\infty}\textrm{He})=1.383763$~a.u. differs from the exact non-relativistic
Hylleraas value of 1.383192 a.u. \cite{yan96a,pachucki01a} at the same level.

A microwave cavity was recently used to measure the refractive index of helium
giving a polarizability of 1.383759(13) a.u. \cite{schmidt07a}.  The best
experiment has an uncertainty of about 10 ppm and is in accord with the most
accurate theory value \cite{lach04a}. Availability of such precise theoretical
and experimental values of He polarizability allows for accurate determinations
of the thermodynamic temperature and may lead to a more accurate value of the
Boltzmann constant \cite{schmidt07a}. This application is discussed in more
detail
 in Section~\ref{thermometry}.

\subsection{Monovalent systems}

The theoretical
\cite{yan96a,hamonou07a,zhang07a,maroulis04a,lim99a,thakkar05a,lim04a,lim05a,derevianko99a,johnson08a,iskrenova07a,iskrenova08a,safronova04b,safronova07a,safronova08b,arora07a,arora07b,jiang09a,pipin93a,zhu04a,tang09b,wang94a,mitroy03f,mitroy09a,mitroy08b,mitroy08c}
and experimental
\cite{molof74a,miffre06a,ekstrom95a,holmgren10a,kadarkallen92a,amini03a,windholz92a,krenn97a,hunter88a,hannaford79a,kawamura09a,tanner88a,lyons98a,vaidyanathan82a,chang83a,gallagher82a,snow05a,snow07b,snow08a}
values of static scalar ($\alpha_0$) and tensor ($\alpha_2$) polarizabilities
of alkali atoms and scalar static polarizabilities of singly ionized monovalent
ions  are compared  in Table~\ref{alkali} and \ref{alkaliions}. The same
designations are used as in the noble gas table. The following additional
method abbreviations are used: EH - $E\text{-}H$ balance or beam-deflection,
sum-rule - hybrid $f$-sum rules with experimental data for primary
contribution, RESIS - resonant excitation Stark ionization spectroscopy,
RLCCSDT - linearized CCSD method with partial triple contributions included.
First, some general remarks are made for monovalent systems, and then Li, Na,
Mg$^+$ and Cs are considered in more detail.

The comparatively simple electronic structure of these atoms render them amenable to accurate calculation by the
coupled-cluster and CICP methods. The sum-rule polarizabilities \cite{derevianko99a} come from a hybrid calculation
that use the RLCCSD calculation as a template. However, the matrix element for the resonance transition has been
replaced by high accuracy experimental matrix elements compiled in \cite{safronova99a}.  The \textit{ab initio} RLCCSD
values are in excellent agreement (better than 1$\%$) with these hybrid recommended values. The semi-empirical CICP
calculations reveal a similar level of accuracy, although there has been some degradation in accuracy for the heavier
Rb system. The  CI calculations with a semi-empirical core potential (CICP) are in excellent agreement with RLCCSDT
calculations and experiment for lighter systems.  The non-relativistic CICP cannot be expected to be particularly
accurate for states with significant spin-orbit splitting, e.g. the $np_J$ states of Rb. The best that can be expected
is that the CICP calculation will do a reasonable job of reproducing the statistically weighted $np_J$ average
polarizability.

The results of the coupled-cluster calculations can be sensitive to particular
contributions that are included, owing to cancellations of various terms (for
example, some triple excitations beyond perturbative treatment may partially
cancel with non-linear single-double terms), leading to some differences
between different coupled-cluster calculations \cite{derevianko-triples}. The
properties involving $nd$ states (i.e. $np$ polarizabilities) are also
sensitive to the number of partial of waves included in the basis sets.
Omission or inadequate inclusion of partial waves with $l>3$ may lead to poor
results for matrix elements involving $nd$ states, and, subsequently, relevant
excited-state polarizabilities.

Some of the  most stringent tests of polarizability calculations of monovalent systems come from  Stark shift
measurements of alkali resonance transitions. Therefore, it is  useful to compare the experimental values for the
polarizability difference obtained from Stark shift measurement directly with theoretical predictions in these cases.
Scalar polarizability differences $\alpha_0(np_J)-\alpha_0(ns)$ (in a.u.) of the alkali transitions derived from Stark
shift measurements are compared with theoretical values in Table~\ref{starkalkali}
\cite{tang10a,mitroy03f,zhang07a,johnson08a,derevianko99a,zhu04a,safronova08b,arora07b,zhu04a,iskrenova07a,hunter91a,miller94a,hunter92a,windholz92a,windholz85a,krenn97a,windholz89a,tanner88a,hunter88a}.
 For the elements heavier that Li, the finite mass effects are smaller than the uncertainty of the calculation.

The tensor polarizability of an open shell atom can be extracted from the difference in polarizabilities between the
different magnetic sub-levels. The scalar and tensor polarizabilities
\cite{ashby03a,tang09a,zhang07a,harvey75a,schmeider71a,arora07a,khadjavi68a,hogevorst75a,hogevorst75a,svanberg72a,mitroy08c,jiang09a}
of some low lying excited states of Li, Na, K, Rb, Ca$^+$ and Sr$^+$  are listed in Table~\ref{alkalitensor}.  There is
a paucity of experimental data for excited states, even for well-studied alkali atoms. The polarizabilities of the
$nd_{5/2}$ states of Ca$^+$ and Sr$^+$ are given owing to their importance for evaluation of the black-body radiation
shifts. Some older and less accurate Stark shifts and tensor polarizabilities are omitted from these Tables.

\subsubsection{Lithium}

The lithium polarizability could assume a pivotal role in polarizability metrology if a multi-species interferometer
can be constructed that is capable of measuring the ratio of the polarizability of other atoms to that of Li to a
relative accuracy of $10^{-4}$  \cite{cronin09a}.  In this case, a measurement of such ratios  in conjunction with
a definitive calculation of the Li $\alpha_0$ could lead to new accuracy benchmarks for the polarizabilities of a number of
 elements.

Correlated basis calculations are possible for  lithium since it only has three electrons.  Consequently it has been
possible to calculate the polarizability to very high precision \cite{yan96a,tang09a}.  The uncertainty in the
experimental value of the polarizability 164.2(11) a.u. \cite{miffre06a} spans all of the theoretical results reported
in Table~\ref{alkali}.

The most recent Hylleraas calculation gave $\alpha_0 = 164.112(1)$ a.u. for
$^{\infty}$Li \cite{tang09a}. Including finite mass effects gave $\alpha_0 =
164.161(1)$ a.u. for $^7$Li.  An approximate treatment of relativistic effects
gave a recommended value of 164.11(3) a.u. \cite{tang10a}. Hylleraas
polarizabilities could also serve as benchmarks for coupled-cluster type
calculations which can be applied to atoms heavier than lithium.

The most stringent test of Li polarizability calculations is presently the Stark shift measurement of the
$2s$-$2p_{1/2}$ transitions  by Hunter {\em et al.} \cite{hunter91a}, which gave a polarizability difference of
$-$37.14(2) a.u.  The current theoretical benchmark is the recent Hylleraas calculations that include finite mass and
relativistic effects \cite{tang09a,tang10a}. The $^7$Li Hylleraas polarizability difference of $-$37.14(4) a.u.
\cite{tang10a} is in excellent agreement with the experimental polarizability difference \cite{hunter91a}. The RLCCSDT
value of $-$37.104 is within 2 standard deviations of the Hunter experiment while the CICP value is 4 standard
deviations too large.  Table~\ref{starkalkali} shows that the Stark shift data offer the most precise information with
which to discriminate between various theoretical calculations.

\begin{figure} [th]
\includegraphics[width=8.40cm,angle=0]{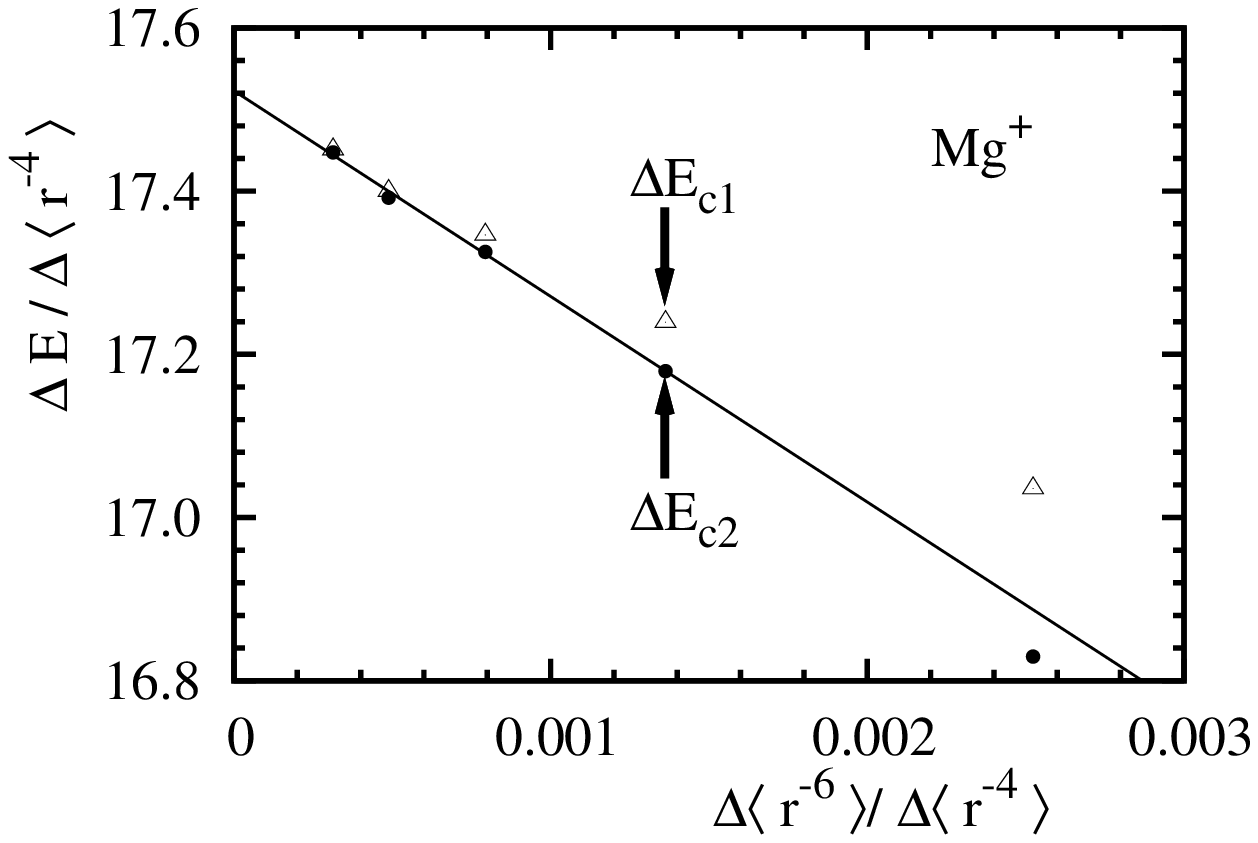}
\vspace{0.02cm} \caption[]{ \label{mg+} The polarization plot of the fine-structure intervals of Mg for the $n = 17$
Rydberg levels.  The $\Delta E_{c1}$ intervals are corrected for relativistic, second-order and Stark shifts.  The
$\Delta E_{c2}$ intervals account for $\langle r^{-7} \rangle$ and $\langle r^{-8} \rangle$ shifts. The linear
regression for the $\Delta E_{\rm c2}$ plot did not include the last point. }
\end{figure}

\subsubsection{Sodium}

A chronological list detailing selected values
\cite{yoshimine64a,muller84,reinsch76a,maeder79a,mitroy03f,hamonou07a,derevianko99a,lim99a,maroulis01a,soldan03a,maroulis04a,thakkar05a,dalgarno59a,hall74a,molof74a,ekstrom95a,holmgren10a}
of the sodium ground state polarizability is presented in Table \ref{sodium}.  The theory values are also sorted by the
type of calculation. The $3s \to 3p$ resonant transition accounts for 98.8$\%$ of the polarizability.

 The most notable
feature of this table is the excellent agreement of the semi-empirical CICP type calculations with the recent
high-precision experimental values of 162.6(3) a.u. \cite{derevianko99a} and 162.7(8) a.u. \cite{ekstrom95a}.  All
three calculations \cite{maeder79a,muller84,mitroy03f}, performed over a period of three decades lie within the
experimental uncertainties.

The coupled-cluster calculations, with the exception of the RLCCSD one \cite{derevianko99a}, tended to give
polarizabilities which were $1 - 2 \%$ larger than experiment until the most recent RCCSDT calculation of Thakkar and
Lupinetti \cite{thakkar05a} which gave 162.9(6) a.u.. The earlier CCSDT calculations tend to overestimate the
polarizability most likely due to basis set issues \cite{lim99a,maroulis01a,soldan03a,maroulis04a}. The same problem
could also be leading to the overestimation of the polarizability by the CI \cite{hamonou07a} and CEPA-PNO
\cite{reinsch76a} calculations.

By way of contrast, the RLCCSD calculation \cite{derevianko99a} gave a polarizability of 163.0 a.u. which is in
agreement with experiment.  We have discussed the differences of the RLCCSD approach from the other coupled-cluster
calculation in Section~\ref{all-order-section}. An important feature here is that this calculation uses a $B$-spline
basis which is effectively complete \cite{derevianko99a,safronova08a}.  As we have discussed on the example of He CCSD
calculation \cite{soldan01a}, polarizability coupled-cluster results vary significantly with the choice of the basis
set if it is not sufficiently saturated. In summary, large (effectively complete) basis sets are needed for precision
polarizability calculations by a coupled-cluster method.

The relativistic correction to the dipole polarizability is about $-1.0$ a.u. \cite{thakkar05a}.  The three
non-relativistic CICP calculations all lie within 0.5$\%$ of the experimental polarizability. As mentioned earlier,
these calculations implicitly include relativistic effects to some extent by tuning the core polarization potential to
the experimental binding energies. The RLCCSD calculation uses a relativistic Hamiltonian and intrinsically includes
relativistic corrections. The recommended value 162.6(3) is based on the RLCCSD calculation with resonant $3s-3p_j$
transition matrix elements replaced by their experimental values.

The Na  polarizability of 162.7(8) \cite{ekstrom95a} obtained by interferometry experiment  served as the reference
polarizability in the determination of the K and Rb polarizabilities by the interferometry ratio approach
\cite{holmgren10a}. Table~\ref{alkali} shows excellent agreement of these values with the hybrid RLCCSD f-sum
polarizabilities of \cite{derevianko99a}.

\subsubsection{Mg$^+$}

We use Mg$^+$ to illustrate the  RESIS experimental approach  owing to recent advances in that area. Both the potential
and the problems of determining the polarizabilities of ions using spectral analysis are evident by contrasting the
different values listed for Mg$^+$ and Ba$^+$. The original analysis of the RESIS data for Mg$^+$ reported a dipole
polarizability of 35.00(5) a.u. \cite{snow08a}. However, the contributions from the $C_7$ and $C_8$ terms of
Eq.~(\ref{DeltaE1}) can possibly corrupt the value of $\alpha_0$ if they are significant as  described in
Section~\ref{resis}. A more detailed analysis of the polarization plot  which explicitly included the $C_7$ and $C_8$
terms was subsequently performed in Ref.~\cite{mitroy09a}.
 This polarization plot is shown in Figure \ref{mg+}.  The data points including the
explicit subtraction of the $C_7$ and $C_8$ terms show a higher degree of linearity.  The revised analysis resulted in
$\alpha_0 = 35.05(3)$ a.u..  This is only 0.15$\%$ larger than the original value and lies within the original error
limits.

The treatment of non-adiabatic corrections is a much more serious issue for the
Ba$^+$ ground state.  Table \ref{alkali} shows that subsequent analysis of the
RESIS data \cite{snow05a,snow07b} do not  lie within their mutual
uncertainties. The most recent analysis of RESIS data gave a polarizability of
123.88(5) a.u. \cite{snow07b}. This analysis explicitly included non-adiabatic
effects from the low-lying $5d$ excitation.  However, non-adiabatic effects
from the $6s$-$6p$ excitation are also significant and need to be included for
a RESIS polarizability to be regarded as definitive.

The influence of the non-adiabatic effects in the $C_7$ and $C_8$ terms of Eq.~({\ref{DeltaE1}) can be minimized by
taking measurements at high values of $L$, e.g. $L \ge 8$.  Unfortunately, as the non-adiabatic corrections diminish
with increasing  $L$, the states with very high $L$ are more sensitive to Stark shifts due to stray electric fields. As
the energy splitting of the Rydberg states gets smaller at higher $L$, the polarizabilities of the $(n,L)$ levels then
get larger due to the very small $(n,L - n,L\pm1)$ energy differences.  To a certain extent one has to choose the
($n$,$L$) states to navigate between the low-$L$ Scylla \cite{homerBC} of non-adiabatic corrections and the high-$L$
Charybdis \cite{homerBC} of Stark shifts \cite{mitroy09a}.

\subsubsection{Cesium}

The Cs atom has been studied extensively owing to the parity-violation experiments on this system \cite{Cs-PNC}.
 A
comprehensive set of Cs scalar and tensor polarizabilities for the $7s-12s$, $7p_{1/2} - 10p_{1/2}$, $7p_{3/2} -
10p_{3/2}$, $5d_{3/2} - 10d_{3/2}$, and $5d_{5/2} - 10d_{5/2}$ states
\cite{bennett99a,amini03a,gunawardena07a,wijngaarden94a,wijngaarden:1994,khvoshtenko68a,fredriksson77a,wessel87a,xia:1997,auznish06a,iskrenova07a,safronova99a,domelunksen83a,khadjavi68a,auznish07a}
taken from Ref.~\cite{iskrenova07a} is given in Table~\ref{cesium}.

The polarizabilities listed in Table~\ref{cesium} are in $10^3$~a.u. since the values range in
size from 300 a.u. to $ 7\times 10^6$ a.u.
 The results of Ref.~\cite{iskrenova07a} are obtained from sum-over-state calculation using the
RLCCSDT matrix elements and experimental energies for a large number of states. The remaining contributions from
highly-excited states were evaluated as well. In a few cases, some of the RLCCSDT matrix elements have been replaced
with matrix elements extracted from experiment \cite{iskrenova07a}. Incorporating such highly-excited states as $12s$
required the use of a very  large $R=220 $~a.u. spherical cavity and large  B-spline basis sets. Extensive tests of
numerical stability of the calculations in such a large cavity have been conducted to verify the accuracy of a finite
basis set representation. All matrix elements used to evaluate dominant polarizability contributions were critically
evaluated for their accuracy based of the size and type of the dominant correlation corrections and semi-empirical
estimates of the omitted correlation terms. Such uncertainty evaluation is discussed in more detail in
Section~\ref{uncertainty}.

  Coulomb approximation (CA)
values \cite{wijngaarden:1994} were also computed with a sum-over-states approach. One interesting feature of
Table~\ref{cesium} is the reasonable level of agreement between the CA and RLCCSDT values for many of the
polarizabilities.  The CA results are computed with wave functions which are tuned to experimental energies.  The
radial matrix elements that arise in the sum-over-states calculation are dominated by the form of the wave function at
large distances. Tuning the wave functions to have the correct energy goes a long way to ensuring that the long-range
part of the wave function has the correct shape.

A number of the  experimental values in Table ~\ref{cesium} were obtained from Stark shift experiments.  In many cases,
the excited state polarizabilities are much larger than the Cs ground state polarizability, so uncertainties in the
ground state have minimal impact on the overall uncertainty. The agreement between the experimental and RLCCSDT
polarizabilities is excellent for the $ns$ states, in most cases the difference between them is
 less than 1$\%$.
The situation is not so clear-cut for the $nd$ states.  Differences between theory and experiment are large in some
cases, but so are the uncertainties of many of the experimental values. However, the RLCCSDT results were found in good
agreement with more recent  experiments \cite{xia:1997,auznish06a,auznish07a}. The RLCCSDT calculation
\cite{iskrenova07a} provided critically evaluated recommended values for a large number of Cs polarizabilities for
which accurate experimental data are not available.

\begin{table}[th]
\caption[]{Static polarizability differences (in a.u.) derived from selected Stark shift measurements. Uncertainties in
the last digits are given in parentheses. References are given in square brackets. $^a$RMBPT, $^b$RLCCSDT, $^c$CICP,
$^d$RCI+MBPT.} \label{starkother}
\begin{ruledtabular}
\begin{tabular}{llrr}
  \multicolumn{1}{l}{Atom} & \multicolumn{1}{l}{State} &  \multicolumn{1}{c}{Experiment}  &  \multicolumn{1}{c}{Theory} \\ \hline
\textbf{Cs} &  $6s$$-$$7s$ &          5837(6) \cite{bennett99a}    & 5834$^a$ \cite{kozlov01a}        \\
                   & & 5709(19) \cite{watts83a}     &                                   \\
 &$6s$$-$$8s$ & 37660(250) \cite{gunawardena07a}   &  37820(290)$^b$   \cite{gunawardena07a}   \\
\textbf{Mg} &$3s^2 - 3s3p$ $^3P^o_1$ & &      \\
  &$m=1$ & 32.1(4.0) \cite{rieger93a} &   37.6$^c$ \cite{mitroy07e}  \\
  &$m=0$ & 15.7(4) \cite{rieger96a} &   16.3$^c$ \cite{mitroy07e}  \\
\textbf{Ca} &$4s^2 - 4s4p$ $^3P^o_1$ &  &   \\
  &    $(m = 0)$      &  90.4(13.5) \cite{morinaga96a}  &  107.5$^c$  \cite{mitroy08g}    \\
  &                 &         98.97(33) \cite{li96a} &         \\
\textbf{Ba} &$6s^2 -6s6p$ $^1P^o_1$ &                     &                          \\
  &    $(m = 0)$      & $-$229.32(48) \cite{li95a}  &    $-$247$^d$ \cite{kozlov99a}    \\
\textbf{Yb} &$6s^2 -6s6p$ $^3P^o_1$  &               &  160(60)$^d$  \cite{porsev99a}       \\
    &   $(m=0)$                       & 123.85(38) \cite{li95b}  &  110(18)$^d$  \cite{porsev99a}    \\
\textbf{Hg} &$6s^2 -6s6p$ $^3P^o_1$ & 26.68(48) \cite{harber01a}  &  26.95$^d$   \cite{hachisu08a}       \\
\textbf{Ga} &$4s^24p_{3/2}$- $4s^25s$&  788(40)   \cite{krenn97a}  &          \\
\textbf{Tl} &$6s^26p_{1/2}$- $6s^27s$ & $-$900(48) \cite{demille94a} &  $-$830$^b$  \cite{safronova06b}          \\
                             &&  $-$829.7(3.1)  \cite{doret02a}  &          \\
\textbf{Tl}&$6p_{1/2}$- $7p_{1/2}$ &  $-$4967(249) \cite{demille94a} & $-$4866$^b$ \cite{safronova06b}   \\
\textbf{Yb}$^+$&$6s$-$5d \ ^2D_{3/2}$  &  $-$41.8(8.5) \cite{schneider05a}      &     \\
\end{tabular}
\end{ruledtabular}
\end{table}

\subsection{Two electron atoms and ions, $ns^2~^1S$ and $nsnp~^3P^o_0$ states}

Table \ref{alkaline} gives the polarizabilities for a number of divalent
species including the alkaline-earth atoms from
Refs.~\cite{komasa02a,tunega97a,archibong91a,bendazzoli04a,hamonou08a,glass87a,muller84,mitroy03f,mitroy04b,mitroy07e,mitroy08a,mitroy08g,mitroy09b,mitroy08k,
mitroy89a,lim04a,sadlej91a,schafer07a,miller76a,schwartz74a,reshetnikov08a,ye08a,hachisu08a,dzuba10a,goebel96b,thierfelder09a,goebel96b,goebel95a,
goebel95b,tang08a,komara05a,porsev99a,porsev06a,porsev06b,porsev08a,kozlov99a}.
The beryllium atom serves as a theoretical benchmark since a very accurate
value has been obtained with a basis of exponentially correlated Gaussians
 (ECG) \cite{komasa02a}. The CICP \cite{mitroy03f} and RCI+MBPT polarizabilities \cite{porsev06a}
lie within 0.2$\%$ of the ECG basis polarizability.

The sub-1$\%$ agreement between the highest quality theory and experiment that
occurred for the alkali atoms is not observed for the alkaline-earth atoms
owing to their more complicated atomic structure and resulting mixing of
configurations. As we have described in Section~\ref{ci+mbpt}, perturbative
methods do not work well for strong valence-valence correlations.  The hybrid
values for Ca and Sr based on the RCI+MBPT calculations with the matrix
elements for the resonance transitions replaced by values derived from
experiments are respectively 1.1$\%$ and 2.5$\%$ smaller than the \textit{ab
initio} RCI+MBPT estimates \cite{porsev06a}. With the exception of Be, our
recommended values for alkaline-earth polarizabilities are those obtained from
the hybrid RCI+MBPT method.   We note very good agreement of the RCCSDT
calculations of Ref.~\cite{lim04a} for the ground state polarizabilities of Ca,
Sr, and Ba with the recommended values in all three cases. One of the problems
of the hybrid approach is the paucity of high-precision experimental data for
divalent atoms. Strontium is the only atom where the polarizability has been
quoted with a precision approaching $0.1\%$ \cite{porsev06a}. This is due to
the availability of a high precision estimate of the resonant oscillator
strength obtained by Yasuda and Katori using photo-association spectroscopy
\cite{yasuda06a}. However, an alternate photo-association experiment
\cite{nagel05a} gave a lifetime 0.8$\%$ smaller than the Yasuda and Katori
value, so it may be over-optimistic to assign an uncertainty of 0.1$\%$ to the
strontium polarizability. Currently the best estimate of the $5s5p$ $^3P^o_0$
excited state polarizability of Sr is accurate to 0.8\% despite the use of the
experimental data. The Sr polarizabilities are discussed in detail in
Ref.~\cite{porsev08a}.

The $^{27}$Al$^{+}$ ion is included in Table \ref{alkaline} since it is being used in
the development of a single ion optical frequency
standard \cite{rosenband08a}. The most reliable calculation of the ground state polarizability $\alpha_0$ is probably
given by the CICP calculation. The only experimental value is of low precision (3$\%$) and was obtained by summing
experimental oscillator strengths. A CICP calculation of the isoelectronic Si$^{2+}$ system gave a polarizability that
was within 0.2$\%$ of the value from a RESIS experiment.

The scatter amongst the different calculations of ytterbium underlines the difficulties of performing calculations in
this system.  The source of the problem lies in the weakly bound $4f^{14}$  core.  There are 20$\%$ differences between
two of the RCI+MBPT calculations that are discussed in recent work by Dzuba and Derevianko \cite{dzuba10a} and are
attributed to the inconsistent use of experimental matrix element for the principal transition in \cite{porsev06b}.
 Yb is of particular
interest for many applications, including ultracold atoms, optical frequency standards, and parity violation
experiments.

There is a significant discrepancy for Cd between the refractive index value of
49.65(1.49) a.u. \cite{goebel95a} and the calculated value of 44.63 a.u. from
the RCICP calculation \cite{ye08a}.  For a number of reasons, including the
measured values of the oscillator strengths for the $5s^2$ $^1S$ $-$ $5s5p$
$^1P^o$ transitions, it has been suggested that the experimental polarizability
might be overestimated \cite{bromley02d}.

The polarizabilities of other excited states, tensor polarizabilities, and Stark shifts  in divalent systems are
discussed in the next subsection.

\subsection{Other data}

Ground state polarizabilities for the other selected  systems from
Refs.~\cite{schwerdtfeger94a,safronova09a,neogrady97a,ilias99a,zhang08d,topcu06a,henderson97a,hamonou07a,mitroy09a,
johnson83a,neogrady96a,snow07a,magnusson77a,lundeen07b,lupinetti05a,kozlov01b,hibbert80a,kozlov01b,fleig05a,
milani90a,guella84a,guella85a,hibbert80a,thierfelder08a,bardon84a,kadarkallen94a}
are given in Table \ref{miscellaneous}. In this review, we list data for
selected systems with a monovalent $ns$ ground state: Cu, Ag, Au, Zn$^+$,
Hg$^+$, and Yb$^+$, and Al$^{2+}$; ions for which recent RESIS experiments have
been performed: Si$^{3+}$ and Kr$^{6+}$; neutral atoms with three and four
valence electrons: Al, Ga, In, Tl, Si, Sn, Pd, and Ir; and U. The reader is
referred to a recent review \cite{schwerdtfeger06a} for atomic ground state
polarizabilities of other systems not listed herein.

One notable discrepancy between theory and experiment occurs for the Al ground state where the best calculations exceed
the experiment value from an EH balance experiment by 25$\%$ \cite{lupinetti05a,milani90a}. The most precise
experimental value in Table \ref{miscellaneous} is the RESIS value for Si$^{3+}$. The final value, 7.433(25) a.u. comes
from a reanalysis of the raw experimental data \cite{komara03a,snow07a} that includes estimates of $r^{-7}$ and
$r^{-8}$ polarization corrections from RLCCSDT and CICP calculations \cite{mitroy09a}.  The agreement between the
RLCCSD polarizability of 7.419 \cite{mitroy09a} and the latest RESIS reanalysis is at the 0.2$\%$ level.

Table \ref{excitedalkaline} shows a number of  measurements and calculations of the tensor polarizability of non-alkali
systems including Ca, Sr, Ba, Zn, Cd, Hg, Tl, Yb, Yb$^+$ from
Refs.~\cite{mitroy08g,kreutztrager73a,mitroy08g,oppen70a,yanagimachi02a,zeiske95a,oppen71a,kreutztrager73a,porsev08a,
oppen69b,kozlov99a,kreutztrager73a,hese77a,vanleeuwen83a,li96a,rinkleff79a,legowski95a,hachisu08a,khadjavi68a,
sandle75a,kaul72a,martin68a,lupinetti05a,fleig05a,kozlov01b,gould76a,porsev99a,rinkleff80a,li95b,kulina82a,schneider05a}.
These systems are the ones under consideration as frequency standards or are being used in atomic parity violation
experiments.  Measurements for some states have been omitted from the Table, and some older or less precise results on
Sc, Y, La and Lu
 \cite{rinkleff94a}, Cd \cite{khadjavi68a},
Ba \cite{kreutztrager73a,vanleeuwen83a,schuh96a}, Hg \cite{khadjavi68a}, Yb
\cite{rinkleff80a}, Sm and Eu \cite{martin68a} have also been omitted.

One feature of Table~\ref{excitedalkaline} is the relatively small number of modern calculations performed.  For
example, the best calculated polarizabilities for the $4s4p$ $^{1,3}P^o_1$ states of Ca are the non-relativistic CICP
calculations.  Another feature is the relatively large uncertainties in many of the experimental values. There are only
five tensor polarizabilities with uncertainties less than 2$\%$.  The most precisely measured $\alpha_2$ of
$-$$43.04(40)$ a.u. occurs for the Ba $6s6p$ $^1P^o_1$ state.  The RCI+MBPT value of $-$51 a.u. is incompatible with
experiment.

The static polarizability differences (in a.u.) for selected transitions in Cs, Mg, Ca, Ba, Yb, Hg, Ga, Tl, and Yb$^+$
derived from Stark shift measurements
\cite{bennett99a,gunawardena07a,rieger93a,rieger96a,morinaga96a,li96a,li95b,harber01a,krenn97a,demille94a,doret02a,schneider05a}
are compared with theoretical calculations \cite{kozlov01a,gunawardena07a,mitroy08g,kozlov99a,hachisu08a,safronova06b}
in Table~\ref{starkother}. Total polarizability differences are given for the cases where $m$ values are listed,
otherwise scalar polarizability differences are listed.

There have been sub-1$\%$ experiments on  four systems, Cs, Ba, Yb, and Hg. The
ability of RCI+MBPT calculations to reproduce experiment for the divalent
systems is mixed.  The agreement for the Hg $6s^2$ $^1S$ - $6s6p$ $^3P^o_1$ is
excellent, but 10$\%$ discrepancies exist for the Ba $6s^2$ $^1S$ - $6s6p$
$^1P^o_1$ and Yb $6s^2$ $^1S$ - $6s6p$ $^3P^o_1$ transitions.  However, the
RCI+MBPT calculations for Ba \cite{kozlov99a} and Yb \cite{porsev99a} were
among the first RCI+MBPT calculations reported.

\section{Evaluating uncertainties of theoretical values}
\label{uncertainty}

\subsection{Sources of theoretical uncertainty}

As illustrated by the tables in the previous section, benchmark comparisons of theory and experiment carry more value
when the theoretical results are accompanied by uncertainty evaluations.  Uncertainty bounds are particularly important
for the recommended values obtained by either high-precision theory methods or by combination of theory values with
experimental data.  The analysis of the theoretical uncertainties has been stimulated by the applications that require
an error bound to be placed on the recommended values.  Such applications include  parity violation, development of the
next-generation frequency standards, ultra-cold atom studies, etc. Analysis of certain experiments requires input of
some data that cannot be easily measured and have to be obtained from theory.  In those cases, the uncertainties of the
theoretical data have to be included in the uncertainty of the final experimental value.  Evaluations of the
theoretical uncertainties are still few and cannot be carried out for all of the methods and in all cases. Here, we
discuss how some theoretical uncertainties may be evaluated.

There are two distinct sources of theoretical uncertainties.  First, there is
an uncertainty associated with the numerical constraints upon the calculations.
Many of the methods that we discussed in this review are computationally very
intensive and restrictions are imposed so that the calculations can be
performed within a reasonable time. Most common numerical uncertainties are
associated with the choice of the basis sets, configuration space, radial grid,
termination of the iterative procedures after achieving the specified
convergence tolerance, etc. Generally, it is possible to at least estimate
uncertainties caused by numerical issues by varying the appropriate parameters
and recording the changes in the results. In many cases, it is possible to
simply continue to change parameters until the change in the resulting values
is sufficiently small or negligible.

For example, it is relatively easy to test the convergence of B-spline basis sets.  The dimensionality of the radial
basis in a RLCCSD calculation for each partial wave (e.g, $ns$, $np_{1/2}$, $np_{3/2}$, $\ldots$ states) is steadily
increased.  The final values of a property like the sodium ground state polarizability do not change, within the quoted
digits whether the $B$-spline basis has a dimension of 40, 50, or 70 orbitals \cite{derevianko99a}. Using only 20
orbitals, however, will lead to change in the final value that is not negligible. Also, truncating partial wave
expansion at $l=3$ will measurably affect the final result, while including all partial waves up to $l=6$ is
sufficiently complete in this case.  Generally, such tests do not have to be carried out at the level of the most
accurate calculation possible and it is sometimes sufficient to study the lowest-order results or low-order MBPT
values. In some cases, it may become necessary to completely repeat the entire calculation. However, such numerical
problems may be studied by well understood conventional methods. In most cases, numerical errors of the theoretical
values can be made small enough not to affect any of the significant figures that are quoted, or can be evaluated and
quoted as uncertainty in the last digit.

Investigations using the Hylleraas method typically perform a series of
calculations of increasing dimension while keeping the non-linear parameters
the same.  The convergence of the data against a value of
the total polynomial power is studied.  The total polynomial
power for a correlated wave function such as Eq.~(\ref{Hyl})
would be
\begin{equation}
\Omega = j_1 + j_2 + j_3 + j_{12} + j_{13} + j_{23}  .
\end{equation}
Most expectation values in a Hylleraas calculation converge as $\sim 1/\Omega^p$. This result is exploited to give
uncertainties in energies, transition matrix elements, polarizabilities, and other quantities \cite{yan96a,tang09a}.

The theoretical uncertainties of the second type are much harder to evaluate.
These are the uncertainties associated with the particular theoretical
methodology, for example,  the uncertainty associated with stopping a
perturbation theory treatment at third order.  Ideally, the total uncertainty
of the theoretical value should give an estimate of how far any value is from
the actual (unknown) exact result. Evaluation of the complete theoretical
uncertainty is non-trivial since it essentially involves the evaluation of a
quantity that is not known beforehand and cannot be determined by the
theoretical methodology adopted.

\subsection{Sources of uncertainties in the sum-over-states polarizability calculations}

It is particularly problematic to evaluate full theoretical uncertainties for
the semi-empirical theoretical methods. In this case, there may be no basis to
make assumptions regarding the missing theory. It may be possible to infer some
information based on the agreement of CICP calculations with quality
experiments for similar states in other members of the same iso-electronic
series.  For example, the CICP ground state polarizability for Al$^+$ of 24.14
a.u. has been assessed at $\pm 0.5\%$ \cite{mitroy09b} based on the 0.3$\%$
agreement between a CICP calculation of the Si$^{2+}$ polarizability and a
RESIS experiment \cite{komara05a,mitroy08k}. The assessment of uncertainties,
for states that lack validating information, as in the case of the $^3P^o_0$
state of Al$^+$ has a larger speculative element \cite{mitroy09b}.

Several strategies exist for uncertainty evaluation for  the \textit{ab initio} MBPT, correlation potential, and
all-order linearized coupled-cluster (RLCCSDT) approaches.  These strategies are illustrated using RLCCSDT method which
utilizes the sum over states algorithm.  For brevity,  we refer to RLCCSDT calculation as ``all-order'' in the text
below.

We use the example discussed in Section~\ref{sumoverstates}, i.e.  the
polarizability of the $5p_{1/2}$ state.  Table~\ref{table-rb} lists a
detailed breakdown of the contributions to this value. There are
three separate contributions:  the main part ($5s - 11s$ and $4d_{3/2}
- 9d_{3/2}$), remainder (all other valence terms), and core
contribution.
The uncertainty in each term of the main part has to be determined.  The
energy levels of low-lying states are generally well known.   Therefore,
the determination of the uncertainty here reduces to the evaluation of
the uncertainty in the corresponding electric-dipole matrix elements.
The relative uncertainty in the polarizability contribution is twice
the relative uncertainty in the matrix element (see Eq.~(\ref{5p})).

The uncertainty of the remainder (higher $n$ contributions) as well as
the uncertainty of the ionic core have to be determined separately.
The uncertainty in the RPA value of the core is estimated from comparison
of the RPA values for noble gases with experiment and precision
coupled-cluster calculations (see Table~\ref{tab:3} and the corresponding
discussion). The evaluation of the uncertainty of the remaining highly-excited
contribution has been discussed in great detail in recent work on
the Sr$^+$ polarizabilities~\cite{jiang09a}.

In most cases, all of the uncertainties are added in quadrature to obtain
the final uncertainty of the polarizability value.

\subsubsection{Determination of the uncertainties in E1 matrix elements}

Ultimately, the theoretical uncertainty estimates in the polarizability need
uncertainties in the E1 matrix elements such as those listed in
Table~\ref{table-rb}. The starting point of relativistic MBPT or all-order
RLCCSD calculations for monovalent systems is a DHF calculation. We refer to
the DHF value as the lowest order. Essentially all corrections to that value
come from Coulomb correlations. Breit interaction corrections to the E1 matrix
elements are generally insignificant at the present level of accuracy
\cite{derevianko02b}, and the relativistic corrections are intrinsically
included due to use of a relativistic Hamiltonian. Therefore, an uncertainty
evaluation requires an estimation of the missing part of the correlation
correction. The strategies to do so include:
\begin{itemize}
\item  approximate evaluation of the size of the correlation correction;
\item evaluation of the size of the higher-order corrections; \item study of the
order-by-order convergence of perturbation theory; \item study of the breakdown
of the various all-order contributions and identification of the most important
terms;
\item semi-empirical determination of dominant missing contributions.
\end{itemize}

\begin{table}

\caption{\label{unc-example1}  Rb electric-dipole matrix elements
(in a.u.) calculated in different approximations
\cite{safronova04a}. The rows labelled ``correlation'' list an
estimate of the correlation contribution, determined as the
relative difference between the lowest-order and the all-order
values. The rows labelled ``higher orders'' list an estimate of the
4$\textrm{th}$ and higher-order contributions, determined as the
relative difference between the third-order and the all-order
values. Absolute values are listed. The negative sign in front of
the lowest-order $6d_{3/2}-6p_{1/2}$ value indicates that the
lowest-order gives incorrect sign for this matrix element.}

\begin{ruledtabular}
\begin{tabular}{lccc}
& $5s-5p_{1/2}$ &$5s-6p_{1/2}$&$6s-5p_{1/2}$\\
\hline
Lowest order& 4.819  & 0.383 &  4.256 \\
 Third order& 4.181  & 0.363 &  4.189 \\
All order   & 4.221  & 0.333 &  4.119 \\
 Correlation&  14\%  &  15\% &  3.3\% \\
Higher  orders& 0.9\% &   9\% &  1.7\% \\
\hline
& $8s-8p_{1/2}$&$4d_{3/2}-5p_{1/2}$&$6d_{3/2}-6p_{1/2}$\\
\hline
Lowest order& 26.817&  9.046&   -0.047\\
 Third order& 25.587&  8.092&   2.184\\
All order   &  25.831&  7.847&   2.974\\
 Correlation&   3.8\% &   15\%&   100\%\\
Higher orders & 0.9\%&  3\%&    27\%\\
\end{tabular}
\end{ruledtabular}
\end{table}

The first three strategies are aimed at providing rough estimate of the matrix element uncertainty. Separate
third-order RMBPT and all-order calculations have to be carried out to evaluate the accuracy of the all-order values
since the extraction of third-order matrix elements from the all-order values is impractical.

The application of the first three strategies are illustrated in Table~\ref{unc-example1} where Rb E1 matrix elements
are listed \cite{safronova04a}.  Three values are given for each matrix element: lowest-order DHF value, the
third-order RMBPT value, and all-order values obtained from an RLCCSD calculation. Third-order values include the
second-order, third-order, and RPA corrections iterated to all-orders (see \cite{johnson96a} for detailed description
of the third-order MBPT calculations}).  The size of the correlation correction is estimated as the relative difference
between the lowest-order and the all-order values.  It is given as a percentage change in the rows labelled
``correlation''. The size of the fourth and higher-order corrections is estimated as a percentage difference between
the third-order and all-order values and listed in the rows labelled ``higher orders''.

Study of the ``correlation'' and ``higher orders'' rows gives some
insight into the accuracy of the final all-order values.  First,
it is noted that the corrections vary significantly among the different
transitions.   Very rough estimate of the uncertainty can be obtained
by assuming that higher-order corrections incorporated into RLCCSD
are smaller than the higher orders that are omitted by RLCCSD.  Thus,
the difference between third and all-orders is taken as the uncertainty.
In most cases, this procedure will significantly overestimate the
uncertainty since Table~\ref{unc-example1} shows that contributions
from all higher orders are lower than the second and third-order in
all cases except the small $5s-6p_{1/2}$ matrix element.  However,
this procedure clearly indicates that while $5s-5p_{1/2}$ matrix element
is probably accurate to better than 1\%, SD all-order $6d_{3/2}-6p_{3/2}$
matrix element may only be accurate to about 25\%.

The last two strategies should be employed if more accurate uncertainty
evaluations are required. This can only be done for certain cases within the
framework of the RLCCSDT method and requires substantial additional
calculations and the careful analysis of all available data. First, the
breakdown of the all-order terms have to be studied. Triple excitations need to
be added at least partially. If certain types of the contributions (associated
with so-called Brueckner orbital terms) are dominant, they may be estimated by
the semi-empirical scaling described, for example, in Ref.~\cite{safronova08a}.
This procedure involves rescaling single-excitation coefficients $\rho_{mv}$
(see Section~\ref{all-order-section}) using experimental energies, and
re-running the entire matrix element calculation with the modified
coefficients. Obviously, this method is only expected to produce more accurate
values if correlation correction is dominated by the terms containing single
valence excitation coefficients. However, this is true in many cases.
Non-linear terms may also be evaluated. The most extensive uncertainty study of
this type has recently been carried out for the atomic quadrupole moments of
Ca$^+$, Sr$^+$, and Ba$^+$ in \cite{jiang08a}.

Detailed studies of the uncertainties of the electric-dipole matrix elements are described, for example, in
Refs.~\cite{arora07b,jiang09a,gunawardena07a}. A brief description is given here for the case of the $4d_{5/2} –-
5p_{3/2}$ matrix element in Sr$^+$ \cite{jiang09a}.  This transition is important in the evaluation of the Sr$^+$ BBR
shift.  Correlation corrections change the matrix element by about $20\%$. Study of the correction breakdown indicates
that the correlation is dominated by a single term that contains single valence excitations. Therefore, we carry out
additional \textit{ab initio} calculations that partially include triple excitations, and also perform scaled  RLCCSD
and RLCCSDT calculations. The results of these four calculations are listed below. All data are in atomic units. The
first line corresponds to the ``all-order'' lines in Table~\ref{unc-example1}.

$$
\begin{array} {lcr}
\textrm{RLCCSD} &&4.150\\
\textrm{RLCCSDT} && 4.198\\
\textrm{RLCCSD scaled} && 4.187\\
\textrm{RLCCSDT scaled} && 4.173\\
\textrm{Final} && 4.187(14)\\
\end{array}
$$

Note that scaled values are much closer together than the SD and SDT \textit{ab initio} values. The final value was
taken to be RLCCSD scaled 4.187(14) result (see, for example, \cite{safronova08a} and references therein for the
discussion of this choice).  The uncertainty of 0.014 is determined as the maximum
difference between the scaled SD values and the \textit{ab initio} SDT and scaled SDT values.

\section{Applications}

\subsection{Parity non-conservation}
\label{PNC}

The goals of the parity nonconservation (PNC) studies in heavy atoms  are  to search for new physics beyond the
standard model of the electroweak interaction by precise evaluation of the weak charge $Q_w$, and to probe parity
violation in the nucleus by evaluating the nuclear anapole moment. The study of PNC in the cesium $6s-7s$ transition
involving both high-precision measurement \cite{Cs-PNC} and several high-precision calculations provided an
atomic-physics test of the standard model of the electroweak interactions \cite{porsev09-pnc}.
  Moreover, accurate determination of the uncertainty in
theoretical values was necessary, leading to detailed studies of parity-conserving quantities in Cs including the
polarizabilities of the $6s$, $6p_J$, and $7s$ states
(see \cite{dzuba97a,safronova99a,vasilyev02a,safronova08a} and references therein).
 The analysis of the Cs experiment
was instrumental in developing methods to evaluate the uncertainties of the theoretical data \cite{blundell92a}.

In the Cs experiment \cite{Cs-PNC}, the PNC amplitude was measured relative to
Stark-induced tensor transition polarizability $\beta_S$ (some works refer to
this quantity as the vector transition polarizability). The DC electric field
mixes states of opposite parity allowing electric-dipole transitions between
$ns$ states. The Stark-induced amplitude is expressed via the Stark-induced
scalar and tensor transition polarizabilities $\alpha_S$ and $\beta_S$. In the
case of the Cs $7s-6s$ transition, they are calculated as sum-over-states using
the expressions \cite{blundell92a}

\begin{eqnarray}
 \alpha_{S}& =& \frac{1}{6} \sum_{n}
   { \langle 7s \| D\| np_{1/2}  \rangle \langle np_{1/2}  \| D\| 6s  \rangle }\nonumber \\
  & \times& \left( \frac{1}{E_{7s}-E_{np_{1/2}}} + \frac{1}{E_{6s}-E_{np_{1/2}}}\right)  \nonumber \\
   &-& \frac{1}{6} \sum_{n} \langle 7s \| D\| np_{3/2}  \rangle \langle np_{3/2}  \| D\| 6s  \rangle \nonumber \\
  & \times& \left( \frac{1}{E_{7s}-E_{np_{3/2}}} + \frac{1}{E_{6s}-E_{np_{3/2}}}\right), \nonumber
\end{eqnarray}

\begin{eqnarray}
 \beta_{S}& =& \frac{1}{6} \sum_{n}
   { \langle 7s \| D\| np_{1/2}  \rangle \langle np_{1/2}  \| D\| 6s  \rangle }\nonumber \\
  & \times& \left( \frac{1}{E_{7s}-E_{np_{1/2}}} - \frac{1}{E_{6s}-E_{np_{1/2}}}\right)  \nonumber \\
   &+& \frac{1}{12} \sum_{n} \langle 7s \| D\| np_{3/2}  \rangle \langle np_{3/2}  \| D\| 6s  \rangle \nonumber \\
  & \times& \left( \frac{1}{E_{7s}-E_{np_{3/2}}} - \frac{1}{E_{6s}-E_{np_{3/2}}}\right).
\end{eqnarray}
These quantities have been extensively studied due to their importance in PNC research,
\cite{dzuba97a,blundell92a,Cs-PNC,cho97a,vasilyev02a}.
  It is more complicated to calculate $\beta_S$ accurately,
in comparison to $\alpha_S$, owing to severe cancellations between different terms contributing to $\beta_S$. The ratio
of $\alpha_S$ and $\beta_S$ has been measured to high precision \cite{cho97a}.
 At the present time, Cs experiment is consistent with the standard model \cite{porsev09-pnc}.

However, the precise
measurement of PNC amplitudes in Cs \cite{Cs-PNC} also led to an experimental value of the small contribution from the
nuclear-spin dependent PNC accurate to 14\%. The constraints on weak nucleon-nucleon coupling constants derived from
this experiment and calculations in Cs were found to be significantly inconsistent with constraints from deep inelastic
scattering and other nuclear experiments \cite{haxton01a,haxton01b,safronova09-pnc}. At the present time, this discrepancy
remains unexplained.

More PNC
experiments in other atomic systems, such as Ra$^+$, Yb, and Fr are currently in progress. Experiments in Pb, Bi, and
Tl have been conducted but theoretical calculations of comparable accuracy are not available to permit precise
comparison of experiments with the standard model. Comparison of theoretical and experimental values of $\alpha_S$ and $\beta_S$
for Tl are given in Ref.~\cite{safronova06b}.

\subsection{Ultracold atoms in optical lattices and quantum computation}
\label{magicwav}

 Quantum computation is a field of research which is aimed at using the quantum nature of matter to produce
fundamentally new methods of computation. There are various approaches to the experimental realization of the quantum
computation. In the quantum computation scheme relevant to this review, the qubits are realized as internal states of
neutral atoms trapped in optical lattices or microtraps. This approach to quantum computation has many advantages, such
as the long decoherence times of the internal states of the atoms, flexibility in controlling atomic interactions,
scalability, possible massive parallelism, and well-developed experimental techniques.

Trapping an atom or a group of atoms in an optical lattice raises the possibility that the laser field used to create
the lattice might shift the energy levels of the lower and upper states by different amounts.  This can result in a
wavelength (and intensity) dependent shift of the clock transition.  This issue was first raised for the atomic clocks
based on neutral atoms trapped in optical lattices.

A solution to this problem was proposed by Katori \textit{et al.}~\cite{katori99b} who suggested that the laser
can be tuned to a magic wavelength $\lambda_{\textrm{magic}}$, where lattice potentials of equal depth are produced for
the two electronic states of the clock transition. At such wavelength,
 the  AC polarizabilities of the two relevant states  satisfy the condition

\begin{equation}
\alpha_{\rm upper}(\lambda) = \alpha_{\rm lower}(\lambda).
\end{equation}

\textit{Ab initio} calculations of the dynamic polarizability are valuable in making an initial estimate of the magic
wavelength prior to construction of the optical lattice.  However, it is possible to make very precise determinations
of the magic wavelength once the lattice has been constructed and atoms have been trapped since the experimental design
is that of a null experiment.  The experimental magic wavelength can be used as a constraint upon the dynamic
polarizability and used to refine the polarizability calculation.

\begin{figure} [h]
\includegraphics[angle=0,width=8.50cm]{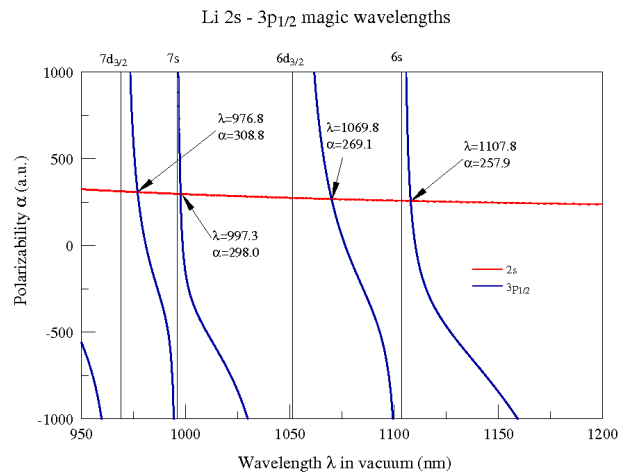}
\caption[]{ \label{magic-fig} Magic wavelengths for the $2s-3p_{1/2}$ transition in Li.
Upper state of the resonant transition is marked on top of the box.}
\end{figure}

Examples of a magic wavelength calculation are depicted in Fig.~\ref{magic-fig}, where polarizabilities of the Li $2s$
and $3p_{1/2}$ states obtained using RLCCSDT method are plotted. The magic wavelengths are located at the crossing
points of the two curves. The ground state polarizability is nearly flat in this wavelength region, while the
$3p_{1/2}$ polarizability has several resonances noted by the vertical lines.

One of the current goals of the quantum information projects is to design an apparatus capable of interconnecting
``flying" and ``stationary" qubits. The ability to trap neutral atoms inside high-$Q$ cavities in the strong coupling
regime is of particular importance for such schemes.  In a far-detuned optical dipole trap, the potential experienced
by an atom can be either attractive or repulsive depending on the sign of the AC Stark shift due to the trap light. The
excited states may experience an AC Stark shift with an opposite sign to the ground state Stark shift which will affect
the fidelity of the experiments. McKeever \textit{et al.} \cite{mckeever03a} demonstrated state-insensitive trapping of
Cs atoms at $\lambda_{\textrm{magic}} = 935$~nm while still maintaining strong coupling for the $6p_{3/2}-6s_{1/2}$
transition.

The magic wavelengths in Na, K, Rb, and Cs atoms for which the $ns$ ground state and either of the first two $np_j$
excited states experience the same optical potential for state-insensitive cooling and trapping were evaluated in
\cite{arora07c}. This was accomplished by matching the dynamic polarizabilities of the atomic $ns$ and $np_j$ states
using extensive relativistic all-order calculations. Uncertainties in the dynamic polarizabilities
were also evaluated.

One requirement for the experimental realization of the scalable quantum computer is the design of a quantum gate with
low error rate which will allow for error correction. Therefore, it is important to study the various decoherence
mechanisms and to search for ways to optimize gate performance.

The issue of the mismatch of the polarizabilities of the ground and excited
states has also arisen in schemes to perform quantum logical operations where
it is a source of decoherence. In the Rydberg gate scheme \cite{Rgate}, the
qubit is based on two ground hyperfine states of neutral atoms confined in an
optical lattice. A two-qubit phase gate may be realized by conditionally
exciting two atoms to relatively low-lying Rydberg states. The choice of this
particular scheme results from its potential for fast (sub-microsecond) gate
operations. Such a gate has been recently experimentally demonstrated
\cite{saffman10}. An atom in a Rydberg state will, in general, move in a
different optical lattice potential than that experienced by the ground state.
Therefore, the vibrational state of the atom in the lattice may change after
the gate operation is completed, leading to decoherence due to motional
heating. The optical potential for a given state depends on its AC
polarizability, so we can seek to minimize this motional heating effect by the
choice of a particular Rydberg state or of the lattice photon frequency
$\omega$. A method for accomplishing this by matching the frequency-dependent
polarizabilities $\alpha(\lambda)$ of the atomic ground state and Rydberg state
is described in \cite{us-gate,saffman05}.

In recent work \cite{rey09},   a novel approach to quantum information
processing, in which multiple qubits can be encoded and manipulated using
electronic and nuclear degrees of freedom associated with individual
alkaline-earth-metal atoms trapped in an optical lattice, was proposed and
analyzed. In this scheme, curves of dynamic polarizabilities are needed for
alkali and group II atom elements to locate the wavelengths where one of the
species can escape or where AC Stark shifts cancel for a specific transition.

\begin{table*}[th]
\caption[]{ The blackbody radiation shifts for a number of proposed optical
frequency standards.  The polarizability difference, $\delta \alpha$, is
negative when the upper state polarizability is smaller than the lower state
polarizability.  A negative polarizability difference means the frequency shift
is positive. All BBR shifts are evaluated at 300 K and values that include the
dynamic shifts are indicated with an asterisk ($^*$). Linewidths are converted
from lifetimes, $\tau$ using $\delta \nu_{\rm nat} = 1/(2\pi\tau)$; natural
linewidths are given for fermionic isotopes for the $ns^2-nsnp$ clock
transitions. Uncertainties in the last digits are given in parentheses.
References are given in square brackets.  The composite CI calculation for
Yb$^+$ is a hybrid calculation that used CI to explicitly allow for core
excitations but also included core polarization using a semi-empirical core
polarization potential. } \label{BBshifts}
\begin{ruledtabular}
\begin{tabular}{lcccccc}
Transition  & $\nu_0$ ($\times 10^{15}$ Hz)   &  $\Delta \nu_{\rm nat}$ (Hz) & $\Delta \alpha$ ($a_0^3$) &  $\Delta
\nu_{\rm BBR}$ (Hz) & $\displaystyle{\bigl| \frac{\Delta \nu_{\rm BBR}}{\nu_0} \bigr|\! \times \! 10^{15} }$ &  Approach  \\
\hline
Ca$^+$($4s_{1/2} - 3d_{5/2}$)    &  0.411 \cite{chwalla09a} &  0.14 \cite{margolis09a}  &  $-$44.1(1.5) &  0.38(1) & 0.925  &  RLCCSDT \cite{arora07a} \\
                        &                &       &  $-$42.8     & 0.369  & 0.895   &  CICP \cite{mitroy08b}    \\
Sr$^+$($5s_{1/2} - 4d_{5/2}$)    &  0.445 \cite{dube07a} & 0.4 \cite{margolis09a} &  $-$29.3(1.1)    & 0.250(9)$^*$ & 0.562$^*$  &  RLCCSDT \cite{jiang09a}   \\
Hg$^+$($5d^{10}6s - 5d^9_{5/2}6s^2$)  &   1.06 \cite{oskay06a} & 1.8 \cite{margolis09a} &    &      &    & Cryogenic \\
Yb$^+$($4f^{14}6s$-$4f^{13}6s^2~^2F_{7/2}$)   &  0.642 \cite{hosaka09a} &  $\sim 10^{-9}$ \cite{margolis09a}   & 11.7  & $-$0.101  & 0.16 & $f$-sum composite CI \cite{biemont98a,lea06a,hosaka09a} \\
                             &           &                        & 6.9(1.8)  & $-$0.057(14)  & 0.089 & $f$-sum (Lifetimes) \cite{lea06a,hosaka09a} \\
Yb$^+$($4f^{14}6s$-$4f^{14}5d~^2D_{3/2}$)   &  0.688 \cite{tamm07a} &  3.1 \cite{margolis09a}   &  42(8) &  $-$0.36(7)  & 0.53(10) & Expt. \cite{tamm07a}  \\
Al$^+$($3s^2 \ ^1S - 3s3p \ ^{3}P^{o}_0$) &  1.12 \cite{rosenband07a} & 0.008 \cite{margolis09a}  & 0.483  &  $-$0.0042(32) & 0.004(3)  &  CICP \cite{mitroy09b} \\
                                           &             &               &   &  $-$0.008(3) & 0.007(3)   &  Expt. \cite{rosenband06a}  \\
In$^+$($5s^2 \ ^1S - 5s5p \ ^{3}P^{o}_0$) & 1.27 \cite{becker01a,eichenseer03a} & 0.8 \cite{margolis09a} & $< 30.7$  & $>$ $-$0.264  & $< 0.20$  & Theory, using $\Delta \alpha$(Cd)    \\
Mg($3s^2 \ ^1S - 3s3p \ ^3P^o_0$)  & 0.655 \cite{nistasd315} &  0.00014 \cite{santra04a}  & 29.9(7) &  $-$0.258(7)$^*$ & 0.394(11)  &  RCI+MBPT \cite{porsev06b} \\
                                     &                     &    & 30.1 &  $-$0.259    & 0.395   & CICP \cite{mitroy07e} \\
Mg($3s^2 \ ^1S - 3s3p \ ^3P^o_1$)  & 0.656 \cite{friebe08a}  & 57 \cite{santra04a} &  30.1  & $-$0.259   &  0.394  & CICP \cite{mitroy07e} \\
Ca($4s^2 \ ^1S - 4s4p \ ^3P^o_0$)  & 0.454 \cite{NIST2}  &  0.0005 \cite{santra04a} &  133.2(2.0)  &   $-$1.171(17)$^*$ & 2.58(4) & RCI+MBPT \cite{porsev06b} \\
                                     &      &     &   135.9  & $-$   &    &  CICP \cite{mitroy08g}  \\
Sr($5s^2 \ ^1S - 5s5p \ ^3P^o_0$)  & 0.429 \cite{campbell08a} & 0.0014 \cite{santra04a} & 261.1(3.6) &    $-$2.354(32)$^*$ & 5.49(7) & RCI+MBPT \cite{porsev06b}    \\
Yb($6s^2 \ ^1S - 6s6p \ ^3P^o_0$)  & 0.518  \cite{poli08a} & 0.008 \cite{porsev04b} &  155(15)  &  $-$1.34(13)$^*$ & 2.6(3) & RCI-MBPT \cite{porsev06b}  \\
                                     &            &   & 161(15)   & $-$1.39(13) & 2.7(3) & Hybrid RCI+MBPT \cite{dzuba10a}  \\
Zn($4s^2 \ ^1S - 4s4p \ ^3P^o_0$)  & 0.969 \cite{nistasd315}  & 0.0025 \cite{wang07b} & 29.57  &   $-$0.255 & 0.263 &  RCICP \cite{ye08a}  \\
Cd($5s^2 \ ^1S - 5s5p \ ^3P^o_0$)  & 0.903 \cite{NIST2} & $\sim 10^{-2}$  & 30.66  &   $-$0.264 & 0.292 &  RCICP \cite{ye08a}  \\
Hg($6s^2 \ ^1S - 6s6p \ ^3P^o_0$)  & 1.13 \cite{hachisu08a} & 0.11 \cite{bigeon67a} &  21.0 & $-$0.181 & 0.160 & RCI+MBPT \cite{hachisu08a}    \\
                                     &                        &  &  24.00 & $-$0.207 &0.183  & RCICP \cite{ye08a} \\
\end{tabular}
\end{ruledtabular}
\end{table*}

\subsection{Atomic clocks}

The current definition of the second in the International System of Units (SI)
is based on the microwave transition between the two hyperfine levels of the
ground state of $^{133}$Cs \cite{NIST330}. The present relative standard
uncertainty of the Cs microwave frequency standard is around $5\times10^{-16}$.
More accurate clocks are needed for a variety of applications. Significant
recent progress in optical spectroscopy and measurement techniques has led to
the achievement of relative standard uncertainties in optical frequency
standards that are comparable to the Cs microwave benchmark. The frequencies of
feasible optical clock transitions are five orders of magnitude greater than
the standard microwave transitions, and so smaller relative uncertainties are
potentially achievable. A list of optical transitions recommended for this
purpose has recently been disseminated by the International Committee for
Weights and Measures \cite{BIPM2006}.

There are two types of optical atomic clocks under active investigation at the
moment. Both types of clocks are based on optical frequency transitions with a
narrow linewidth. The narrow linewidth mandates that the upper state of the
clock transition be a long-lived metastable state. One type of clock is
implemented using a group of cold atoms trapped in an optical lattice.  The
second consists of a single laser cooled ion. With extremely low systematic
perturbations and better stability and accuracy, such optical frequency
standards should exceed the performance of the existing Cs standard. A commonly
quoted target for the new generation of optical frequency standards is a
fractional uncertainty of $\Delta \nu / \nu_0 = 10^{-18}$
\cite{becker01a,udem02a,katori03a,takamoto05a}.

There are two main interconnecting areas of theoretic atomic clock research:
prediction of atomic properties required for new clock proposals and
determination of quantities contributing to the uncertainty budget. New clock
proposals require estimates of the atomic properties for details of the
proposals (transition rates, lifetimes, branching ratios, magic wavelengths,
scattering rates, etc.) and evaluation of the systematic shifts (Zeeman shift,
electric quadruple shift, blackbody radiation shift, AC Stark shifts due to
various laser fields, etc.). While a large fraction of these quantities may be
eventually measured, lack of knowledge of some of these properties may delay
experimental realization of new proposals. In the case of well-developed
proposals, one of the main uncertainty issues is the blackbody radiation (BBR)
shift. The operation of atomic clocks is generally carried out at room
temperature, whereas the definition of the second refers to the clock
transition in an atom at absolute zero. This implies that the clock transition
frequency should be corrected for effects of finite temperature, of which the
leading contributor is the blackbody radiation shift.  The BBR shift is looming
as a major component in the uncertainty budget of the optical frequency
standards.  Table \ref{UncertaintyBudget} shows the fractional uncertainty
budget for a $^{87}$Sr optical frequency standard \cite{ludlow08a}.  The BBR
shift is by far the largest source of uncertainty in the uncertainty budget. It
is noteworthy that the second largest source of uncertainty is the AC stark
shift caused by the optical lattice (this estimate did not take into account
possible corrections due to M1 and E2 multipoles caused by spatial
inhomogeneities of the lattice field). Experimental measurements of BBR shifts
are difficult and high-precision theoretical calculations are presently needed.

\begin{table}[th]
\caption[]{ Fractional uncertainty budget for the $^{87}$Sr atomic frequency standard \cite{ludlow08a}.
The BBR shifts are evaluated at 296 K.  Corrections that include knowledge of
polarizabilities are preceded by an asterisk (*). }
\label{UncertaintyBudget}
\begin{ruledtabular}
\begin{tabular}{lcc}
Effect  & Correction   &  Uncertainty   \\
        & ($\times 10^{16}$)   &  ($\times 10^{16}$)   \\
\hline
*Lattice Stark shifts  & $-$6.5 & 0.5   \\
*Lattice hyperpolarizability & & \\
Stark shifts  & $-$0.2 & 0.2   \\
*BBR shifts  & 52.1 & 1.0   \\
*Probe laser Stark shifts  & 0.2 & 0.1   \\
1st order Zeeman  & 0.2 & 0.2   \\
2nd order Zeeman  & 0.2 & 0.02   \\
Collisional shift & 8.9 & 0.8   \\
Line pulling & 0.0 & 0.2   \\
Servo error shift & 8.9 & 0.5   \\
2nd order Doppler shift& 8.9 & $< 0.01$   \\
Totals & 54.9 & 1.5   \\
\end{tabular}
\end{ruledtabular}
\end{table}

\subsubsection{Black body radiation shifts}

The BBR shift is the AC Stark shift resulting from the ambient blackbody
radiation field surrounding the atom.  The BBR energy shift of an atomic state
can be approximately calculated as \cite{porsev06a}
\begin{equation}
\Delta E = -\frac{2}{15} (\alpha \pi )^3 \alpha_0(0) T^4 (1 + \eta) \ , \label{BBR1}
\end{equation}
where $\alpha$ is the fine structure constant. The static scalar polarizability
$\alpha_0(0)$ and energy shift $\Delta E$ in Eq.~(\ref{BBR1}) are in atomic
units. In this expression, the temperature in K is multiplied by $3.166 \ 8153
\times 10^{-6}$.  This is converted to Hz by multiplying by $6.579684 \times
10^{15}$. The factor $\eta$ is a correction factor that allows for the
frequency dependence of the polarizability when the blackbody integral is
performed \cite{itano82a,porsev06b,mitroy09b}.  The factor $\eta$, referred to
as the dynamic shift, is most conveniently written as
\cite{porsev06b,mitroy09b}
\begin{equation}
\eta \approx  -\frac{40 \pi^2 T^2}{21 \alpha_d(0)} S(-4) \ .
\label{BBS2}
\end{equation}
The dynamic shift is largest when the excitation energies of the states
that make the largest contribution to the polarizabilities are small.  The
dynamic shift is largest for strontium and increases the BBR shift by
2.7$\%$ \cite{porsev06b}.

Under most circumstances, the energy shift of an atomic level by a radiation
field is dominated by the dipole component.  However, other multipoles might
make a contribution when the atomic level is part of a spin-orbit multiplet
\cite{porsev06b}. The $nsnp$ $^3$P$_J$ levels of the alkaline-earth
 atoms have relatively small energy
splittings.  The frequency shift due to magnetic dipole (M1) transitions could become important at the 10$^{-18}$ level
of accuracy. The M1 frequency shift for Sr has been estimated at $2.4 \times 10^{-5}$ Hz \cite{porsev06b}.
The frequency shifts for other alkaline earths can be estimated using the approximate
result
$\delta \nu_{\rm X} \approx \delta \nu_{\rm Sr} \delta E_{\rm Sr}(^3P^o_1-{^3}P^o_0)/\delta E_{\rm X}(^3P^o_1 - {^3}P^o_0)$
since the magnetic dipole matrix elements between the two members of the triplet show little
variation between different species.

Table \ref{BBshifts} lists the frequencies, linewidths, and blackbody radiation shifts for a number of potential
optical frequency standards from
Refs.~\cite{chwalla09a,dube07a,oskay06a,hosaka09a,tamm07a,rosenband07a,nistasd315,friebe08a,wilpers07a,campbell08a,poli08a,hachisu08a,margolis09a,santra04a,porsev04b,bigeon67a,wang07b,arora07a,mitroy08b,jiang09a,lea06a,mitroy09b,rosenband06a,porsev06b,mitroy07e,mitroy08g,dzuba10a,ye08a}.
  The polarizability difference,
$\Delta \alpha$, is negative when the upper state polarizability is smaller than the lower state polarizability.  All
BBR shifts are evaluated at 300 K.
Linewidths are converted from lifetimes, $\tau$ using $\delta \nu_{\rm nat} = 1/(2\pi\tau)$; natural linewidths are
given for fermionic isotopes for the $ns^2-nsnp$ clock transitions.
It is immediately apparent that the proposed ion clocks generally have smaller polarizability differences
than the electrically neutral atoms in the lattice clocks.

All of the proposed frequency standards, with the exception of the
Al$^+$($3s^2$-$3s3p \ ^{3}P^{o}_0$) transition have $T = 300$ K
fractional shifts of $10^{-16}$ or higher.  The frequency shifts
of the neutrals are generally larger than the singly charged ions.

\subsubsection{Optical lattice clocks}

Many of the issues that impact on the optimal choice for an optical frequency
standard are present in the proposed strontium $^1$$S$-$^3$$P$$^o_0$ optical
frequency standard \cite{boyd07a,campbell08a}. While Sr might be a desirable
atom from the perspective of practical experimentation, it is the most
susceptible to BBR shifts since the polarizability difference between the two
states is 259.8 a.u., giving a BBR shift of $2.35$ Hz at $T = 300$ K
\cite{porsev08a}.

Assuming that the polarizability difference can be determined to 0.2$\%$ accuracy, the resulting BBR uncertainty would
be $0.0047$ Hz which corresponds to a fractional uncertainty of $1 \times 10^{-17}$. Achieving such a level of
precision requires experimental determination of the five most important transitions in the oscillator strength
sum-over-states to a precision of 0.1$\%$ \cite{porsev08a}.

Another problem associated with large polarizability differences is the enhanced sensitivity with respect to variations
in temperature. A 1.0 K uncertainty in the temperature at 300 K would lead to a frequency uncertainty of $\Delta \nu =
0.064$ Hz.  The large BBR shift makes a Sr standard particularly sensitive to an imprecisely known temperature.  These
problems can be reduced by running the clock at lower temperatures.  For example, the BBR uncertainties stated above
can be reduced in size by a factor of more than 200 by maintaining the clock at liquid nitrogen temperatures.

Sensitivity to BBR fields has resulted in a proposal that mercury would be a superior candidate for an optical
frequency standard \cite{hachisu08a} despite the inconvenience of much shorter optical lattice wavelengths.  Cadmium and
zinc have also been identified as candidates with reduced BBR shifts \cite{wang07b,ye08a}.  The drawback of the group
IIB atoms are the greater uncertainties in the determination of the polarizabilities.   The underlying $(nd)^{10}$
shell of the group IIB atoms implies large core polarizabilities, stronger valence-core correlations and valence
expectation values that are slower to converge.  In addition, the resonant oscillator strength for these atoms is about
1.4, as opposed to 1.7-1.8 for the group II atoms.  Consequently the use of a high precision resonant transition matrix
element from a photo-association experiment would do less to minimize the uncertainty than in a group II atom.

Ytterbium has also been the subject of increased experimental interest \cite{poli08a}. This system also suffers from the
drawback that it has a large polarizability. Furthermore,  a first-principles calculation of the polarizability to a
guaranteed accuracy of even 1$\%$ is a very difficult proposition.  The most weakly bound core shell is the $(4f)^{14}$
shell and the Yb$^{2+}$ polarizability is $~\sim 9$ a.u. \cite{safronova09a}.

The atoms that have so far been used in most experiments, are those that are
amenable to cooling and trapping.  The lighter group II atoms, Be and Mg, have
the disadvantage that they are difficult to cool, but have the advantage of
much smaller BBR shifts \cite{friebe08a}. Further, it would be easier to
compensate for the effect of the BBR shift in Be and Mg than in most other
atoms. Besides having  smaller shifts, these are relatively light atoms with
small core polarizabilities, so the uncertainties associated with any
calculation will be smaller than those of other lattice clock.  These
considerations apply most strongly to beryllium.  In this case, the
polarizability difference of the clock atom states is only 1.8 a.u. (Table
\ref{alkaline}).  Beryllium has only four electrons, so calculations with ECGs
are possible and should ultimately be able to achieve a precision approaching
0.01 a.u.

The dynamic correction to the BBR shift makes a finite contribution when the precision reaches the $10^{-18}$ Hz level
\cite{porsev06a,jiang09a} but should not lead to a significant increase in the BBR shift uncertainty.  The sum rule for
evaluation of $S(-4)$ is more strongly dominated by a few major transitions than $\alpha_0$ and the relative
uncertainty in $S(-4)$ will not be any larger than that of $\alpha_0$.  Further, the dynamic contribution will be small
so the need for a precise evaluation is reduced.

One recent complication has been the recent realization that higher order
multipoles could have an impact upon the magic wavelength.  The inhomogenous
spatial distributions of the electric and magnetic fields in the standing wave
patterns that define the lattice can lead to energy shifts in the atomic
vibrational motion \cite{taichenachev08a}. This requires the definition of a
motion insensitive magic wavelength which requires knowledge of the frequency
dependent electric quadrupole and magnetic dipole polarizabilities
\cite{katori09a}.

\subsubsection{Ion clocks}

\begin{table*}[th]
\caption[]{ Summary of the recent theoretical calculations of the Stark shift coefficient $k$ in
$10^{-10}$~Hz/(V/m)$^2$ and the BBR radiation shift parameter $\beta$ for transitions between the ground hyperfine
states and comparison with experiment.  All BBR shifts are evaluated at 300~K. Uncertainties in the last digits are
given in parentheses. References are given in square brackets.} \label{BBRMicrowave}
\begin{ruledtabular}
\begin{tabular}{lcccc}
 Atom         & Transition                    &  Method          & $k$       &    $\beta$ \\ \hline
$^7$Li        & $2s~ (F=2 \leftrightarrow F=1)$& RLCCSDT \cite{johnson08a}          &  $-$0.05824 &
$-$0.5017$\times 10^{-14}$\\
              &            &    Expt.             \cite{mowat72a}       &  $-$0.061(2)& \\
$^{23}$Na    & $3s~ (F=2 \leftrightarrow F=1)$& RLCCSDT \cite{safronova10a}     &  $-$0.1285 &  $-$0.5019$\times
10^{-14}$\\
        &                               &Expt.             \cite{mowat72a}    &  $-$0.124(3)&
           \\ \hline
$^{39}$K       & $4s~ (F=2 \leftrightarrow F=1)$ RLCCSDT  \cite{safronova08b}           &  $-$0.0746  & $-$1.118$\times 1
                                              0^{-14}$\\
         &            &Expt.          \cite{mowat72a}       &  $-$0.071(2)&
          \\ \hline
$^{87}$Rb      & $5s~ (F=2 \leftrightarrow F=1)$& RLCCSDT  \cite{safronova10b}    &  $-$1.272*  & $-$$1.287 \times 10^{-14}$\\
               &                               & RCI+MBPT   \cite{angstmann06b}      &  $-$1.24(1) &  $-$1.26(1)$\times 10^{-14}$              \\
         &      &Expt.             \cite{mowat72a}       &  $-$1.23(3) &        \\
$^{133}$Cs     & $6s~ (F=4 \leftrightarrow F=3)$& RLCCSDT  \cite{beloy06a}       &  $-$2.271(8)& $-$1.710(6)$\times 10^{-14}$\\
              &             &Theory, PTSCI   \cite{angstmann06a}  &  $-$2.26(2) &  $-$1.70(2)$\times 10^{-14}$\\
              &                               &Expt.            \cite{simon98a} &  $-$2.271(4)& $-$1.710(3)$\times 10^{-14}$\\
  &              &Expt.           \cite{godone05a}&  $-$2.05(4) &  $-$1.54(4)$\times 10^{-14}$\\ \hline
$^{137}$Ba$^+$ & $6s~ (F=2 \leftrightarrow F=1)$&PTSCI  \cite{angstmann06b}      &  $-$0.284(3)& $-$0.245(2)$\times 10^{-14}$\\
$^{171}$Yb$^+$ & $6s~ (F=1 \leftrightarrow F=0)$&RMBPT3  \cite{safronova09a}      &  $-$0.1796  &   $-$0.0983$\times 10^{-14}$\\
        &     &Theory, PTSCI \cite{angstmann06b}      &  $-$0.171(9)& $-$0.094(5)$\times 10^{-14}$\\
                                                                           \hline
$^{199}$Hg$^+$  & $6s~ (F=1 \leftrightarrow F=0)$&PTSCI \cite{angstmann06b}   &  $-$0.060(3)& $-$0.0102(5)$\times 10^{-14}$ \\
\end{tabular}
\end{ruledtabular}
*Preliminary value
\end{table*}

Ion state polarizabilities are generally smaller than those for neutral atoms
because
 the electrons are more tightly bound.
None of the ion clocks have polarizability differences that exceed 50 a.u..

The Al$^+$($3s^2$-$3s3p \ ^{3}P^{o}_0$) transition has the smallest BBR shift of any ion clock due to the fortuitous
near equality of polarizabilities of the two states in the clock transition.  The CICP BBR shift is only $-$0.0042(32) Hz
\cite{mitroy09b} while experiment gave $-$0.008(3) Hz \cite{rosenband06a}. However, the technical requirements for
construction of an Al$^+$ clock are much more demanding since the clock transition and cooling laser are in the
ultraviolet \cite{rosenband07a}.

While the Ca$^+$ system is monovalent, calculation of its polarizabilities using RLCCSDT method
 is a more difficult
proposition than for the iso-electronic neutral potassium \cite{arora07a}. The
difficulties lie in the determination of the  $3d$ state polarizability. First,
the $3d$ state is quite compact and its charge distribution does perturb the
charge distribution of the $3s^23p^6$ core.  This leads to a more slowly
convergent perturbation theory or CI expansion. Second, about 30$\%$ of the
polarizability comes from the $3d \to nf$ excitations.  The sum-over-states in
this case is not dominated by a single transition, so discrete excitations up
to the $12f$ have to be included. Furthermore, the continuum contribution is
significant. Including states up to $12f$ means a much larger $B$-spline basis
needs to be used, which in turn makes the calculation more exacting. Similar
considerations impact the BBR shift calculation for  Sr$^+$  \cite{jiang09a}.

The relative uncertainties associated with the determination of the BBR shifts for the two proposed Yb$^+$ standards
are also large due to the underlying $4f^{14}$ core.  This is partly mitigated by the small size of the BBR shifts. The
BBR shift for In$^+$ was determined by assuming the polarizability difference would be smaller than that of cadmium.
This estimate will be an overestimate since the In$^+$ ion will have smaller polarizabilities than cadmium.

\subsubsection{Experimental possibilities}

So far discussions have focused largely on theory-based approaches to the
determination of the relevant polarizabilities.  However, experimental avenues
do exist.  For example, the polarizability of the Si$^{2+}$ ground state, an
ion iso-electronic with Al$^+$ has been determined by RESIS to an accuracy of
better than 0.1$\%$.   A RESIS experiment on Al$^+$ should be able to achieve a
similar precision.  Similarly, a RESIS experiment should be able to determine
the In$^+$ ground state polarizability to an accuracy of 0.1$\%$.  However, an
improved theoretical analysis would be needed to get RESIS polarizabilities for
Ca$^+$, Sr$^+$ and Yb$^+$. Application of RESIS to excited parent ions also
remains a challenge.

The  actual knowledge of the ground and excited state polarizabilities is
 mainly important because it enables the
determination of the BBR Stark shift.  Direct Stark shift experiments, on the
other hand, might ultimately give the most accurate polarizability differences.
Table \ref{starkalkali} shows that experiments on the $ns$-$np_{1/2}$
transitions of the alkali atoms have yielded polarizability differences with
uncertainties less than 0.1 a.u..

Photo-association (PA) experiment lifetimes have been utilized in estimating
polarizabilities with sub-$1\%$ precision. However, PA spectroscopy has never
been applied to measure transitions to an excited state,  and  excited state
polarizabilities have significant contributions from more than one transition.

\subsubsection{BBR shifts in microwave frequency standards}
\label{BBR-mic}
A BBR shift also exists for the different hyperfine states involved in microwave frequency standards. In the case of
the optical transitions, the lowest (second) order polarizabilities of the clock states are different. In the case of
the ground-state hyperfine  microwave frequency standards, the lowest (second) order polarizabilities of the clock
states are identical and the lowest-order BBR shift vanishes. To evaluate the BBR shift,  third-order $F$-dependent
polarizabilities must be calculated.

 The third-order
$F$-dependent ($F$ is the angular momentum of the hyperfine state) static polarizability, $\alpha_F$ can be written
\cite{beloy06a}
\begin{equation}
\alpha_F = A g_I \mu_n (2T + C + R),
\end{equation}
where $A$ is an angular coefficient, $G_I$ is the nuclear gyromagnetic ratio,
and $\mu_n$ is the nuclear magneton. The quantities $T$, $C$ and $R$ arise from
third-order perturbation theory and typically involve two electric-dipole
matrix elements $\langle i\|D\|j \rangle$ and a one matrix element involving
the  magnetic hyperfine operator $\mathcal{T}^{(1)}$. For example, term $T$ is
given by \cite{beloy06a}
\begin{equation}
T = \sum_{m\neq v} \sum_{n\neq v} A_1 \delta_{j_n j_v} \frac{\langle v \|D \| m\rangle \langle m \|D \| n\rangle
\langle n \| \mathcal{T}^{(1)}\| v\rangle}{\left( E_m - E_v\right)\left( E_n - E_v\right)}.
 \nonumber
\end{equation}
  Here, $A_1$ is the angular coefficient and sums over $m$, $n$ run over all possible states allowed
  by the selection rules.

 The BBR shift at room temperature
effecting the Cs microwave frequency standard has been calculated to high accuracy (0.35\% and 1\%) in
Refs.~\cite{beloy06a,angstmann06a}, respectively, implying a $6\times10^{-17}$ fractional uncertainty. These
calculations are in agreement with a 0.2\% measurement \cite{simon98a}.

A summary of recent theoretical calculations
\cite{johnson08a,safronova10a,safronova10b,angstmann06b,beloy06a,safronova09a}
of the Stark shift coefficient $k$ in $10^{-10}$~Hz/(V/m)$^2$ and the BBR
radiation shift parameter $\beta$ for transitions between the ground hyperfine
states and comparison with experiment \cite{mowat72a,simon98a} is given in
Table~\ref{BBRMicrowave}.  All BBR shifts are evaluated at 300~K. The Stark
coefficient $k$ is defined as
\begin{equation}
\delta \nu =k E^2, \label{ee3}
\end{equation}
where $\delta \nu$ is the frequency shift in the static electric field. The Stark coefficient for the transition
between states $F$ and $I$ is related to the polarizability as
\begin{equation}
 k = -\frac{1}{2}[\alpha_0(F)-\alpha_0(I)].
 \label{ee4}
\end{equation}
The parameter $\beta$ of the relative temperature-dependent BBR shift of the microwave frequency standard
 is  defined as
\begin{equation}
\frac{\delta \nu}{\nu_0} = \beta \left( \frac{T(K)}{T_0} \right)^4 \left(1+\epsilon
\left(\frac{T(K)}{T_0}\right)^2\right),
\end{equation}
where $T_0$ is generally taken to be room temperature, $300K$, $\epsilon$ parameterizes the lowest-order (in T)
contribution to the dynamic correction $\eta$ in Eq.~(\ref{BBR1}), and $\nu_0$ is clock transition frequency. The
parameter $\beta$ is calculated directly from the Stark-shift coefficient $k$ defined by Eqs.~(\ref{ee3}-\ref{ee4}) as
\begin{equation}
\beta = \frac{k}{\nu_0} \left( 831.9~\textrm{V/m} \right)^2.
\end{equation}

\begin{table*}[th]
\caption[]{ The lowest order dispersion coefficient, $C_6$ for homo-nuclear atom-atom pairs.  The Hybrid-RLCCSD
 replaces the calculated matrix element for the resonance transition
with an experimental value.  } \label{C6}
\begin{ruledtabular}
\begin{tabular}{lcccccc}
Method & $^{\infty}$Li  &  Na  & K  &  Rb  & Cs    &Fr \\ \hline
Hylleraas \cite{tang09a} & 1393.42(5) &     &   &   & &  \\
Model Potential \cite{marinescu94a} & 1388 & 1472 & 3813 & 4426 & 6331& \\
CICP \cite{mitroy03f} & 1394.6 & 1561 & 3905 & 4635  & & \\
RLCCSD \cite{derevianko99a} &    &  1564 & 3867 & 4628 &  6899  &  5174 \\
Hybrid-RLCCSD \cite{derevianko99a,derevianko01a} & 1390(2) & 1556(4) & 3897(15) & 4691(23) & 6851(74) & 5256(89)\\
Expt.                                           &      &      &  3921  \cite{pashov08a}  &  4698(4) \cite{vankempen02a}  &  6877(24)  \cite{amini03a} &     \\
Expt.                                           &      &      &    &   &  6860(25)\cite{tiesinga04} &     \\
\end{tabular}
\end{ruledtabular}
\end{table*}

\subsection{Long-range interatomic potentials}

The long-range dispersion interaction between two spherically
symmetric atoms has the form
\begin{equation}
V_{\rm disp} = - \frac{C_6}{R^6} - \frac{C_8}{R^8} - \frac{C_{10}}{R^{10}} \ldots
\label{dispersion}
\end{equation}
where the $C_n$ coefficients are called the dispersion coefficients. The calculation of the dispersion interaction is
closely related to polarizability calculations.  For example, the $C_6$ parameter for two atoms, $a$ and $b$, in states
$m$ and $n$, can be evaluated using the  oscillator strengths as
\begin{equation}
C_6 = \frac{3}{2} \sum_{ij} \frac{f_{mi}f_{nj}}
{\Delta E_{mi}\Delta E_{nj}(\Delta E_{mi}+\Delta E_{nj})}  \ .
\label{C6v1}
\end{equation}
The equation is reminiscent of Eq.~(\ref{alpha1}) and any calculation using Eq.~(\ref{C6v1}) automatically generates
the necessary information to generate the dipole polarizability.  The dispersion coefficients can also be directly
evaluated from the polarizability at imaginary frequencies as
\begin{equation}
C_6 = \frac{3}{2} \int_{0}^{\infty} \alpha_{a,0}(i\omega) \ \alpha_{b,0}(i\omega) \ d\omega. \label{C6v2}
\end{equation}
The polarizability of state $n$ at imaginary frequencies is written
\begin{equation}
\alpha_0(i\omega) = \sum_{i} \frac{f_{ni}}{(\Delta E^2_{ni}+\omega^2)}. \label{alphaiw}
\end{equation}

Equation (\ref{dispersion}) and subsequent expressions given by
Eqs.~(\ref{C6v1}), or (\ref{C6v2}) are the best way to evaluate long-range
atom-atom interactions.  Orthodox quantum chemistry techniques are not well
suited to determining the very small energies of the long range potential.

The importance of a good description of the long-range atom-atom interaction
increases at very low energies.  Determination of the dissociation energy
for many molecules often involves an extrapolation from the rovibrational
energy levels of the highest vibrational states \cite{leroy71a}.  This has
been accomplished in the semi-classical (WKB) LeRoy-Bernstein procedure
\cite{leroy71a}.  Similarly, the determination of the scattering length in
cold-atom collisions often requires knowledge of the dispersion parameters
\cite{marte02a}.

Better information about the specific values of the dispersion coefficients for
many atoms has become available primarily because of the importance of such
data for the field of cold-atom physics.  There have been the near exact
non-relativistic calculation by Yan and co-workers on H, He and Li using
Hylleraas basis sets
\cite{yan96a,yan00a,zhang05a,zhang06a,zhang06b,zhang07b,tang09a}. An important
series of calculations on the ground and excited states of the alkali atoms
were reported by Marinescu and co-workers
\cite{marinescu94a,marinescu95a,marinescu96a,marinescu97b,marinescu99a}.
However, these calculations were performed with a model potential approach that
omitted some dynamical features (e.g. transitions from the core) that should be
included.  Later calculations with semi-empirical Hamiltonians by Mitroy and
co-workers
\cite{mitroy03f,zhang07a,mitroy07d,mitroy07e,zhang07c,mitroy08a,mitroy08g} and
RLCCSD/RCI+MBPT calculations by Derevianko and co-workers
\cite{derevianko99a,safronova99a,porsev02a,porsev03a,zhu04a,porsev06a} should
be preferred since the underlying atomic structure descriptions are superior.
These calculations encompass both the alkali and alkaline-earth atoms.  Table
\ref{C6} shows that CICP and RLCCSDT calculations of $C_6$ for homo-nuclear
pairs of alkali atoms agree at the 1$\%$ level.  This agreement extends to
hetero-nuclear pairs of alkali atoms \cite{mitroy03f} and to alkaline-earth
atoms \cite{porsev06a}.

\subsection{Thermometry and other macroscopic standards}
\label{thermometry}

 The present definition of temperature is based on the triple point of water which is set to 273.16
K.  An alternative approach would be to fix the Boltzmann constant, $k_B$ and then measure the thermometric properties
of a substance which depend on the product $k_BT$.  At present, the best estimate of the Boltzmann constant was
determined by the speed of sound in helium gas.  Acoustic gas thermometry (AGT) has resulted in a value of $k_B$
accurate to 1.8 ppm \cite{mohr05a,fellmuth06a}.

The speed of sound is not the only thermometric property that can be used to
determine $k_B$.  Two other properties are the dielectric constant for helium
gas and the refractive index for helium gas
\cite{stone04a,fellmuth06a,schmidt07a}. The most recent refractive index
experiment using a microwave cavity \cite{schmidt07a} has given the dipole
polarizability to an accuracy of 9.3 ppm.  If the $^4$He polarizability is
taken as a known quantity from theory, then the microwave cavity experiment
admits other interpretations. Taking the polarizability and diamagnetic
susceptibility as known quantities, the refractive index experiment yields a
value for the universal gas constant, $R = 8.314487(76)$, which is not far
removed in precision from the recommended value of 8.314 472(15)
\cite{mohr05a}. Boltzmann's constant, the definition of the mol and the
universal gas constant are all inter-related through the identity, $R = k_B
N_A$.

\subsection{Atomic transition rate determinations}

The sum-over-states approach described in Section~\ref{sumoverstates} is generally used to determine the
polarizabilities from  calculated or experimental oscillator strengths or E1 matrix elements. It is possible to reverse
the process for systems which have a precisely known polarizability that is dominated by a single strong transition.  A
good example occurs for the cesium atom \cite{derevianko02a} where the dipole polarizability \cite{amini03a} and line
strength ratio \cite{rafac98a} have been measured to high accuracy.

The ground state static polarizability $\alpha_0$ can be written as
\begin{equation}
\alpha_0 = \alpha_{6p} + \alpha_{v}^{\prime} + \alpha_{\rm core} ,
\end{equation}
where $\alpha_{6p}$ is the contribution of the resonance excitations to the polarizability, i.e. from $6s-6p_{1/2}$ and
$6s-6p_{3/2}$ transitions, and  $\alpha_{\rm v}^{\prime}$ includes contributions from all other excited states .
Rearranging and expressing $\alpha_{6p}$ in terms of the $6s \to 6p_{1/2}$ line strength gives
\begin{equation}
S_{6s  -  6p_{1/2}} =  \frac{  \alpha_0-\alpha_{v}^{\prime} - \alpha_{\rm core}  } { \displaystyle{ \frac{1}{3\Delta
E_{6s\!-\!6p_{1/2}}} + \frac{R}{3\Delta E_{6s\!-\!6p_{3/2}}}   } }  \ . \label{S6p}
\end{equation}
The factor $R$ is the ratio of the line strengths of the spin-orbit doublet. Using $\alpha_0 = 401.0(6)$
\cite{amini03a}, $R = 1.9809(9)$ \cite{rafac98a}, ionic core polarizability of $15.644(4)$ \cite{zhou89a} that needs to
be corrected for the presence of the valence electron by the term  $\alpha_{cv} = -0.72$, $\alpha_{v}^{\prime} = 1.81$
 yields $S_{6s - 6p_{1/2}} = 20.308(42)$ and $S_{6s - 6p_{3/2}} = 40.227(84)$. The corresponding values for the
reduced  matrix elements in atomic units are 4.510(4) and 6.347(5) for the
$6s-6p_{1/2}$ and $6s-6p_{3/2}$ transitions respectively. The uncertainties of
these values are dominated by the uncertainty in the experimental value of
$\alpha_0$.

A similar approach has been used to determine the multiplet strengths for the resonance transitions in Mg$^+$,
Si$^{3+}$ \cite{mitroy09a} and Si$^{2+}$ \cite{mitroy08k} from RESIS experimental data \cite{snow08a,komara05a}.

Stark shifts for the $ns$-$np_{1/2}$ transition \cite{miller94a} have also been used to derive estimates for the
$S(np_{1/2}-(n\!-\!1)d_{3/2})$ line strengths with a precision of about 1$\%$ for potassium and rubidium \cite{arora07b}.
This analysis relied on the result that 80-90$\%$ of the $np_{1/2}$ polarizability comes from the excitation to the
$(n-1)d_{3/2}$ state. These values were also used to determine the magic wavelengths for the $np-ns$
transitions in these alkali
atoms \cite{arora07c}. Such determination of matrix elements permitted benchmark comparisons of
theory and experiment \cite{arora07b}.

The procedures described above also permit the crosschecking of results from completely different types of experiment.
The domination of the $6p$ Cs scalar polarizabilities by the $5d-6p$ dipole matrix elements facilitated an exacting
consistency check of the $5d$ lifetime with $6p$ polarizability data \cite{safronova04b}. In that work, $5d-6p$ matrix
elements obtained from experimental Stark shift data were compared with the values extracted from the $5d$ lifetimes.
The experimental measurements of the $5d$ lifetime and $6$p scalar polarizabilities were found to be inconsistent
within the uncertainties quoted by the experimental groups \cite{safronova04b}. Theoretical RLCCSDT matrix
elements~\cite{safronova04b} were found to be in agreement with the Stark shift experiments but not with the lifetime
measurements.

\section{Conclusions}

The advent of cold-atom physics owes its existence to the ability to
manipulate groups of atoms with electromagnetic fields.  Consequently,
many topics in the area of field-atom interactions have recently been
the subject of considerable interest and heightened importance.  This
applies to a quantity like the dipole polarizability which governs the
first-order response of an atom to an applied electric field and the
preceding few years have seen many calculations of atomic
polarizabilities for a variety of systems.

The aim of the present review has been to provide a reasonably comprehensive
treatment of polarizability related issues as they relate to topics of
contemporary importance.  However, our treatment is not exhaustive. The
polarizabilities of many atoms such as the halogens have been omitted.  The
reader is referred to the broader treatment in \cite{schwerdtfeger06a}.
Similarly, the treatment of DC and AC Stark shift data is better described as
selective as opposed to exhaustive.

Part of the motivation for this review has been the importance in developing
new atomic based standards of time \cite{BIPM2006}, and corresponding  need for
precise knowledge of the blackbody radiation shifts. The
 primary requirement for the BBR
application is for polarizabilities and Stark shifts to be known with a
precision of 0.1$\%$ or better. Much of the existing body of experimental data
is an order of magnitude less precise. Direct measurements of clock transition
Stark shifts would be helpful in reducing the BBR shift uncertainties.

One area where theory might be useful in this endeavor would be in the
development of atom based polarizability standards.  Such a standard is already
in existence for helium where theoretical and experimental polarizabilities
have uncertainties of 0.17 ppm and 9.1 ppm \cite{schmidt07a}, respectively.
These results are not relevant to atomic clock research and another atom needs
to serve as a standard. Hylleraas calculations on lithium could yet serve to
provide a theoretical reference point for Stark shift experiments.  At the
moment the uncertainties in the best calculation and best experiment are $0.11
\%$ \cite{tang10a} and $0.07 \%$ \cite{hunter91a}.
 A better treatment of relativistic
effects should result in uncertainty in the Hylleraas calculation decreasing
to the 0.01$\%$ level of precision.

One possible avenue for improvement could be in the development of hybrid
theoretical approaches combining the best features of different methods. For
example, orbital-based approaches cannot match the extreme accuracies
achievable with correlated basis sets.  Direct incorporation of the
 the Dirac Hamiltonian in orbital-based
calculations is now relatively routine, but this is not the case for
calculations with correlated basis sets. Perhaps, comparisons of correlated
basis calculations with non-relativistic orbital-based calculations and with
relativistic orbital-based calculations could be used to estimate relativistic
corrections to Hylleraas calculations or correlation corrections to
orbital-based calculations.

It is likely that the determination of polarizabilities will become
increasingly important in the future.  As experiments become capable of greater
precision, it will become necessary to make more detailed corrections of the
effects of electromagnetic fields that are used for  manipulation and
investigation of atoms.

\section{Acknowledgments}
 The authors express their appreciation to Dr. U. I. Safronova for useful
  discussions and comments on the manuscript, and
 in particularly for her  thorough check of most of the tables.
This work was partly supported by the National Science
Foundation under Physics Frontiers Center grant PHY-0822671 to the University
of Maryland.
 This research has made extensive use of NASA's Astrophysics Data System.
 The work of MSS was
supported in part by US National Science Foundation Grant  No.\ PHY-07-58088.
The work of JM was supported in part by the Australian Research Council
Discovery Project DP-1092620.  This research was performed in part under the
sponsorship of the US Department of Commerce, National Institute of Standards
and Technology.

\begin{thebibliography}{383}
\expandafter\ifx\csname natexlab\endcsname\relax\def\natexlab#1{#1}\fi
\expandafter\ifx\csname bibnamefont\endcsname\relax
  \def\bibnamefont#1{#1}\fi
\expandafter\ifx\csname bibfnamefont\endcsname\relax
  \def\bibfnamefont#1{#1}\fi
\expandafter\ifx\csname citenamefont\endcsname\relax
  \def\citenamefont#1{#1}\fi
\expandafter\ifx\csname url\endcsname\relax
  \def\url#1{\texttt{#1}}\fi
\expandafter\ifx\csname urlprefix\endcsname\relax\def\urlprefix{URL }\fi
\providecommand{\bibinfo}[2]{#2} \providecommand{\eprint}[2][]{\url{#2}}

\bibitem[{\citenamefont{Maxwell}(1864)}]{maxwell1864a}
\bibinfo{author}{\bibfnamefont{J.~C.} \bibnamefont{Maxwell}},
  \bibinfo{journal}{Phil.~Trans.~R.~Soc. (London)}
  \textbf{\bibinfo{volume}{155}}, \bibinfo{pages}{459} (\bibinfo{year}{1864}).

\bibitem[{\citenamefont{Edmonds}(1996)}]{edmonds96}
\bibinfo{author}{\bibfnamefont{A.~R.} \bibnamefont{Edmonds}},
  \emph{\bibinfo{title}{Angular Momentum in Quantum Mechanics}}
  (\bibinfo{publisher}{Princeton University Press},
  \bibinfo{address}{Princeton, NJ}, \bibinfo{year}{1996}).

\bibitem[{\citenamefont{Schr{\"o}dinger}(1926)}]{schrodinger26b}
\bibinfo{author}{\bibfnamefont{E.}~\bibnamefont{Schr{\"o}dinger}},
  \bibinfo{journal}{Annalen.~der Physik} \textbf{\bibinfo{volume}{385}},
  \bibinfo{pages}{437} (\bibinfo{year}{1926}).

\bibitem[{\citenamefont{{Madej} and {Bernard}}(2001)}]{madej01a}
\bibinfo{author}{\bibfnamefont{A.~A.} \bibnamefont{{Madej}}} \bibnamefont{and}
  \bibinfo{author}{\bibfnamefont{J.~E.} \bibnamefont{{Bernard}}}, in
  \emph{\bibinfo{booktitle}{Frequency Measurement and Control, Edited by Andre
  N. Luiten, Topics in Applied Physics, vol. 79, pp.153-195}}, edited by
  \bibinfo{editor}{\bibfnamefont{A.~N.} \bibnamefont{{Luiten}}}
  (\bibinfo{year}{2001}), pp. \bibinfo{pages}{153--195}.

\bibitem[{\citenamefont{Udem et~al.}(2002)\citenamefont{Udem, Holzwarth, and
  Hansch}}]{udem02a}
\bibinfo{author}{\bibfnamefont{T.}~\bibnamefont{Udem}},
  \bibinfo{author}{\bibfnamefont{R.}~\bibnamefont{Holzwarth}},
  \bibnamefont{and} \bibinfo{author}{\bibfnamefont{T.~W.}
  \bibnamefont{Hansch}}, \bibinfo{journal}{Nature}
  \textbf{\bibinfo{volume}{416}}, \bibinfo{pages}{233} (\bibinfo{year}{2002}).

\bibitem[{\citenamefont{{Diddams} et~al.}(2004)\citenamefont{{Diddams},
  {Bergquist}, {Jefferts}, and {Oates}}}]{diddams04a}
\bibinfo{author}{\bibfnamefont{S.~A.} \bibnamefont{{Diddams}}},
  \bibinfo{author}{\bibfnamefont{J.~C.} \bibnamefont{{Bergquist}}},
  \bibinfo{author}{\bibfnamefont{S.~R.} \bibnamefont{{Jefferts}}},
  \bibnamefont{and} \bibinfo{author}{\bibfnamefont{C.~W.}
  \bibnamefont{{Oates}}}, \bibinfo{journal}{Science}
  \textbf{\bibinfo{volume}{306}}, \bibinfo{pages}{1318} (\bibinfo{year}{2004}).

\bibitem[{\citenamefont{{Gill} et~al.}(2003)\citenamefont{{Gill}, {Barwood},
  {Klein}, {Huang}, {Webster}, {Blythe}, {Hosaka}, {Lea}, and
  {Margolis}}}]{gill03a}
\bibinfo{author}{\bibfnamefont{P.}~\bibnamefont{{Gill}}},
  \bibinfo{author}{\bibfnamefont{G.~P.} \bibnamefont{{Barwood}}},
  \bibinfo{author}{\bibfnamefont{H.~A.} \bibnamefont{{Klein}}},
  \bibinfo{author}{\bibfnamefont{G.}~\bibnamefont{{Huang}}},
  \bibinfo{author}{\bibfnamefont{S.~A.} \bibnamefont{{Webster}}},
  \bibinfo{author}{\bibfnamefont{P.~J.} \bibnamefont{{Blythe}}},
  \bibinfo{author}{\bibfnamefont{K.}~\bibnamefont{{Hosaka}}},
  \bibinfo{author}{\bibfnamefont{S.~N.} \bibnamefont{{Lea}}}, \bibnamefont{and}
  \bibinfo{author}{\bibfnamefont{H.~S.} \bibnamefont{{Margolis}}},
  \bibinfo{journal}{Meas.~Sci.~Technol.} \textbf{\bibinfo{volume}{14}},
  \bibinfo{pages}{1174} (\bibinfo{year}{2003}).

\bibitem[{\citenamefont{{Gill}}(2005)}]{gill05a}
\bibinfo{author}{\bibfnamefont{P.}~\bibnamefont{{Gill}}},
  \bibinfo{journal}{Metrologia} \textbf{\bibinfo{volume}{42}},
  \bibinfo{pages}{S125} (\bibinfo{year}{2005}).

\bibitem[{\citenamefont{Margolis}(2009)}]{margolis09a}
\bibinfo{author}{\bibfnamefont{H.~S.} \bibnamefont{Margolis}},
  \bibinfo{journal}{J.~Phys.~B} \textbf{\bibinfo{volume}{41}},
  \bibinfo{pages}{154017} (\bibinfo{year}{2009}).

\bibitem[{\citenamefont{{Gallagher} and {Cooke}}(1979)}]{gallagher79a}
\bibinfo{author}{\bibfnamefont{T.~F.} \bibnamefont{{Gallagher}}}
  \bibnamefont{and} \bibinfo{author}{\bibfnamefont{W.~E.}
  \bibnamefont{{Cooke}}}, \bibinfo{journal}{Phys.~Rev.~Lett.}
  \textbf{\bibinfo{volume}{42}}, \bibinfo{pages}{835} (\bibinfo{year}{1979}).

\bibitem[{\citenamefont{Itano et~al.}(1982)\citenamefont{Itano, Lewis, and
  Wineland}}]{itano82a}
\bibinfo{author}{\bibfnamefont{W.~M.} \bibnamefont{Itano}},
  \bibinfo{author}{\bibfnamefont{L.~L.} \bibnamefont{Lewis}}, \bibnamefont{and}
  \bibinfo{author}{\bibfnamefont{D.~J.} \bibnamefont{Wineland}},
  \bibinfo{journal}{Phys. Rev. A} \textbf{\bibinfo{volume}{25}},
  \bibinfo{pages}{1233} (\bibinfo{year}{1982}).

\bibitem[{\citenamefont{{Hollberg} and {Hall}}(1984)}]{hollberg84a}
\bibinfo{author}{\bibfnamefont{L.}~\bibnamefont{{Hollberg}}} \bibnamefont{and}
  \bibinfo{author}{\bibfnamefont{J.~L.} \bibnamefont{{Hall}}},
  \bibinfo{journal}{Phys.~Rev.~Lett.} \textbf{\bibinfo{volume}{53}},
  \bibinfo{pages}{230} (\bibinfo{year}{1984}).

\bibitem[{\citenamefont{Porsev and Derevianko}(2006{\natexlab{a}})}]{porsev06b}
\bibinfo{author}{\bibfnamefont{S.~G.} \bibnamefont{Porsev}} \bibnamefont{and}
  \bibinfo{author}{\bibfnamefont{A.}~\bibnamefont{Derevianko}},
  \bibinfo{journal}{Phys.~Rev.~A} \textbf{\bibinfo{volume}{74}},
  \bibinfo{pages}{020502(R)} (\bibinfo{year}{2006}{\natexlab{a}}).

\bibitem[{\citenamefont{Dalgarno}(1962)}]{dalgarno62a}
\bibinfo{author}{\bibfnamefont{A.}~\bibnamefont{Dalgarno}},
  \bibinfo{journal}{Advan.~Phys.} \textbf{\bibinfo{volume}{11}},
  \bibinfo{pages}{281} (\bibinfo{year}{1962}).

\bibitem[{\citenamefont{{Teachout} and {Pack}}(1971)}]{teachout71a}
\bibinfo{author}{\bibfnamefont{R.~R.} \bibnamefont{{Teachout}}}
  \bibnamefont{and} \bibinfo{author}{\bibfnamefont{R.~T.}
  \bibnamefont{{Pack}}}, \bibinfo{journal}{Atomic Data}
  \textbf{\bibinfo{volume}{3}}, \bibinfo{pages}{195} (\bibinfo{year}{1971}).

\bibitem[{\citenamefont{Miller and Bederson}(1977)}]{miller77a}
\bibinfo{author}{\bibfnamefont{T.~M.} \bibnamefont{Miller}} \bibnamefont{and}
  \bibinfo{author}{\bibfnamefont{B.}~\bibnamefont{Bederson}},
  \bibinfo{journal}{Adv.~At.~Mol.~Phys.} \textbf{\bibinfo{volume}{13}},
  \bibinfo{pages}{1} (\bibinfo{year}{1977}).

\bibitem[{\citenamefont{Miller}(1988)}]{miller88a}
\bibinfo{author}{\bibfnamefont{T.~M.} \bibnamefont{Miller}},
  \bibinfo{journal}{Adv.~At.~Mol.~Phys.} \textbf{\bibinfo{volume}{25}},
  \bibinfo{pages}{37} (\bibinfo{year}{1988}).

\bibitem[{\citenamefont{van Wijngaarden W~A}(1996)}]{wijngaarden96a}
\bibinfo{author}{\bibnamefont{van Wijngaarden W~A}},
  \bibinfo{journal}{Adv.~At.~Mol.~Opt.~Phys.} \textbf{\bibinfo{volume}{36}},
  \bibinfo{pages}{141} (\bibinfo{year}{1996}).

\bibitem[{\citenamefont{{Bonin} and Kresin}(1997)}]{bonin97a}
\bibinfo{author}{\bibfnamefont{K.~D.} \bibnamefont{{Bonin}}} \bibnamefont{and}
  \bibinfo{author}{\bibfnamefont{V.~V.} \bibnamefont{Kresin}},
  \emph{\bibinfo{title}{Electric dipole polarizabilities of atoms, molecules
  and clusters}} (\bibinfo{publisher}{World Scientific},
  \bibinfo{address}{Singapore}, \bibinfo{year}{1997}).

\bibitem[{\citenamefont{{Delone} and {Krainov}}(1999)}]{delone99a}
\bibinfo{author}{\bibfnamefont{N.~B.} \bibnamefont{{Delone}}} \bibnamefont{and}
  \bibinfo{author}{\bibfnamefont{V.~P.} \bibnamefont{{Krainov}}},
  \bibinfo{journal}{Physics Uspekhi} \textbf{\bibinfo{volume}{42}},
  \bibinfo{pages}{669} (\bibinfo{year}{1999}).

\bibitem[{\citenamefont{Schwerdtfeger}(2006)}]{schwerdtfeger06a}
\bibinfo{author}{\bibfnamefont{P.}~\bibnamefont{Schwerdtfeger}},
  \emph{\bibinfo{title}{Atomic Static Dipole Polarizabilities}},
  chap.~\bibinfo{chapter}{1}, p.~\bibinfo{pages}{1}, in  \cite{maroulis06a}
  (\bibinfo{year}{2006}).

\bibitem[{\citenamefont{Gould and Miller}(2005)}]{gould05a}
\bibinfo{author}{\bibfnamefont{H.}~\bibnamefont{Gould}} \bibnamefont{and}
  \bibinfo{author}{\bibfnamefont{T.~M.} \bibnamefont{Miller}},
  \bibinfo{journal}{Adv.~At.~Mol.~Opt.~Phys.} \textbf{\bibinfo{volume}{51}},
  \bibinfo{pages}{343} (\bibinfo{year}{2005}).

\bibitem[{\citenamefont{Lundeen}(2005)}]{lundeen05a}
\bibinfo{author}{\bibfnamefont{S.~R.} \bibnamefont{Lundeen}},
  \bibinfo{journal}{Adv.~At.~Mol.~Opt.~Phys.} \textbf{\bibinfo{volume}{52}},
  \bibinfo{pages}{161} (\bibinfo{year}{2005}).

\bibitem[{\citenamefont{Miller}(2007)}]{miller07a}
\bibinfo{author}{\bibfnamefont{T.~M.} \bibnamefont{Miller}},
  \emph{\bibinfo{title}{Atomic and Molecular Polarizabilities}},
  chap.~\bibinfo{chapter}{10}, pp. \bibinfo{pages}{10--192},
  vol.~\bibinfo{volume}{88} of  \cite{crc07a} (\bibinfo{year}{2007}).

\bibitem[{\citenamefont{Lupinetti and Thakkar}(2006)}]{lupinetti06b}
\bibinfo{author}{\bibfnamefont{C.}~\bibnamefont{Lupinetti}} \bibnamefont{and}
  \bibinfo{author}{\bibfnamefont{A.~J.} \bibnamefont{Thakkar}}, in
  \cite{maroulis06a}, p. \bibinfo{pages}{505}.

\bibitem[{\citenamefont{Gallagher}(2005)}]{gallagher05a}
\bibinfo{author}{\bibfnamefont{T.~F.} \bibnamefont{Gallagher}},
  \emph{\bibinfo{title}{Rydberg Atoms}} (\bibinfo{publisher}{Cambridge
  University Press}, \bibinfo{address}{Cambridge}, \bibinfo{year}{2005}).

\bibitem[{\citenamefont{Fuhr and Wiese}(1995)}]{fuhr95a}
\bibinfo{author}{\bibfnamefont{J.~R.} \bibnamefont{Fuhr}} \bibnamefont{and}
  \bibinfo{author}{\bibfnamefont{W.~L.} \bibnamefont{Wiese}},
  \emph{\bibinfo{title}{Atomic Transition Probabilities}}
  (\bibinfo{publisher}{CRC Press}, \bibinfo{address}{Boca Raton, Florida},
  \bibinfo{year}{1995}), vol.~\bibinfo{volume}{76},
  chap.~\bibinfo{chapter}{10}, p. \bibinfo{pages}{128}.

\bibitem[{\citenamefont{Mitroy and Bromley}(2004)}]{mitroy04b}
\bibinfo{author}{\bibfnamefont{J.}~\bibnamefont{Mitroy}} \bibnamefont{and}
  \bibinfo{author}{\bibfnamefont{M.~W.~J.} \bibnamefont{Bromley}},
  \bibinfo{journal}{Phys.~Rev.~A} \textbf{\bibinfo{volume}{70}},
  \bibinfo{pages}{052503} (\bibinfo{year}{2004}).

\bibitem[{\citenamefont{Arora et~al.}(2007)\citenamefont{Arora, Safronova, and
  Clark}}]{arora07c}
\bibinfo{author}{\bibfnamefont{B.}~\bibnamefont{Arora}},
  \bibinfo{author}{\bibfnamefont{M.}~\bibnamefont{Safronova}},
  \bibnamefont{and} \bibinfo{author}{\bibfnamefont{C.~W.} \bibnamefont{Clark}},
  \bibinfo{journal}{Phys. Rev. A} \textbf{\bibinfo{volume}{76}},
  \bibinfo{pages}{052509} (\bibinfo{year}{2007}).

\bibitem[{\citenamefont{{Arora}
  et~al.}(2007{\natexlab{a}})\citenamefont{{Arora}, {Safronova}, and
  {Clark}}}]{arora07b}
\bibinfo{author}{\bibfnamefont{B.}~\bibnamefont{{Arora}}},
  \bibinfo{author}{\bibfnamefont{M.~S.} \bibnamefont{{Safronova}}},
  \bibnamefont{and} \bibinfo{author}{\bibfnamefont{C.~W.}
  \bibnamefont{{Clark}}}, \bibinfo{journal}{Phys. Rev. A}
  \textbf{\bibinfo{volume}{76}}, \bibinfo{pages}{052516}
  (\bibinfo{year}{2007}{\natexlab{a}}).

\bibitem[{\citenamefont{Safronova et~al.}(1999)\citenamefont{Safronova,
  Johnson, and Derevianko}}]{safronova99a}
\bibinfo{author}{\bibfnamefont{M.~S.} \bibnamefont{Safronova}},
  \bibinfo{author}{\bibfnamefont{W.~R.} \bibnamefont{Johnson}},
  \bibnamefont{and}
  \bibinfo{author}{\bibfnamefont{A.}~\bibnamefont{Derevianko}},
  \bibinfo{journal}{Phys.~Rev.~A} \textbf{\bibinfo{volume}{60}},
  \bibinfo{pages}{4476} (\bibinfo{year}{1999}).

\bibitem[{\citenamefont{Volz and Schmoranzer}(1996)}]{volz96a}
\bibinfo{author}{\bibfnamefont{U.}~\bibnamefont{Volz}} \bibnamefont{and}
  \bibinfo{author}{\bibfnamefont{H.}~\bibnamefont{Schmoranzer}},
  \bibinfo{journal}{Phys.~Scr.} \textbf{\bibinfo{volume}{T65}},
  \bibinfo{pages}{48} (\bibinfo{year}{1996}).

\bibitem[{\citenamefont{Sansonetti et~al.}(2005)\citenamefont{Sansonetti,
  Martin, and Young}}]{NIST2}
\bibinfo{author}{\bibfnamefont{J.}~\bibnamefont{Sansonetti}},
  \bibinfo{author}{\bibfnamefont{W.}~\bibnamefont{Martin}}, \bibnamefont{and}
  \bibinfo{author}{\bibfnamefont{S.}~\bibnamefont{Young}},
  \emph{\bibinfo{title}{Handbook of basic atomic spectroscopic data}}
  (\bibinfo{year}{2005}), \bibinfo{note}{(version 1.1.2). [Online] Available:
  http://physics.nist.gov/Handbook [2007, August 29]. National Institute of
  Standards and Technology, Gaithersburg, MD}.

\bibitem[{\citenamefont{Moore}(1971)}]{moore71b}
\bibinfo{author}{\bibfnamefont{C.~E.} \bibnamefont{Moore}},
  \emph{\bibinfo{title}{Atomic Energy Levels (Chromium-Niobium NSRDS-NBS 35)}},
  vol.~\bibinfo{volume}{2} (\bibinfo{publisher}{US GPO},
  \bibinfo{address}{Washington DC}, \bibinfo{year}{1971}).

\bibitem[{\citenamefont{Lim and Schwerdfeger}(2004)}]{lim04a}
\bibinfo{author}{\bibfnamefont{I.~S.} \bibnamefont{Lim}} \bibnamefont{and}
  \bibinfo{author}{\bibfnamefont{P.}~\bibnamefont{Schwerdfeger}},
  \bibinfo{journal}{Phys.~Rev.~A} \textbf{\bibinfo{volume}{70}},
  \bibinfo{pages}{062501} (\bibinfo{year}{2004}).

\bibitem[{\citenamefont{{Lupinetti} and {Thakkar}}(2005)}]{lupinetti05a}
\bibinfo{author}{\bibfnamefont{C.}~\bibnamefont{{Lupinetti}}} \bibnamefont{and}
  \bibinfo{author}{\bibfnamefont{A.~J.} \bibnamefont{{Thakkar}}},
  \bibinfo{journal}{J.~Chem.~Phys.} \textbf{\bibinfo{volume}{122}},
  \bibinfo{pages}{044301} (\bibinfo{year}{2005}).

\bibitem[{\citenamefont{Safronova and Johnson}(2008)}]{safronova08a}
\bibinfo{author}{\bibfnamefont{M.~S.} \bibnamefont{Safronova}}
  \bibnamefont{and} \bibinfo{author}{\bibfnamefont{W.~R.}
  \bibnamefont{Johnson}}, \bibinfo{journal}{Adv.~At.~Mol.~Opt.~Phys.}
  \textbf{\bibinfo{volume}{55}}, \bibinfo{pages}{191} (\bibinfo{year}{2008}).

\bibitem[{\citenamefont{{Hibbert} et~al.}(1977)\citenamefont{{Hibbert},
  {LeDourneuf}, and {Lan}}}]{hibbert77a}
\bibinfo{author}{\bibfnamefont{A.}~\bibnamefont{{Hibbert}}},
  \bibinfo{author}{\bibfnamefont{M.}~\bibnamefont{{LeDourneuf}}},
  \bibnamefont{and} \bibinfo{author}{\bibfnamefont{V.~K.} \bibnamefont{{Lan}}},
  \bibinfo{journal}{J.~Phys.~B} \textbf{\bibinfo{volume}{10}},
  \bibinfo{pages}{1015} (\bibinfo{year}{1977}).

\bibitem[{\citenamefont{Dalgarno and Lewis}(1955)}]{dalgarno55a}
\bibinfo{author}{\bibfnamefont{A.}~\bibnamefont{Dalgarno}} \bibnamefont{and}
  \bibinfo{author}{\bibfnamefont{J.~T.} \bibnamefont{Lewis}},
  \bibinfo{journal}{Proc.~R.~Soc.~London~Ser.~A}
  \textbf{\bibinfo{volume}{233}}, \bibinfo{pages}{70} (\bibinfo{year}{1955}).

\bibitem[{\citenamefont{Epstein}(1926)}]{epstein26}
\bibinfo{author}{\bibfnamefont{P.}~\bibnamefont{Epstein}},
  \bibinfo{journal}{Phys. Rev.} \textbf{\bibinfo{volume}{28}},
  \bibinfo{pages}{695} (\bibinfo{year}{1926}).

\bibitem[{\citenamefont{Waller}(1926)}]{waller26a}
\bibinfo{author}{\bibfnamefont{I.}~\bibnamefont{Waller}}, \bibinfo{journal}{Z.
  Phys.} \textbf{\bibinfo{volume}{38}}, \bibinfo{pages}{635}
  (\bibinfo{year}{1926}).

\bibitem[{\citenamefont{Wentzel}(1926)}]{wentzel26}
\bibinfo{author}{\bibfnamefont{G.}~\bibnamefont{Wentzel}}, \bibinfo{journal}{Z.
  Phys.} \textbf{\bibinfo{volume}{38}}, \bibinfo{pages}{527}
  (\bibinfo{year}{1926}).

\bibitem[{\citenamefont{Yakhontov}(2003)}]{yakhontov03}
\bibinfo{author}{\bibfnamefont{V.}~\bibnamefont{Yakhontov}},
  \bibinfo{journal}{Phys. Rev. Lett.} \textbf{\bibinfo{volume}{91}},
  \bibinfo{pages}{093001} (\bibinfo{year}{2003}).

\bibitem[{\citenamefont{Szmytkowski and Mielewczyk}(2004)}]{szmytkowski04}
\bibinfo{author}{\bibfnamefont{R.}~\bibnamefont{Szmytkowski}} \bibnamefont{and}
  \bibinfo{author}{\bibfnamefont{K.}~\bibnamefont{Mielewczyk}},
  \bibinfo{journal}{J. Phys. B: At. Mol. Opt. Phys.}
  \textbf{\bibinfo{volume}{37}}, \bibinfo{pages}{3961} (\bibinfo{year}{2004}).

\bibitem[{\citenamefont{Szmytkowski}(2006)}]{szmytkowski06}
\bibinfo{author}{\bibfnamefont{R.}~\bibnamefont{Szmytkowski}},
  \bibinfo{journal}{Chem. Phys. Lett.} \textbf{\bibinfo{volume}{419}},
  \bibinfo{pages}{537} (\bibinfo{year}{2006}).

\bibitem[{\citenamefont{Jentschura and Haas}(2008)}]{jentschura08}
\bibinfo{author}{\bibfnamefont{U.~D.} \bibnamefont{Jentschura}}
  \bibnamefont{and} \bibinfo{author}{\bibfnamefont{M.}~\bibnamefont{Haas}},
  \bibinfo{journal}{Phys. Rev. A} \textbf{\bibinfo{volume}{78}},
  \bibinfo{pages}{042504} (\bibinfo{year}{2008}).

\bibitem[{\citenamefont{Bethe and Salpeter}(1977)}]{bethe77a}
\bibinfo{author}{\bibfnamefont{H.~A.} \bibnamefont{Bethe}} \bibnamefont{and}
  \bibinfo{author}{\bibfnamefont{E.~E.} \bibnamefont{Salpeter}},
  \emph{\bibinfo{title}{Quantum mechanics of one- and two-electron atoms}}
  (\bibinfo{publisher}{Plenum}, \bibinfo{address}{New York},
  \bibinfo{year}{1977}).

\bibitem[{\citenamefont{Fano and Cooper}(1968)}]{fano68a}
\bibinfo{author}{\bibfnamefont{U.}~\bibnamefont{Fano}} \bibnamefont{and}
  \bibinfo{author}{\bibfnamefont{J.~W.} \bibnamefont{Cooper}},
  \bibinfo{journal}{Rev.~Mod.~Phys.} \textbf{\bibinfo{volume}{40}},
  \bibinfo{pages}{441} (\bibinfo{year}{1968}).

\bibitem[{\citenamefont{Clark}(1990)}]{clark90}
\bibinfo{author}{\bibfnamefont{C.~W.} \bibnamefont{Clark}},
  \bibinfo{journal}{J. Opt. Soc. Am. B} \textbf{\bibinfo{volume}{7}},
  \bibinfo{pages}{488} (\bibinfo{year}{1990}).

\bibitem[{\citenamefont{Johnson et~al.}(1983)\citenamefont{Johnson, Kolb, and
  Huang}}]{johnson83a}
\bibinfo{author}{\bibfnamefont{W.~R.} \bibnamefont{Johnson}},
  \bibinfo{author}{\bibfnamefont{D.}~\bibnamefont{Kolb}}, \bibnamefont{and}
  \bibinfo{author}{\bibfnamefont{K.}~\bibnamefont{Huang}},
  \bibinfo{journal}{At.~Data~Nucl.~Data~Tables} \textbf{\bibinfo{volume}{28}},
  \bibinfo{pages}{333} (\bibinfo{year}{1983}).

\bibitem[{\citenamefont{Johnson et~al.}(1996)\citenamefont{Johnson, Liu, and
  Sapirstein}}]{johnson96a}
\bibinfo{author}{\bibfnamefont{W.~R.} \bibnamefont{Johnson}},
  \bibinfo{author}{\bibfnamefont{Z.~W.} \bibnamefont{Liu}}, \bibnamefont{and}
  \bibinfo{author}{\bibfnamefont{J.}~\bibnamefont{Sapirstein}},
  \bibinfo{journal}{At.~Data and Nuclear Data Tables}
  \textbf{\bibinfo{volume}{64}}, \bibinfo{pages}{279} (\bibinfo{year}{1996}).

\bibitem[{\citenamefont{Johnson et~al.}(1988)\citenamefont{Johnson, Blundell,
  and Sapirstein}}]{johnson88a}
\bibinfo{author}{\bibfnamefont{W.~R.} \bibnamefont{Johnson}},
  \bibinfo{author}{\bibfnamefont{S.~A.} \bibnamefont{Blundell}},
  \bibnamefont{and}
  \bibinfo{author}{\bibfnamefont{J.}~\bibnamefont{Sapirstein}},
  \bibinfo{journal}{Phys.~Rev.~A} \textbf{\bibinfo{volume}{37}},
  \bibinfo{pages}{307} (\bibinfo{year}{1988}).

\bibitem[{\citenamefont{{Kozlov} and {Porsev}}(1999)}]{kozlov99a}
\bibinfo{author}{\bibfnamefont{M.~G.} \bibnamefont{{Kozlov}}} \bibnamefont{and}
  \bibinfo{author}{\bibfnamefont{S.~G.} \bibnamefont{{Porsev}}},
  \bibinfo{journal}{Eur.~Phys.~J.~D} \textbf{\bibinfo{volume}{5}},
  \bibinfo{pages}{59} (\bibinfo{year}{1999}).

\bibitem[{\citenamefont{{Safronova} et~al.}(2006)\citenamefont{{Safronova},
  {Arora}, and {Clark}}}]{safronova06c}
\bibinfo{author}{\bibfnamefont{M.~S.} \bibnamefont{{Safronova}}},
  \bibinfo{author}{\bibfnamefont{B.}~\bibnamefont{{Arora}}}, \bibnamefont{and}
  \bibinfo{author}{\bibfnamefont{C.~W.} \bibnamefont{{Clark}}},
  \bibinfo{journal}{Phys.~Rev.~A} \textbf{\bibinfo{volume}{73}},
  \bibinfo{pages}{022505} (\bibinfo{year}{2006}).

\bibitem[{\citenamefont{{Katori} et~al.}(1999)\citenamefont{{Katori}, {Ido},
  and {Kuwata-Gonokami}}}]{katori99b}
\bibinfo{author}{\bibfnamefont{H.}~\bibnamefont{{Katori}}},
  \bibinfo{author}{\bibfnamefont{T.}~\bibnamefont{{Ido}}}, \bibnamefont{and}
  \bibinfo{author}{\bibfnamefont{M.}~\bibnamefont{{Kuwata-Gonokami}}},
  \bibinfo{journal}{J.~Phys.~Soc.~Japan} \textbf{\bibinfo{volume}{68}},
  \bibinfo{pages}{2479} (\bibinfo{year}{1999}).

\bibitem[{\citenamefont{Derevianko et~al.}(1999)\citenamefont{Derevianko,
  Johnson, Safronova, and Babb}}]{derevianko99a}
\bibinfo{author}{\bibfnamefont{A.}~\bibnamefont{Derevianko}},
  \bibinfo{author}{\bibfnamefont{W.~R.} \bibnamefont{Johnson}},
  \bibinfo{author}{\bibfnamefont{M.~S.} \bibnamefont{Safronova}},
  \bibnamefont{and} \bibinfo{author}{\bibfnamefont{J.~F.} \bibnamefont{Babb}},
  \bibinfo{journal}{Phys.~Rev.~Lett.} \textbf{\bibinfo{volume}{82}},
  \bibinfo{pages}{3589} (\bibinfo{year}{1999}).

\bibitem[{\citenamefont{Porsev and Derevianko}(2006{\natexlab{b}})}]{porsev06a}
\bibinfo{author}{\bibfnamefont{S.~G.} \bibnamefont{Porsev}} \bibnamefont{and}
  \bibinfo{author}{\bibfnamefont{A.}~\bibnamefont{Derevianko}},
  \bibinfo{journal}{JETP} \textbf{\bibinfo{volume}{102}}, \bibinfo{pages}{195}
  (\bibinfo{year}{2006}{\natexlab{b}}).

\bibitem[{\citenamefont{{Bouloufa} et~al.}(2009)\citenamefont{{Bouloufa},
  {Crubellier}, and {Dulieu}}}]{bouloufa09a}
\bibinfo{author}{\bibfnamefont{N.}~\bibnamefont{{Bouloufa}}},
  \bibinfo{author}{\bibfnamefont{A.}~\bibnamefont{{Crubellier}}},
  \bibnamefont{and} \bibinfo{author}{\bibfnamefont{O.}~\bibnamefont{{Dulieu}}},
  \bibinfo{journal}{Physica Scripta Volume T} \textbf{\bibinfo{volume}{134}},
  \bibinfo{pages}{014014} (\bibinfo{year}{2009}).

\bibitem[{\citenamefont{Kleinman et~al.}(1968)\citenamefont{Kleinman, Hahn, and
  Spruch}}]{kleinman68a}
\bibinfo{author}{\bibfnamefont{C.~J.} \bibnamefont{Kleinman}},
  \bibinfo{author}{\bibfnamefont{Y.}~\bibnamefont{Hahn}}, \bibnamefont{and}
  \bibinfo{author}{\bibfnamefont{L.}~\bibnamefont{Spruch}},
  \bibinfo{journal}{Phys. Rev.} \textbf{\bibinfo{volume}{165}},
  \bibinfo{pages}{53} (\bibinfo{year}{1968}).

\bibitem[{\citenamefont{Dalgarno et~al.}(1968)\citenamefont{Dalgarno, Drake,
  and Victor}}]{dalgarno68a}
\bibinfo{author}{\bibfnamefont{A.~D.} \bibnamefont{Dalgarno}},
  \bibinfo{author}{\bibfnamefont{G.~W.~F.} \bibnamefont{Drake}},
  \bibnamefont{and} \bibinfo{author}{\bibfnamefont{G.~A.}
  \bibnamefont{Victor}}, \bibinfo{journal}{Phys.~Rev.}
  \textbf{\bibinfo{volume}{176}}, \bibinfo{pages}{194} (\bibinfo{year}{1968}).

\bibitem[{\citenamefont{Dalgarno and Kingston}(1960)}]{dalgarno60a}
\bibinfo{author}{\bibfnamefont{A.}~\bibnamefont{Dalgarno}} \bibnamefont{and}
  \bibinfo{author}{\bibfnamefont{A.~E.} \bibnamefont{Kingston}},
  \bibinfo{journal}{Proc.~R.~Soc.~London~A} \textbf{\bibinfo{volume}{259}},
  \bibinfo{pages}{424} (\bibinfo{year}{1960}).

\bibitem[{\citenamefont{Langhoff and Karplus}(1969)}]{langhoff69a}
\bibinfo{author}{\bibfnamefont{P.~W.} \bibnamefont{Langhoff}} \bibnamefont{and}
  \bibinfo{author}{\bibfnamefont{M.}~\bibnamefont{Karplus}},
  \bibinfo{journal}{J.~Opt.~Soc.~Am.} \textbf{\bibinfo{volume}{59}},
  \bibinfo{pages}{863} (\bibinfo{year}{1969}).

\bibitem[{\citenamefont{{Schmidt} et~al.}(2007)\citenamefont{{Schmidt},
  {Gavioso}, {May}, and {Moldover}}}]{schmidt07a}
\bibinfo{author}{\bibfnamefont{J.~W.} \bibnamefont{{Schmidt}}},
  \bibinfo{author}{\bibfnamefont{R.~M.} \bibnamefont{{Gavioso}}},
  \bibinfo{author}{\bibfnamefont{E.~F.} \bibnamefont{{May}}}, \bibnamefont{and}
  \bibinfo{author}{\bibfnamefont{M.~R.} \bibnamefont{{Moldover}}},
  \bibinfo{journal}{Phys.~Rev.~Lett.} \textbf{\bibinfo{volume}{98}},
  \bibinfo{pages}{254504} (\bibinfo{year}{2007}).

\bibitem[{\citenamefont{Goebel and Hohm}(1995)}]{goebel95a}
\bibinfo{author}{\bibfnamefont{D.}~\bibnamefont{Goebel}} \bibnamefont{and}
  \bibinfo{author}{\bibfnamefont{U.}~\bibnamefont{Hohm}},
  \bibinfo{journal}{Phys.~Rev.~A} \textbf{\bibinfo{volume}{52}},
  \bibinfo{pages}{3691} (\bibinfo{year}{1995}).

\bibitem[{\citenamefont{Goebel and Hohm}(1996)}]{goebel96a}
\bibinfo{author}{\bibfnamefont{D.}~\bibnamefont{Goebel}} \bibnamefont{and}
  \bibinfo{author}{\bibfnamefont{U.}~\bibnamefont{Hohm}},
  \bibinfo{journal}{J.~Phys.~Chem.} \textbf{\bibinfo{volume}{100}},
  \bibinfo{pages}{7710} (\bibinfo{year}{1996}).

\bibitem[{\citenamefont{{Hall} and {Zorn}}(1974)}]{hall74a}
\bibinfo{author}{\bibfnamefont{W.~D.} \bibnamefont{{Hall}}} \bibnamefont{and}
  \bibinfo{author}{\bibfnamefont{J.~C.} \bibnamefont{{Zorn}}},
  \bibinfo{journal}{Phys.~Rev.~A} \textbf{\bibinfo{volume}{10}},
  \bibinfo{pages}{1141} (\bibinfo{year}{1974}).

\bibitem[{\citenamefont{Molof et~al.}(1974)\citenamefont{Molof, Schwartz,
  Miller, and Bederson}}]{molof74a}
\bibinfo{author}{\bibfnamefont{R.~W.} \bibnamefont{Molof}},
  \bibinfo{author}{\bibfnamefont{H.~L.} \bibnamefont{Schwartz}},
  \bibinfo{author}{\bibfnamefont{T.~M.} \bibnamefont{Miller}},
  \bibnamefont{and} \bibinfo{author}{\bibfnamefont{B.}~\bibnamefont{Bederson}},
  \bibinfo{journal}{Phys.~Rev.~A} \textbf{\bibinfo{volume}{10}},
  \bibinfo{pages}{1131} (\bibinfo{year}{1974}).

\bibitem[{\citenamefont{Miller and Bederson}(1976)}]{miller76a}
\bibinfo{author}{\bibfnamefont{T.~M.} \bibnamefont{Miller}} \bibnamefont{and}
  \bibinfo{author}{\bibfnamefont{B.}~\bibnamefont{Bederson}},
  \bibinfo{journal}{Phys.~Rev.~A} \textbf{\bibinfo{volume}{14}},
  \bibinfo{pages}{1572} (\bibinfo{year}{1976}).

\bibitem[{\citenamefont{Schwartz et~al.}(1974)\citenamefont{Schwartz, Miller,
  and Bederson}}]{schwartz74a}
\bibinfo{author}{\bibfnamefont{H.~L.} \bibnamefont{Schwartz}},
  \bibinfo{author}{\bibfnamefont{T.~M.} \bibnamefont{Miller}},
  \bibnamefont{and} \bibinfo{author}{\bibfnamefont{B.}~\bibnamefont{Bederson}},
  \bibinfo{journal}{Phys.~Rev.~A} \textbf{\bibinfo{volume}{10}},
  \bibinfo{pages}{1924} (\bibinfo{year}{1974}).

\bibitem[{\citenamefont{{Cronin} et~al.}(2009)\citenamefont{{Cronin},
  {Schmiedmayer}, and {Pritchard}}}]{cronin09a}
\bibinfo{author}{\bibfnamefont{A.~D.} \bibnamefont{{Cronin}}},
  \bibinfo{author}{\bibfnamefont{J.}~\bibnamefont{{Schmiedmayer}}},
  \bibnamefont{and} \bibinfo{author}{\bibfnamefont{D.~E.}
  \bibnamefont{{Pritchard}}}, \bibinfo{journal}{Rev.~Mod.~Phys.}
  \textbf{\bibinfo{volume}{81}}, \bibinfo{pages}{1051} (\bibinfo{year}{2009}).

\bibitem[{\citenamefont{Miffre et~al.}(2006)\citenamefont{Miffre, Jacquet,
  Buchner, Trenec, and Vigue}}]{miffre06a}
\bibinfo{author}{\bibfnamefont{A.}~\bibnamefont{Miffre}},
  \bibinfo{author}{\bibfnamefont{M.}~\bibnamefont{Jacquet}},
  \bibinfo{author}{\bibfnamefont{M.}~\bibnamefont{Buchner}},
  \bibinfo{author}{\bibfnamefont{G.}~\bibnamefont{Trenec}}, \bibnamefont{and}
  \bibinfo{author}{\bibfnamefont{J.}~\bibnamefont{Vigue}},
  \bibinfo{journal}{Eur.~Phys.~J.~D} \textbf{\bibinfo{volume}{38}},
  \bibinfo{pages}{353} (\bibinfo{year}{2006}).

\bibitem[{\citenamefont{Ekstrom et~al.}(1995)\citenamefont{Ekstrom,
  Schmiedmayer, Chapman, Hammond, and Pritchard}}]{ekstrom95a}
\bibinfo{author}{\bibfnamefont{C.~R.} \bibnamefont{Ekstrom}},
  \bibinfo{author}{\bibfnamefont{J.}~\bibnamefont{Schmiedmayer}},
  \bibinfo{author}{\bibfnamefont{M.~S.} \bibnamefont{Chapman}},
  \bibinfo{author}{\bibfnamefont{T.~D.} \bibnamefont{Hammond}},
  \bibnamefont{and} \bibinfo{author}{\bibfnamefont{D.~E.}
  \bibnamefont{Pritchard}}, \bibinfo{journal}{Phys. Rev. A}
  \textbf{\bibinfo{volume}{51}}, \bibinfo{pages}{3883} (\bibinfo{year}{1995}).

\bibitem[{\citenamefont{Holmgren et~al.}(2010)\citenamefont{Holmgren, Revelle,
  Lonij, and Cronin}}]{holmgren10a}
\bibinfo{author}{\bibfnamefont{W.~F.} \bibnamefont{Holmgren}},
  \bibinfo{author}{\bibfnamefont{M.~C.} \bibnamefont{Revelle}},
  \bibinfo{author}{\bibfnamefont{V.~P.~A.} \bibnamefont{Lonij}},
  \bibnamefont{and} \bibinfo{author}{\bibfnamefont{A.}~\bibnamefont{Cronin}}
  (\bibinfo{year}{2010}), \eprint{1001.3888v1}.

\bibitem[{\citenamefont{{Amini} and {Gould}}(2003)}]{amini03a}
\bibinfo{author}{\bibfnamefont{J.~M.} \bibnamefont{{Amini}}} \bibnamefont{and}
  \bibinfo{author}{\bibfnamefont{H.}~\bibnamefont{{Gould}}},
  \bibinfo{journal}{Phys.~Rev.~Lett.} \textbf{\bibinfo{volume}{91}},
  \bibinfo{pages}{153001} (\bibinfo{year}{2003}).

\bibitem[{\citenamefont{Kadar-Kallen and Bonin}(1992)}]{kadarkallen92a}
\bibinfo{author}{\bibfnamefont{M.~A.} \bibnamefont{Kadar-Kallen}}
  \bibnamefont{and} \bibinfo{author}{\bibfnamefont{K.~D.} \bibnamefont{Bonin}},
  \bibinfo{journal}{Phys. Rev. Lett.} \textbf{\bibinfo{volume}{68}},
  \bibinfo{pages}{2015} (\bibinfo{year}{1992}).

\bibitem[{\citenamefont{Kadar-Kallen and Bonin}(1994)}]{kadarkallen94a}
\bibinfo{author}{\bibfnamefont{M.~A.} \bibnamefont{Kadar-Kallen}}
  \bibnamefont{and} \bibinfo{author}{\bibfnamefont{K.~D.} \bibnamefont{Bonin}},
  \bibinfo{journal}{Phys. Rev. Lett.} \textbf{\bibinfo{volume}{72}},
  \bibinfo{pages}{828} (\bibinfo{year}{1994}).

\bibitem[{\citenamefont{Hu and Kusse}(2002)}]{hu02a}
\bibinfo{author}{\bibfnamefont{M.}~\bibnamefont{Hu}} \bibnamefont{and}
  \bibinfo{author}{\bibfnamefont{B.~R.} \bibnamefont{Kusse}},
  \bibinfo{journal}{Phys. Rev. A} \textbf{\bibinfo{volume}{66}},
  \bibinfo{pages}{062506} (\bibinfo{year}{2002}).

\bibitem[{\citenamefont{Sarkisov et~al.}(2006)\citenamefont{Sarkisov, Beigman,
  Shevelko, and Struve}}]{sarkisov06a}
\bibinfo{author}{\bibfnamefont{G.~S.} \bibnamefont{Sarkisov}},
  \bibinfo{author}{\bibfnamefont{I.~L.} \bibnamefont{Beigman}},
  \bibinfo{author}{\bibfnamefont{V.~P.} \bibnamefont{Shevelko}},
  \bibnamefont{and} \bibinfo{author}{\bibfnamefont{K.~W.}
  \bibnamefont{Struve}}, \bibinfo{journal}{Phys.~Rev.~E}
  \textbf{\bibinfo{volume}{73}}, \bibinfo{pages}{042501}
  (\bibinfo{year}{2006}).

\bibitem[{\citenamefont{Born and Heisenberg}(1924)}]{born24a}
\bibinfo{author}{\bibfnamefont{M.}~\bibnamefont{Born}} \bibnamefont{and}
  \bibinfo{author}{\bibfnamefont{W.}~\bibnamefont{Heisenberg}},
  \bibinfo{journal}{Z. Phys.} \textbf{\bibinfo{volume}{407}},
  \bibinfo{pages}{407} (\bibinfo{year}{1924}).

\bibitem[{\citenamefont{Mayer and Mayer}(1933)}]{mayer33a}
\bibinfo{author}{\bibfnamefont{J.~E.} \bibnamefont{Mayer}} \bibnamefont{and}
  \bibinfo{author}{\bibfnamefont{M.~G.} \bibnamefont{Mayer}},
  \bibinfo{journal}{Phys. Rev.} \textbf{\bibinfo{volume}{43}},
  \bibinfo{pages}{605} (\bibinfo{year}{1933}).

\bibitem[{\citenamefont{Dalgarno and Lewis}(1956)}]{dalgarno56a}
\bibinfo{author}{\bibfnamefont{A.}~\bibnamefont{Dalgarno}} \bibnamefont{and}
  \bibinfo{author}{\bibfnamefont{J.~T.} \bibnamefont{Lewis}},
  \bibinfo{journal}{Proc.~Phys.~Soc.~London~Ser.~A}
  \textbf{\bibinfo{volume}{69}}, \bibinfo{pages}{57} (\bibinfo{year}{1956}).

\bibitem[{\citenamefont{Drachman}(1982)}]{drachman82b}
\bibinfo{author}{\bibfnamefont{R.~J.} \bibnamefont{Drachman}},
  \bibinfo{journal}{Phys.~Rev.~A} \textbf{\bibinfo{volume}{26}},
  \bibinfo{pages}{1228} (\bibinfo{year}{1982}).

\bibitem[{\citenamefont{Drachman and Bhatia}(1995)}]{drachman95a}
\bibinfo{author}{\bibfnamefont{R.~J.} \bibnamefont{Drachman}} \bibnamefont{and}
  \bibinfo{author}{\bibfnamefont{A.~K.} \bibnamefont{Bhatia}},
  \bibinfo{journal}{Phys.~Rev.~A} \textbf{\bibinfo{volume}{51}},
  \bibinfo{pages}{2926} (\bibinfo{year}{1995}).

\bibitem[{\citenamefont{Mitroy and Safronova}(2009)}]{mitroy09a}
\bibinfo{author}{\bibfnamefont{J.}~\bibnamefont{Mitroy}} \bibnamefont{and}
  \bibinfo{author}{\bibfnamefont{M.~S.} \bibnamefont{Safronova}},
  \bibinfo{journal}{Phys.~Rev.~A} \textbf{\bibinfo{volume}{79}},
  \bibinfo{pages}{012513} (\bibinfo{year}{2009}).

\bibitem[{\citenamefont{{Bockasten}}(1974)}]{bockasten74a}
\bibinfo{author}{\bibfnamefont{K.}~\bibnamefont{{Bockasten}}},
  \bibinfo{journal}{Phys.~Rev.~A} \textbf{\bibinfo{volume}{9}},
  \bibinfo{pages}{1087} (\bibinfo{year}{1974}).

\bibitem[{\citenamefont{{Drake} and {Swainson}}(1991)}]{drake91a}
\bibinfo{author}{\bibfnamefont{G.~W.~F.} \bibnamefont{{Drake}}}
  \bibnamefont{and} \bibinfo{author}{\bibfnamefont{R.~A.}
  \bibnamefont{{Swainson}}}, \bibinfo{journal}{Phys.~Rev.~A}
  \textbf{\bibinfo{volume}{44}}, \bibinfo{pages}{5448} (\bibinfo{year}{1991}).

\bibitem[{\citenamefont{{Swainson} and {Drake}}(1992)}]{swainson92b}
\bibinfo{author}{\bibfnamefont{R.~A.} \bibnamefont{{Swainson}}}
  \bibnamefont{and} \bibinfo{author}{\bibfnamefont{G.~W.~F.}
  \bibnamefont{{Drake}}}, \bibinfo{journal}{Can.~J.~Phys.}
  \textbf{\bibinfo{volume}{70}}, \bibinfo{pages}{187} (\bibinfo{year}{1992}).

\bibitem[{\citenamefont{Mitroy}(2008)}]{mitroy08k}
\bibinfo{author}{\bibfnamefont{J.}~\bibnamefont{Mitroy}},
  \bibinfo{journal}{Phys.~Rev.~A} \textbf{\bibinfo{volume}{78}},
  \bibinfo{pages}{052515} (\bibinfo{year}{2008}).

\bibitem[{\citenamefont{Stark}(1913)}]{stark13a}
\bibinfo{author}{\bibfnamefont{J.}~\bibnamefont{Stark}},
  \bibinfo{journal}{Annalen.~der Physik} \textbf{\bibinfo{volume}{43}},
  \bibinfo{pages}{965} (\bibinfo{year}{1913}).

\bibitem[{\citenamefont{{Hunter} et~al.}(1988)\citenamefont{{Hunter}, {Krause},
  {Murthy}, and {Sung}}}]{hunter88a}
\bibinfo{author}{\bibfnamefont{L.~R.} \bibnamefont{{Hunter}}},
  \bibinfo{author}{\bibfnamefont{D.}~\bibnamefont{{Krause}},
  \bibfnamefont{Jr.}},
  \bibinfo{author}{\bibfnamefont{S.}~\bibnamefont{{Murthy}}}, \bibnamefont{and}
  \bibinfo{author}{\bibfnamefont{T.~W.} \bibnamefont{{Sung}}},
  \bibinfo{journal}{Phys.~Rev.~A} \textbf{\bibinfo{volume}{37}},
  \bibinfo{pages}{3283} (\bibinfo{year}{1988}).

\bibitem[{\citenamefont{Hunter et~al.}(1991)\citenamefont{Hunter, Krause,
  Berkeland, and Boshier}}]{hunter91a}
\bibinfo{author}{\bibfnamefont{L.~R.} \bibnamefont{Hunter}},
  \bibinfo{author}{\bibfnamefont{D.}~\bibnamefont{Krause}},
  \bibinfo{author}{\bibfnamefont{D.~J.} \bibnamefont{Berkeland}},
  \bibnamefont{and} \bibinfo{author}{\bibfnamefont{M.~G.}
  \bibnamefont{Boshier}}, \bibinfo{journal}{Phys. Rev. A}
  \textbf{\bibinfo{volume}{44}}, \bibinfo{pages}{6140} (\bibinfo{year}{1991}).

\bibitem[{\citenamefont{{Hunter} et~al.}(1992)\citenamefont{{Hunter}, {Krause},
  {Miller}, {Berkeland}, and {Boshier}}}]{hunter92a}
\bibinfo{author}{\bibfnamefont{L.~R.} \bibnamefont{{Hunter}}},
  \bibinfo{author}{\bibfnamefont{D.}~\bibnamefont{{Krause}}},
  \bibinfo{author}{\bibfnamefont{K.~E.} \bibnamefont{{Miller}}},
  \bibinfo{author}{\bibfnamefont{D.~J.} \bibnamefont{{Berkeland}}},
  \bibnamefont{and} \bibinfo{author}{\bibfnamefont{M.~G.}
  \bibnamefont{{Boshier}}}, \bibinfo{journal}{Opt.~Commun.}
  \textbf{\bibinfo{volume}{94}}, \bibinfo{pages}{210} (\bibinfo{year}{1992}).

\bibitem[{\citenamefont{Bennett et~al.}(1999)\citenamefont{Bennett, Roberts,
  and Wieman}}]{bennett99a}
\bibinfo{author}{\bibfnamefont{S.~C.} \bibnamefont{Bennett}},
  \bibinfo{author}{\bibfnamefont{J.~L.} \bibnamefont{Roberts}},
  \bibnamefont{and} \bibinfo{author}{\bibfnamefont{C.~E.}
  \bibnamefont{Wieman}}, \bibinfo{journal}{Phys. Rev. A}
  \textbf{\bibinfo{volume}{59}}, \bibinfo{pages}{R16} (\bibinfo{year}{1999}).

\bibitem[{\citenamefont{{van Wijngaarden}}(1999)}]{wijngaarden99a}
\bibinfo{author}{\bibfnamefont{W.~A.} \bibnamefont{{van Wijngaarden}}}, in
  \emph{\bibinfo{booktitle}{American Institute of Physics Conference Series}},
  edited by \bibinfo{editor}{\bibfnamefont{W.~E.} \bibnamefont{{Baylis}}}
  \bibnamefont{and} \bibinfo{editor}{\bibfnamefont{G.~W.~F.}
  \bibnamefont{{Drake}}} (\bibinfo{year}{1999}), vol. \bibinfo{volume}{477} of
  \emph{\bibinfo{series}{American Institute of Physics Conference Series}}, pp.
  \bibinfo{pages}{305--321}.

\bibitem[{\citenamefont{Sherman et~al.}(2005)\citenamefont{Sherman, Koerber,
  Markhotok, Nagourney, and Fortson}}]{sherman05a}
\bibinfo{author}{\bibfnamefont{J.~A.} \bibnamefont{Sherman}},
  \bibinfo{author}{\bibfnamefont{T.~W.} \bibnamefont{Koerber}},
  \bibinfo{author}{\bibfnamefont{A.}~\bibnamefont{Markhotok}},
  \bibinfo{author}{\bibfnamefont{W.}~\bibnamefont{Nagourney}},
  \bibnamefont{and} \bibinfo{author}{\bibfnamefont{E.~N.}
  \bibnamefont{Fortson}}, \bibinfo{journal}{Phys. Rev. Lett.}
  \textbf{\bibinfo{volume}{94}}, \bibinfo{pages}{243001}
  (\bibinfo{year}{2005}).

\bibitem[{\citenamefont{{Iskrenova-Tchoukova} and
  {Safronova}}(2008)}]{iskrenova08a}
\bibinfo{author}{\bibfnamefont{E.}~\bibnamefont{{Iskrenova-Tchoukova}}}
  \bibnamefont{and} \bibinfo{author}{\bibfnamefont{M.~S.}
  \bibnamefont{{Safronova}}}, \bibinfo{journal}{Phys.~Rev.~A}
  \textbf{\bibinfo{volume}{78}}, \bibinfo{pages}{012508}
  (\bibinfo{year}{2008}).

\bibitem[{\citenamefont{Rosenband et~al.}(2006)\citenamefont{Rosenband, Itano,
  Schmidt, Hume, Koelemeij, Bergquist, and Wineland}}]{rosenband06a}
\bibinfo{author}{\bibfnamefont{T.}~\bibnamefont{Rosenband}},
  \bibinfo{author}{\bibfnamefont{W.~M.} \bibnamefont{Itano}},
  \bibinfo{author}{\bibfnamefont{P.~O.} \bibnamefont{Schmidt}},
  \bibinfo{author}{\bibfnamefont{D.~B.} \bibnamefont{Hume}},
  \bibinfo{author}{\bibfnamefont{J.~C.~J.} \bibnamefont{Koelemeij}},
  \bibinfo{author}{\bibfnamefont{J.~C.} \bibnamefont{Bergquist}},
  \bibnamefont{and} \bibinfo{author}{\bibfnamefont{D.~J.}
  \bibnamefont{Wineland}}, \bibinfo{journal}{Proceedings of the 20th European
  Frequency and Time Forum, PTB Braunschweig, Germany, 2006.} p.
  \bibinfo{pages}{289} (\bibinfo{year}{2006}).

\bibitem[{\citenamefont{{S{\'a}nchez} et~al.}(2009)\citenamefont{{S{\'a}nchez},
  {{\v Z}{\'a}kov{\'a}}, {Andjelkovic}, {Bushaw}, {Dasgupta}, {Ewald},
  {Geppert}, {Kluge}, {Kr{\"a}mer}, {Nothhelfer} et~al.}}]{sanchez09a}
\bibinfo{author}{\bibfnamefont{R.}~\bibnamefont{{S{\'a}nchez}}},
  \bibinfo{author}{\bibfnamefont{M.}~\bibnamefont{{{\v Z}{\'a}kov{\'a}}}},
  \bibinfo{author}{\bibfnamefont{Z.}~\bibnamefont{{Andjelkovic}}},
  \bibinfo{author}{\bibfnamefont{B.~A.} \bibnamefont{{Bushaw}}},
  \bibinfo{author}{\bibfnamefont{K.}~\bibnamefont{{Dasgupta}}},
  \bibinfo{author}{\bibfnamefont{G.}~\bibnamefont{{Ewald}}},
  \bibinfo{author}{\bibfnamefont{C.}~\bibnamefont{{Geppert}}},
  \bibinfo{author}{\bibfnamefont{H.}~\bibnamefont{{Kluge}}},
  \bibinfo{author}{\bibfnamefont{J.}~\bibnamefont{{Kr{\"a}mer}}},
  \bibinfo{author}{\bibfnamefont{M.}~\bibnamefont{{Nothhelfer}}},
  \bibnamefont{et~al.}, \bibinfo{journal}{New Journal of Physics}
  \textbf{\bibinfo{volume}{11}}, \bibinfo{pages}{073016}
  (\bibinfo{year}{2009}).

\bibitem[{\citenamefont{Hibbert}(1975)}]{hibbert75a}
\bibinfo{author}{\bibfnamefont{A.}~\bibnamefont{Hibbert}},
  \bibinfo{journal}{Rep.~Prog.~Phys.} \textbf{\bibinfo{volume}{38}},
  \bibinfo{pages}{1217} (\bibinfo{year}{1975}).

\bibitem[{\citenamefont{Dzuba et~al.}(1996)\citenamefont{Dzuba, Flambaum, and
  Kozlov}}]{dzuba96b}
\bibinfo{author}{\bibfnamefont{V.~A.} \bibnamefont{Dzuba}},
  \bibinfo{author}{\bibfnamefont{V.~V.} \bibnamefont{Flambaum}},
  \bibnamefont{and} \bibinfo{author}{\bibfnamefont{M.~G.}
  \bibnamefont{Kozlov}}, \bibinfo{journal}{Phys.\ Rev.\ A}
  \textbf{\bibinfo{volume}{54}}, \bibinfo{pages}{3948} (\bibinfo{year}{1996}).

\bibitem[{\citenamefont{{Froese Fischer} et~al.}(1997)\citenamefont{{Froese
  Fischer}, Brage, and {J{\"o}nsson}}}]{fischer97a}
\bibinfo{author}{\bibfnamefont{C.}~\bibnamefont{{Froese Fischer}}},
  \bibinfo{author}{\bibfnamefont{T.}~\bibnamefont{Brage}}, \bibnamefont{and}
  \bibinfo{author}{\bibfnamefont{P.}~\bibnamefont{{J{\"o}nsson}}},
  \emph{\bibinfo{title}{Computational Atomic Structure (An MCHF Approach)}}
  (\bibinfo{publisher}{Institute of Phyics Publishing},
  \bibinfo{address}{Bristol}, \bibinfo{year}{1997}).

\bibitem[{\citenamefont{Grant}(2007)}]{grant07}
\bibinfo{author}{\bibfnamefont{I.~P.} \bibnamefont{Grant}},
  \emph{\bibinfo{title}{Relativistic Quantum Theory of Atoms and Molecules
  Theory and Computation}} (\bibinfo{publisher}{Springer},
  \bibinfo{address}{New York}, \bibinfo{year}{2007}).

\bibitem[{\citenamefont{Laughlin and Victor}(1988)}]{laughlin88a}
\bibinfo{author}{\bibfnamefont{C.}~\bibnamefont{Laughlin}} \bibnamefont{and}
  \bibinfo{author}{\bibfnamefont{G.~A.} \bibnamefont{Victor}},
  \bibinfo{journal}{Adv.~At.~Mol.~Phys.} \textbf{\bibinfo{volume}{25}},
  \bibinfo{pages}{163} (\bibinfo{year}{1988}).

\bibitem[{\citenamefont{M{\"u}ller et~al.}(1984)\citenamefont{M{\"u}ller,
  Flesch, and Meyer}}]{muller84}
\bibinfo{author}{\bibfnamefont{W.}~\bibnamefont{M{\"u}ller}},
  \bibinfo{author}{\bibfnamefont{J.}~\bibnamefont{Flesch}}, \bibnamefont{and}
  \bibinfo{author}{\bibfnamefont{W.}~\bibnamefont{Meyer}},
  \bibinfo{journal}{J.~Chem.~Phys.} \textbf{\bibinfo{volume}{80}},
  \bibinfo{pages}{3297} (\bibinfo{year}{1984}).

\bibitem[{\citenamefont{Mitroy and Bromley}(2003)}]{mitroy03f}
\bibinfo{author}{\bibfnamefont{J.}~\bibnamefont{Mitroy}} \bibnamefont{and}
  \bibinfo{author}{\bibfnamefont{M.~W.~J.} \bibnamefont{Bromley}},
  \bibinfo{journal}{Phys.~Rev.~A} \textbf{\bibinfo{volume}{68}},
  \bibinfo{pages}{052714} (\bibinfo{year}{2003}).

\bibitem[{\citenamefont{Victor et~al.}(1976)\citenamefont{Victor, Stewart, and
  Laughlin}}]{victor76}
\bibinfo{author}{\bibfnamefont{G.~A.} \bibnamefont{Victor}},
  \bibinfo{author}{\bibfnamefont{R.~F.} \bibnamefont{Stewart}},
  \bibnamefont{and} \bibinfo{author}{\bibfnamefont{C.}~\bibnamefont{Laughlin}},
  \bibinfo{journal}{Astrophys.~J.~Suppl.~Ser.} \textbf{\bibinfo{volume}{31}},
  \bibinfo{pages}{237} (\bibinfo{year}{1976}).

\bibitem[{\citenamefont{Norcross and Seaton}(1976)}]{norcross76a}
\bibinfo{author}{\bibfnamefont{D.~W.} \bibnamefont{Norcross}} \bibnamefont{and}
  \bibinfo{author}{\bibfnamefont{M.~J.} \bibnamefont{Seaton}},
  \bibinfo{journal}{J.~Phys.~B} \textbf{\bibinfo{volume}{9}},
  \bibinfo{pages}{2983} (\bibinfo{year}{1976}).

\bibitem[{\citenamefont{Santra et~al.}(2004)\citenamefont{Santra, Christ, and
  Greene}}]{santra04a}
\bibinfo{author}{\bibfnamefont{R.}~\bibnamefont{Santra}},
  \bibinfo{author}{\bibfnamefont{K.~V.} \bibnamefont{Christ}},
  \bibnamefont{and} \bibinfo{author}{\bibfnamefont{C.~H.}
  \bibnamefont{Greene}}, \bibinfo{journal}{Phys.~Rev.~A}
  \textbf{\bibinfo{volume}{69}}, \bibinfo{pages}{042510}
  (\bibinfo{year}{2004}).

\bibitem[{\citenamefont{{Migdalek} and {Baylis}}(1978)}]{migdalek78a}
\bibinfo{author}{\bibfnamefont{J.}~\bibnamefont{{Migdalek}}} \bibnamefont{and}
  \bibinfo{author}{\bibfnamefont{W.~E.} \bibnamefont{{Baylis}}},
  \bibinfo{journal}{J. Phys. B} \textbf{\bibinfo{volume}{11}},
  \bibinfo{pages}{L497} (\bibinfo{year}{1978}).

\bibitem[{\citenamefont{Hameed et~al.}(1968)\citenamefont{Hameed, Herzenberg,
  and James}}]{hameed68a}
\bibinfo{author}{\bibfnamefont{S.}~\bibnamefont{Hameed}},
  \bibinfo{author}{\bibfnamefont{A.}~\bibnamefont{Herzenberg}},
  \bibnamefont{and} \bibinfo{author}{\bibfnamefont{M.~G.} \bibnamefont{James}},
  \bibinfo{journal}{J.~Phys.~B} \textbf{\bibinfo{volume}{1}},
  \bibinfo{pages}{822} (\bibinfo{year}{1968}).

\bibitem[{\citenamefont{Hameed}(1972)}]{hameed72a}
\bibinfo{author}{\bibfnamefont{S.}~\bibnamefont{Hameed}},
  \bibinfo{journal}{J.~Phys.~B} \textbf{\bibinfo{volume}{5}},
  \bibinfo{pages}{746} (\bibinfo{year}{1972}).

\bibitem[{\citenamefont{Zhou and Norcross}(1989)}]{zhou89a}
\bibinfo{author}{\bibfnamefont{H.~L.} \bibnamefont{Zhou}} \bibnamefont{and}
  \bibinfo{author}{\bibfnamefont{D.~W.} \bibnamefont{Norcross}},
  \bibinfo{journal}{Phys. Rev. A} \textbf{\bibinfo{volume}{40}},
  \bibinfo{pages}{5048} (\bibinfo{year}{1989}).

\bibitem[{\citenamefont{{Ye} and {Wang}}(2008)}]{ye08a}
\bibinfo{author}{\bibfnamefont{A.}~\bibnamefont{{Ye}}} \bibnamefont{and}
  \bibinfo{author}{\bibfnamefont{G.}~\bibnamefont{{Wang}}},
  \bibinfo{journal}{Phys.~Rev.~A} \textbf{\bibinfo{volume}{78}},
  \bibinfo{pages}{014502} (\bibinfo{year}{2008}).

\bibitem[{\citenamefont{Doolen}(1984)}]{doolen84a}
\bibinfo{author}{\bibfnamefont{G.~D.} \bibnamefont{Doolen}},
  \emph{\bibinfo{title}{Crc handbook of chemistry and physics}}
  (\bibinfo{year}{1984}), \bibinfo{note}{unpublished, referenced in
  \cite{miller07a}}.

\bibitem[{\citenamefont{Chu and Dalgarno}(2004)}]{chu04a}
\bibinfo{author}{\bibfnamefont{X.}~\bibnamefont{Chu}} \bibnamefont{and}
  \bibinfo{author}{\bibfnamefont{A.}~\bibnamefont{Dalgarno}},
  \bibinfo{journal}{J.~Chem.~Phys.} \textbf{\bibinfo{volume}{121}},
  \bibinfo{pages}{4083} (\bibinfo{year}{2004}).

\bibitem[{\citenamefont{{Pachucki} and {Sapirstein}}(2001)}]{pachucki01a}
\bibinfo{author}{\bibfnamefont{K.}~\bibnamefont{{Pachucki}}} \bibnamefont{and}
  \bibinfo{author}{\bibfnamefont{J.}~\bibnamefont{{Sapirstein}}},
  \bibinfo{journal}{Phys.~Rev.~A} \textbf{\bibinfo{volume}{63}},
  \bibinfo{pages}{012504} (\bibinfo{year}{2001}).

\bibitem[{\citenamefont{Yan et~al.}(1996)\citenamefont{Yan, Babb, Dalgarno, and
  Drake}}]{yan96a}
\bibinfo{author}{\bibfnamefont{Z.~C.} \bibnamefont{Yan}},
  \bibinfo{author}{\bibfnamefont{J.~F.} \bibnamefont{Babb}},
  \bibinfo{author}{\bibfnamefont{A.}~\bibnamefont{Dalgarno}}, \bibnamefont{and}
  \bibinfo{author}{\bibfnamefont{G.~W.~F.} \bibnamefont{Drake}},
  \bibinfo{journal}{Phys.~Rev.~A} \textbf{\bibinfo{volume}{54}},
  \bibinfo{pages}{2824} (\bibinfo{year}{1996}).

\bibitem[{\citenamefont{{Tang} et~al.}(2009{\natexlab{a}})\citenamefont{{Tang},
  {Yan}, {Shi}, and {Babb}}}]{tang09a}
\bibinfo{author}{\bibfnamefont{L.-Y.} \bibnamefont{{Tang}}},
  \bibinfo{author}{\bibfnamefont{Z.-C.} \bibnamefont{{Yan}}},
  \bibinfo{author}{\bibfnamefont{T.-Y.} \bibnamefont{{Shi}}}, \bibnamefont{and}
  \bibinfo{author}{\bibfnamefont{J.~F.} \bibnamefont{{Babb}}},
  \bibinfo{journal}{Phys.~Rev.~A} \textbf{\bibinfo{volume}{79}},
  \bibinfo{pages}{062712} (\bibinfo{year}{2009}{\natexlab{a}}).

\bibitem[{\citenamefont{{{\L}ach} et~al.}(2004)\citenamefont{{{\L}ach},
  {Jeziorski}, and {Szalewicz}}}]{lach04a}
\bibinfo{author}{\bibfnamefont{G.}~\bibnamefont{{{\L}ach}}},
  \bibinfo{author}{\bibfnamefont{B.}~\bibnamefont{{Jeziorski}}},
  \bibnamefont{and}
  \bibinfo{author}{\bibfnamefont{K.}~\bibnamefont{{Szalewicz}}},
  \bibinfo{journal}{Phys.~Rev.~Lett.} \textbf{\bibinfo{volume}{92}},
  \bibinfo{pages}{233001} (\bibinfo{year}{2004}).

\bibitem[{\citenamefont{Suzuki and Varga}(1998)}]{suzuki98a}
\bibinfo{author}{\bibfnamefont{Y.}~\bibnamefont{Suzuki}} \bibnamefont{and}
  \bibinfo{author}{\bibfnamefont{K.}~\bibnamefont{Varga}},
  \emph{\bibinfo{title}{Stochastic variational Approach to Quantum-Mechanical
  Few-Body Problems}}, 172 (\bibinfo{publisher}{Springer},
  \bibinfo{address}{New York}, \bibinfo{year}{1998}).

\bibitem[{\citenamefont{Komasa}(2002)}]{komasa02a}
\bibinfo{author}{\bibfnamefont{J.}~\bibnamefont{Komasa}},
  \bibinfo{journal}{Phys.~Rev.~A} \textbf{\bibinfo{volume}{65}},
  \bibinfo{pages}{012506} (\bibinfo{year}{2002}).

\bibitem[{\citenamefont{Brown and Ravenhall}(1951)}]{brown51a}
\bibinfo{author}{\bibfnamefont{G.~E.} \bibnamefont{Brown}} \bibnamefont{and}
  \bibinfo{author}{\bibfnamefont{D.}~\bibnamefont{Ravenhall}},
  \bibinfo{journal}{Proc.~R.~Soc.~London, Ser. A}
  \textbf{\bibinfo{volume}{208}}, \bibinfo{pages}{552} (\bibinfo{year}{1951}).

\bibitem[{\citenamefont{Cannon and Derevianko}(2004)}]{cannon04a}
\bibinfo{author}{\bibfnamefont{C.}~\bibnamefont{Cannon}} \bibnamefont{and}
  \bibinfo{author}{\bibfnamefont{A.}~\bibnamefont{Derevianko}},
  \bibinfo{journal}{Phys. Rev. A} \textbf{\bibinfo{volume}{69}},
  \bibinfo{pages}{030502(R)} (\bibinfo{year}{2004}).

\bibitem[{\citenamefont{{Safronova} and {Safronova}}(2009)}]{safronova09a}
\bibinfo{author}{\bibfnamefont{U.~I.} \bibnamefont{{Safronova}}}
  \bibnamefont{and} \bibinfo{author}{\bibfnamefont{M.~S.}
  \bibnamefont{{Safronova}}}, \bibinfo{journal}{Phys.~Rev.~A}
  \textbf{\bibinfo{volume}{79}}, \bibinfo{pages}{022512}
  (\bibinfo{year}{2009}).

\bibitem[{\citenamefont{Dzuba et~al.}(1989)\citenamefont{Dzuba, Flambaum, and
  Sushkov}}]{dzuba89b}
\bibinfo{author}{\bibfnamefont{V.~A.} \bibnamefont{Dzuba}},
  \bibinfo{author}{\bibfnamefont{V.~V.} \bibnamefont{Flambaum}},
  \bibnamefont{and} \bibinfo{author}{\bibfnamefont{O.~P.}
  \bibnamefont{Sushkov}}, \bibinfo{journal}{Phys.~Lett.~A}
  \textbf{\bibinfo{volume}{140}}, \bibinfo{pages}{493} (\bibinfo{year}{1989}).

\bibitem[{\citenamefont{Dzuba et~al.}(2001)\citenamefont{Dzuba, Flambaum, and
  Ginges}}]{dzuba01a}
\bibinfo{author}{\bibfnamefont{V.~A.} \bibnamefont{Dzuba}},
  \bibinfo{author}{\bibfnamefont{V.~V.} \bibnamefont{Flambaum}},
  \bibnamefont{and} \bibinfo{author}{\bibfnamefont{J.~S.~M.}
  \bibnamefont{Ginges}}, \bibinfo{journal}{Phys. Rev. A}
  \textbf{\bibinfo{volume}{63}}, \bibinfo{pages}{062101}
  (\bibinfo{year}{2001}).

\bibitem[{\citenamefont{Dzuba et~al.}(2002)\citenamefont{Dzuba, Flambaum, and
  Ginges}}]{dzuba02a}
\bibinfo{author}{\bibfnamefont{V.~A.} \bibnamefont{Dzuba}},
  \bibinfo{author}{\bibfnamefont{V.~V.} \bibnamefont{Flambaum}},
  \bibnamefont{and} \bibinfo{author}{\bibfnamefont{J.~S.~M.}
  \bibnamefont{Ginges}}, \bibinfo{journal}{Phys. Rev. D}
  \textbf{\bibinfo{volume}{66}}, \bibinfo{pages}{076013}
  (\bibinfo{year}{2002}).

\bibitem[{\citenamefont{Dzuba}(2008)}]{dzuba08a}
\bibinfo{author}{\bibfnamefont{V.~A.} \bibnamefont{Dzuba}},
  \bibinfo{journal}{Phys. Rev. A} \textbf{\bibinfo{volume}{78}},
  \bibinfo{pages}{042502} (\bibinfo{year}{2008}).

\bibitem[{\citenamefont{{Angstmann}
  et~al.}(2006{\natexlab{a}})\citenamefont{{Angstmann}, {Dzuba}, and
  {Flambaum}}}]{angstmann06a}
\bibinfo{author}{\bibfnamefont{E.~J.} \bibnamefont{{Angstmann}}},
  \bibinfo{author}{\bibfnamefont{V.~A.} \bibnamefont{{Dzuba}}},
  \bibnamefont{and} \bibinfo{author}{\bibfnamefont{V.~V.}
  \bibnamefont{{Flambaum}}}, \bibinfo{journal}{Phys.~Rev.~Lett.}
  \textbf{\bibinfo{volume}{97}}, \bibinfo{pages}{040802}
  (\bibinfo{year}{2006}{\natexlab{a}}).

\bibitem[{\citenamefont{{Angstmann}
  et~al.}(2006{\natexlab{b}})\citenamefont{{Angstmann}, {Dzuba}, and
  {Flambaum}}}]{angstmann06b}
\bibinfo{author}{\bibfnamefont{E.~J.} \bibnamefont{{Angstmann}}},
  \bibinfo{author}{\bibfnamefont{V.~A.} \bibnamefont{{Dzuba}}},
  \bibnamefont{and} \bibinfo{author}{\bibfnamefont{V.~V.}
  \bibnamefont{{Flambaum}}}, \bibinfo{journal}{Phys.~Rev.~A}
  \textbf{\bibinfo{volume}{74}}, \bibinfo{pages}{023405}
  (\bibinfo{year}{2006}{\natexlab{b}}).

\bibitem[{\citenamefont{Coester and K{\"{u}}mmel}(1960)}]{coester60a}
\bibinfo{author}{\bibfnamefont{F.}~\bibnamefont{Coester}} \bibnamefont{and}
  \bibinfo{author}{\bibfnamefont{H.}~\bibnamefont{K{\"{u}}mmel}},
  \bibinfo{journal}{Nucl.~Phys.} \textbf{\bibinfo{volume}{17}},
  \bibinfo{pages}{477} (\bibinfo{year}{1960}).

\bibitem[{\citenamefont{{Lim} et~al.}(2005)\citenamefont{{Lim},
  {Schwerdtfeger}, {Metz}, and {Stoll}}}]{lim05a}
\bibinfo{author}{\bibfnamefont{I.~S.} \bibnamefont{{Lim}}},
  \bibinfo{author}{\bibfnamefont{P.}~\bibnamefont{{Schwerdtfeger}}},
  \bibinfo{author}{\bibfnamefont{B.}~\bibnamefont{{Metz}}}, \bibnamefont{and}
  \bibinfo{author}{\bibfnamefont{H.}~\bibnamefont{{Stoll}}},
  \bibinfo{journal}{J.~Chem.~Phys.} \textbf{\bibinfo{volume}{122}},
  \bibinfo{pages}{104103} (\bibinfo{year}{2005}).

\bibitem[{\citenamefont{{Sadlej} et~al.}(1991)\citenamefont{{Sadlej}, {Urban},
  and {Gropen}}}]{sadlej91a}
\bibinfo{author}{\bibfnamefont{A.~J.} \bibnamefont{{Sadlej}}},
  \bibinfo{author}{\bibfnamefont{M.}~\bibnamefont{{Urban}}}, \bibnamefont{and}
  \bibinfo{author}{\bibfnamefont{O.}~\bibnamefont{{Gropen}}},
  \bibinfo{journal}{Phys.~Rev.~A} \textbf{\bibinfo{volume}{44}},
  \bibinfo{pages}{5547} (\bibinfo{year}{1991}).

\bibitem[{\citenamefont{{Sch{\"a}fer} et~al.}(2007)\citenamefont{{Sch{\"a}fer},
  {Mehring}, {Sch{\"a}fer}, and {Schwerdtfeger}}}]{schafer07a}
\bibinfo{author}{\bibfnamefont{S.}~\bibnamefont{{Sch{\"a}fer}}},
  \bibinfo{author}{\bibfnamefont{M.}~\bibnamefont{{Mehring}}},
  \bibinfo{author}{\bibfnamefont{R.}~\bibnamefont{{Sch{\"a}fer}}},
  \bibnamefont{and}
  \bibinfo{author}{\bibfnamefont{P.}~\bibnamefont{{Schwerdtfeger}}},
  \bibinfo{journal}{Phys.~Rev.~A} \textbf{\bibinfo{volume}{76}},
  \bibinfo{pages}{052515} (\bibinfo{year}{2007}).

\bibitem[{\citenamefont{Blundell
  et~al.}(1989{\natexlab{a}})\citenamefont{Blundell, Johnson, Liu, and
  Sapirstein}}]{blundell89b}
\bibinfo{author}{\bibfnamefont{S.~A.} \bibnamefont{Blundell}},
  \bibinfo{author}{\bibfnamefont{W.~R.} \bibnamefont{Johnson}},
  \bibinfo{author}{\bibfnamefont{Z.~W.} \bibnamefont{Liu}}, \bibnamefont{and}
  \bibinfo{author}{\bibfnamefont{J.}~\bibnamefont{Sapirstein}},
  \bibinfo{journal}{Phys. Rev. A} \textbf{\bibinfo{volume}{39}},
  \bibinfo{pages}{3768} (\bibinfo{year}{1989}{\natexlab{a}}).

\bibitem[{\citenamefont{Blundell
  et~al.}(1989{\natexlab{b}})\citenamefont{Blundell, Johnson, Liu, and
  Sapirstein}}]{blundell89a}
\bibinfo{author}{\bibfnamefont{S.~A.} \bibnamefont{Blundell}},
  \bibinfo{author}{\bibfnamefont{W.~R.} \bibnamefont{Johnson}},
  \bibinfo{author}{\bibfnamefont{Z.~W.} \bibnamefont{Liu}}, \bibnamefont{and}
  \bibinfo{author}{\bibfnamefont{J.}~\bibnamefont{Sapirstein}},
  \bibinfo{journal}{Phys. Rev. A} \textbf{\bibinfo{volume}{40}},
  \bibinfo{pages}{2233} (\bibinfo{year}{1989}{\natexlab{b}}).

\bibitem[{\citenamefont{{Pal} et~al.}(2007)\citenamefont{{Pal}, {Safronova},
  {Johnson}, {Derevianko}, and {Porsev}}}]{pal07a}
\bibinfo{author}{\bibfnamefont{R.}~\bibnamefont{{Pal}}},
  \bibinfo{author}{\bibfnamefont{M.~S.} \bibnamefont{{Safronova}}},
  \bibinfo{author}{\bibfnamefont{W.~R.} \bibnamefont{{Johnson}}},
  \bibinfo{author}{\bibfnamefont{A.}~\bibnamefont{{Derevianko}}},
  \bibnamefont{and} \bibinfo{author}{\bibfnamefont{S.~G.}
  \bibnamefont{{Porsev}}}, \bibinfo{journal}{Phys.~Rev.~A}
  \textbf{\bibinfo{volume}{75}}, \bibinfo{pages}{042515}
  (\bibinfo{year}{2007}).

\bibitem[{\citenamefont{Porsev and Derevianko}(2006{\natexlab{c}})}]{porsev06c}
\bibinfo{author}{\bibfnamefont{S.~G.} \bibnamefont{Porsev}} \bibnamefont{and}
  \bibinfo{author}{\bibfnamefont{A.}~\bibnamefont{Derevianko}},
  \bibinfo{journal}{Phys. Rev. A} \textbf{\bibinfo{volume}{73}},
  \bibinfo{pages}{012501} (\bibinfo{year}{2006}{\natexlab{c}}).

\bibitem[{\citenamefont{Safronova
  et~al.}(2009{\natexlab{a}})\citenamefont{Safronova, Kozlov, Johnson, and
  Jiang}}]{safronova09c}
\bibinfo{author}{\bibfnamefont{M.~S.} \bibnamefont{Safronova}},
  \bibinfo{author}{\bibfnamefont{M.~G.} \bibnamefont{Kozlov}},
  \bibinfo{author}{\bibfnamefont{W.~R.} \bibnamefont{Johnson}},
  \bibnamefont{and} \bibinfo{author}{\bibfnamefont{D.}~\bibnamefont{Jiang}},
  \bibinfo{journal}{Phys.~Rev.~A} \textbf{\bibinfo{volume}{80}},
  \bibinfo{pages}{012516} (\bibinfo{year}{2009}{\natexlab{a}}).

\bibitem[{\citenamefont{Porsev et~al.}(2001)\citenamefont{Porsev, Kozlov,
  Rakhlina, and Derevianko}}]{porsev01a}
\bibinfo{author}{\bibfnamefont{S.~G.} \bibnamefont{Porsev}},
  \bibinfo{author}{\bibfnamefont{M.~G.} \bibnamefont{Kozlov}},
  \bibinfo{author}{\bibfnamefont{Y.~G.} \bibnamefont{Rakhlina}},
  \bibnamefont{and}
  \bibinfo{author}{\bibfnamefont{A.}~\bibnamefont{Derevianko}},
  \bibinfo{journal}{Phys.~Rev.~A} \textbf{\bibinfo{volume}{64}},
  \bibinfo{pages}{012508} (\bibinfo{year}{2001}).

\bibitem[{\citenamefont{{Porsev} et~al.}(2008)\citenamefont{{Porsev}, {Ludlow},
  {Boyd}, and {Ye}}}]{porsev08a}
\bibinfo{author}{\bibfnamefont{S.~G.} \bibnamefont{{Porsev}}},
  \bibinfo{author}{\bibfnamefont{A.~D.} \bibnamefont{{Ludlow}}},
  \bibinfo{author}{\bibfnamefont{M.~M.} \bibnamefont{{Boyd}}},
  \bibnamefont{and} \bibinfo{author}{\bibfnamefont{J.}~\bibnamefont{{Ye}}},
  \bibinfo{journal}{Phys.~Rev.~A} \textbf{\bibinfo{volume}{78}},
  \bibinfo{pages}{032508} (\bibinfo{year}{2008}).

\bibitem[{\citenamefont{Dzuba and Ginges}(2006)}]{dzuba06a}
\bibinfo{author}{\bibfnamefont{V.~A.} \bibnamefont{Dzuba}} \bibnamefont{and}
  \bibinfo{author}{\bibfnamefont{J.~S.} \bibnamefont{Ginges}},
  \bibinfo{journal}{Phys. Rev. A} \textbf{\bibinfo{volume}{73}},
  \bibinfo{pages}{032503} (\bibinfo{year}{2006}).

\bibitem[{\citenamefont{{Dzuba} and {Derevianko}}(2010)}]{dzuba10a}
\bibinfo{author}{\bibfnamefont{V.~A.} \bibnamefont{{Dzuba}}} \bibnamefont{and}
  \bibinfo{author}{\bibfnamefont{A.}~\bibnamefont{{Derevianko}}},
  \bibinfo{journal}{J.~Phys.~B} \textbf{\bibinfo{volume}{43}},
  \bibinfo{pages}{074011} (\bibinfo{year}{2010}).

\bibitem[{\citenamefont{Dzuba et~al.}(1998)\citenamefont{Dzuba, Kozlov, Porsev,
  and Flambaum}}]{JETP}
\bibinfo{author}{\bibfnamefont{V.}~\bibnamefont{Dzuba}},
  \bibinfo{author}{\bibfnamefont{M.}~\bibnamefont{Kozlov}},
  \bibinfo{author}{\bibfnamefont{S.}~\bibnamefont{Porsev}}, \bibnamefont{and}
  \bibinfo{author}{\bibfnamefont{V.}~\bibnamefont{Flambaum}},
  \bibinfo{journal}{Zh. Eksp. Theor. Fiz.} \textbf{\bibinfo{volume}{114}},
  \bibinfo{pages}{1636} (\bibinfo{year}{1998}).

\bibitem[{\citenamefont{{Hachisu} et~al.}(2008)\citenamefont{{Hachisu},
  {Miyagishi}, {Porsev}, {Derevianko}, {Ovsiannikov}, {Pal'Chikov}, {Takamoto},
  and {Katori}}}]{hachisu08a}
\bibinfo{author}{\bibfnamefont{H.}~\bibnamefont{{Hachisu}}},
  \bibinfo{author}{\bibfnamefont{K.}~\bibnamefont{{Miyagishi}}},
  \bibinfo{author}{\bibfnamefont{S.~G.} \bibnamefont{{Porsev}}},
  \bibinfo{author}{\bibfnamefont{A.}~\bibnamefont{{Derevianko}}},
  \bibinfo{author}{\bibfnamefont{V.~D.} \bibnamefont{{Ovsiannikov}}},
  \bibinfo{author}{\bibfnamefont{V.~G.} \bibnamefont{{Pal'Chikov}}},
  \bibinfo{author}{\bibfnamefont{M.}~\bibnamefont{{Takamoto}}},
  \bibnamefont{and} \bibinfo{author}{\bibfnamefont{H.}~\bibnamefont{{Katori}}},
  \bibinfo{journal}{Phys.~Rev.~Lett.} \textbf{\bibinfo{volume}{100}},
  \bibinfo{pages}{053001} (\bibinfo{year}{2008}), \eprint{0711.4638}.

\bibitem[{\citenamefont{Nikoli\'{c} and Lindroth}(2004)}]{lindroth04}
\bibinfo{author}{\bibfnamefont{D.}~\bibnamefont{Nikoli\'{c}}} \bibnamefont{and}
  \bibinfo{author}{\bibfnamefont{E.}~\bibnamefont{Lindroth}},
  \bibinfo{journal}{J. Phys. B} \textbf{\bibinfo{volume}{37}},
  \bibinfo{pages}{L285} (\bibinfo{year}{2004}).

\bibitem[{\citenamefont{{Sold{\'a}n} et~al.}(2001)\citenamefont{{Sold{\'a}n},
  {Lee}, and {Wright}}}]{soldan01a}
\bibinfo{author}{\bibfnamefont{P.}~\bibnamefont{{Sold{\'a}n}}},
  \bibinfo{author}{\bibfnamefont{E.~P.~F.} \bibnamefont{{Lee}}},
  \bibnamefont{and} \bibinfo{author}{\bibfnamefont{T.~G.}
  \bibnamefont{{Wright}}}, \bibinfo{journal}{Physical Chemistry Chemical
  Physics (Incorporating Faraday Transactions)} \textbf{\bibinfo{volume}{3}},
  \bibinfo{pages}{4661} (\bibinfo{year}{2001}).

\bibitem[{\citenamefont{Nakajima and Hirao}(2001)}]{nakajima01a}
\bibinfo{author}{\bibfnamefont{T.}~\bibnamefont{Nakajima}} \bibnamefont{and}
  \bibinfo{author}{\bibfnamefont{K.}~\bibnamefont{Hirao}},
  \bibinfo{journal}{Chemistry Lett.} \textbf{\bibinfo{volume}{30}},
  \bibinfo{pages}{706} (\bibinfo{year}{2001}).

\bibitem[{\citenamefont{{Hald} et~al.}(2003)\citenamefont{{Hald},
  {Paw{\l}owski}, {J{\o}rgensen}, and {H{\"a}ttig}}}]{hald03a}
\bibinfo{author}{\bibfnamefont{K.}~\bibnamefont{{Hald}}},
  \bibinfo{author}{\bibfnamefont{F.}~\bibnamefont{{Paw{\l}owski}}},
  \bibinfo{author}{\bibfnamefont{P.}~\bibnamefont{{J{\o}rgensen}}},
  \bibnamefont{and}
  \bibinfo{author}{\bibfnamefont{C.}~\bibnamefont{{H{\"a}ttig}}},
  \bibinfo{journal}{J.~Chem.~Phys.} \textbf{\bibinfo{volume}{118}},
  \bibinfo{pages}{1292} (\bibinfo{year}{2003}).

\bibitem[{\citenamefont{Thakkar et~al.}(1992)\citenamefont{Thakkar, Hettema,
  and Wormer}}]{thakkar92a}
\bibinfo{author}{\bibfnamefont{A.~J.} \bibnamefont{Thakkar}},
  \bibinfo{author}{\bibfnamefont{H.}~\bibnamefont{Hettema}}, \bibnamefont{and}
  \bibinfo{author}{\bibfnamefont{P.~E.~S.} \bibnamefont{Wormer}},
  \bibinfo{journal}{J.~Chem.~Phys.} \textbf{\bibinfo{volume}{97}},
  \bibinfo{pages}{3252} (\bibinfo{year}{1992}).

\bibitem[{\citenamefont{{Franke} et~al.}(2001)\citenamefont{{Franke},
  {M{\"u}ller}, and {Noga}}}]{franke01a}
\bibinfo{author}{\bibfnamefont{R.}~\bibnamefont{{Franke}}},
  \bibinfo{author}{\bibfnamefont{H.}~\bibnamefont{{M{\"u}ller}}},
  \bibnamefont{and} \bibinfo{author}{\bibfnamefont{J.}~\bibnamefont{{Noga}}},
  \bibinfo{journal}{J.~Chem.~Phys.} \textbf{\bibinfo{volume}{114}},
  \bibinfo{pages}{7746} (\bibinfo{year}{2001}).

\bibitem[{\citenamefont{{Gugan} and {Michel}}(1980{\natexlab{a}})}]{gugan80a}
\bibinfo{author}{\bibfnamefont{D.}~\bibnamefont{{Gugan}}} \bibnamefont{and}
  \bibinfo{author}{\bibfnamefont{G.~W.} \bibnamefont{{Michel}}},
  \bibinfo{journal}{Molec.~Phys.~} \textbf{\bibinfo{volume}{39}},
  \bibinfo{pages}{783} (\bibinfo{year}{1980}{\natexlab{a}}).

\bibitem[{\citenamefont{{Gugan} and {Michel}}(1980{\natexlab{b}})}]{gugan80b}
\bibinfo{author}{\bibfnamefont{D.}~\bibnamefont{{Gugan}}} \bibnamefont{and}
  \bibinfo{author}{\bibfnamefont{G.~W.} \bibnamefont{{Michel}}},
  \bibinfo{journal}{Metrologia} \textbf{\bibinfo{volume}{16}},
  \bibinfo{pages}{149} (\bibinfo{year}{1980}{\natexlab{b}}).

\bibitem[{\citenamefont{{Orcutt} and {Cole}}(1967)}]{orcutt67a}
\bibinfo{author}{\bibfnamefont{R.~H.} \bibnamefont{{Orcutt}}} \bibnamefont{and}
  \bibinfo{author}{\bibfnamefont{R.~H.} \bibnamefont{{Cole}}},
  \bibinfo{journal}{J.~Chem.~Phys.} \textbf{\bibinfo{volume}{46}},
  \bibinfo{pages}{697} (\bibinfo{year}{1967}).

\bibitem[{\citenamefont{Newell and Baird}(1965)}]{newell65a}
\bibinfo{author}{\bibfnamefont{A.~C.} \bibnamefont{Newell}} \bibnamefont{and}
  \bibinfo{author}{\bibfnamefont{R.~C.} \bibnamefont{Baird}},
  \bibinfo{journal}{J.~Appl.~Phys.} \textbf{\bibinfo{volume}{36}},
  \bibinfo{pages}{3751} (\bibinfo{year}{1965}).

\bibitem[{\citenamefont{Bhatia and Drachman}(1997)}]{bhatia97a}
\bibinfo{author}{\bibfnamefont{A.~K.} \bibnamefont{Bhatia}} \bibnamefont{and}
  \bibinfo{author}{\bibfnamefont{R.~J.} \bibnamefont{Drachman}},
  \bibinfo{journal}{Can.~J.~Phys.} \textbf{\bibinfo{volume}{75}},
  \bibinfo{pages}{11} (\bibinfo{year}{1997}).

\bibitem[{\citenamefont{{Johnson} and {Cheng}}(1996)}]{johnson96b}
\bibinfo{author}{\bibfnamefont{W.~R.} \bibnamefont{{Johnson}}}
  \bibnamefont{and} \bibinfo{author}{\bibfnamefont{K.~T.}
  \bibnamefont{{Cheng}}}, \bibinfo{journal}{Phys.~Rev.~A}
  \textbf{\bibinfo{volume}{53}}, \bibinfo{pages}{1375} (\bibinfo{year}{1996}).

\bibitem[{\citenamefont{{Lim} et~al.}(2002)\citenamefont{{Lim}, {Laerdahl}, and
  {Schwerdtfeger}}}]{lim02a}
\bibinfo{author}{\bibfnamefont{I.~S.} \bibnamefont{{Lim}}},
  \bibinfo{author}{\bibfnamefont{J.~K.} \bibnamefont{{Laerdahl}}},
  \bibnamefont{and}
  \bibinfo{author}{\bibfnamefont{P.}~\bibnamefont{{Schwerdtfeger}}},
  \bibinfo{journal}{J.~Chem.~Phys.} \textbf{\bibinfo{volume}{116}},
  \bibinfo{pages}{172} (\bibinfo{year}{2002}).

\bibitem[{\citenamefont{{Cooke} et~al.}(1977)\citenamefont{{Cooke},
  {Gallagher}, {Hill}, and {Edelstein}}}]{cooke77a}
\bibinfo{author}{\bibfnamefont{W.~E.} \bibnamefont{{Cooke}}},
  \bibinfo{author}{\bibfnamefont{T.~F.} \bibnamefont{{Gallagher}}},
  \bibinfo{author}{\bibfnamefont{R.~M.} \bibnamefont{{Hill}}},
  \bibnamefont{and} \bibinfo{author}{\bibfnamefont{S.~A.}
  \bibnamefont{{Edelstein}}}, \bibinfo{journal}{Phys.~Rev.~A}
  \textbf{\bibinfo{volume}{16}}, \bibinfo{pages}{1141} (\bibinfo{year}{1977}).

\bibitem[{\citenamefont{{\"{O}}pik}(1967)}]{opik67a}
\bibinfo{author}{\bibfnamefont{U.}~\bibnamefont{{\"{O}}pik}},
  \bibinfo{journal}{Proc.~Phys.~Soc.~London} \textbf{\bibinfo{volume}{92}},
  \bibinfo{pages}{566} (\bibinfo{year}{1967}).

\bibitem[{\citenamefont{Johansson}(1960)}]{johansson60a}
\bibinfo{author}{\bibfnamefont{I.}~\bibnamefont{Johansson}},
  \bibinfo{journal}{Ark.~Fys.} \textbf{\bibinfo{volume}{20}},
  \bibinfo{pages}{135} (\bibinfo{year}{1960}).

\bibitem[{\citenamefont{{Safinya} et~al.}(1980)\citenamefont{{Safinya},
  {Gallagher}, and {Sandner}}}]{safinya80a}
\bibinfo{author}{\bibfnamefont{K.~A.} \bibnamefont{{Safinya}}},
  \bibinfo{author}{\bibfnamefont{T.~F.} \bibnamefont{{Gallagher}}},
  \bibnamefont{and}
  \bibinfo{author}{\bibfnamefont{W.}~\bibnamefont{{Sandner}}},
  \bibinfo{journal}{Phys.~Rev.~A} \textbf{\bibinfo{volume}{22}},
  \bibinfo{pages}{2672} (\bibinfo{year}{1980}).

\bibitem[{\citenamefont{Freeman and Kleppner}(1976)}]{freeman76a}
\bibinfo{author}{\bibfnamefont{R.~R.} \bibnamefont{Freeman}} \bibnamefont{and}
  \bibinfo{author}{\bibfnamefont{D.}~\bibnamefont{Kleppner}},
  \bibinfo{journal}{Phys. Rev. A} \textbf{\bibinfo{volume}{14}},
  \bibinfo{pages}{1614} (\bibinfo{year}{1976}).

\bibitem[{\citenamefont{{Curtis} and Ramanujam}(1981)}]{curtis81b}
\bibinfo{author}{\bibfnamefont{L.~J.} \bibnamefont{{Curtis}}} \bibnamefont{and}
  \bibinfo{author}{\bibfnamefont{P.~S.} \bibnamefont{Ramanujam}},
  \bibinfo{journal}{J.~Opt.~Soc.~Am.~B} \textbf{\bibinfo{volume}{71}},
  \bibinfo{pages}{1315} (\bibinfo{year}{1981}).

\bibitem[{\citenamefont{{Gray} et~al.}(1988)\citenamefont{{Gray}, {Sun}, and
  {MacAdam}}}]{gray88a}
\bibinfo{author}{\bibfnamefont{L.~G.} \bibnamefont{{Gray}}},
  \bibinfo{author}{\bibfnamefont{X.}~\bibnamefont{{Sun}}}, \bibnamefont{and}
  \bibinfo{author}{\bibfnamefont{K.~B.} \bibnamefont{{MacAdam}}},
  \bibinfo{journal}{Phys.~Rev.~A} \textbf{\bibinfo{volume}{38}},
  \bibinfo{pages}{4985} (\bibinfo{year}{1988}).

\bibitem[{\citenamefont{{Weber} and {Sansonetti}}(1987)}]{weber87a}
\bibinfo{author}{\bibfnamefont{K.-H.} \bibnamefont{{Weber}}} \bibnamefont{and}
  \bibinfo{author}{\bibfnamefont{C.~J.} \bibnamefont{{Sansonetti}}},
  \bibinfo{journal}{Phys.~Rev.~A} \textbf{\bibinfo{volume}{35}},
  \bibinfo{pages}{4650} (\bibinfo{year}{1987}).

\bibitem[{\citenamefont{{Bockasten}}(1956)}]{bockasten56a}
\bibinfo{author}{\bibfnamefont{K.}~\bibnamefont{{Bockasten}}},
  \bibinfo{journal}{Phys.~Rev.~} \textbf{\bibinfo{volume}{102}},
  \bibinfo{pages}{729} (\bibinfo{year}{1956}).

\bibitem[{\citenamefont{Hamonou and Hibbert}(2007)}]{hamonou07a}
\bibinfo{author}{\bibfnamefont{L.}~\bibnamefont{Hamonou}} \bibnamefont{and}
  \bibinfo{author}{\bibfnamefont{A.}~\bibnamefont{Hibbert}},
  \bibinfo{journal}{J.~Phys.~B} \textbf{\bibinfo{volume}{40}},
  \bibinfo{pages}{3555} (\bibinfo{year}{2007}).

\bibitem[{\citenamefont{Safronova and Clark}(2004)}]{safronova04b}
\bibinfo{author}{\bibfnamefont{M.~S.} \bibnamefont{Safronova}}
  \bibnamefont{and} \bibinfo{author}{\bibfnamefont{C.~W.} \bibnamefont{Clark}},
  \bibinfo{journal}{Phys. Rev. A} \textbf{\bibinfo{volume}{69}},
  \bibinfo{pages}{040501(R)} (\bibinfo{year}{2004}).

\bibitem[{\citenamefont{{Tang} et~al.}(2010)\citenamefont{{Tang}, {Yan}, {Shi},
  and {Mitroy}}}]{tang10a}
\bibinfo{author}{\bibfnamefont{L.-Y.} \bibnamefont{{Tang}}},
  \bibinfo{author}{\bibfnamefont{Z.-C.} \bibnamefont{{Yan}}},
  \bibinfo{author}{\bibfnamefont{T.-Y.} \bibnamefont{{Shi}}}, \bibnamefont{and}
  \bibinfo{author}{\bibfnamefont{J.}~\bibnamefont{{Mitroy}}},
  \bibinfo{journal}{Phys.~Rev.~A} \textbf{\bibinfo{volume}{10X}},
  \bibinfo{pages}{"in press"} (\bibinfo{year}{2010}), \eprint{1001.4116v1}.

\bibitem[{\citenamefont{Zhang et~al.}(2007{\natexlab{a}})\citenamefont{Zhang,
  Mitroy, and Bromley}}]{zhang07a}
\bibinfo{author}{\bibfnamefont{J.~Y.} \bibnamefont{Zhang}},
  \bibinfo{author}{\bibfnamefont{J.}~\bibnamefont{Mitroy}}, \bibnamefont{and}
  \bibinfo{author}{\bibfnamefont{M.~W.~J.} \bibnamefont{Bromley}},
  \bibinfo{journal}{Phys.~Rev.~A.} \textbf{\bibinfo{volume}{75}},
  \bibinfo{pages}{042509} (\bibinfo{year}{2007}{\natexlab{a}}).

\bibitem[{\citenamefont{{Maroulis}}(2004)}]{maroulis04a}
\bibinfo{author}{\bibfnamefont{G.}~\bibnamefont{{Maroulis}}},
  \bibinfo{journal}{J.~Chem.~Phys.} \textbf{\bibinfo{volume}{121}},
  \bibinfo{pages}{10519} (\bibinfo{year}{2004}).

\bibitem[{\citenamefont{{Lim} et~al.}(1999)\citenamefont{{Lim}, {Pernpointner},
  {Seth}, {Laerdahl}, {Schwerdtfeger}, {Neogrady}, and {Urban}}}]{lim99a}
\bibinfo{author}{\bibfnamefont{I.~S.} \bibnamefont{{Lim}}},
  \bibinfo{author}{\bibfnamefont{M.}~\bibnamefont{{Pernpointner}}},
  \bibinfo{author}{\bibfnamefont{M.}~\bibnamefont{{Seth}}},
  \bibinfo{author}{\bibfnamefont{J.~K.} \bibnamefont{{Laerdahl}}},
  \bibinfo{author}{\bibfnamefont{P.}~\bibnamefont{{Schwerdtfeger}}},
  \bibinfo{author}{\bibfnamefont{P.}~\bibnamefont{{Neogrady}}},
  \bibnamefont{and} \bibinfo{author}{\bibfnamefont{M.}~\bibnamefont{{Urban}}},
  \bibinfo{journal}{Phys.~Rev.~A} \textbf{\bibinfo{volume}{60}},
  \bibinfo{pages}{2822} (\bibinfo{year}{1999}).

\bibitem[{\citenamefont{{Thakkar} and {Lupinetti}}(2005)}]{thakkar05a}
\bibinfo{author}{\bibfnamefont{A.~J.} \bibnamefont{{Thakkar}}}
  \bibnamefont{and}
  \bibinfo{author}{\bibfnamefont{C.}~\bibnamefont{{Lupinetti}}},
  \bibinfo{journal}{Chem.~Phys.~Lett.} \textbf{\bibinfo{volume}{402}},
  \bibinfo{pages}{270} (\bibinfo{year}{2005}).

\bibitem[{\citenamefont{{Johnson} et~al.}(2008)\citenamefont{{Johnson},
  {Safronova}, {Derevianko}, and {Safronova}}}]{johnson08a}
\bibinfo{author}{\bibfnamefont{W.~R.} \bibnamefont{{Johnson}}},
  \bibinfo{author}{\bibfnamefont{U.~I.} \bibnamefont{{Safronova}}},
  \bibinfo{author}{\bibfnamefont{A.}~\bibnamefont{{Derevianko}}},
  \bibnamefont{and} \bibinfo{author}{\bibfnamefont{M.~S.}
  \bibnamefont{{Safronova}}}, \bibinfo{journal}{Phys.~Rev.~A}
  \textbf{\bibinfo{volume}{77}}, \bibinfo{pages}{022510}
  (\bibinfo{year}{2008}).

\bibitem[{\citenamefont{{Safronova} and {Safronova}}(2008)}]{safronova08b}
\bibinfo{author}{\bibfnamefont{U.~I.} \bibnamefont{{Safronova}}}
  \bibnamefont{and} \bibinfo{author}{\bibfnamefont{M.~S.}
  \bibnamefont{{Safronova}}}, \bibinfo{journal}{Phys.~Rev.~A}
  \textbf{\bibinfo{volume}{78}}, \bibinfo{pages}{052504}
  (\bibinfo{year}{2008}).

\bibitem[{\citenamefont{{Iskrenova-Tchoukova}
  et~al.}(2007)\citenamefont{{Iskrenova-Tchoukova}, {Safronova}, and
  {Safronova}}}]{iskrenova07a}
\bibinfo{author}{\bibfnamefont{E.}~\bibnamefont{{Iskrenova-Tchoukova}}},
  \bibinfo{author}{\bibfnamefont{M.~S.} \bibnamefont{{Safronova}}},
  \bibnamefont{and} \bibinfo{author}{\bibfnamefont{U.~I.}
  \bibnamefont{{Safronova}}}, \bibinfo{journal}{J.~Comput.~Methods~Sci.~Eng.}
  \textbf{\bibinfo{volume}{7}}, \bibinfo{pages}{521} (\bibinfo{year}{2007}).

\bibitem[{\citenamefont{{Safronova} et~al.}(2007)\citenamefont{{Safronova},
  {Johnson}, and {Safronova}}}]{safronova07a}
\bibinfo{author}{\bibfnamefont{U.~I.} \bibnamefont{{Safronova}}},
  \bibinfo{author}{\bibfnamefont{W.~R.} \bibnamefont{{Johnson}}},
  \bibnamefont{and} \bibinfo{author}{\bibfnamefont{M.~S.}
  \bibnamefont{{Safronova}}}, \bibinfo{journal}{Phys.~Rev.~A}
  \textbf{\bibinfo{volume}{76}}, \bibinfo{pages}{042504}
  (\bibinfo{year}{2007}).

\bibitem[{\citenamefont{Zhang and Mitroy}(2007)}]{zhang07c}
\bibinfo{author}{\bibfnamefont{J.~Y.} \bibnamefont{Zhang}} \bibnamefont{and}
  \bibinfo{author}{\bibfnamefont{J.}~\bibnamefont{Mitroy}},
  \bibinfo{journal}{Phys.~Rev.~A} \textbf{\bibinfo{volume}{76}},
  \bibinfo{pages}{022705} (\bibinfo{year}{2007}).

\bibitem[{\citenamefont{Zhu et~al.}(2004)\citenamefont{Zhu, Dalgarno, Porsev,
  and Derevianko}}]{zhu04a}
\bibinfo{author}{\bibfnamefont{C.}~\bibnamefont{Zhu}},
  \bibinfo{author}{\bibfnamefont{A.}~\bibnamefont{Dalgarno}},
  \bibinfo{author}{\bibfnamefont{S.~G.} \bibnamefont{Porsev}},
  \bibnamefont{and}
  \bibinfo{author}{\bibfnamefont{A.}~\bibnamefont{Derevianko}},
  \bibinfo{journal}{Phys.~Rev.~A.} \textbf{\bibinfo{volume}{70}},
  \bibinfo{pages}{032722} (\bibinfo{year}{2004}).

\bibitem[{\citenamefont{Windholz et~al.}(1992)\citenamefont{Windholz, Musso,
  Zerza, and Jager}}]{windholz92a}
\bibinfo{author}{\bibfnamefont{L.}~\bibnamefont{Windholz}},
  \bibinfo{author}{\bibfnamefont{M.}~\bibnamefont{Musso}},
  \bibinfo{author}{\bibfnamefont{G.}~\bibnamefont{Zerza}}, \bibnamefont{and}
  \bibinfo{author}{\bibfnamefont{H.}~\bibnamefont{Jager}},
  \bibinfo{journal}{Phys.~Rev.~A} \textbf{\bibinfo{volume}{46}},
  \bibinfo{pages}{5812} (\bibinfo{year}{1992}).

\bibitem[{\citenamefont{Windholz and Musso}(1989)}]{windholz89a}
\bibinfo{author}{\bibfnamefont{L.}~\bibnamefont{Windholz}} \bibnamefont{and}
  \bibinfo{author}{\bibfnamefont{M.}~\bibnamefont{Musso}},
  \bibinfo{journal}{Phys.~Rev.~A} \textbf{\bibinfo{volume}{39}},
  \bibinfo{pages}{2472} (\bibinfo{year}{1989}).

\bibitem[{\citenamefont{Krenn et~al.}(1997)\citenamefont{Krenn, Scherf, Khait,
  Musso, and Windholz}}]{krenn97a}
\bibinfo{author}{\bibfnamefont{C.}~\bibnamefont{Krenn}},
  \bibinfo{author}{\bibfnamefont{W.}~\bibnamefont{Scherf}},
  \bibinfo{author}{\bibfnamefont{O.}~\bibnamefont{Khait}},
  \bibinfo{author}{\bibfnamefont{M.}~\bibnamefont{Musso}}, \bibnamefont{and}
  \bibinfo{author}{\bibfnamefont{L.}~\bibnamefont{Windholz}},
  \bibinfo{journal}{Z.~Phys.~D} \textbf{\bibinfo{volume}{41}},
  \bibinfo{pages}{229} (\bibinfo{year}{1997}).

\bibitem[{\citenamefont{{Hannaford} et~al.}(1979)\citenamefont{{Hannaford},
  {MacGillivray}, and {Standage}}}]{hannaford79a}
\bibinfo{author}{\bibfnamefont{P.}~\bibnamefont{{Hannaford}}},
  \bibinfo{author}{\bibfnamefont{W.~R.} \bibnamefont{{MacGillivray}}},
  \bibnamefont{and} \bibinfo{author}{\bibfnamefont{M.~C.}
  \bibnamefont{{Standage}}}, \bibinfo{journal}{J.~Phys.~B}
  \textbf{\bibinfo{volume}{12}}, \bibinfo{pages}{4033} (\bibinfo{year}{1979}).

\bibitem[{\citenamefont{Kawamura et~al.}(2009)\citenamefont{Kawamura, Jin,
  Takahasi, and Minowi}}]{kawamura09a}
\bibinfo{author}{\bibfnamefont{M.}~\bibnamefont{Kawamura}},
  \bibinfo{author}{\bibfnamefont{W.}~\bibnamefont{Jin}},
  \bibinfo{author}{\bibfnamefont{N.}~\bibnamefont{Takahasi}}, \bibnamefont{and}
  \bibinfo{author}{\bibfnamefont{T.}~\bibnamefont{Minowi}},
  \bibinfo{journal}{J.~Phys.~Soc.~Japan} \textbf{\bibinfo{volume}{78}},
  \bibinfo{pages}{034301} (\bibinfo{year}{2009}).

\bibitem[{\citenamefont{Tanner and Wieman}(1988)}]{tanner88a}
\bibinfo{author}{\bibfnamefont{C.~E.} \bibnamefont{Tanner}} \bibnamefont{and}
  \bibinfo{author}{\bibfnamefont{C.}~\bibnamefont{Wieman}},
  \bibinfo{journal}{Phys. Rev. A} \textbf{\bibinfo{volume}{38}},
  \bibinfo{pages}{162} (\bibinfo{year}{1988}).

\bibitem[{\citenamefont{Werner and Meyer}(1976)}]{werner76a}
\bibinfo{author}{\bibfnamefont{H.-J.} \bibnamefont{Werner}} \bibnamefont{and}
  \bibinfo{author}{\bibfnamefont{W.}~\bibnamefont{Meyer}},
  \bibinfo{journal}{Phys. Rev. A} \textbf{\bibinfo{volume}{13}},
  \bibinfo{pages}{13} (\bibinfo{year}{1976}).

\bibitem[{\citenamefont{{Tang} et~al.}(2009{\natexlab{b}})\citenamefont{{Tang},
  {Zhang}, {Yan}, {Shi}, {Babb}, and {Mitroy}}}]{tang09b}
\bibinfo{author}{\bibfnamefont{L.-Y.} \bibnamefont{{Tang}}},
  \bibinfo{author}{\bibfnamefont{J.-Y.} \bibnamefont{{Zhang}}},
  \bibinfo{author}{\bibfnamefont{Z.-C.} \bibnamefont{{Yan}}},
  \bibinfo{author}{\bibfnamefont{T.-Y.} \bibnamefont{{Shi}}},
  \bibinfo{author}{\bibfnamefont{J.~F.} \bibnamefont{{Babb}}},
  \bibnamefont{and} \bibinfo{author}{\bibfnamefont{J.}~\bibnamefont{{Mitroy}}},
  \bibinfo{journal}{Phys.~Rev.~A} \textbf{\bibinfo{volume}{80}},
  \bibinfo{pages}{042511} (\bibinfo{year}{2009}{\natexlab{b}}).

\bibitem[{\citenamefont{{Wang} and {Chung}}(1994)}]{wang94a}
\bibinfo{author}{\bibfnamefont{Z.-W.} \bibnamefont{{Wang}}} \bibnamefont{and}
  \bibinfo{author}{\bibfnamefont{K.~T.} \bibnamefont{{Chung}}},
  \bibinfo{journal}{J.~Phys.~B} \textbf{\bibinfo{volume}{27}},
  \bibinfo{pages}{855} (\bibinfo{year}{1994}).

\bibitem[{\citenamefont{Mitroy and Zhang}(2008{\natexlab{a}})}]{mitroy08b}
\bibinfo{author}{\bibfnamefont{J.}~\bibnamefont{Mitroy}} \bibnamefont{and}
  \bibinfo{author}{\bibfnamefont{J.~Y.} \bibnamefont{Zhang}},
  \bibinfo{journal}{Eur.~Phys.~J D} \textbf{\bibinfo{volume}{46}},
  \bibinfo{pages}{415} (\bibinfo{year}{2008}{\natexlab{a}}).

\bibitem[{\citenamefont{Mitroy et~al.}(2008)\citenamefont{Mitroy, Zhang, and
  Bromley}}]{mitroy08c}
\bibinfo{author}{\bibfnamefont{J.}~\bibnamefont{Mitroy}},
  \bibinfo{author}{\bibfnamefont{J.~Y.} \bibnamefont{Zhang}}, \bibnamefont{and}
  \bibinfo{author}{\bibfnamefont{M.~W.~J.} \bibnamefont{Bromley}},
  \bibinfo{journal}{Phys.~Rev.~A} \textbf{\bibinfo{volume}{77}},
  \bibinfo{pages}{032512} (\bibinfo{year}{2008}).

\bibitem[{\citenamefont{{Arora}
  et~al.}(2007{\natexlab{b}})\citenamefont{{Arora}, {Safronova}, and
  {Clark}}}]{arora07a}
\bibinfo{author}{\bibfnamefont{B.}~\bibnamefont{{Arora}}},
  \bibinfo{author}{\bibfnamefont{M.~S.} \bibnamefont{{Safronova}}},
  \bibnamefont{and} \bibinfo{author}{\bibfnamefont{C.~W.}
  \bibnamefont{{Clark}}}, \bibinfo{journal}{Phys. Rev. A}
  \textbf{\bibinfo{volume}{76}}, \bibinfo{pages}{064501}
  (\bibinfo{year}{2007}{\natexlab{b}}).

\bibitem[{\citenamefont{{Jiang} et~al.}(2009)\citenamefont{{Jiang}, {Arora},
  {Safronova}, and {Clark}}}]{jiang09a}
\bibinfo{author}{\bibfnamefont{D.}~\bibnamefont{{Jiang}}},
  \bibinfo{author}{\bibfnamefont{B.}~\bibnamefont{{Arora}}},
  \bibinfo{author}{\bibfnamefont{M.~S.} \bibnamefont{{Safronova}}},
  \bibnamefont{and} \bibinfo{author}{\bibfnamefont{C.~W.}
  \bibnamefont{{Clark}}}, \bibinfo{journal}{J.~Phys.~B}
  \textbf{\bibinfo{volume}{42}}, \bibinfo{pages}{150420}
  (\bibinfo{year}{2009}).

\bibitem[{\citenamefont{Lyons and Gallagher}(1998)}]{lyons98a}
\bibinfo{author}{\bibfnamefont{B.~J.} \bibnamefont{Lyons}} \bibnamefont{and}
  \bibinfo{author}{\bibfnamefont{T.~F.} \bibnamefont{Gallagher}},
  \bibinfo{journal}{Phys.~ Rev.~A} \textbf{\bibinfo{volume}{57}},
  \bibinfo{pages}{2426} (\bibinfo{year}{1998}).

\bibitem[{\citenamefont{{Chang}}(1983)}]{chang83a}
\bibinfo{author}{\bibfnamefont{E.~S.} \bibnamefont{{Chang}}},
  \bibinfo{journal}{J.~Phys.~B} \textbf{\bibinfo{volume}{16}},
  \bibinfo{pages}{L539} (\bibinfo{year}{1983}).

\bibitem[{\citenamefont{{Gallagher} et~al.}(1982)\citenamefont{{Gallagher},
  {Kachru}, and {Tran}}}]{gallagher82a}
\bibinfo{author}{\bibfnamefont{T.~F.} \bibnamefont{{Gallagher}}},
  \bibinfo{author}{\bibfnamefont{R.}~\bibnamefont{{Kachru}}}, \bibnamefont{and}
  \bibinfo{author}{\bibfnamefont{N.~H.} \bibnamefont{{Tran}}},
  \bibinfo{journal}{Phys.~Rev.~A} \textbf{\bibinfo{volume}{26}},
  \bibinfo{pages}{2611} (\bibinfo{year}{1982}).

\bibitem[{\citenamefont{{Snow} et~al.}(2005)\citenamefont{{Snow}, {Gearba},
  {Komara}, {Lundeen}, and {Sturrus}}}]{snow05a}
\bibinfo{author}{\bibfnamefont{E.~L.} \bibnamefont{{Snow}}},
  \bibinfo{author}{\bibfnamefont{M.~A.} \bibnamefont{{Gearba}}},
  \bibinfo{author}{\bibfnamefont{R.~A.} \bibnamefont{{Komara}}},
  \bibinfo{author}{\bibfnamefont{S.~R.} \bibnamefont{{Lundeen}}},
  \bibnamefont{and} \bibinfo{author}{\bibfnamefont{W.~G.}
  \bibnamefont{{Sturrus}}}, \bibinfo{journal}{Phys.~Rev.~A}
  \textbf{\bibinfo{volume}{71}}, \bibinfo{pages}{022510}
  (\bibinfo{year}{2005}).

\bibitem[{\citenamefont{{Snow} and {Lundeen}}(2008)}]{snow08a}
\bibinfo{author}{\bibfnamefont{E.~L.} \bibnamefont{{Snow}}} \bibnamefont{and}
  \bibinfo{author}{\bibfnamefont{S.~R.} \bibnamefont{{Lundeen}}},
  \bibinfo{journal}{Phys.~Rev.~A} \textbf{\bibinfo{volume}{77}},
  \bibinfo{pages}{052501} (\bibinfo{year}{2008}).

\bibitem[{\citenamefont{{Snow} and {Lundeen}}(2007{\natexlab{a}})}]{snow07b}
\bibinfo{author}{\bibfnamefont{E.~L.} \bibnamefont{{Snow}}} \bibnamefont{and}
  \bibinfo{author}{\bibfnamefont{S.~R.} \bibnamefont{{Lundeen}}},
  \bibinfo{journal}{Phys.~Rev.~A} \textbf{\bibinfo{volume}{76}},
  \bibinfo{pages}{052505} (\bibinfo{year}{2007}{\natexlab{a}}).

\bibitem[{\citenamefont{Theodosiou et~al.}(1995)\citenamefont{Theodosiou,
  Curtis, and Nicolaides}}]{theodosiou95a}
\bibinfo{author}{\bibfnamefont{C.~E.} \bibnamefont{Theodosiou}},
  \bibinfo{author}{\bibfnamefont{L.~J.} \bibnamefont{Curtis}},
  \bibnamefont{and} \bibinfo{author}{\bibfnamefont{C.~A.}
  \bibnamefont{Nicolaides}}, \bibinfo{journal}{Phys.~Rev.~A}
  \textbf{\bibinfo{volume}{52}}, \bibinfo{pages}{3677} (\bibinfo{year}{1995}).

\bibitem[{\citenamefont{Miller et~al.}(1994)\citenamefont{Miller, Krause~Jr.,
  and Hunter}}]{miller94a}
\bibinfo{author}{\bibfnamefont{K.~E.} \bibnamefont{Miller}},
  \bibinfo{author}{\bibfnamefont{D.}~\bibnamefont{Krause~Jr.}},
  \bibnamefont{and} \bibinfo{author}{\bibfnamefont{L.~R.}
  \bibnamefont{Hunter}}, \bibinfo{journal}{Phys.~Rev.~A}
  \textbf{\bibinfo{volume}{49}}, \bibinfo{pages}{5128} (\bibinfo{year}{1994}).

\bibitem[{\citenamefont{{Windholz} and {Neureiter}}(1985)}]{windholz85a}
\bibinfo{author}{\bibfnamefont{L.}~\bibnamefont{{Windholz}}} \bibnamefont{and}
  \bibinfo{author}{\bibfnamefont{C.}~\bibnamefont{{Neureiter}}},
  \bibinfo{journal}{Phys.~Lett.~A} \textbf{\bibinfo{volume}{109}},
  \bibinfo{pages}{155} (\bibinfo{year}{1985}).

\bibitem[{\citenamefont{Ashby et~al.}(2003)\citenamefont{Ashby, Clarke, and van
  Wijngaarden}}]{ashby03a}
\bibinfo{author}{\bibfnamefont{R.}~\bibnamefont{Ashby}},
  \bibinfo{author}{\bibfnamefont{J.~J.} \bibnamefont{Clarke}},
  \bibnamefont{and} \bibinfo{author}{\bibfnamefont{W.~A.} \bibnamefont{van
  Wijngaarden}}, \bibinfo{journal}{Eur.~Phys.~J.~D}
  \textbf{\bibinfo{volume}{23}}, \bibinfo{pages}{327} (\bibinfo{year}{2003}).

\bibitem[{\citenamefont{Harvey et~al.}(1975)\citenamefont{Harvey, Hawkins,
  Meisel, and Schawlow}}]{harvey75a}
\bibinfo{author}{\bibfnamefont{K.~C.} \bibnamefont{Harvey}},
  \bibinfo{author}{\bibfnamefont{R.~T.} \bibnamefont{Hawkins}},
  \bibinfo{author}{\bibfnamefont{G.}~\bibnamefont{Meisel}}, \bibnamefont{and}
  \bibinfo{author}{\bibfnamefont{A.~L.} \bibnamefont{Schawlow}},
  \bibinfo{journal}{Phys. Rev. Lett.} \textbf{\bibinfo{volume}{34}},
  \bibinfo{pages}{1073} (\bibinfo{year}{1975}).

\bibitem[{\citenamefont{{Schmieder} et~al.}(1971)\citenamefont{{Schmieder},
  {Lurio}, and {Happer}}}]{schmeider71a}
\bibinfo{author}{\bibfnamefont{R.~W.} \bibnamefont{{Schmieder}}},
  \bibinfo{author}{\bibfnamefont{A.}~\bibnamefont{{Lurio}}}, \bibnamefont{and}
  \bibinfo{author}{\bibfnamefont{W.}~\bibnamefont{{Happer}}},
  \bibinfo{journal}{Phys.~Rev.~A} \textbf{\bibinfo{volume}{3}},
  \bibinfo{pages}{1209} (\bibinfo{year}{1971}).

\bibitem[{\citenamefont{Khadjavi et~al.}(1968)\citenamefont{Khadjavi, Lurio,
  and Happer}}]{khadjavi68a}
\bibinfo{author}{\bibfnamefont{A.}~\bibnamefont{Khadjavi}},
  \bibinfo{author}{\bibfnamefont{A.}~\bibnamefont{Lurio}}, \bibnamefont{and}
  \bibinfo{author}{\bibfnamefont{W.}~\bibnamefont{Happer}},
  \bibinfo{journal}{Phys. Rev.} \textbf{\bibinfo{volume}{167}},
  \bibinfo{pages}{128} (\bibinfo{year}{1968}).

\bibitem[{\citenamefont{{van Wijngaarden}}(1997)}]{wijngaarden97a}
\bibinfo{author}{\bibfnamefont{W.}~\bibnamefont{{van Wijngaarden}}},
  \bibinfo{journal}{J. Quant. Spect. Rad. Transf.}
  \textbf{\bibinfo{volume}{57}}, \bibinfo{pages}{275} (\bibinfo{year}{1997}).

\bibitem[{\citenamefont{{Hogervorst} and {Svanberg}}(1975)}]{hogevorst75a}
\bibinfo{author}{\bibfnamefont{W.}~\bibnamefont{{Hogervorst}}}
  \bibnamefont{and}
  \bibinfo{author}{\bibfnamefont{S.}~\bibnamefont{{Svanberg}}},
  \bibinfo{journal}{Phys.~Src.} \textbf{\bibinfo{volume}{12}},
  \bibinfo{pages}{67} (\bibinfo{year}{1975}).

\bibitem[{\citenamefont{{Svanberg}}(1972)}]{svanberg72a}
\bibinfo{author}{\bibfnamefont{S.}~\bibnamefont{{Svanberg}}},
  \bibinfo{journal}{Phys.~Scr.} \textbf{\bibinfo{volume}{5}},
  \bibinfo{pages}{132} (\bibinfo{year}{1972}).

\bibitem[{\citenamefont{{Yoshimine} and {Hurst}}(1964)}]{yoshimine64a}
\bibinfo{author}{\bibfnamefont{M.}~\bibnamefont{{Yoshimine}}} \bibnamefont{and}
  \bibinfo{author}{\bibfnamefont{R.~P.} \bibnamefont{{Hurst}}},
  \bibinfo{journal}{Phys.~Rev.} \textbf{\bibinfo{volume}{135}},
  \bibinfo{pages}{612} (\bibinfo{year}{1964}).

\bibitem[{\citenamefont{Reinsch and Meyer}(1976)}]{reinsch76a}
\bibinfo{author}{\bibfnamefont{E.~A.} \bibnamefont{Reinsch}} \bibnamefont{and}
  \bibinfo{author}{\bibfnamefont{W.}~\bibnamefont{Meyer}},
  \bibinfo{journal}{Phys.~Rev.~A} \textbf{\bibinfo{volume}{14}},
  \bibinfo{pages}{915} (\bibinfo{year}{1976}).

\bibitem[{\citenamefont{Maeder and Kutzelnigg}(1979)}]{maeder79a}
\bibinfo{author}{\bibfnamefont{F.}~\bibnamefont{Maeder}} \bibnamefont{and}
  \bibinfo{author}{\bibfnamefont{W.}~\bibnamefont{Kutzelnigg}},
  \bibinfo{journal}{Chem.~Phys.} \textbf{\bibinfo{volume}{42}},
  \bibinfo{pages}{95} (\bibinfo{year}{1979}).

\bibitem[{\citenamefont{{Maroulis}}(2001)}]{maroulis01a}
\bibinfo{author}{\bibfnamefont{G.}~\bibnamefont{{Maroulis}}},
  \bibinfo{journal}{Chem.~Phys.~Lett.} \textbf{\bibinfo{volume}{334}},
  \bibinfo{pages}{207} (\bibinfo{year}{2001}).

\bibitem[{\citenamefont{{Sold{\'a}n} et~al.}(2003)\citenamefont{{Sold{\'a}n},
  {Cvita{\v s}}, and {Hutson}}}]{soldan03a}
\bibinfo{author}{\bibfnamefont{P.}~\bibnamefont{{Sold{\'a}n}}},
  \bibinfo{author}{\bibfnamefont{M.~T.} \bibnamefont{{Cvita{\v s}}}},
  \bibnamefont{and} \bibinfo{author}{\bibfnamefont{J.~M.}
  \bibnamefont{{Hutson}}}, \bibinfo{journal}{Phys.~Rev.~A}
  \textbf{\bibinfo{volume}{67}}, \bibinfo{pages}{054702}
  (\bibinfo{year}{2003}).

\bibitem[{\citenamefont{Dalgarno and Kingston}(1959)}]{dalgarno59a}
\bibinfo{author}{\bibfnamefont{A.}~\bibnamefont{Dalgarno}} \bibnamefont{and}
  \bibinfo{author}{\bibfnamefont{A.~E.} \bibnamefont{Kingston}},
  \bibinfo{journal}{Proc.~Phys.~Soc.~London} \textbf{\bibinfo{volume}{73}},
  \bibinfo{pages}{455} (\bibinfo{year}{1959}).

\bibitem[{\citenamefont{{Gunawardena} et~al.}(2007)\citenamefont{{Gunawardena},
  {Elliott}, {Safronova}, and {Safronova}}}]{gunawardena07a}
\bibinfo{author}{\bibfnamefont{M.}~\bibnamefont{{Gunawardena}}},
  \bibinfo{author}{\bibfnamefont{D.~S.} \bibnamefont{{Elliott}}},
  \bibinfo{author}{\bibfnamefont{M.~S.} \bibnamefont{{Safronova}}},
  \bibnamefont{and}
  \bibinfo{author}{\bibfnamefont{U.}~\bibnamefont{{Safronova}}},
  \bibinfo{journal}{Phys.~Rev.~A} \textbf{\bibinfo{volume}{75}},
  \bibinfo{pages}{022507} (\bibinfo{year}{2007}).

\bibitem[{\citenamefont{{van Wijngaarden} et~al.}(1994)\citenamefont{{van
  Wijngaarden}, {Hessels}, {Li}, and {Rothery}}}]{wijngaarden94a}
\bibinfo{author}{\bibfnamefont{W.~A.} \bibnamefont{{van Wijngaarden}}},
  \bibinfo{author}{\bibfnamefont{E.~A.} \bibnamefont{{Hessels}}},
  \bibinfo{author}{\bibfnamefont{J.}~\bibnamefont{{Li}}}, \bibnamefont{and}
  \bibinfo{author}{\bibfnamefont{N.~E.} \bibnamefont{{Rothery}}},
  \bibinfo{journal}{Phys.~Rev.~A} \textbf{\bibinfo{volume}{49}},
  \bibinfo{pages}{2220} (\bibinfo{year}{1994}).

\bibitem[{\citenamefont{van Wijngaarden and Li}(1994)}]{wijngaarden:1994}
\bibinfo{author}{\bibfnamefont{W.}~\bibnamefont{van Wijngaarden}}
  \bibnamefont{and} \bibinfo{author}{\bibfnamefont{J.}~\bibnamefont{Li}},
  \bibinfo{journal}{J. Quant. Spect. Rad. Transf.}
  \textbf{\bibinfo{volume}{52}}, \bibinfo{pages}{555} (\bibinfo{year}{1994}).

\bibitem[{\citenamefont{Khvoshtenko and Chaika}(1968)}]{khvoshtenko68a}
\bibinfo{author}{\bibfnamefont{G.}~\bibnamefont{Khvoshtenko}} \bibnamefont{and}
  \bibinfo{author}{\bibfnamefont{M.}~\bibnamefont{Chaika}},
  \bibinfo{journal}{Optica.~Spectrosk.} \textbf{\bibinfo{volume}{25}},
  \bibinfo{pages}{246} (\bibinfo{year}{1968}).

\bibitem[{\citenamefont{{Fredriksson} and {Svanberg}}(1977)}]{fredriksson77a}
\bibinfo{author}{\bibfnamefont{K.}~\bibnamefont{{Fredriksson}}}
  \bibnamefont{and}
  \bibinfo{author}{\bibfnamefont{S.}~\bibnamefont{{Svanberg}}},
  \bibinfo{journal}{Z.~Physik.} \textbf{\bibinfo{volume}{281}},
  \bibinfo{pages}{189} (\bibinfo{year}{1977}).

\bibitem[{\citenamefont{Wessel and Cooper}(1987)}]{wessel87a}
\bibinfo{author}{\bibfnamefont{J.~E.} \bibnamefont{Wessel}} \bibnamefont{and}
  \bibinfo{author}{\bibfnamefont{D.~E.} \bibnamefont{Cooper}},
  \bibinfo{journal}{Phys. Rev. A} \textbf{\bibinfo{volume}{35}},
  \bibinfo{pages}{1621} (\bibinfo{year}{1987}).

\bibitem[{\citenamefont{Xia et~al.}(1997)\citenamefont{Xia, Clarke, Li, and van
  Wijngaarden}}]{xia:1997}
\bibinfo{author}{\bibfnamefont{J.}~\bibnamefont{Xia}},
  \bibinfo{author}{\bibfnamefont{J.}~\bibnamefont{Clarke}},
  \bibinfo{author}{\bibfnamefont{J.}~\bibnamefont{Li}}, \bibnamefont{and}
  \bibinfo{author}{\bibfnamefont{W.}~\bibnamefont{van Wijngaarden}},
  \bibinfo{journal}{Phys. Rev. A} \textbf{\bibinfo{volume}{56}},
  \bibinfo{pages}{5176} (\bibinfo{year}{1997}).

\bibitem[{\citenamefont{{Auzinsh} et~al.}(2006)\citenamefont{{Auzinsh},
  {Blushs}, {Ferber}, {Gahbauer}, {Jarmola}, and {Tamanis}}}]{auznish06a}
\bibinfo{author}{\bibfnamefont{M.}~\bibnamefont{{Auzinsh}}},
  \bibinfo{author}{\bibfnamefont{K.}~\bibnamefont{{Blushs}}},
  \bibinfo{author}{\bibfnamefont{R.}~\bibnamefont{{Ferber}}},
  \bibinfo{author}{\bibfnamefont{F.}~\bibnamefont{{Gahbauer}}},
  \bibinfo{author}{\bibfnamefont{A.}~\bibnamefont{{Jarmola}}},
  \bibnamefont{and}
  \bibinfo{author}{\bibfnamefont{M.}~\bibnamefont{{Tamanis}}},
  \bibinfo{journal}{Opt.~Commun.} \textbf{\bibinfo{volume}{264}},
  \bibinfo{pages}{333} (\bibinfo{year}{2006}).

\bibitem[{\citenamefont{Domelunksen}(1983)}]{domelunksen83a}
\bibinfo{author}{\bibfnamefont{V.}~\bibnamefont{Domelunksen}},
  \bibinfo{journal}{Optica.~Spectrosk.} \textbf{\bibinfo{volume}{54}},
  \bibinfo{pages}{950} (\bibinfo{year}{1983}).

\bibitem[{\citenamefont{{Auzinsh} et~al.}(2007)\citenamefont{{Auzinsh},
  {Bluss}, {Ferber}, {Gahbauer}, {Jarmola}, {Safronova}, {Safronova}, and
  {Tamanis}}}]{auznish07a}
\bibinfo{author}{\bibfnamefont{M.}~\bibnamefont{{Auzinsh}}},
  \bibinfo{author}{\bibfnamefont{K.}~\bibnamefont{{Bluss}}},
  \bibinfo{author}{\bibfnamefont{R.}~\bibnamefont{{Ferber}}},
  \bibinfo{author}{\bibfnamefont{F.}~\bibnamefont{{Gahbauer}}},
  \bibinfo{author}{\bibfnamefont{A.}~\bibnamefont{{Jarmola}}},
  \bibinfo{author}{\bibfnamefont{M.~S.} \bibnamefont{{Safronova}}},
  \bibinfo{author}{\bibfnamefont{U.~I.} \bibnamefont{{Safronova}}},
  \bibnamefont{and}
  \bibinfo{author}{\bibfnamefont{M.}~\bibnamefont{{Tamanis}}},
  \bibinfo{journal}{Phys.~Rev.~A} \textbf{\bibinfo{volume}{75}},
  \bibinfo{pages}{022502} (\bibinfo{year}{2007}).

\bibitem[{\citenamefont{{Tunega}}(1997)}]{tunega97a}
\bibinfo{author}{\bibfnamefont{D.}~\bibnamefont{{Tunega}}},
  \bibinfo{journal}{Chem.~Phys.~Lett.} \textbf{\bibinfo{volume}{269}},
  \bibinfo{pages}{435} (\bibinfo{year}{1997}).

\bibitem[{\citenamefont{Archibong and Thakkar}(1991)}]{archibong91a}
\bibinfo{author}{\bibfnamefont{E.~F.} \bibnamefont{Archibong}}
  \bibnamefont{and} \bibinfo{author}{\bibfnamefont{A.~J.}
  \bibnamefont{Thakkar}}, \bibinfo{journal}{Phys.~Rev.~A}
  \textbf{\bibinfo{volume}{44}}, \bibinfo{pages}{5478} (\bibinfo{year}{1991}).

\bibitem[{\citenamefont{{Bendazzoli} and {Monari}}(2004)}]{bendazzoli04a}
\bibinfo{author}{\bibfnamefont{G.~L.} \bibnamefont{{Bendazzoli}}}
  \bibnamefont{and} \bibinfo{author}{\bibfnamefont{A.}~\bibnamefont{{Monari}}},
  \bibinfo{journal}{Chem.~Phys.} \textbf{\bibinfo{volume}{306}},
  \bibinfo{pages}{153} (\bibinfo{year}{2004}).

\bibitem[{\citenamefont{Hamonou and Hibbert}(2008)}]{hamonou08a}
\bibinfo{author}{\bibfnamefont{L.}~\bibnamefont{Hamonou}} \bibnamefont{and}
  \bibinfo{author}{\bibfnamefont{A.}~\bibnamefont{Hibbert}},
  \bibinfo{journal}{J.~Phys.~B} \textbf{\bibinfo{volume}{41}},
  \bibinfo{pages}{245004} (\bibinfo{year}{2008}).

\bibitem[{\citenamefont{{Glass}}(1987)}]{glass87a}
\bibinfo{author}{\bibfnamefont{R.}~\bibnamefont{{Glass}}},
  \bibinfo{journal}{J.~Phys.~B} \textbf{\bibinfo{volume}{20}},
  \bibinfo{pages}{4649} (\bibinfo{year}{1987}).

\bibitem[{\citenamefont{Reshetnikov et~al.}(2008)\citenamefont{Reshetnikov,
  Curtis, Brown, and Irving}}]{reshetnikov08a}
\bibinfo{author}{\bibfnamefont{N.}~\bibnamefont{Reshetnikov}},
  \bibinfo{author}{\bibfnamefont{L.~J.} \bibnamefont{Curtis}},
  \bibinfo{author}{\bibfnamefont{M.~S.} \bibnamefont{Brown}}, \bibnamefont{and}
  \bibinfo{author}{\bibfnamefont{R.~E.} \bibnamefont{Irving}},
  \bibinfo{journal}{Phys.~Scr.} \textbf{\bibinfo{volume}{77}},
  \bibinfo{pages}{015301} (\bibinfo{year}{2008}).

\bibitem[{\citenamefont{Mitroy and Zhang}(2007{\natexlab{a}})}]{mitroy07e}
\bibinfo{author}{\bibfnamefont{J.}~\bibnamefont{Mitroy}} \bibnamefont{and}
  \bibinfo{author}{\bibfnamefont{J.~Y.} \bibnamefont{Zhang}},
  \bibinfo{journal}{Phys.~Rev.~A} \textbf{\bibinfo{volume}{76}},
  \bibinfo{pages}{062703} (\bibinfo{year}{2007}{\natexlab{a}}).

\bibitem[{\citenamefont{Mitroy and Zhang}(2008{\natexlab{b}})}]{mitroy08a}
\bibinfo{author}{\bibfnamefont{J.}~\bibnamefont{Mitroy}} \bibnamefont{and}
  \bibinfo{author}{\bibfnamefont{J.~Y.} \bibnamefont{Zhang}},
  \bibinfo{journal}{Mol.~Phys.} \textbf{\bibinfo{volume}{106}},
  \bibinfo{pages}{127} (\bibinfo{year}{2008}{\natexlab{b}}).

\bibitem[{\citenamefont{Mitroy and Zhang}(2008{\natexlab{c}})}]{mitroy08g}
\bibinfo{author}{\bibfnamefont{J.}~\bibnamefont{Mitroy}} \bibnamefont{and}
  \bibinfo{author}{\bibfnamefont{J.~Y.} \bibnamefont{Zhang}},
  \bibinfo{journal}{J.~Chem.~Phys.} \textbf{\bibinfo{volume}{128}},
  \bibinfo{pages}{134305} (\bibinfo{year}{2008}{\natexlab{c}}).

\bibitem[{\citenamefont{Mitroy et~al.}(2009)\citenamefont{Mitroy, Zhang,
  Bromley, and Rollin}}]{mitroy09b}
\bibinfo{author}{\bibfnamefont{J.}~\bibnamefont{Mitroy}},
  \bibinfo{author}{\bibfnamefont{J.~Y.} \bibnamefont{Zhang}},
  \bibinfo{author}{\bibfnamefont{M.~W.~J.} \bibnamefont{Bromley}},
  \bibnamefont{and} \bibinfo{author}{\bibfnamefont{K.~G.}
  \bibnamefont{Rollin}}, \bibinfo{journal}{Eur.~Phys.~J.~D}
  \textbf{\bibinfo{volume}{53}}, \bibinfo{pages}{15} (\bibinfo{year}{2009}).

\bibitem[{\citenamefont{Goebel et~al.}(1996)\citenamefont{Goebel, Hohm, and
  Maroulis}}]{goebel96b}
\bibinfo{author}{\bibfnamefont{D.}~\bibnamefont{Goebel}},
  \bibinfo{author}{\bibfnamefont{U.}~\bibnamefont{Hohm}}, \bibnamefont{and}
  \bibinfo{author}{\bibfnamefont{G.}~\bibnamefont{Maroulis}},
  \bibinfo{journal}{Phys.~Rev.~A} \textbf{\bibinfo{volume}{54}},
  \bibinfo{pages}{1973} (\bibinfo{year}{1996}).

\bibitem[{\citenamefont{Thierfelder and
  {Schwerdtfeger}}(2009)}]{thierfelder09a}
\bibinfo{author}{\bibfnamefont{C.}~\bibnamefont{Thierfelder}} \bibnamefont{and}
  \bibinfo{author}{\bibfnamefont{P.}~\bibnamefont{{Schwerdtfeger}}},
  \bibinfo{journal}{Phys.~Rev.~A} \textbf{\bibinfo{volume}{79}},
  \bibinfo{pages}{032512} (\bibinfo{year}{2009}).

\bibitem[{\citenamefont{{Tang} and {Toennies}}(2008)}]{tang08a}
\bibinfo{author}{\bibfnamefont{K.~T.} \bibnamefont{{Tang}}} \bibnamefont{and}
  \bibinfo{author}{\bibfnamefont{J.~P.} \bibnamefont{{Toennies}}},
  \bibinfo{journal}{Mol.~Phys.~} \textbf{\bibinfo{volume}{106}},
  \bibinfo{pages}{1645} (\bibinfo{year}{2008}).

\bibitem[{\citenamefont{{Komara} et~al.}(2005)\citenamefont{{Komara}, {Gearba},
  {Fehrenbach}, and {Lundeen}}}]{komara05a}
\bibinfo{author}{\bibfnamefont{R.~A.} \bibnamefont{{Komara}}},
  \bibinfo{author}{\bibfnamefont{M.~A.} \bibnamefont{{Gearba}}},
  \bibinfo{author}{\bibfnamefont{C.~W.} \bibnamefont{{Fehrenbach}}},
  \bibnamefont{and} \bibinfo{author}{\bibfnamefont{S.~R.}
  \bibnamefont{{Lundeen}}}, \bibinfo{journal}{J.~Phys.~B}
  \textbf{\bibinfo{volume}{38}}, \bibinfo{pages}{S87} (\bibinfo{year}{2005}).

\bibitem[{\citenamefont{Porsev et~al.}(1999)\citenamefont{Porsev, Rakhlina, and
  Kozlov}}]{porsev99a}
\bibinfo{author}{\bibfnamefont{S.~G.} \bibnamefont{Porsev}},
  \bibinfo{author}{\bibfnamefont{Y.~G.} \bibnamefont{Rakhlina}},
  \bibnamefont{and} \bibinfo{author}{\bibfnamefont{M.~G.}
  \bibnamefont{Kozlov}}, \bibinfo{journal}{Phys. Rev. A}
  \textbf{\bibinfo{volume}{60}}, \bibinfo{pages}{2781} (\bibinfo{year}{1999}).

\bibitem[{\citenamefont{{Schwerdtfeger} and
  {Bowmaker}}(1994)}]{schwerdtfeger94a}
\bibinfo{author}{\bibfnamefont{P.}~\bibnamefont{{Schwerdtfeger}}}
  \bibnamefont{and} \bibinfo{author}{\bibfnamefont{G.~A.}
  \bibnamefont{{Bowmaker}}}, \bibinfo{journal}{J.~Chem.~Phys.~}
  \textbf{\bibinfo{volume}{100}}, \bibinfo{pages}{4487} (\bibinfo{year}{1994}).

\bibitem[{\citenamefont{Neogrady et~al.}(1997)\citenamefont{Neogrady, Kello,
  Urban, and Sadlej}}]{neogrady97a}
\bibinfo{author}{\bibfnamefont{P.}~\bibnamefont{Neogrady}},
  \bibinfo{author}{\bibfnamefont{V.}~\bibnamefont{Kello}},
  \bibinfo{author}{\bibfnamefont{M.}~\bibnamefont{Urban}}, \bibnamefont{and}
  \bibinfo{author}{\bibfnamefont{A.~J.} \bibnamefont{Sadlej}},
  \bibinfo{journal}{Int.~J.~Quant.~Chem.} \textbf{\bibinfo{volume}{63}},
  \bibinfo{pages}{557} (\bibinfo{year}{1997}).

\bibitem[{\citenamefont{{Ilias} and Neogrady}(1999)}]{ilias99a}
\bibinfo{author}{\bibfnamefont{M.}~\bibnamefont{{Ilias}}} \bibnamefont{and}
  \bibinfo{author}{\bibfnamefont{P.}~\bibnamefont{Neogrady}},
  \bibinfo{journal}{Chem.~Phys.~Lett.} \textbf{\bibinfo{volume}{309}},
  \bibinfo{pages}{441} (\bibinfo{year}{1999}).

\bibitem[{\citenamefont{Zhang et~al.}(2008)\citenamefont{Zhang, Mitroy,
  Sadeghpour, and Bromley}}]{zhang08d}
\bibinfo{author}{\bibfnamefont{J.~Y.} \bibnamefont{Zhang}},
  \bibinfo{author}{\bibfnamefont{J.}~\bibnamefont{Mitroy}},
  \bibinfo{author}{\bibfnamefont{H.~R.} \bibnamefont{Sadeghpour}},
  \bibnamefont{and} \bibinfo{author}{\bibfnamefont{M.~W.~J.}
  \bibnamefont{Bromley}}, \bibinfo{journal}{Phys.~Rev.~A}
  \textbf{\bibinfo{volume}{78}}, \bibinfo{pages}{062710}
  (\bibinfo{year}{2008}).

\bibitem[{\citenamefont{{Henderson} et~al.}(1997)\citenamefont{{Henderson},
  {Curtis}, {Matulioniene}, {Ellis}, and {Theodosiou}}}]{henderson97a}
\bibinfo{author}{\bibfnamefont{M.}~\bibnamefont{{Henderson}}},
  \bibinfo{author}{\bibfnamefont{L.~J.} \bibnamefont{{Curtis}}},
  \bibinfo{author}{\bibfnamefont{R.}~\bibnamefont{{Matulioniene}}},
  \bibinfo{author}{\bibfnamefont{D.~G.} \bibnamefont{{Ellis}}},
  \bibnamefont{and} \bibinfo{author}{\bibfnamefont{C.~E.}
  \bibnamefont{{Theodosiou}}}, \bibinfo{journal}{Phys.~Rev.~A}
  \textbf{\bibinfo{volume}{56}}, \bibinfo{pages}{1872} (\bibinfo{year}{1997}).

\bibitem[{\citenamefont{{Curtis} and {Theodosiou}}(1995)}]{curtis95a}
\bibinfo{author}{\bibfnamefont{L.~J.} \bibnamefont{{Curtis}}} \bibnamefont{and}
  \bibinfo{author}{\bibfnamefont{C.~E.} \bibnamefont{{Theodosiou}}},
  \bibinfo{journal}{J.~Opt.~Soc.~Am.~B} \textbf{\bibinfo{volume}{12}},
  \bibinfo{pages}{175} (\bibinfo{year}{1995}).

\bibitem[{\citenamefont{Neogrady et~al.}(1996)\citenamefont{Neogrady, Kello,
  Urban, and Sadlej}}]{neogrady96a}
\bibinfo{author}{\bibfnamefont{P.}~\bibnamefont{Neogrady}},
  \bibinfo{author}{\bibfnamefont{V.}~\bibnamefont{Kello}},
  \bibinfo{author}{\bibfnamefont{M.}~\bibnamefont{Urban}}, \bibnamefont{and}
  \bibinfo{author}{\bibfnamefont{A.~J.} \bibnamefont{Sadlej}},
  \bibinfo{journal}{Theor.~Chim.~Acta} \textbf{\bibinfo{volume}{93}},
  \bibinfo{pages}{101} (\bibinfo{year}{1996}).

\bibitem[{\citenamefont{{Snow} and {Lundeen}}(2007{\natexlab{b}})}]{snow07a}
\bibinfo{author}{\bibfnamefont{E.~L.} \bibnamefont{{Snow}}} \bibnamefont{and}
  \bibinfo{author}{\bibfnamefont{S.~R.} \bibnamefont{{Lundeen}}},
  \bibinfo{journal}{Phys.~Rev.~A} \textbf{\bibinfo{volume}{75}},
  \bibinfo{pages}{062512} (\bibinfo{year}{2007}{\natexlab{b}}).

\bibitem[{\citenamefont{{Lundeen} and {Fehrenbach}}(2007)}]{lundeen07b}
\bibinfo{author}{\bibfnamefont{S.~R.} \bibnamefont{{Lundeen}}}
  \bibnamefont{and} \bibinfo{author}{\bibfnamefont{C.~W.}
  \bibnamefont{{Fehrenbach}}}, \bibinfo{journal}{Phys.~Rev.~A}
  \textbf{\bibinfo{volume}{75}}, \bibinfo{pages}{032523}
  (\bibinfo{year}{2007}).

\bibitem[{\citenamefont{Magnusson and Zetterberg}(1977)}]{magnusson77a}
\bibinfo{author}{\bibfnamefont{C.~E.} \bibnamefont{Magnusson}}
  \bibnamefont{and} \bibinfo{author}{\bibfnamefont{P.~O.}
  \bibnamefont{Zetterberg}}, \bibinfo{journal}{Phys.~Scr.}
  \textbf{\bibinfo{volume}{15}}, \bibinfo{pages}{237} (\bibinfo{year}{1977}).

\bibitem[{\citenamefont{Kozlov et~al.}(2001{\natexlab{a}})\citenamefont{Kozlov,
  Porsev, and Johnson}}]{kozlov01b}
\bibinfo{author}{\bibfnamefont{M.~G.} \bibnamefont{Kozlov}},
  \bibinfo{author}{\bibfnamefont{S.~G.} \bibnamefont{Porsev}},
  \bibnamefont{and} \bibinfo{author}{\bibfnamefont{W.~R.}
  \bibnamefont{Johnson}}, \bibinfo{journal}{Phys. Rev. A}
  \textbf{\bibinfo{volume}{64}}, \bibinfo{pages}{052107}
  (\bibinfo{year}{2001}{\natexlab{a}}).

\bibitem[{\citenamefont{{Hibbert}}(1980)}]{hibbert80a}
\bibinfo{author}{\bibfnamefont{A.}~\bibnamefont{{Hibbert}}},
  \bibinfo{journal}{J.~Phys.~B} \textbf{\bibinfo{volume}{13}},
  \bibinfo{pages}{3725} (\bibinfo{year}{1980}).

\bibitem[{\citenamefont{{Fleig}}(2005)}]{fleig05a}
\bibinfo{author}{\bibfnamefont{T.}~\bibnamefont{{Fleig}}},
  \bibinfo{journal}{Phys.~Rev.~A} \textbf{\bibinfo{volume}{72}},
  \bibinfo{pages}{052506} (\bibinfo{year}{2005}).

\bibitem[{\citenamefont{{Milani} et~al.}(1990)\citenamefont{{Milani},
  {Moullet}, and {de Heer}}}]{milani90a}
\bibinfo{author}{\bibfnamefont{P.}~\bibnamefont{{Milani}}},
  \bibinfo{author}{\bibfnamefont{I.}~\bibnamefont{{Moullet}}},
  \bibnamefont{and} \bibinfo{author}{\bibfnamefont{W.~A.} \bibnamefont{{de
  Heer}}}, \bibinfo{journal}{Phys.~Rev.~A} \textbf{\bibinfo{volume}{42}},
  \bibinfo{pages}{5150} (\bibinfo{year}{1990}).

\bibitem[{\citenamefont{{Guella} et~al.}(1984)\citenamefont{{Guella}, {Miller},
  {Bederson}, {Stockdale}, and {Jaduszliwer}}}]{guella84a}
\bibinfo{author}{\bibfnamefont{T.~P.} \bibnamefont{{Guella}}},
  \bibinfo{author}{\bibfnamefont{T.~M.} \bibnamefont{{Miller}}},
  \bibinfo{author}{\bibfnamefont{B.}~\bibnamefont{{Bederson}}},
  \bibinfo{author}{\bibfnamefont{J.~A.~D.} \bibnamefont{{Stockdale}}},
  \bibnamefont{and}
  \bibinfo{author}{\bibfnamefont{B.}~\bibnamefont{{Jaduszliwer}}},
  \bibinfo{journal}{Phys.~Rev.~A} \textbf{\bibinfo{volume}{29}},
  \bibinfo{pages}{2977} (\bibinfo{year}{1984}).

\bibitem[{\citenamefont{{Guella}}(1985)}]{guella85a}
\bibinfo{author}{\bibfnamefont{T.~P.} \bibnamefont{{Guella}}}, Ph.D. thesis,
  \bibinfo{school}{New York University} (\bibinfo{year}{1985}).

\bibitem[{\citenamefont{{Thierfelder} et~al.}(2008)\citenamefont{{Thierfelder},
  {Assadollahzadeh}, {Schwerdtfeger}, {Sch{\"a}fer}, and
  {Sch{\"a}fer}}}]{thierfelder08a}
\bibinfo{author}{\bibfnamefont{C.}~\bibnamefont{{Thierfelder}}},
  \bibinfo{author}{\bibfnamefont{B.}~\bibnamefont{{Assadollahzadeh}}},
  \bibinfo{author}{\bibfnamefont{P.}~\bibnamefont{{Schwerdtfeger}}},
  \bibinfo{author}{\bibfnamefont{S.}~\bibnamefont{{Sch{\"a}fer}}},
  \bibnamefont{and}
  \bibinfo{author}{\bibfnamefont{R.}~\bibnamefont{{Sch{\"a}fer}}},
  \bibinfo{journal}{Phys.~Rev.~A} \textbf{\bibinfo{volume}{78}},
  \bibinfo{pages}{052506} (\bibinfo{year}{2008}).

\bibitem[{\citenamefont{Bardon and Auduffren}(1984)}]{bardon84a}
\bibinfo{author}{\bibfnamefont{J.}~\bibnamefont{Bardon}} \bibnamefont{and}
  \bibinfo{author}{\bibfnamefont{M.}~\bibnamefont{Auduffren}},
  \bibinfo{journal}{J.~Physique.~(Colloq)} \textbf{\bibinfo{volume}{C9}},
  \bibinfo{pages}{245} (\bibinfo{year}{1984}).

\bibitem[{\citenamefont{{Kreutztr{\"a}ger} and
  {v.~Oppen}}(1973)}]{kreutztrager73a}
\bibinfo{author}{\bibfnamefont{A.}~\bibnamefont{{Kreutztr{\"a}ger}}}
  \bibnamefont{and}
  \bibinfo{author}{\bibfnamefont{G.}~\bibnamefont{{v.~Oppen}}},
  \bibinfo{journal}{Z.~Phys.} \textbf{\bibinfo{volume}{265}},
  \bibinfo{pages}{421} (\bibinfo{year}{1973}).

\bibitem[{\citenamefont{{von Oppen}}(1970)}]{oppen70a}
\bibinfo{author}{\bibfnamefont{G.}~\bibnamefont{{von Oppen}}},
  \bibinfo{journal}{Z.~Phys.} \textbf{\bibinfo{volume}{232}},
  \bibinfo{pages}{473} (\bibinfo{year}{1970}).

\bibitem[{\citenamefont{Yanagimachi et~al.}(2002)\citenamefont{Yanagimachi,
  Kajiro, Machiya, and Morinaga}}]{yanagimachi02a}
\bibinfo{author}{\bibfnamefont{S.}~\bibnamefont{Yanagimachi}},
  \bibinfo{author}{\bibfnamefont{M.}~\bibnamefont{Kajiro}},
  \bibinfo{author}{\bibfnamefont{M.}~\bibnamefont{Machiya}}, \bibnamefont{and}
  \bibinfo{author}{\bibfnamefont{A.}~\bibnamefont{Morinaga}},
  \bibinfo{journal}{Phys.~Rev.~A} \textbf{\bibinfo{volume}{65}},
  \bibinfo{pages}{042104} (\bibinfo{year}{2002}).

\bibitem[{\citenamefont{Zeiske et~al.}(1995)\citenamefont{Zeiske, Zinner,
  Riehle, and Helmcke}}]{zeiske95a}
\bibinfo{author}{\bibfnamefont{K.}~\bibnamefont{Zeiske}},
  \bibinfo{author}{\bibfnamefont{G.}~\bibnamefont{Zinner}},
  \bibinfo{author}{\bibfnamefont{F.}~\bibnamefont{Riehle}}, \bibnamefont{and}
  \bibinfo{author}{\bibfnamefont{J.}~\bibnamefont{Helmcke}},
  \bibinfo{journal}{Appl.~Phys.~B} \textbf{\bibinfo{volume}{60}},
  \bibinfo{pages}{205} (\bibinfo{year}{1995}).

\bibitem[{\citenamefont{{von Oppen}}(1971)}]{oppen71a}
\bibinfo{author}{\bibfnamefont{G.}~\bibnamefont{{von Oppen}}},
  \bibinfo{journal}{Z.~Phys.} \textbf{\bibinfo{volume}{248}},
  \bibinfo{pages}{41} (\bibinfo{year}{1971}).

\bibitem[{\citenamefont{{von Oppen}}(1969)}]{oppen69b}
\bibinfo{author}{\bibfnamefont{G.}~\bibnamefont{{von Oppen}}},
  \bibinfo{journal}{Z.~Phys.~} \textbf{\bibinfo{volume}{227}},
  \bibinfo{pages}{207} (\bibinfo{year}{1969}).

\bibitem[{\citenamefont{{Hese} et~al.}(1977)\citenamefont{{Hese}, {Renn}, and
  {Schweda}}}]{hese77a}
\bibinfo{author}{\bibfnamefont{A.}~\bibnamefont{{Hese}}},
  \bibinfo{author}{\bibfnamefont{A.}~\bibnamefont{{Renn}}}, \bibnamefont{and}
  \bibinfo{author}{\bibfnamefont{H.~S.} \bibnamefont{{Schweda}}},
  \bibinfo{journal}{Opt.~Commun.} \textbf{\bibinfo{volume}{20}},
  \bibinfo{pages}{385} (\bibinfo{year}{1977}).

\bibitem[{\citenamefont{{van Leeuwen} and {Hogervorst}}(1983)}]{vanleeuwen83a}
\bibinfo{author}{\bibfnamefont{K.~A.~H.} \bibnamefont{{van Leeuwen}}}
  \bibnamefont{and}
  \bibinfo{author}{\bibfnamefont{W.}~\bibnamefont{{Hogervorst}}},
  \bibinfo{journal}{Z.~Phys.} \textbf{\bibinfo{volume}{310}},
  \bibinfo{pages}{37} (\bibinfo{year}{1983}).

\bibitem[{\citenamefont{{Li} et~al.}(2004)\citenamefont{{Li}, {Rochester},
  {Kozlov}, and {Budker}}}]{li04b}
\bibinfo{author}{\bibfnamefont{C.-H.} \bibnamefont{{Li}}},
  \bibinfo{author}{\bibfnamefont{S.~M.} \bibnamefont{{Rochester}}},
  \bibinfo{author}{\bibfnamefont{M.~G.} \bibnamefont{{Kozlov}}},
  \bibnamefont{and} \bibinfo{author}{\bibfnamefont{D.}~\bibnamefont{{Budker}}},
  \bibinfo{journal}{Phys.~Rev.~A} \textbf{\bibinfo{volume}{69}},
  \bibinfo{pages}{042507} (\bibinfo{year}{2004}).

\bibitem[{\citenamefont{{Rinkleff}}(1979)}]{rinkleff79a}
\bibinfo{author}{\bibfnamefont{R.-H.} \bibnamefont{{Rinkleff}}},
  \bibinfo{journal}{Z.~Phys.} \textbf{\bibinfo{volume}{291}},
  \bibinfo{pages}{23} (\bibinfo{year}{1979}).

\bibitem[{\citenamefont{Robinson}(1969)}]{robinson69a}
\bibinfo{author}{\bibfnamefont{E.~J.} \bibnamefont{Robinson}},
  \bibinfo{journal}{J.~Opt.~Soc.~Am.} \textbf{\bibinfo{volume}{59}},
  \bibinfo{pages}{782} (\bibinfo{year}{1969}).

\bibitem[{\citenamefont{{Legowski} et~al.}(1995)\citenamefont{{Legowski},
  {Molhem}, {Osi{\'n}ski}, and {Rudecki}}}]{legowski95a}
\bibinfo{author}{\bibfnamefont{S.}~\bibnamefont{{Legowski}}},
  \bibinfo{author}{\bibfnamefont{A.}~\bibnamefont{{Molhem}}},
  \bibinfo{author}{\bibfnamefont{G.}~\bibnamefont{{Osi{\'n}ski}}},
  \bibnamefont{and}
  \bibinfo{author}{\bibfnamefont{P.}~\bibnamefont{{Rudecki}}},
  \bibinfo{journal}{Z.~Phys.~D} \textbf{\bibinfo{volume}{35}},
  \bibinfo{pages}{101} (\bibinfo{year}{1995}).

\bibitem[{\citenamefont{{Sandle} et~al.}(1975)\citenamefont{{Sandle},
  {Standage}, and {Warrington}}}]{sandle75a}
\bibinfo{author}{\bibfnamefont{W.~J.} \bibnamefont{{Sandle}}},
  \bibinfo{author}{\bibfnamefont{M.~C.} \bibnamefont{{Standage}}},
  \bibnamefont{and} \bibinfo{author}{\bibfnamefont{D.~M.}
  \bibnamefont{{Warrington}}}, \bibinfo{journal}{J.~Phys.~B}
  \textbf{\bibinfo{volume}{8}}, \bibinfo{pages}{1203} (\bibinfo{year}{1975}).

\bibitem[{\citenamefont{{Kaul} and {Latshaw}}(1972)}]{kaul72a}
\bibinfo{author}{\bibfnamefont{R.~D.} \bibnamefont{{Kaul}}} \bibnamefont{and}
  \bibinfo{author}{\bibfnamefont{W.~S.} \bibnamefont{{Latshaw}}},
  \bibinfo{journal}{J.~Opt.~Soc.~America} \textbf{\bibinfo{volume}{62}},
  \bibinfo{pages}{615} (\bibinfo{year}{1972}).

\bibitem[{\citenamefont{{Martin} et~al.}(1968)\citenamefont{{Martin},
  {Sandars}, and {Woodgate}}}]{martin68a}
\bibinfo{author}{\bibfnamefont{N.~J.} \bibnamefont{{Martin}}},
  \bibinfo{author}{\bibfnamefont{P.~G.~H.} \bibnamefont{{Sandars}}},
  \bibnamefont{and} \bibinfo{author}{\bibfnamefont{G.~K.}
  \bibnamefont{{Woodgate}}}, \bibinfo{journal}{Proc.~R.~Soc.~London~A}
  \textbf{\bibinfo{volume}{305}}, \bibinfo{pages}{139} (\bibinfo{year}{1968}).

\bibitem[{\citenamefont{Petersen et~al.}(1968)\citenamefont{Petersen, Palmer,
  and Shirly}}]{peterson68a}
\bibinfo{author}{\bibfnamefont{F.~R.} \bibnamefont{Petersen}},
  \bibinfo{author}{\bibfnamefont{H.~G.} \bibnamefont{Palmer}},
  \bibnamefont{and} \bibinfo{author}{\bibfnamefont{J.~H.}
  \bibnamefont{Shirly}}, \bibinfo{journal}{Bull.~Am.~Phys.~Soc.}
  \textbf{\bibinfo{volume}{13}}, \bibinfo{pages}{1674} (\bibinfo{year}{1968}).

\bibitem[{\citenamefont{Gould}(1976)}]{gould76a}
\bibinfo{author}{\bibfnamefont{H.}~\bibnamefont{Gould}},
  \bibinfo{journal}{Phys. Rev. A} \textbf{\bibinfo{volume}{14}},
  \bibinfo{pages}{922} (\bibinfo{year}{1976}).

\bibitem[{\citenamefont{{Rinkleff}}(1980)}]{rinkleff80a}
\bibinfo{author}{\bibfnamefont{R.-H.} \bibnamefont{{Rinkleff}}},
  \bibinfo{journal}{Z.~Phys.} \textbf{\bibinfo{volume}{296}},
  \bibinfo{pages}{101} (\bibinfo{year}{1980}).

\bibitem[{\citenamefont{{Kulina} and {Rinkleff}}(1982)}]{kulina82a}
\bibinfo{author}{\bibfnamefont{P.}~\bibnamefont{{Kulina}}} \bibnamefont{and}
  \bibinfo{author}{\bibfnamefont{R.-H.} \bibnamefont{{Rinkleff}}},
  \bibinfo{journal}{Z.~Phys.} \textbf{\bibinfo{volume}{304}},
  \bibinfo{pages}{371} (\bibinfo{year}{1982}).

\bibitem[{\citenamefont{{Li} and {Van Wijngaarden}}(1995)}]{li95b}
\bibinfo{author}{\bibfnamefont{J.}~\bibnamefont{{Li}}} \bibnamefont{and}
  \bibinfo{author}{\bibfnamefont{W.~A.} \bibnamefont{{Van Wijngaarden}}},
  \bibinfo{journal}{J.~Phys.~B} \textbf{\bibinfo{volume}{28}},
  \bibinfo{pages}{2559} (\bibinfo{year}{1995}).

\bibitem[{\citenamefont{{Schneider} et~al.}(2005)\citenamefont{{Schneider},
  {Peik}, and {Tamm}}}]{schneider05a}
\bibinfo{author}{\bibfnamefont{T.}~\bibnamefont{{Schneider}}},
  \bibinfo{author}{\bibfnamefont{E.}~\bibnamefont{{Peik}}}, \bibnamefont{and}
  \bibinfo{author}{\bibfnamefont{C.}~\bibnamefont{{Tamm}}},
  \bibinfo{journal}{Phys.~Rev.~Lett.} \textbf{\bibinfo{volume}{94}},
  \bibinfo{pages}{230801} (\bibinfo{year}{2005}).

\bibitem[{\citenamefont{Bhatia and Drachman}(1992)}]{bhatia92a}
\bibinfo{author}{\bibfnamefont{A.~K.} \bibnamefont{Bhatia}} \bibnamefont{and}
  \bibinfo{author}{\bibfnamefont{R.~J.} \bibnamefont{Drachman}},
  \bibinfo{journal}{Phys.~Rev.~A} \textbf{\bibinfo{volume}{45}},
  \bibinfo{pages}{7752} (\bibinfo{year}{1992}).

\bibitem[{\citenamefont{Korobov}(2000)}]{korobov01a}
\bibinfo{author}{\bibfnamefont{V.~I.} \bibnamefont{Korobov}},
  \bibinfo{journal}{Phys. Rev. A} \textbf{\bibinfo{volume}{61}},
  \bibinfo{pages}{064503} (\bibinfo{year}{2000}).

\bibitem[{\citenamefont{Cencek et~al.}(2001)\citenamefont{Cencek, Szalewicz,
  and Jeziorski}}]{cencek01a}
\bibinfo{author}{\bibfnamefont{W.}~\bibnamefont{Cencek}},
  \bibinfo{author}{\bibfnamefont{K.}~\bibnamefont{Szalewicz}},
  \bibnamefont{and}
  \bibinfo{author}{\bibfnamefont{B.}~\bibnamefont{Jeziorski}},
  \bibinfo{journal}{Phys. Rev. Lett.} \textbf{\bibinfo{volume}{86}},
  \bibinfo{pages}{5675} (\bibinfo{year}{2001}).

\bibitem[{\citenamefont{{Pipin} and {Bishop}}(1993)}]{pipin93a}
\bibinfo{author}{\bibfnamefont{J.}~\bibnamefont{{Pipin}}} \bibnamefont{and}
  \bibinfo{author}{\bibfnamefont{D.~M.} \bibnamefont{{Bishop}}},
  \bibinfo{journal}{Phys.~Rev.A} \textbf{\bibinfo{volume}{47}},
  \bibinfo{pages}{4571} (\bibinfo{year}{1993}).

\bibitem[{\citenamefont{{Vaidyanathan}
  et~al.}(1982)\citenamefont{{Vaidyanathan}, {Spencer}, {Rubbmark}, {Kuiper},
  {Fabre}, {Kleppner}, and {Ducas}}}]{vaidyanathan82a}
\bibinfo{author}{\bibfnamefont{A.~G.} \bibnamefont{{Vaidyanathan}}},
  \bibinfo{author}{\bibfnamefont{W.~P.} \bibnamefont{{Spencer}}},
  \bibinfo{author}{\bibfnamefont{J.~R.} \bibnamefont{{Rubbmark}}},
  \bibinfo{author}{\bibfnamefont{H.}~\bibnamefont{{Kuiper}}},
  \bibinfo{author}{\bibfnamefont{C.}~\bibnamefont{{Fabre}}},
  \bibinfo{author}{\bibfnamefont{D.}~\bibnamefont{{Kleppner}}},
  \bibnamefont{and} \bibinfo{author}{\bibfnamefont{T.~W.}
  \bibnamefont{{Ducas}}}, \bibinfo{journal}{Phys.~Rev.~A}
  \textbf{\bibinfo{volume}{26}}, \bibinfo{pages}{3346} (\bibinfo{year}{1982}).

\bibitem[{\citenamefont{Porsev and
  Derevianko}(2006{\natexlab{d}})}]{derevianko-triples}
\bibinfo{author}{\bibfnamefont{S.~G.} \bibnamefont{Porsev}} \bibnamefont{and}
  \bibinfo{author}{\bibfnamefont{A.}~\bibnamefont{Derevianko}},
  \bibinfo{journal}{Phys.~Rev.~A} \textbf{\bibinfo{volume}{73}},
  \bibinfo{pages}{012501} (\bibinfo{year}{2006}{\natexlab{d}}).

\bibitem[{\citenamefont{Homer}(circa 900BC)}]{homerBC}
\bibinfo{author}{\bibnamefont{Homer}}, \emph{\bibinfo{title}{The Odyssey}}
  (\bibinfo{year}{circa 900BC}).

\bibitem[{\citenamefont{Wood et~al.}(1997)\citenamefont{Wood, Bennett, Cho,
  Masterson, Roberts, Tanner, and Wieman}}]{Cs-PNC}
\bibinfo{author}{\bibfnamefont{C.~S.} \bibnamefont{Wood}},
  \bibinfo{author}{\bibfnamefont{S.~C.} \bibnamefont{Bennett}},
  \bibinfo{author}{\bibfnamefont{D.}~\bibnamefont{Cho}},
  \bibinfo{author}{\bibfnamefont{B.~P.} \bibnamefont{Masterson}},
  \bibinfo{author}{\bibfnamefont{J.~L.} \bibnamefont{Roberts}},
  \bibinfo{author}{\bibfnamefont{C.~E.} \bibnamefont{Tanner}},
  \bibnamefont{and} \bibinfo{author}{\bibfnamefont{C.~E.}
  \bibnamefont{Wieman}}, \bibinfo{journal}{Science}
  \textbf{\bibinfo{volume}{275}}, \bibinfo{pages}{1759} (\bibinfo{year}{1997}).

\bibitem[{\citenamefont{Kozlov et~al.}(2001{\natexlab{b}})\citenamefont{Kozlov,
  Porsev, and Tupitsyn}}]{kozlov01a}
\bibinfo{author}{\bibfnamefont{M.~G.} \bibnamefont{Kozlov}},
  \bibinfo{author}{\bibfnamefont{S.~G.} \bibnamefont{Porsev}},
  \bibnamefont{and} \bibinfo{author}{\bibfnamefont{I.~I.}
  \bibnamefont{Tupitsyn}}, \bibinfo{journal}{Phys. Rev. Lett.}
  \textbf{\bibinfo{volume}{86}}, \bibinfo{pages}{3260}
  (\bibinfo{year}{2001}{\natexlab{b}}).

\bibitem[{\citenamefont{{Watts} et~al.}(1983)\citenamefont{{Watts}, {Gilbert},
  and {Wieman}}}]{watts83a}
\bibinfo{author}{\bibfnamefont{R.~N.} \bibnamefont{{Watts}}},
  \bibinfo{author}{\bibfnamefont{S.~L.} \bibnamefont{{Gilbert}}},
  \bibnamefont{and} \bibinfo{author}{\bibfnamefont{C.~E.}
  \bibnamefont{{Wieman}}}, \bibinfo{journal}{Phys.~Rev.~A}
  \textbf{\bibinfo{volume}{27}}, \bibinfo{pages}{2769} (\bibinfo{year}{1983}).

\bibitem[{\citenamefont{{Rieger} et~al.}(1993)\citenamefont{{Rieger},
  {Sengstock}, {Sterr}, {M{\"u}ller}, and {Ertmer}}}]{rieger93a}
\bibinfo{author}{\bibfnamefont{V.}~\bibnamefont{{Rieger}}},
  \bibinfo{author}{\bibfnamefont{K.}~\bibnamefont{{Sengstock}}},
  \bibinfo{author}{\bibfnamefont{U.}~\bibnamefont{{Sterr}}},
  \bibinfo{author}{\bibfnamefont{J.~H.} \bibnamefont{{M{\"u}ller}}},
  \bibnamefont{and} \bibinfo{author}{\bibfnamefont{W.}~\bibnamefont{{Ertmer}}},
  \bibinfo{journal}{Opt.~Commun.} \textbf{\bibinfo{volume}{99}},
  \bibinfo{pages}{172} (\bibinfo{year}{1993}).

\bibitem[{\citenamefont{Rieger}(1996)}]{rieger96a}
\bibinfo{author}{\bibfnamefont{V.}~\bibnamefont{Rieger}}, Ph.D. thesis,
  \bibinfo{school}{Univerisitat Hannover} (\bibinfo{year}{1996}).

\bibitem[{\citenamefont{Morinaga et~al.}(1996)\citenamefont{Morinaga, Nakamura,
  Kurosu, and Ito}}]{morinaga96a}
\bibinfo{author}{\bibfnamefont{A.}~\bibnamefont{Morinaga}},
  \bibinfo{author}{\bibfnamefont{M.}~\bibnamefont{Nakamura}},
  \bibinfo{author}{\bibfnamefont{T.}~\bibnamefont{Kurosu}}, \bibnamefont{and}
  \bibinfo{author}{\bibfnamefont{N.}~\bibnamefont{Ito}},
  \bibinfo{journal}{Phys.~Rev.~A} \textbf{\bibinfo{volume}{54}},
  \bibinfo{pages}{21(R)} (\bibinfo{year}{1996}).

\bibitem[{\citenamefont{Li and van Wijngaarden}(1996)}]{li96a}
\bibinfo{author}{\bibfnamefont{J.}~\bibnamefont{Li}} \bibnamefont{and}
  \bibinfo{author}{\bibfnamefont{W.~A.} \bibnamefont{van Wijngaarden}},
  \bibinfo{journal}{Phys.~Rev.~A} \textbf{\bibinfo{volume}{53}},
  \bibinfo{pages}{604} (\bibinfo{year}{1996}).

\bibitem[{\citenamefont{{Li} and {van Wijngaarden}}(1995)}]{li95a}
\bibinfo{author}{\bibfnamefont{J.}~\bibnamefont{{Li}}} \bibnamefont{and}
  \bibinfo{author}{\bibfnamefont{W.~A.} \bibnamefont{{van Wijngaarden}}},
  \bibinfo{journal}{Phys.~Rev.~A} \textbf{\bibinfo{volume}{51}},
  \bibinfo{pages}{3560} (\bibinfo{year}{1995}).

\bibitem[{\citenamefont{{Harber} and {Romalis}}(2001)}]{harber01a}
\bibinfo{author}{\bibfnamefont{D.~M.} \bibnamefont{{Harber}}} \bibnamefont{and}
  \bibinfo{author}{\bibfnamefont{M.~V.} \bibnamefont{{Romalis}}},
  \bibinfo{journal}{Phys.~Rev.~A} \textbf{\bibinfo{volume}{63}},
  \bibinfo{pages}{013402} (\bibinfo{year}{2001}).

\bibitem[{\citenamefont{DeMille et~al.}(1994)\citenamefont{DeMille, Budker, and
  Commins}}]{demille94a}
\bibinfo{author}{\bibfnamefont{D.}~\bibnamefont{DeMille}},
  \bibinfo{author}{\bibfnamefont{D.}~\bibnamefont{Budker}}, \bibnamefont{and}
  \bibinfo{author}{\bibfnamefont{E.~D.} \bibnamefont{Commins}},
  \bibinfo{journal}{Phys. Rev. A} \textbf{\bibinfo{volume}{50}},
  \bibinfo{pages}{4657} (\bibinfo{year}{1994}).

\bibitem[{\citenamefont{Safronova et~al.}(2006)\citenamefont{Safronova,
  Johnson, Safronova, and Cowan}}]{safronova06b}
\bibinfo{author}{\bibfnamefont{M.~S.} \bibnamefont{Safronova}},
  \bibinfo{author}{\bibfnamefont{W.~R.} \bibnamefont{Johnson}},
  \bibinfo{author}{\bibfnamefont{U.~I.} \bibnamefont{Safronova}},
  \bibnamefont{and} \bibinfo{author}{\bibfnamefont{T.~E.} \bibnamefont{Cowan}},
  \bibinfo{journal}{Phys.~Rev.~A} \textbf{\bibinfo{volume}{74}},
  \bibinfo{pages}{022504} (\bibinfo{year}{2006}).

\bibitem[{\citenamefont{Doret et~al.}(2002)\citenamefont{Doret, Friedberg,
  Speck, Richardson, and Majumder}}]{doret02a}
\bibinfo{author}{\bibfnamefont{S.~C.} \bibnamefont{Doret}},
  \bibinfo{author}{\bibfnamefont{P.~D.} \bibnamefont{Friedberg}},
  \bibinfo{author}{\bibfnamefont{A.~J.} \bibnamefont{Speck}},
  \bibinfo{author}{\bibfnamefont{D.~S.} \bibnamefont{Richardson}},
  \bibnamefont{and} \bibinfo{author}{\bibfnamefont{P.~K.}
  \bibnamefont{Majumder}}, \bibinfo{journal}{Phys. Rev. A}
  \textbf{\bibinfo{volume}{66}}, \bibinfo{pages}{052504}
  (\bibinfo{year}{2002}).

\bibitem[{\citenamefont{Mitroy and Norcross}(1989)}]{mitroy89a}
\bibinfo{author}{\bibfnamefont{J.}~\bibnamefont{Mitroy}} \bibnamefont{and}
  \bibinfo{author}{\bibfnamefont{D.~W.} \bibnamefont{Norcross}},
  \bibinfo{journal}{Phys.~Rev.~A} \textbf{\bibinfo{volume}{39}},
  \bibinfo{pages}{537} (\bibinfo{year}{1989}).

\bibitem[{\citenamefont{Goebel et~al.}(1995)\citenamefont{Goebel, Hohm, and
  Kerl}}]{goebel95b}
\bibinfo{author}{\bibfnamefont{D.}~\bibnamefont{Goebel}},
  \bibinfo{author}{\bibfnamefont{U.}~\bibnamefont{Hohm}}, \bibnamefont{and}
  \bibinfo{author}{\bibfnamefont{K.}~\bibnamefont{Kerl}},
  \bibinfo{journal}{J.~Mol.~Struct.} \textbf{\bibinfo{volume}{349}},
  \bibinfo{pages}{253} (\bibinfo{year}{1995}).

\bibitem[{\citenamefont{{Yasuda} et~al.}(2006)\citenamefont{{Yasuda},
  {Kishimoto}, {Takamoto}, and {Katori}}}]{yasuda06a}
\bibinfo{author}{\bibfnamefont{M.}~\bibnamefont{{Yasuda}}},
  \bibinfo{author}{\bibfnamefont{T.}~\bibnamefont{{Kishimoto}}},
  \bibinfo{author}{\bibfnamefont{M.}~\bibnamefont{{Takamoto}}},
  \bibnamefont{and} \bibinfo{author}{\bibfnamefont{H.}~\bibnamefont{{Katori}}},
  \bibinfo{journal}{Phys.~Rev.~A} \textbf{\bibinfo{volume}{73}},
  \bibinfo{pages}{011403} (\bibinfo{year}{2006}).

\bibitem[{\citenamefont{{Nagel} et~al.}(2005)\citenamefont{{Nagel},
  {Mickelson}, {Saenz}, {Martinez}, {Chen}, {Killian}, {Pellegrini}, and
  {C{\^o}t{\'e}}}}]{nagel05a}
\bibinfo{author}{\bibfnamefont{S.~B.} \bibnamefont{{Nagel}}},
  \bibinfo{author}{\bibfnamefont{P.~G.} \bibnamefont{{Mickelson}}},
  \bibinfo{author}{\bibfnamefont{A.~D.} \bibnamefont{{Saenz}}},
  \bibinfo{author}{\bibfnamefont{Y.~N.} \bibnamefont{{Martinez}}},
  \bibinfo{author}{\bibfnamefont{Y.~C.} \bibnamefont{{Chen}}},
  \bibinfo{author}{\bibfnamefont{T.~C.} \bibnamefont{{Killian}}},
  \bibinfo{author}{\bibfnamefont{P.}~\bibnamefont{{Pellegrini}}},
  \bibnamefont{and}
  \bibinfo{author}{\bibfnamefont{R.}~\bibnamefont{{C{\^o}t{\'e}}}},
  \bibinfo{journal}{Phys.~Rev.~Lett.} \textbf{\bibinfo{volume}{94}},
  \bibinfo{pages}{083004} (\bibinfo{year}{2005}).

\bibitem[{\citenamefont{Rosenband et~al.}(2008)\citenamefont{Rosenband, Hume,
  Brusch, Lorini, Oskay, Drullinger, Fortier, Stalnaker, Diddams, Swann
  et~al.}}]{rosenband08a}
\bibinfo{author}{\bibfnamefont{T.}~\bibnamefont{Rosenband}},
  \bibinfo{author}{\bibfnamefont{C.~W.} \bibnamefont{Hume},
  \bibfnamefont{D~B~Chou}},
  \bibinfo{author}{\bibfnamefont{A.}~\bibnamefont{Brusch}},
  \bibinfo{author}{\bibfnamefont{L.}~\bibnamefont{Lorini}},
  \bibinfo{author}{\bibfnamefont{W.~H.} \bibnamefont{Oskay}},
  \bibinfo{author}{\bibfnamefont{R.~E.} \bibnamefont{Drullinger}},
  \bibinfo{author}{\bibfnamefont{T.~M.} \bibnamefont{Fortier}},
  \bibinfo{author}{\bibfnamefont{J.~E.} \bibnamefont{Stalnaker}},
  \bibinfo{author}{\bibfnamefont{S.~A.} \bibnamefont{Diddams}},
  \bibinfo{author}{\bibfnamefont{W.~C.} \bibnamefont{Swann}},
  \bibnamefont{et~al.}, \bibinfo{journal}{Science}
  \textbf{\bibinfo{volume}{319}}, \bibinfo{pages}{1808} (\bibinfo{year}{2008}).

\bibitem[{\citenamefont{Bromley and Mitroy}(2002)}]{bromley02d}
\bibinfo{author}{\bibfnamefont{M.~W.~J.} \bibnamefont{Bromley}}
  \bibnamefont{and} \bibinfo{author}{\bibfnamefont{J.}~\bibnamefont{Mitroy}},
  \bibinfo{journal}{Phys.~Rev.~A} \textbf{\bibinfo{volume}{65}},
  \bibinfo{pages}{062506} (\bibinfo{year}{2002}).

\bibitem[{\citenamefont{{Topcu} et~al.}(2006)\citenamefont{{Topcu}, {Nasser},
  {Daku}, and {Fritzsche}}}]{topcu06a}
\bibinfo{author}{\bibfnamefont{S.}~\bibnamefont{{Topcu}}},
  \bibinfo{author}{\bibfnamefont{J.}~\bibnamefont{{Nasser}}},
  \bibinfo{author}{\bibfnamefont{L.~M.~L.} \bibnamefont{{Daku}}},
  \bibnamefont{and}
  \bibinfo{author}{\bibfnamefont{S.}~\bibnamefont{{Fritzsche}}},
  \bibinfo{journal}{Phys.~Rev.~A} \textbf{\bibinfo{volume}{73}},
  \bibinfo{pages}{042503} (\bibinfo{year}{2006}).

\bibitem[{\citenamefont{{Komara} et~al.}(2003)\citenamefont{{Komara}, {Gearba},
  {Lundeen}, and {Fehrenbach}}}]{komara03a}
\bibinfo{author}{\bibfnamefont{R.~A.} \bibnamefont{{Komara}}},
  \bibinfo{author}{\bibfnamefont{M.~A.} \bibnamefont{{Gearba}}},
  \bibinfo{author}{\bibfnamefont{S.~R.} \bibnamefont{{Lundeen}}},
  \bibnamefont{and} \bibinfo{author}{\bibfnamefont{C.~W.}
  \bibnamefont{{Fehrenbach}}}, \bibinfo{journal}{Phys.~Rev.~A}
  \textbf{\bibinfo{volume}{67}}, \bibinfo{pages}{062502}
  (\bibinfo{year}{2003}).

\bibitem[{\citenamefont{{Rinkleff} and {Thorn}}(1994)}]{rinkleff94a}
\bibinfo{author}{\bibfnamefont{R.-H.} \bibnamefont{{Rinkleff}}}
  \bibnamefont{and} \bibinfo{author}{\bibfnamefont{F.}~\bibnamefont{{Thorn}}},
  \bibinfo{journal}{Z.~Phys.~D} \textbf{\bibinfo{volume}{31}},
  \bibinfo{pages}{31} (\bibinfo{year}{1994}).

\bibitem[{\citenamefont{{Schuh} et~al.}(1996)\citenamefont{{Schuh},
  {Neureiter}, {J{\"a}ger}, and {Windholz}}}]{schuh96a}
\bibinfo{author}{\bibfnamefont{B.}~\bibnamefont{{Schuh}}},
  \bibinfo{author}{\bibfnamefont{C.}~\bibnamefont{{Neureiter}}},
  \bibinfo{author}{\bibfnamefont{H.}~\bibnamefont{{J{\"a}ger}}},
  \bibnamefont{and}
  \bibinfo{author}{\bibfnamefont{L.}~\bibnamefont{{Windholz}}},
  \bibinfo{journal}{Z.~Phys.~D} \textbf{\bibinfo{volume}{37}},
  \bibinfo{pages}{149} (\bibinfo{year}{1996}).

\bibitem[{\citenamefont{Derevianko}(2002)}]{derevianko02b}
\bibinfo{author}{\bibfnamefont{A.}~\bibnamefont{Derevianko}},
  \bibinfo{journal}{Phys. Rev. A} \textbf{\bibinfo{volume}{65}},
  \bibinfo{pages}{012106} (\bibinfo{year}{2002}).

\bibitem[{\citenamefont{Safronova et~al.}(2004)\citenamefont{Safronova,
  Williams, and Clark}}]{safronova04a}
\bibinfo{author}{\bibfnamefont{M.~S.} \bibnamefont{Safronova}},
  \bibinfo{author}{\bibfnamefont{C.~J.} \bibnamefont{Williams}},
  \bibnamefont{and} \bibinfo{author}{\bibfnamefont{C.~W.} \bibnamefont{Clark}},
  \bibinfo{journal}{Phys.~Rev.~A} \textbf{\bibinfo{volume}{69}},
  \bibinfo{pages}{022509} (\bibinfo{year}{2004}).

\bibitem[{\citenamefont{Jiang et~al.}(2008)\citenamefont{Jiang, Arora, and
  Safronova}}]{jiang08a}
\bibinfo{author}{\bibfnamefont{D.}~\bibnamefont{Jiang}},
  \bibinfo{author}{\bibfnamefont{B.}~\bibnamefont{Arora}}, \bibnamefont{and}
  \bibinfo{author}{\bibfnamefont{M.~S.} \bibnamefont{Safronova}},
  \bibinfo{journal}{Phys. Rev. A} \textbf{\bibinfo{volume}{78}},
  \bibinfo{pages}{022514} (\bibinfo{year}{2008}).

\bibitem[{\citenamefont{Porsev et~al.}(2009)\citenamefont{Porsev, Beloy, and
  Derevianko}}]{porsev09-pnc}
\bibinfo{author}{\bibfnamefont{S.~G.} \bibnamefont{Porsev}},
  \bibinfo{author}{\bibfnamefont{K.}~\bibnamefont{Beloy}}, \bibnamefont{and}
  \bibinfo{author}{\bibfnamefont{A.}~\bibnamefont{Derevianko}},
  \bibinfo{journal}{Phys. Rev. Lett.} \textbf{\bibinfo{volume}{102}},
  \bibinfo{pages}{181601} (\bibinfo{year}{2009}).

\bibitem[{\citenamefont{Dzuba et~al.}(1997)\citenamefont{Dzuba, Flambaum, and
  Sushkov}}]{dzuba97a}
\bibinfo{author}{\bibfnamefont{V.~A.} \bibnamefont{Dzuba}},
  \bibinfo{author}{\bibfnamefont{V.~V.} \bibnamefont{Flambaum}},
  \bibnamefont{and} \bibinfo{author}{\bibfnamefont{O.~P.}
  \bibnamefont{Sushkov}}, \bibinfo{journal}{Phys. Rev. A}
  \textbf{\bibinfo{volume}{56}}, \bibinfo{pages}{R4357} (\bibinfo{year}{1997}).

\bibitem[{\citenamefont{Vasilyev et~al.}(2002)\citenamefont{Vasilyev, Savukov,
  Safronova, and Berry}}]{vasilyev02a}
\bibinfo{author}{\bibfnamefont{A.~A.} \bibnamefont{Vasilyev}},
  \bibinfo{author}{\bibfnamefont{I.~M.} \bibnamefont{Savukov}},
  \bibinfo{author}{\bibfnamefont{M.~S.} \bibnamefont{Safronova}},
  \bibnamefont{and} \bibinfo{author}{\bibfnamefont{H.}~\bibnamefont{Berry}},
  \bibinfo{journal}{Phys. Rev. A} \textbf{\bibinfo{volume}{66}},
  \bibinfo{pages}{020101R} (\bibinfo{year}{2002}).

\bibitem[{\citenamefont{Blundell et~al.}(1992)\citenamefont{Blundell,
  Sapirstein, and Johnson}}]{blundell92a}
\bibinfo{author}{\bibfnamefont{S.~A.} \bibnamefont{Blundell}},
  \bibinfo{author}{\bibfnamefont{J.}~\bibnamefont{Sapirstein}},
  \bibnamefont{and} \bibinfo{author}{\bibfnamefont{W.~R.}
  \bibnamefont{Johnson}}, \bibinfo{journal}{Phys. Rev. D}
  \textbf{\bibinfo{volume}{45}}, \bibinfo{pages}{1602} (\bibinfo{year}{1992}).

\bibitem[{\citenamefont{Cho et~al.}(1997)\citenamefont{Cho, Wood, Bennett,
  Roberts, and Wieman}}]{cho97a}
\bibinfo{author}{\bibfnamefont{D.}~\bibnamefont{Cho}},
  \bibinfo{author}{\bibfnamefont{C.~S.} \bibnamefont{Wood}},
  \bibinfo{author}{\bibfnamefont{S.~C.} \bibnamefont{Bennett}},
  \bibinfo{author}{\bibfnamefont{J.~L.} \bibnamefont{Roberts}},
  \bibnamefont{and} \bibinfo{author}{\bibfnamefont{C.~E.}
  \bibnamefont{Wieman}}, \bibinfo{journal}{Phys. Rev. A}
  \textbf{\bibinfo{volume}{55}}, \bibinfo{pages}{1007} (\bibinfo{year}{1997}).

\bibitem[{\citenamefont{Haxton and Wieman}(2001)}]{haxton01a}
\bibinfo{author}{\bibfnamefont{W.~C.} \bibnamefont{Haxton}} \bibnamefont{and}
  \bibinfo{author}{\bibfnamefont{C.~E.} \bibnamefont{Wieman}},
  \bibinfo{journal}{Ann.\ Rev.\ Nucl.\ Part.\ Sci.}
  \textbf{\bibinfo{volume}{51}}, \bibinfo{pages}{261} (\bibinfo{year}{2001}).

\bibitem[{\citenamefont{Haxton et~al.}(2001)\citenamefont{Haxton, Liu, and
  Ramsey-Musolf}}]{haxton01b}
\bibinfo{author}{\bibfnamefont{W.~C.} \bibnamefont{Haxton}},
  \bibinfo{author}{\bibfnamefont{C.-P.} \bibnamefont{Liu}}, \bibnamefont{and}
  \bibinfo{author}{\bibfnamefont{M.~J.} \bibnamefont{Ramsey-Musolf}},
  \bibinfo{journal}{Phys. Rev. Lett.} \textbf{\bibinfo{volume}{86}},
  \bibinfo{pages}{5247} (\bibinfo{year}{2001}).

\bibitem[{\citenamefont{Safronova
  et~al.}(2009{\natexlab{b}})\citenamefont{Safronova, Pal, Jiang, Kozlov,
  Johnson, and Safronova}}]{safronova09-pnc}
\bibinfo{author}{\bibfnamefont{M.~S.} \bibnamefont{Safronova}},
  \bibinfo{author}{\bibfnamefont{R.}~\bibnamefont{Pal}},
  \bibinfo{author}{\bibfnamefont{D.}~\bibnamefont{Jiang}},
  \bibinfo{author}{\bibfnamefont{M.}~\bibnamefont{Kozlov}},
  \bibinfo{author}{\bibfnamefont{W.}~\bibnamefont{Johnson}}, \bibnamefont{and}
  \bibinfo{author}{\bibfnamefont{U.~I.} \bibnamefont{Safronova}},
  \bibinfo{journal}{Nucl. Phys. A} \textbf{\bibinfo{volume}{827}},
  \bibinfo{pages}{411c} (\bibinfo{year}{2009}{\natexlab{b}}).

\bibitem[{\citenamefont{McKeever et~al.}(2003)\citenamefont{McKeever, Buck,
  Boozer, Kuzmich, Nagerl, Stamper-Kurn, and Kimble}}]{mckeever03a}
\bibinfo{author}{\bibfnamefont{J.}~\bibnamefont{McKeever}},
  \bibinfo{author}{\bibfnamefont{J.~R.} \bibnamefont{Buck}},
  \bibinfo{author}{\bibfnamefont{A.~D.} \bibnamefont{Boozer}},
  \bibinfo{author}{\bibfnamefont{A.}~\bibnamefont{Kuzmich}},
  \bibinfo{author}{\bibfnamefont{H.-C.} \bibnamefont{Nagerl}},
  \bibinfo{author}{\bibfnamefont{D.~M.} \bibnamefont{Stamper-Kurn}},
  \bibnamefont{and} \bibinfo{author}{\bibfnamefont{H.~J.}
  \bibnamefont{Kimble}}, \bibinfo{journal}{Phys. Rev. Lett.}
  \textbf{\bibinfo{volume}{90}}, \bibinfo{pages}{133602}
  (\bibinfo{year}{2003}).

\bibitem[{\citenamefont{Jaksch et~al.}(1999)\citenamefont{Jaksch, Briegel,
  Cirac, Gardiner, and Zoller}}]{Rgate}
\bibinfo{author}{\bibfnamefont{D.}~\bibnamefont{Jaksch}},
  \bibinfo{author}{\bibfnamefont{H.-J.} \bibnamefont{Briegel}},
  \bibinfo{author}{\bibfnamefont{J.~I.} \bibnamefont{Cirac}},
  \bibinfo{author}{\bibfnamefont{C.~W.} \bibnamefont{Gardiner}},
  \bibnamefont{and} \bibinfo{author}{\bibfnamefont{P.}~\bibnamefont{Zoller}},
  \bibinfo{journal}{Phys. Rev. Lett.} \textbf{\bibinfo{volume}{82}},
  \bibinfo{pages}{1975} (\bibinfo{year}{1999}).

\bibitem[{\citenamefont{Isenhower et~al.}(2010)\citenamefont{Isenhower, Urban,
  Zhang, Gill, Henage, Johnson, Walker, and Saffman}}]{saffman10}
\bibinfo{author}{\bibfnamefont{L.}~\bibnamefont{Isenhower}},
  \bibinfo{author}{\bibfnamefont{E.}~\bibnamefont{Urban}},
  \bibinfo{author}{\bibfnamefont{X.~L.} \bibnamefont{Zhang}},
  \bibinfo{author}{\bibfnamefont{A.~T.} \bibnamefont{Gill}},
  \bibinfo{author}{\bibfnamefont{T.}~\bibnamefont{Henage}},
  \bibinfo{author}{\bibfnamefont{T.~A.} \bibnamefont{Johnson}},
  \bibinfo{author}{\bibfnamefont{T.~G.} \bibnamefont{Walker}},
  \bibnamefont{and} \bibinfo{author}{\bibfnamefont{M.}~\bibnamefont{Saffman}},
  \bibinfo{journal}{Phys. Rev. Lett.} \textbf{\bibinfo{volume}{104}},
  \bibinfo{pages}{010503} (\bibinfo{year}{2010}).

\bibitem[{\citenamefont{Safronova et~al.}(2003)\citenamefont{Safronova,
  Williams, and Clark}}]{us-gate}
\bibinfo{author}{\bibfnamefont{M.}~\bibnamefont{Safronova}},
  \bibinfo{author}{\bibfnamefont{C.~J.} \bibnamefont{Williams}},
  \bibnamefont{and} \bibinfo{author}{\bibfnamefont{C.~W.} \bibnamefont{Clark}},
  \bibinfo{journal}{Phys. Rev. A} \textbf{\bibinfo{volume}{67}},
  \bibinfo{pages}{040303} (\bibinfo{year}{2003}).

\bibitem[{\citenamefont{Saffman and Walker}(2005)}]{saffman05}
\bibinfo{author}{\bibfnamefont{M.}~\bibnamefont{Saffman}} \bibnamefont{and}
  \bibinfo{author}{\bibfnamefont{T.~G.} \bibnamefont{Walker}},
  \bibinfo{journal}{Phys. Rev. A} \textbf{\bibinfo{volume}{72}},
  \bibinfo{pages}{22347} (\bibinfo{year}{2005}).

\bibitem[{\citenamefont{Gorshkov et~al.}(2009)\citenamefont{Gorshkov, Rey,
  Daley, Boyd, Ye, Zoller, and Lukin}}]{rey09}
\bibinfo{author}{\bibfnamefont{A.~V.} \bibnamefont{Gorshkov}},
  \bibinfo{author}{\bibfnamefont{A.~M.} \bibnamefont{Rey}},
  \bibinfo{author}{\bibfnamefont{A.~J.} \bibnamefont{Daley}},
  \bibinfo{author}{\bibfnamefont{M.~M.} \bibnamefont{Boyd}},
  \bibinfo{author}{\bibfnamefont{J.}~\bibnamefont{Ye}},
  \bibinfo{author}{\bibfnamefont{P.}~\bibnamefont{Zoller}}, \bibnamefont{and}
  \bibinfo{author}{\bibfnamefont{M.~D.} \bibnamefont{Lukin}},
  \bibinfo{journal}{Phys. Rev. Lett.} \textbf{\bibinfo{volume}{102}},
  \bibinfo{pages}{110503} (\bibinfo{year}{2009}).

\bibitem[{\citenamefont{{Chwalla} et~al.}(2009)\citenamefont{{Chwalla},
  {Benhelm}, {Kim}, {Kirchmair}, {Monz}, {Riebe}, {Schindler}, {Villar},
  {H{\"a}nsel}, {Roos} et~al.}}]{chwalla09a}
\bibinfo{author}{\bibfnamefont{M.}~\bibnamefont{{Chwalla}}},
  \bibinfo{author}{\bibfnamefont{J.}~\bibnamefont{{Benhelm}}},
  \bibinfo{author}{\bibfnamefont{K.}~\bibnamefont{{Kim}}},
  \bibinfo{author}{\bibfnamefont{G.}~\bibnamefont{{Kirchmair}}},
  \bibinfo{author}{\bibfnamefont{T.}~\bibnamefont{{Monz}}},
  \bibinfo{author}{\bibfnamefont{M.}~\bibnamefont{{Riebe}}},
  \bibinfo{author}{\bibfnamefont{P.}~\bibnamefont{{Schindler}}},
  \bibinfo{author}{\bibfnamefont{A.~S.} \bibnamefont{{Villar}}},
  \bibinfo{author}{\bibfnamefont{W.}~\bibnamefont{{H{\"a}nsel}}},
  \bibinfo{author}{\bibfnamefont{C.~F.} \bibnamefont{{Roos}}},
  \bibnamefont{et~al.}, \bibinfo{journal}{Phys.~Rev.~Lett.}
  \textbf{\bibinfo{volume}{102}}, \bibinfo{pages}{023002}
  (\bibinfo{year}{2009}).

\bibitem[{\citenamefont{{Dub{\'e}} et~al.}(2007)\citenamefont{{Dub{\'e}},
  {Madej}, {Bernard}, and {Shiner}}}]{dube07a}
\bibinfo{author}{\bibfnamefont{P.}~\bibnamefont{{Dub{\'e}}}},
  \bibinfo{author}{\bibfnamefont{A.~A.} \bibnamefont{{Madej}}},
  \bibinfo{author}{\bibfnamefont{J.~E.} \bibnamefont{{Bernard}}},
  \bibnamefont{and} \bibinfo{author}{\bibfnamefont{A.~D.}
  \bibnamefont{{Shiner}}}, in \emph{\bibinfo{booktitle}{Society of
  Photo-Optical Instrumentation Engineers (SPIE) Conference Series}}
  (\bibinfo{year}{2007}), vol. \bibinfo{volume}{6673} of
  \emph{\bibinfo{series}{Society of Photo-Optical Instrumentation Engineers
  (SPIE) Conference Series}}.

\bibitem[{\citenamefont{{Oskay} et~al.}(2006)\citenamefont{{Oskay}, {Diddams},
  {Donley}, {Fortier}, {Heavner}, {Hollberg}, {Itano}, {Jefferts}, {Delaney},
  {Kim} et~al.}}]{oskay06a}
\bibinfo{author}{\bibfnamefont{W.~H.} \bibnamefont{{Oskay}}},
  \bibinfo{author}{\bibfnamefont{S.~A.} \bibnamefont{{Diddams}}},
  \bibinfo{author}{\bibfnamefont{E.~A.} \bibnamefont{{Donley}}},
  \bibinfo{author}{\bibfnamefont{T.~M.} \bibnamefont{{Fortier}}},
  \bibinfo{author}{\bibfnamefont{T.~P.} \bibnamefont{{Heavner}}},
  \bibinfo{author}{\bibfnamefont{L.}~\bibnamefont{{Hollberg}}},
  \bibinfo{author}{\bibfnamefont{W.~M.} \bibnamefont{{Itano}}},
  \bibinfo{author}{\bibfnamefont{S.~R.} \bibnamefont{{Jefferts}}},
  \bibinfo{author}{\bibfnamefont{M.~J.} \bibnamefont{{Delaney}}},
  \bibinfo{author}{\bibfnamefont{K.}~\bibnamefont{{Kim}}},
  \bibnamefont{et~al.}, \bibinfo{journal}{Phys.~Rev.~Lett.}
  \textbf{\bibinfo{volume}{97}}, \bibinfo{pages}{020801}
  (\bibinfo{year}{2006}).

\bibitem[{\citenamefont{{Hosaka} et~al.}(2009)\citenamefont{{Hosaka},
  {Webster}, {Stannard}, {Walton}, {Margolis}, and {Gill}}}]{hosaka09a}
\bibinfo{author}{\bibfnamefont{K.}~\bibnamefont{{Hosaka}}},
  \bibinfo{author}{\bibfnamefont{S.~A.} \bibnamefont{{Webster}}},
  \bibinfo{author}{\bibfnamefont{A.}~\bibnamefont{{Stannard}}},
  \bibinfo{author}{\bibfnamefont{B.~R.} \bibnamefont{{Walton}}},
  \bibinfo{author}{\bibfnamefont{H.~S.} \bibnamefont{{Margolis}}},
  \bibnamefont{and} \bibinfo{author}{\bibfnamefont{P.}~\bibnamefont{{Gill}}},
  \bibinfo{journal}{Phys.~Rev.~A} \textbf{\bibinfo{volume}{79}},
  \bibinfo{pages}{033403} (\bibinfo{year}{2009}).

\bibitem[{\citenamefont{{Bi{\'e}mont} et~al.}(1998)\citenamefont{{Bi{\'e}mont},
  {Dutrieux}, {Martin}, and {Quinet}}}]{biemont98a}
\bibinfo{author}{\bibfnamefont{E.}~\bibnamefont{{Bi{\'e}mont}}},
  \bibinfo{author}{\bibfnamefont{J.}~\bibnamefont{{Dutrieux}}},
  \bibinfo{author}{\bibfnamefont{I.}~\bibnamefont{{Martin}}}, \bibnamefont{and}
  \bibinfo{author}{\bibfnamefont{P.}~\bibnamefont{{Quinet}}},
  \bibinfo{journal}{J.~Phys.~B} \textbf{\bibinfo{volume}{31}},
  \bibinfo{pages}{3321} (\bibinfo{year}{1998}).

\bibitem[{\citenamefont{Lea et~al.}(2006)\citenamefont{Lea, Webster, and
  Barwood}}]{lea06a}
\bibinfo{author}{\bibfnamefont{S.~N.} \bibnamefont{Lea}},
  \bibinfo{author}{\bibfnamefont{S.~A.} \bibnamefont{Webster}},
  \bibnamefont{and} \bibinfo{author}{\bibfnamefont{G.~P.}
  \bibnamefont{Barwood}}, \bibinfo{journal}{Proceedings of the 20th European
  Frequency and Time Form, PTB Braunschweig, Germany}
  \textbf{\bibinfo{volume}{85}}, \bibinfo{pages}{302} (\bibinfo{year}{2006}).

\bibitem[{\citenamefont{Tamm et~al.}(2007)\citenamefont{Tamm, Lipphardt,
  Schnatz, Wyands, Weyers, Schneider, and Peik}}]{tamm07a}
\bibinfo{author}{\bibfnamefont{C.}~\bibnamefont{Tamm}},
  \bibinfo{author}{\bibfnamefont{B.}~\bibnamefont{Lipphardt}},
  \bibinfo{author}{\bibfnamefont{H.}~\bibnamefont{Schnatz}},
  \bibinfo{author}{\bibfnamefont{R.}~\bibnamefont{Wyands}},
  \bibinfo{author}{\bibfnamefont{S.}~\bibnamefont{Weyers}},
  \bibinfo{author}{\bibfnamefont{T.}~\bibnamefont{Schneider}},
  \bibnamefont{and} \bibinfo{author}{\bibfnamefont{E.}~\bibnamefont{Peik}},
  \bibinfo{journal}{IEEE Trans.~Instrum.~Measur.}
  \textbf{\bibinfo{volume}{56}}, \bibinfo{pages}{601} (\bibinfo{year}{2007}).

\bibitem[{\citenamefont{Rosenband et~al.}(2007)\citenamefont{Rosenband,
  Schmidt, Hume, Itano, Fortier, Stalnaker, Kim, Diddams, Koelemeij, Bergquist
  et~al.}}]{rosenband07a}
\bibinfo{author}{\bibfnamefont{T.}~\bibnamefont{Rosenband}},
  \bibinfo{author}{\bibfnamefont{P.~O.} \bibnamefont{Schmidt}},
  \bibinfo{author}{\bibfnamefont{D.~B.} \bibnamefont{Hume}},
  \bibinfo{author}{\bibfnamefont{W.~M.} \bibnamefont{Itano}},
  \bibinfo{author}{\bibfnamefont{T.~M.} \bibnamefont{Fortier}},
  \bibinfo{author}{\bibfnamefont{J.~E.} \bibnamefont{Stalnaker}},
  \bibinfo{author}{\bibfnamefont{K.}~\bibnamefont{Kim}},
  \bibinfo{author}{\bibfnamefont{S.~A.} \bibnamefont{Diddams}},
  \bibinfo{author}{\bibfnamefont{J.~C.~J.} \bibnamefont{Koelemeij}},
  \bibinfo{author}{\bibfnamefont{J.~C.} \bibnamefont{Bergquist}},
  \bibnamefont{et~al.}, \bibinfo{journal}{Phys.~Rev.~Lett.}
  \textbf{\bibinfo{volume}{98}}, \bibinfo{pages}{220801}
  (\bibinfo{year}{2007}).

\bibitem[{\citenamefont{{Becker} et~al.}(2001)\citenamefont{{Becker},
  {Zanthier}, {Nevsky}, {Schwedes}, {Skvortsov}, {Walther}, and
  {Peik}}}]{becker01a}
\bibinfo{author}{\bibfnamefont{T.}~\bibnamefont{{Becker}}},
  \bibinfo{author}{\bibfnamefont{J.~V.} \bibnamefont{{Zanthier}}},
  \bibinfo{author}{\bibfnamefont{A.~Y.} \bibnamefont{{Nevsky}}},
  \bibinfo{author}{\bibfnamefont{C.}~\bibnamefont{{Schwedes}}},
  \bibinfo{author}{\bibfnamefont{M.~N.} \bibnamefont{{Skvortsov}}},
  \bibinfo{author}{\bibfnamefont{H.}~\bibnamefont{{Walther}}},
  \bibnamefont{and} \bibinfo{author}{\bibfnamefont{E.}~\bibnamefont{{Peik}}},
  \bibinfo{journal}{Phys.~Rev.~A} \textbf{\bibinfo{volume}{63}},
  \bibinfo{pages}{051802} (\bibinfo{year}{2001}).

\bibitem[{\citenamefont{{Eichenseer} et~al.}(2003)\citenamefont{{Eichenseer},
  {Nevsky}, {Schwedes}, {von Zanthier}, and {Walther}}}]{eichenseer03a}
\bibinfo{author}{\bibfnamefont{M.}~\bibnamefont{{Eichenseer}}},
  \bibinfo{author}{\bibfnamefont{A.~Y.} \bibnamefont{{Nevsky}}},
  \bibinfo{author}{\bibfnamefont{C.}~\bibnamefont{{Schwedes}}},
  \bibinfo{author}{\bibfnamefont{J.}~\bibnamefont{{von Zanthier}}},
  \bibnamefont{and}
  \bibinfo{author}{\bibfnamefont{H.}~\bibnamefont{{Walther}}},
  \bibinfo{journal}{J.~Phys.~B} \textbf{\bibinfo{volume}{36}},
  \bibinfo{pages}{553} (\bibinfo{year}{2003}).

\bibitem[{\citenamefont{Ralchenko et~al.}(2008)\citenamefont{Ralchenko,
  Kramida, Reader, and {NIST ASD Team}}}]{nistasd315}
\bibinfo{author}{\bibfnamefont{Y.}~\bibnamefont{Ralchenko}},
  \bibinfo{author}{\bibfnamefont{A.}~\bibnamefont{Kramida}},
  \bibinfo{author}{\bibfnamefont{J.}~\bibnamefont{Reader}}, \bibnamefont{and}
  \bibinfo{author}{\bibnamefont{{NIST ASD Team}}}, \emph{\bibinfo{title}{{NIST
  Atomic Spectra Database Version 3.1.5}}} (\bibinfo{year}{2008}),
  \urlprefix\url{http://physics.nist.gov/asd3}.

\bibitem[{\citenamefont{{Friebe} et~al.}(2008)\citenamefont{{Friebe}, {Pape},
  {Riedmann}, {Moldenhauer}, {Mehlst{\"a}ubler}, {Rehbein}, {Lisdat}, {Rasel},
  {Ertmer}, {Schnatz} et~al.}}]{friebe08a}
\bibinfo{author}{\bibfnamefont{J.}~\bibnamefont{{Friebe}}},
  \bibinfo{author}{\bibfnamefont{A.}~\bibnamefont{{Pape}}},
  \bibinfo{author}{\bibfnamefont{M.}~\bibnamefont{{Riedmann}}},
  \bibinfo{author}{\bibfnamefont{K.}~\bibnamefont{{Moldenhauer}}},
  \bibinfo{author}{\bibfnamefont{T.}~\bibnamefont{{Mehlst{\"a}ubler}}},
  \bibinfo{author}{\bibfnamefont{N.}~\bibnamefont{{Rehbein}}},
  \bibinfo{author}{\bibfnamefont{C.}~\bibnamefont{{Lisdat}}},
  \bibinfo{author}{\bibfnamefont{E.~M.} \bibnamefont{{Rasel}}},
  \bibinfo{author}{\bibfnamefont{W.}~\bibnamefont{{Ertmer}}},
  \bibinfo{author}{\bibfnamefont{H.}~\bibnamefont{{Schnatz}}},
  \bibnamefont{et~al.}, \bibinfo{journal}{Phys.~Rev.~A}
  \textbf{\bibinfo{volume}{78}}, \bibinfo{pages}{033830}
  (\bibinfo{year}{2008}).

\bibitem[{\citenamefont{{Campbell} et~al.}(2008)\citenamefont{{Campbell},
  {Ludlow}, {Blatt}, {Thomsen}, {Martin}, {de Miranda}, {Zelevinsky}, {Boyd},
  {Ye}, {Diddams} et~al.}}]{campbell08a}
\bibinfo{author}{\bibfnamefont{G.~K.} \bibnamefont{{Campbell}}},
  \bibinfo{author}{\bibfnamefont{A.~D.} \bibnamefont{{Ludlow}}},
  \bibinfo{author}{\bibfnamefont{S.}~\bibnamefont{{Blatt}}},
  \bibinfo{author}{\bibfnamefont{J.~W.} \bibnamefont{{Thomsen}}},
  \bibinfo{author}{\bibfnamefont{M.~J.} \bibnamefont{{Martin}}},
  \bibinfo{author}{\bibfnamefont{M.~H.~G.} \bibnamefont{{de Miranda}}},
  \bibinfo{author}{\bibfnamefont{T.}~\bibnamefont{{Zelevinsky}}},
  \bibinfo{author}{\bibfnamefont{M.~M.} \bibnamefont{{Boyd}}},
  \bibinfo{author}{\bibfnamefont{J.}~\bibnamefont{{Ye}}},
  \bibinfo{author}{\bibfnamefont{S.~A.} \bibnamefont{{Diddams}}},
  \bibnamefont{et~al.}, \bibinfo{journal}{Metrologia}
  \textbf{\bibinfo{volume}{45}}, \bibinfo{pages}{539} (\bibinfo{year}{2008}).

\bibitem[{\citenamefont{{Poli} et~al.}(2008)\citenamefont{{Poli}, {Barber},
  {Lemke}, {Oates}, {Ma}, {Stalnaker}, {Fortier}, {Diddams}, {Hollberg},
  {Bergquist} et~al.}}]{poli08a}
\bibinfo{author}{\bibfnamefont{N.}~\bibnamefont{{Poli}}},
  \bibinfo{author}{\bibfnamefont{Z.~W.} \bibnamefont{{Barber}}},
  \bibinfo{author}{\bibfnamefont{N.~D.} \bibnamefont{{Lemke}}},
  \bibinfo{author}{\bibfnamefont{C.~W.} \bibnamefont{{Oates}}},
  \bibinfo{author}{\bibfnamefont{L.~S.} \bibnamefont{{Ma}}},
  \bibinfo{author}{\bibfnamefont{J.~E.} \bibnamefont{{Stalnaker}}},
  \bibinfo{author}{\bibfnamefont{T.~M.} \bibnamefont{{Fortier}}},
  \bibinfo{author}{\bibfnamefont{S.~A.} \bibnamefont{{Diddams}}},
  \bibinfo{author}{\bibfnamefont{L.}~\bibnamefont{{Hollberg}}},
  \bibinfo{author}{\bibfnamefont{J.~C.} \bibnamefont{{Bergquist}}},
  \bibnamefont{et~al.}, \bibinfo{journal}{Phys.~Rev.~A}
  \textbf{\bibinfo{volume}{77}}, \bibinfo{pages}{050501}
  (\bibinfo{year}{2008}).

\bibitem[{\citenamefont{Porsev and Derevianko}(2004)}]{porsev04b}
\bibinfo{author}{\bibfnamefont{S.~G.} \bibnamefont{Porsev}} \bibnamefont{and}
  \bibinfo{author}{\bibfnamefont{A.}~\bibnamefont{Derevianko}},
  \bibinfo{journal}{Phys. Rev. A} \textbf{\bibinfo{volume}{69}},
  \bibinfo{pages}{042506} (\bibinfo{year}{2004}).

\bibitem[{\citenamefont{{Wang} and {Ye}}(2007)}]{wang07b}
\bibinfo{author}{\bibfnamefont{G.}~\bibnamefont{{Wang}}} \bibnamefont{and}
  \bibinfo{author}{\bibfnamefont{A.}~\bibnamefont{{Ye}}},
  \bibinfo{journal}{Phys.~Rev.~A} \textbf{\bibinfo{volume}{76}},
  \bibinfo{pages}{043409} (\bibinfo{year}{2007}).

\bibitem[{\citenamefont{Bigeon}(1967)}]{bigeon67a}
\bibinfo{author}{\bibfnamefont{M.~C.} \bibnamefont{Bigeon}},
  \bibinfo{journal}{J.~Phys. France} \textbf{\bibinfo{volume}{28}},
  \bibinfo{pages}{51} (\bibinfo{year}{1967}).

\bibitem[{\citenamefont{Taylor and Thompson}(2008)}]{NIST330}
\bibinfo{author}{\bibfnamefont{B.~N.} \bibnamefont{Taylor}} \bibnamefont{and}
  \bibinfo{author}{\bibfnamefont{A.}~\bibnamefont{Thompson}},
  \emph{\bibinfo{title}{The International System of Units (SI)}}
  (\bibinfo{publisher}{National Institute of Standards and Technology, Special
  Publication 330}, \bibinfo{address}{Gaithersburg, MD}, \bibinfo{year}{2008}).

\bibitem[{\citenamefont{Wallard}(2006)}]{BIPM2006}
\bibinfo{author}{\bibfnamefont{A.}~\bibnamefont{Wallard}},
  \bibinfo{journal}{Metrologia} \textbf{\bibinfo{volume}{43}},
  \bibinfo{pages}{175} (\bibinfo{year}{2006}).

\bibitem[{\citenamefont{Katori et~al.}(2003)\citenamefont{Katori, Takamoto,
  Pal'chikov, and Ovsiannikov}}]{katori03a}
\bibinfo{author}{\bibfnamefont{H.}~\bibnamefont{Katori}},
  \bibinfo{author}{\bibfnamefont{M.}~\bibnamefont{Takamoto}},
  \bibinfo{author}{\bibfnamefont{V.~G.} \bibnamefont{Pal'chikov}},
  \bibnamefont{and} \bibinfo{author}{\bibfnamefont{V.~D.}
  \bibnamefont{Ovsiannikov}}, \bibinfo{journal}{Phys.~Rev.~Lett.}
  \textbf{\bibinfo{volume}{91}}, \bibinfo{pages}{173005}
  (\bibinfo{year}{2003}).

\bibitem[{\citenamefont{{Takamoto} et~al.}(2005)\citenamefont{{Takamoto},
  {Hong}, {Higashi}, and {Katori}}}]{takamoto05a}
\bibinfo{author}{\bibfnamefont{M.}~\bibnamefont{{Takamoto}}},
  \bibinfo{author}{\bibfnamefont{F.-L.} \bibnamefont{{Hong}}},
  \bibinfo{author}{\bibfnamefont{R.}~\bibnamefont{{Higashi}}},
  \bibnamefont{and} \bibinfo{author}{\bibfnamefont{H.}~\bibnamefont{{Katori}}},
  \bibinfo{journal}{Nature} \textbf{\bibinfo{volume}{435}},
  \bibinfo{pages}{321} (\bibinfo{year}{2005}).

\bibitem[{\citenamefont{{Ludlow} et~al.}(2008)\citenamefont{{Ludlow},
  {Zelevinsky}, {Campbell}, {Blatt}, {Boyd}, {de Miranda}, {Martin}, {Thomsen},
  {Foreman}, {Ye} et~al.}}]{ludlow08a}
\bibinfo{author}{\bibfnamefont{A.~D.} \bibnamefont{{Ludlow}}},
  \bibinfo{author}{\bibfnamefont{T.}~\bibnamefont{{Zelevinsky}}},
  \bibinfo{author}{\bibfnamefont{G.~K.} \bibnamefont{{Campbell}}},
  \bibinfo{author}{\bibfnamefont{S.}~\bibnamefont{{Blatt}}},
  \bibinfo{author}{\bibfnamefont{M.~M.} \bibnamefont{{Boyd}}},
  \bibinfo{author}{\bibfnamefont{M.~H.~G.} \bibnamefont{{de Miranda}}},
  \bibinfo{author}{\bibfnamefont{M.~J.} \bibnamefont{{Martin}}},
  \bibinfo{author}{\bibfnamefont{J.~W.} \bibnamefont{{Thomsen}}},
  \bibinfo{author}{\bibfnamefont{S.~M.} \bibnamefont{{Foreman}}},
  \bibinfo{author}{\bibfnamefont{J.}~\bibnamefont{{Ye}}}, \bibnamefont{et~al.},
  \bibinfo{journal}{Science} \textbf{\bibinfo{volume}{319}},
  \bibinfo{pages}{1805} (\bibinfo{year}{2008}), \eprint{0801.4344}.

\bibitem[{\citenamefont{{Wilpers} et~al.}(2007)\citenamefont{{Wilpers},
  {Oates}, {Diddams}, {Bartels}, {Fortier}, {Oskay}, {Bergquist}, {Jefferts},
  {Heavner}, {Parker} et~al.}}]{wilpers07a}
\bibinfo{author}{\bibfnamefont{G.}~\bibnamefont{{Wilpers}}},
  \bibinfo{author}{\bibfnamefont{C.~W.} \bibnamefont{{Oates}}},
  \bibinfo{author}{\bibfnamefont{S.~A.} \bibnamefont{{Diddams}}},
  \bibinfo{author}{\bibfnamefont{A.}~\bibnamefont{{Bartels}}},
  \bibinfo{author}{\bibfnamefont{T.~M.} \bibnamefont{{Fortier}}},
  \bibinfo{author}{\bibfnamefont{W.~H.} \bibnamefont{{Oskay}}},
  \bibinfo{author}{\bibfnamefont{J.~C.} \bibnamefont{{Bergquist}}},
  \bibinfo{author}{\bibfnamefont{S.~R.} \bibnamefont{{Jefferts}}},
  \bibinfo{author}{\bibfnamefont{T.~P.} \bibnamefont{{Heavner}}},
  \bibinfo{author}{\bibfnamefont{T.~E.} \bibnamefont{{Parker}}},
  \bibnamefont{et~al.}, \bibinfo{journal}{Metrologia}
  \textbf{\bibinfo{volume}{44}}, \bibinfo{pages}{146} (\bibinfo{year}{2007}).

\bibitem[{\citenamefont{Boyd et~al.}(2007)\citenamefont{Boyd, Ludlow, Blatt,
  Foreman, Ito, Zelevinsky, and Ye}}]{boyd07a}
\bibinfo{author}{\bibfnamefont{M.~M.} \bibnamefont{Boyd}},
  \bibinfo{author}{\bibfnamefont{A.~D.} \bibnamefont{Ludlow}},
  \bibinfo{author}{\bibfnamefont{S.}~\bibnamefont{Blatt}},
  \bibinfo{author}{\bibfnamefont{S.~M.} \bibnamefont{Foreman}},
  \bibinfo{author}{\bibfnamefont{T.}~\bibnamefont{Ito}},
  \bibinfo{author}{\bibfnamefont{T.}~\bibnamefont{Zelevinsky}},
  \bibnamefont{and} \bibinfo{author}{\bibfnamefont{J.}~\bibnamefont{Ye}},
  \bibinfo{journal}{Phys.~Rev.~Lett..} \textbf{\bibinfo{volume}{98}},
  \bibinfo{pages}{083002} (\bibinfo{year}{2007}).

\bibitem[{\citenamefont{{Taichenachev}
  et~al.}(2008)\citenamefont{{Taichenachev}, {Yudin}, {Ovsiannikov},
  {Pal'Chikov}, and {Oates}}}]{taichenachev08a}
\bibinfo{author}{\bibfnamefont{A.~V.} \bibnamefont{{Taichenachev}}},
  \bibinfo{author}{\bibfnamefont{V.~I.} \bibnamefont{{Yudin}}},
  \bibinfo{author}{\bibfnamefont{V.~D.} \bibnamefont{{Ovsiannikov}}},
  \bibinfo{author}{\bibfnamefont{V.~G.} \bibnamefont{{Pal'Chikov}}},
  \bibnamefont{and} \bibinfo{author}{\bibfnamefont{C.~W.}
  \bibnamefont{{Oates}}}, \bibinfo{journal}{Phys.~Rev.~Lett.~}
  \textbf{\bibinfo{volume}{101}}, \bibinfo{pages}{193601}
  (\bibinfo{year}{2008}).

\bibitem[{\citenamefont{{Katori} et~al.}(2009)\citenamefont{{Katori},
  {Hashiguchi}, {Il'Inova}, and {Ovsiannikov}}}]{katori09a}
\bibinfo{author}{\bibfnamefont{H.}~\bibnamefont{{Katori}}},
  \bibinfo{author}{\bibfnamefont{K.}~\bibnamefont{{Hashiguchi}}},
  \bibinfo{author}{\bibfnamefont{E.~Y.} \bibnamefont{{Il'Inova}}},
  \bibnamefont{and} \bibinfo{author}{\bibfnamefont{V.~D.}
  \bibnamefont{{Ovsiannikov}}}, \bibinfo{journal}{Phys.~Rev.~Lett.~}
  \textbf{\bibinfo{volume}{103}}, \bibinfo{pages}{153004}
  (\bibinfo{year}{2009}).

\bibitem[{\citenamefont{Mowat}(1972)}]{mowat72a}
\bibinfo{author}{\bibfnamefont{J.~R.} \bibnamefont{Mowat}},
  \bibinfo{journal}{Phys. Rev. A} \textbf{\bibinfo{volume}{5}},
  \bibinfo{pages}{1059} (\bibinfo{year}{1972}).

\bibitem[{\citenamefont{{Safronova} et~al.}(2010)\citenamefont{{Safronova},
  Jiang, Arora, Clark, Kozlov, {Safronova}, and Johnson}}]{safronova10a}
\bibinfo{author}{\bibfnamefont{M.~S.} \bibnamefont{{Safronova}}},
  \bibinfo{author}{\bibfnamefont{D.}~\bibnamefont{Jiang}},
  \bibinfo{author}{\bibfnamefont{B.}~\bibnamefont{Arora}},
  \bibinfo{author}{\bibfnamefont{C.~W.} \bibnamefont{Clark}},
  \bibinfo{author}{\bibfnamefont{M.~G.} \bibnamefont{Kozlov}},
  \bibinfo{author}{\bibfnamefont{U.~I.} \bibnamefont{{Safronova}}},
  \bibnamefont{and} \bibinfo{author}{\bibfnamefont{W.~R.}
  \bibnamefont{Johnson}}, \bibinfo{journal}{IEEE Trans. Ultrason.
  Ferroelectrics and Frequency Control} \textbf{\bibinfo{volume}{57}},
  \bibinfo{pages}{94} (\bibinfo{year}{2010}).

\bibitem[{\citenamefont{Safronova and Safronova}(2010)}]{safronova10b}
\bibinfo{author}{\bibfnamefont{M.~S.} \bibnamefont{Safronova}}
  \bibnamefont{and} \bibinfo{author}{\bibfnamefont{U.~I.}
  \bibnamefont{Safronova}} (\bibinfo{year}{2010}), \bibinfo{note}{to be
  submitted to Phys. Rev. A (2010)}.

\bibitem[{\citenamefont{Beloy et~al.}(2006)\citenamefont{Beloy, Safronova, and
  Derevianko}}]{beloy06a}
\bibinfo{author}{\bibfnamefont{K.}~\bibnamefont{Beloy}},
  \bibinfo{author}{\bibfnamefont{U.~I.} \bibnamefont{Safronova}},
  \bibnamefont{and}
  \bibinfo{author}{\bibfnamefont{A.}~\bibnamefont{Derevianko}},
  \bibinfo{journal}{Phys. Rev.Lett.} \textbf{\bibinfo{volume}{97}},
  \bibinfo{pages}{040801} (\bibinfo{year}{2006}).

\bibitem[{\citenamefont{{Simon} et~al.}(1998)\citenamefont{{Simon}, {Laurent},
  and {Clairon}}}]{simon98a}
\bibinfo{author}{\bibfnamefont{E.}~\bibnamefont{{Simon}}},
  \bibinfo{author}{\bibfnamefont{P.}~\bibnamefont{{Laurent}}},
  \bibnamefont{and}
  \bibinfo{author}{\bibfnamefont{A.}~\bibnamefont{{Clairon}}},
  \bibinfo{journal}{Phys.~Rev.~A} \textbf{\bibinfo{volume}{57}},
  \bibinfo{pages}{436} (\bibinfo{year}{1998}).

\bibitem[{\citenamefont{{Godone} et~al.}(2005)\citenamefont{{Godone},
  {Calonico}, {Levi}, {Micalizio}, and {Calosso}}}]{godone05a}
\bibinfo{author}{\bibfnamefont{A.}~\bibnamefont{{Godone}}},
  \bibinfo{author}{\bibfnamefont{D.}~\bibnamefont{{Calonico}}},
  \bibinfo{author}{\bibfnamefont{F.}~\bibnamefont{{Levi}}},
  \bibinfo{author}{\bibfnamefont{S.}~\bibnamefont{{Micalizio}}},
  \bibnamefont{and}
  \bibinfo{author}{\bibfnamefont{C.}~\bibnamefont{{Calosso}}},
  \bibinfo{journal}{Phys.~Rev.~A} \textbf{\bibinfo{volume}{71}},
  \bibinfo{pages}{063401} (\bibinfo{year}{2005}).

\bibitem[{\citenamefont{Marinescu et~al.}(1994)\citenamefont{Marinescu,
  Sadeghpour, and Dalgarno}}]{marinescu94a}
\bibinfo{author}{\bibfnamefont{M.}~\bibnamefont{Marinescu}},
  \bibinfo{author}{\bibfnamefont{H.~R.} \bibnamefont{Sadeghpour}},
  \bibnamefont{and} \bibinfo{author}{\bibfnamefont{A.}~\bibnamefont{Dalgarno}},
  \bibinfo{journal}{Phys.~Rev.~A} \textbf{\bibinfo{volume}{49}},
  \bibinfo{pages}{982} (\bibinfo{year}{1994}).

\bibitem[{\citenamefont{Derevianko et~al.}(2001)\citenamefont{Derevianko, Babb,
  and Dalgarno}}]{derevianko01a}
\bibinfo{author}{\bibfnamefont{A.}~\bibnamefont{Derevianko}},
  \bibinfo{author}{\bibfnamefont{J.~F.} \bibnamefont{Babb}}, \bibnamefont{and}
  \bibinfo{author}{\bibfnamefont{A.}~\bibnamefont{Dalgarno}},
  \bibinfo{journal}{Phys.~Rev.~A.} \textbf{\bibinfo{volume}{63}},
  \bibinfo{pages}{052704} (\bibinfo{year}{2001}).

\bibitem[{\citenamefont{{Pashov} et~al.}(2008)\citenamefont{{Pashov}, {Popov},
  {Knockel}, and {Tiemann}}}]{pashov08a}
\bibinfo{author}{\bibfnamefont{A.}~\bibnamefont{{Pashov}}},
  \bibinfo{author}{\bibfnamefont{P.}~\bibnamefont{{Popov}}},
  \bibinfo{author}{\bibfnamefont{H.}~\bibnamefont{{Knockel}}},
  \bibnamefont{and}
  \bibinfo{author}{\bibfnamefont{E.}~\bibnamefont{{Tiemann}}},
  \bibinfo{journal}{Eur.~Phys.~J. D} \textbf{\bibinfo{volume}{46}},
  \bibinfo{pages}{241} (\bibinfo{year}{2008}).

\bibitem[{\citenamefont{van Kempen et~al.}(2002)\citenamefont{van Kempen,
  Kokkelmans, Heinzen, and Verhaar}}]{vankempen02a}
\bibinfo{author}{\bibfnamefont{E.~G.~M.} \bibnamefont{van Kempen}},
  \bibinfo{author}{\bibfnamefont{S.~J. J. M.~F.} \bibnamefont{Kokkelmans}},
  \bibinfo{author}{\bibfnamefont{D.~J.} \bibnamefont{Heinzen}},
  \bibnamefont{and} \bibinfo{author}{\bibfnamefont{B.~J.}
  \bibnamefont{Verhaar}}, \bibinfo{journal}{Phys.~Rev.~Lett.}
  \textbf{\bibinfo{volume}{88}}, \bibinfo{pages}{093201}
  (\bibinfo{year}{2002}).

\bibitem[{\citenamefont{Chin et~al.}(2004)\citenamefont{Chin, Vuleti{\'c},
  Kerman, Chu, Tiesinga, Leo, and Williams}}]{tiesinga04}
\bibinfo{author}{\bibfnamefont{C.}~\bibnamefont{Chin}},
  \bibinfo{author}{\bibfnamefont{V.}~\bibnamefont{Vuleti{\'c}}},
  \bibinfo{author}{\bibfnamefont{A.~J.} \bibnamefont{Kerman}},
  \bibinfo{author}{\bibfnamefont{S.}~\bibnamefont{Chu}},
  \bibinfo{author}{\bibfnamefont{E.}~\bibnamefont{Tiesinga}},
  \bibinfo{author}{\bibfnamefont{P.~J.} \bibnamefont{Leo}}, \bibnamefont{and}
  \bibinfo{author}{\bibfnamefont{C.~J.} \bibnamefont{Williams}},
  \bibinfo{journal}{Phys. Rev. A} \textbf{\bibinfo{volume}{70}},
  \bibinfo{pages}{032701} (\bibinfo{year}{2004}).

\bibitem[{\citenamefont{{Le Roy} and {Bernstein}}(1971)}]{leroy71a}
\bibinfo{author}{\bibfnamefont{R.~J.} \bibnamefont{{Le Roy}}} \bibnamefont{and}
  \bibinfo{author}{\bibfnamefont{R.~B.} \bibnamefont{{Bernstein}}},
  \bibinfo{journal}{J.~Mol.~Spectrosc.} \textbf{\bibinfo{volume}{37}},
  \bibinfo{pages}{109} (\bibinfo{year}{1971}).

\bibitem[{\citenamefont{Marte et~al.}(2002)\citenamefont{Marte, Volz, Schuster,
  Durr, Rempe, van Kempen, and Verhaar}}]{marte02a}
\bibinfo{author}{\bibfnamefont{A.}~\bibnamefont{Marte}},
  \bibinfo{author}{\bibfnamefont{T.}~\bibnamefont{Volz}},
  \bibinfo{author}{\bibfnamefont{J.}~\bibnamefont{Schuster}},
  \bibinfo{author}{\bibfnamefont{S.}~\bibnamefont{Durr}},
  \bibinfo{author}{\bibfnamefont{G.}~\bibnamefont{Rempe}},
  \bibinfo{author}{\bibfnamefont{E.~G.~M.} \bibnamefont{van Kempen}},
  \bibnamefont{and} \bibinfo{author}{\bibfnamefont{B.~J.}
  \bibnamefont{Verhaar}}, \bibinfo{journal}{Phys.~Rev.~Lett.}
  \textbf{\bibinfo{volume}{89}}, \bibinfo{pages}{283202}
  (\bibinfo{year}{2002}).

\bibitem[{\citenamefont{Yan et~al.}(2000)\citenamefont{Yan, Zhu, and
  Zhou}}]{yan00a}
\bibinfo{author}{\bibfnamefont{Z.~C.} \bibnamefont{Yan}},
  \bibinfo{author}{\bibfnamefont{J.~M.} \bibnamefont{Zhu}}, \bibnamefont{and}
  \bibinfo{author}{\bibfnamefont{B.~L.} \bibnamefont{Zhou}},
  \bibinfo{journal}{Phys.~Rev.~A} \textbf{\bibinfo{volume}{62}},
  \bibinfo{pages}{034501} (\bibinfo{year}{2000}).

\bibitem[{\citenamefont{Zhang et~al.}(2005)\citenamefont{Zhang, Yan, Vrinceanu,
  and Sadeghpour}}]{zhang05a}
\bibinfo{author}{\bibfnamefont{J.~Y.} \bibnamefont{Zhang}},
  \bibinfo{author}{\bibfnamefont{Z.~C.} \bibnamefont{Yan}},
  \bibinfo{author}{\bibfnamefont{D.}~\bibnamefont{Vrinceanu}},
  \bibnamefont{and} \bibinfo{author}{\bibfnamefont{H.~R.}
  \bibnamefont{Sadeghpour}}, \bibinfo{journal}{Phys.~Rev.~A.}
  \textbf{\bibinfo{volume}{71}}, \bibinfo{pages}{032712}
  (\bibinfo{year}{2005}).

\bibitem[{\citenamefont{Zhang et~al.}(2006{\natexlab{a}})\citenamefont{Zhang,
  Yan, Vrinceanu, Babb, and Sadeghpour}}]{zhang06a}
\bibinfo{author}{\bibfnamefont{J.~Y.} \bibnamefont{Zhang}},
  \bibinfo{author}{\bibfnamefont{Z.~C.} \bibnamefont{Yan}},
  \bibinfo{author}{\bibfnamefont{D.}~\bibnamefont{Vrinceanu}},
  \bibinfo{author}{\bibfnamefont{J.~F.} \bibnamefont{Babb}}, \bibnamefont{and}
  \bibinfo{author}{\bibfnamefont{H.~R.} \bibnamefont{Sadeghpour}},
  \bibinfo{journal}{Phys.~Rev.~A.} \textbf{\bibinfo{volume}{73}},
  \bibinfo{pages}{022710} (\bibinfo{year}{2006}{\natexlab{a}}).

\bibitem[{\citenamefont{Zhang et~al.}(2006{\natexlab{b}})\citenamefont{Zhang,
  Yan, Vrinceanu, Babb, and Sadeghpour}}]{zhang06b}
\bibinfo{author}{\bibfnamefont{J.~Y.} \bibnamefont{Zhang}},
  \bibinfo{author}{\bibfnamefont{Z.~C.} \bibnamefont{Yan}},
  \bibinfo{author}{\bibfnamefont{D.}~\bibnamefont{Vrinceanu}},
  \bibinfo{author}{\bibfnamefont{J.~F.} \bibnamefont{Babb}}, \bibnamefont{and}
  \bibinfo{author}{\bibfnamefont{H.~R.} \bibnamefont{Sadeghpour}},
  \bibinfo{journal}{Phys.~Rev.~A.} \textbf{\bibinfo{volume}{74}},
  \bibinfo{pages}{014704} (\bibinfo{year}{2006}{\natexlab{b}}).

\bibitem[{\citenamefont{Zhang et~al.}(2007{\natexlab{b}})\citenamefont{Zhang,
  Yan, Vrinceanu, Babb, and Sadeghpour}}]{zhang07b}
\bibinfo{author}{\bibfnamefont{J.~Y.} \bibnamefont{Zhang}},
  \bibinfo{author}{\bibfnamefont{Z.~C.} \bibnamefont{Yan}},
  \bibinfo{author}{\bibfnamefont{D.}~\bibnamefont{Vrinceanu}},
  \bibinfo{author}{\bibfnamefont{J.~F.} \bibnamefont{Babb}}, \bibnamefont{and}
  \bibinfo{author}{\bibfnamefont{H.~R.} \bibnamefont{Sadeghpour}},
  \bibinfo{journal}{Phys.~Rev.~A} \textbf{\bibinfo{volume}{76}},
  \bibinfo{pages}{012723} (\bibinfo{year}{2007}{\natexlab{b}}).

\bibitem[{\citenamefont{Marinescu and Dalgarno}(1995)}]{marinescu95a}
\bibinfo{author}{\bibfnamefont{M.}~\bibnamefont{Marinescu}} \bibnamefont{and}
  \bibinfo{author}{\bibfnamefont{A.}~\bibnamefont{Dalgarno}},
  \bibinfo{journal}{Phys.~Rev.~A} \textbf{\bibinfo{volume}{52}},
  \bibinfo{pages}{311} (\bibinfo{year}{1995}).

\bibitem[{\citenamefont{Marinescu and Dalgarno}(1996)}]{marinescu96a}
\bibinfo{author}{\bibfnamefont{M.}~\bibnamefont{Marinescu}} \bibnamefont{and}
  \bibinfo{author}{\bibfnamefont{A.}~\bibnamefont{Dalgarno}},
  \bibinfo{journal}{Z. Phys.\ D} \textbf{\bibinfo{volume}{36}},
  \bibinfo{pages}{239} (\bibinfo{year}{1996}).

\bibitem[{\citenamefont{Marinescu}(1997)}]{marinescu97b}
\bibinfo{author}{\bibfnamefont{M.}~\bibnamefont{Marinescu}},
  \bibinfo{journal}{Phys.~Rev.~A} \textbf{\bibinfo{volume}{56}},
  \bibinfo{pages}{4764} (\bibinfo{year}{1997}).

\bibitem[{\citenamefont{Marinescu and Sadeghpour}(1999)}]{marinescu99a}
\bibinfo{author}{\bibfnamefont{M.}~\bibnamefont{Marinescu}} \bibnamefont{and}
  \bibinfo{author}{\bibfnamefont{H.~R.} \bibnamefont{Sadeghpour}},
  \bibinfo{journal}{Phys.~Rev.~A} \textbf{\bibinfo{volume}{59}},
  \bibinfo{pages}{390} (\bibinfo{year}{1999}).

\bibitem[{\citenamefont{Mitroy and Zhang}(2007{\natexlab{b}})}]{mitroy07d}
\bibinfo{author}{\bibfnamefont{J.}~\bibnamefont{Mitroy}} \bibnamefont{and}
  \bibinfo{author}{\bibfnamefont{J.~Y.} \bibnamefont{Zhang}},
  \bibinfo{journal}{Phys.~Rev.~A} \textbf{\bibinfo{volume}{76}},
  \bibinfo{pages}{032706} (\bibinfo{year}{2007}{\natexlab{b}}).

\bibitem[{\citenamefont{Porsev and Derevianko}(2002)}]{porsev02a}
\bibinfo{author}{\bibfnamefont{S.~G.} \bibnamefont{Porsev}} \bibnamefont{and}
  \bibinfo{author}{\bibfnamefont{A.}~\bibnamefont{Derevianko}},
  \bibinfo{journal}{Phys.~Rev.~A} \textbf{\bibinfo{volume}{65}},
  \bibinfo{pages}{020701(R)} (\bibinfo{year}{2002}).

\bibitem[{\citenamefont{Porsev and Derevianko}(2003)}]{porsev03a}
\bibinfo{author}{\bibfnamefont{S.~G.} \bibnamefont{Porsev}} \bibnamefont{and}
  \bibinfo{author}{\bibfnamefont{A.}~\bibnamefont{Derevianko}},
  \bibinfo{journal}{J.~Chem.~Phys.} \textbf{\bibinfo{volume}{119}},
  \bibinfo{pages}{844} (\bibinfo{year}{2003}).

\bibitem[{\citenamefont{{Mohr} and {Taylor}}(2005)}]{mohr05a}
\bibinfo{author}{\bibfnamefont{P.~J.} \bibnamefont{{Mohr}}} \bibnamefont{and}
  \bibinfo{author}{\bibfnamefont{B.~N.} \bibnamefont{{Taylor}}},
  \bibinfo{journal}{Rev.~Mod.~Phys.} \textbf{\bibinfo{volume}{77}},
  \bibinfo{pages}{1} (\bibinfo{year}{2005}).

\bibitem[{\citenamefont{{Fellmuth} et~al.}(2006)\citenamefont{{Fellmuth},
  {Gaiser}, and {Fischer}}}]{fellmuth06a}
\bibinfo{author}{\bibfnamefont{B.}~\bibnamefont{{Fellmuth}}},
  \bibinfo{author}{\bibfnamefont{C.}~\bibnamefont{{Gaiser}}}, \bibnamefont{and}
  \bibinfo{author}{\bibfnamefont{J.}~\bibnamefont{{Fischer}}},
  \bibinfo{journal}{Measurement Science and Technology}
  \textbf{\bibinfo{volume}{17}}, \bibinfo{pages}{145} (\bibinfo{year}{2006}).

\bibitem[{\citenamefont{{Stone} and {Stejskal}}(2004)}]{stone04a}
\bibinfo{author}{\bibfnamefont{J.~A.} \bibnamefont{{Stone}}} \bibnamefont{and}
  \bibinfo{author}{\bibfnamefont{A.}~\bibnamefont{{Stejskal}}},
  \bibinfo{journal}{Metrologia} \textbf{\bibinfo{volume}{41}},
  \bibinfo{pages}{189} (\bibinfo{year}{2004}).

\bibitem[{\citenamefont{{Derevianko} and {Porsev}}(2002)}]{derevianko02a}
\bibinfo{author}{\bibfnamefont{A.}~\bibnamefont{{Derevianko}}}
  \bibnamefont{and} \bibinfo{author}{\bibfnamefont{S.~G.}
  \bibnamefont{{Porsev}}}, \bibinfo{journal}{Phys.~Rev.~A}
  \textbf{\bibinfo{volume}{65}}, \bibinfo{pages}{053403}
  (\bibinfo{year}{2002}).

\bibitem[{\citenamefont{Rafac and Tanner}(1998)}]{rafac98a}
\bibinfo{author}{\bibfnamefont{R.~J.} \bibnamefont{Rafac}} \bibnamefont{and}
  \bibinfo{author}{\bibfnamefont{C.~E.} \bibnamefont{Tanner}},
  \bibinfo{journal}{Phys. Rev. A} \textbf{\bibinfo{volume}{58}},
  \bibinfo{pages}{1087} (\bibinfo{year}{1998}).

\bibitem[{\citenamefont{Lide and Frederikse}(2007)}]{crc07a}
\bibinfo{editor}{\bibfnamefont{D.~R.} \bibnamefont{Lide}} \bibnamefont{and}
  \bibinfo{editor}{\bibfnamefont{H.~P.~R.} \bibnamefont{Frederikse}}, eds.,
  \emph{\bibinfo{title}{CRC Handbook of Chemistry and Physics}},
  vol.~\bibinfo{volume}{88} (\bibinfo{publisher}{CRC Press},
  \bibinfo{address}{Boca Raton, Florida}, \bibinfo{year}{2007}).

\bibitem[{\citenamefont{Maroulis}(2006)}]{maroulis06a}
\bibinfo{editor}{\bibfnamefont{G.}~\bibnamefont{Maroulis}}, ed.,
  \emph{\bibinfo{title}{Atoms, Molecules and Clusters in Electric Fields:
  Theoretical Approaches to the calculation of electric polarizability}}
  (\bibinfo{publisher}{World Scientific}, \bibinfo{address}{London},
  \bibinfo{year}{2006}).

\end{thebibliography}

\end{document}